# Self-Aware and Self-Adaptive Autoscaling for Cloud-Based Services

by

# Tao Chen

A thesis submitted to
the University of Birmingham
for the degree of
Doctor of Philosophy

School of Computer Science
College of Engineering and Physical Sciences
University of Birmingham
November 2015

# ABSTRACT

Modern Internet services are increasingly leveraging on cloud computing for flexible, elastic and on-demand provision. Typically, Quality of Service (QoS) of cloud-based services can be tuned using different underlying cloud configurations and resources, e.g., number of threads, CPU and memory etc., which are shared, leased and priced as utilities. This benefit is fundamentally grounded by autoscaling: an automatic and elastic process that adapts cloud configurations on-demand according to time-varying workloads. This thesis proposes a holistic cloud autoscaling framework to effectively and seamlessly address existing challenges related to different logical aspects of autoscaling, including architecting autoscaling system, modelling the QoS of cloud-based service, determining the granularity of control and deciding trade-off autoscaling decisions. The framework takes advantages of the principles of self-awareness and the related algorithms to adaptively handle the dynamics, uncertainties, QoS interference and trade-offs on objectives that are exhibited in the cloud. The major benefit is that, by leveraging the framework, cloud autoscaling can be effectively achieved without heavy human analysis and design time knowledge. Through conducting various experiments using RUBiS benchmark and realistic workload on real cloud setting, this thesis evaluates the effectiveness of the framework based on various quality indicators and compared with other state-of-the-art approaches.

# ACKNOWLEDGEMENT

Firstly, I would like to express my most sincere gratitude to my supervisor, Dr. Rami Bahsoon, for his continuous, helpful and endless supports on my PhD work and research, for his patience, motivation, and immense knowledge. His guidance has provided me with great helps throughout my PhD and every aspects of my research work. I have been extremely lucky to have a supervisor like him, who cares so much about my work, and who is able to discuss research related issues with me properly and promptly. His outlook and comments inspired me a lot and gave me confidence.

Beside my supervisor, other members from my research monitoring group have also provided me with great helps. I would like to thank Prof. Xin Yao for his insightful comments and encouragements on my research; thank Dr. Mirco Musolesi for his timely, but hard questions which incented me to improve my work.

My sincere thanks also go to the other team members from the EPiCS project: Dr. Peter Lewis, Dr. Leandro L. Minku and Dr. Funmilade Faniyi, for the stimulating discussion, for the remarkable collaborations, and for the busy days that we were working together before deadlines. Also, I thank the members from the Software Engineering Research Group: Dr. Sarah al Azzani, Faisal Alrebeish, Dr. Esra Zaghoul, Dr. Shehnila Zardari, Giannis Tziakouris, Abdessalam Elhabbash, Maria Salama and Alexandros Evangelidis, for the interesting and inspired discussion during meetings and talks. In addition, I thank my friend Yiteng Ma and Shanhua Chen for the kind proof-reading of my work.

Last but not the least, I would like to thank my farther and mother for their continuous and endless supports, both financially and spiritually, on me to complete my PhD, which is also the last wish of my farther, who passed away during my PhD study; I hope that he would be aware of my PhD completion in heaven. Also, I thank my wife, Xiao Liang, and mother in law for their supports and care for my life in general. Thank my lovely, newly born daughter Yvonne Yuhan Chen for all the happiness and emotional encouragements that she brought to me.

# CONTENTS

# List of Figures

# LIST OF TABLES

# CHAPTER 1

# INTRODUCTION

Modern IT companies, from large enterprises to small business, are increasingly leveraging on cloud computing to improve their profits and reduce the costs. Unlike many other similar computing paradigms, such as Cluster computing and Grid computing, cloud computing provides on-demand access to virtual computing resources, software stacks, applications and services through the principle of shared infrastructure. Typically, existing cloud providers offer three hierarchical layers according to the levels of abstraction [73], these layers are: Software as-a-Service (SaaS), Platform as-a-Service (PaaS) and Infrastructure as-a-Service (IaaS), as shown in Figure 1.1. In the SaaS layer, the end-users are allowed to access to various cloud-based services, which are often well-deployed and readily available to use. These cloud-based services can refer to the entire application and system, e.g., Gmail; or any conceptual part within an application, e.g., a payroll reporting service in a large human-resource management application. In contrast, the PaaS layer focuses on providing a shared platform that allowing consumers to develop, run and manage application or services without the complexity of building and maintaining the infrastructure. This is achieved by offering a software stack, which consists of various configurable software to support the entire cloud-based service life cycle. Examples of these software include application server where there are configurable number of max threads and session life time; or database server that permits configurable number of max connections and buffer size. The IaaS layer offers a shared infrastructure, where the concern is on provisioning hardware resources, including CPU, memory and bandwidth etc. These resources are usually controlled by virtulization and packed in a Virtual Machine (VM). Clearly, the software configurations at PaaS layer and the hardware resources at IaaS layer serve as the fundamental elements that support the running cloud-based services at SaaS layer. However, this comes with costs, providing that those software and resources are priced by the PaaS and IaaS providers as utilities and, therefore, they need to be leased in a pay-as-you-go manner.

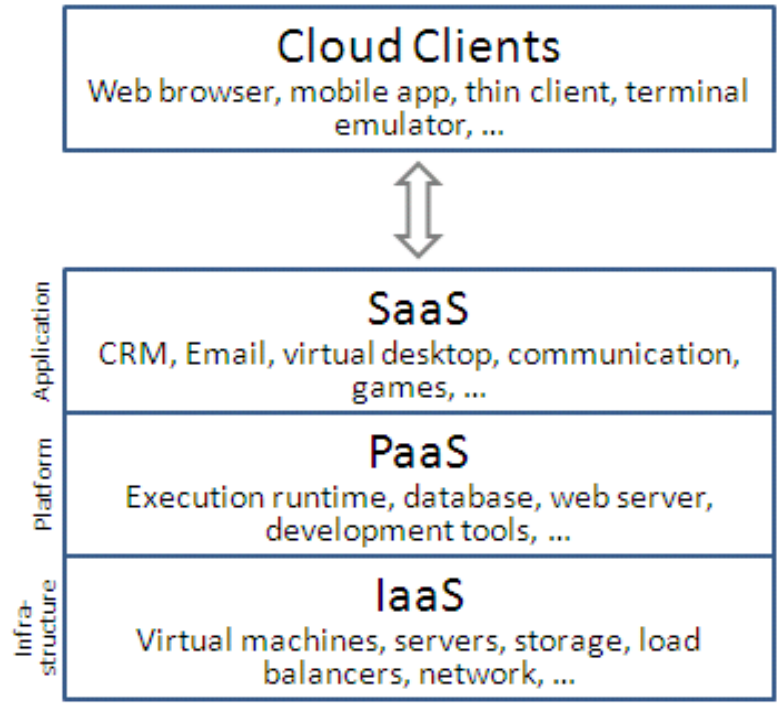

Figure 1.1: The Standard Three Layers Cloud.

## 1.1 Research Storyline

One of the pronounced benefits of the cloud, regardless of the levels of abstraction, is elasticity. In the context of cloud computing, Herbst et al. [73] have defined elasticity as:

*Elasticity measures the degree to which a system is able to adapt to workload changes by provisioning and de-provisioning resources in an autonomic manner, such that at each point in time the available resources match the current demand as closely as possible.*

whereby the demand is measured according to certain Quality of Service (QoS) attributes or cost. The QoS, for example, can be response time, throughput or any other non-functional attribute experienced by the end-users. Herbst et al's definition reflects the on-demand nature of elasticity in the cloud; however, they only focus on elasticity at the IaaS layer where the concern has been on hardware resources (a.k.a. resource elasticity). As discussed by Dustdar et al. [49], understanding elasticity purely from a viewpoint of resource is rather restrictive. Thus, in turn, they claim that elasticity in the cloud should cover two additional dimensions where the focus are the changes in QoS attributes and cost, which are formally termed as QoS elasticity and cost elasticity respectively.

To ensure high level QoS and cost requirements at the SaaS, we argue that elasticity should cover all the three layers in the cloud, especially when the software configurations at PaaS layer can cause considerable interplay and effects on the required hardware resources at IaaS layer, as evident by our early study [31] and much recent work [26] [139] [99] [49]. Indeed, when the cloud-based services at SaaS layer suffer dynamic environmental conditions, such as workloads and the size of the incoming jobs, governing their QoS and the incurred costs is among the primary concerns of IaaS providers, PaaS providers and SaaS providers (a.k.a. service owners). This, in turn, requires QoS models that capture the dynamic, and possibly uncertain sensitivity of QoS to software configurations, hardware resource and environmental conditions. However, as shown in a recent survey [13], existing work on QoS modelling is often coarse-grained, focusing on either application or VM; in addition, the software configurations are rarely considered in conjunction with the hardware resource and changes to the environmental conditions. In general, there is a lack of systematic solution for selecting the most relevant software configurations, hardware resource and environmental conditions that can significantly influence the QoS. To study QoS sensitivity for each individual cloud-based service across all the cloud layers, this thesis looks into a fine-grained, dynamic and online QoS modelling approach to select the relevant inputs of the QoS and model their magnitude in correlation to QoS [31][38][33]. Our experimental analysis reveal that, (i) only the important inputs are useful in QoS modelling; (ii) the performance of algorithms used to build the QoS models vary depending on scenarios. These observations have motivated us to further investigate a QoS modelling approach that is not only able to select the important inputs, but also to identify the best algorithm out of a set of candidates for a given scenario [38][33].

When many cloud-based services run on the shared infrastructure, their QoS can be interfered with by the dynamic and uncertain resources contention caused by the services' neighbours, including the co-located services on a VM [112] and the co-hosted VMs on a Physical Machine (PM) [116]. This phenomenon, referred to as **_QoS interference_**, has become a non-trivial and challenging issue in the cloud. However, QoS interference, especially at the co-located services level, has been rarely considered in state-of-the-art elasticity management approaches. To this end, this thesis explicitly captures the information of QoS interference at both service- and VM-levels, as well as their interplay with the other inputs (i.e., software configurations, hardware resources and environment conditions of a cloud-based service) in the QoS modelling [31][38][33].

In certain predictable scenarios where the environmental condition has strong seasonality and the QoS interference is minimal, the configurations and resources can be approximately predefined. Nevertheless, for many other cases, for examples, spiked workloads and uncertain QoS interference, elasticity can be only enabled by runtime automatic scaling, or simply autoscaling: _an automatic and elastic process, typically running on a PM, that adapts software configurations and hardware resources provisioning on-demand_ *according to the changing environmental conditions.* The concrete scaling actions can be either vertical scaling or horizontal scaling: the former refers to change the configurations and resources within a PM; the later refers to boots up/shutdown VMs on the other PMs via migration or replication. The ultimate goal of autoscaling is *to continually optimise the QoS and cost objectives for all cloud-based services; thus their Service Level Agreement (SLA) and budget requirements can be better complied with.* We term the situation as **globally-optimal benefit** in the cloud when the objectives of all cloud-based services reach their optimal results. However, achieving globally-optimal benefit depends on the right granularity of control that is difficult to ensure in cloud autoscaling. This is because the QoS sensitivity and the possible interference are dynamic and uncertain in nature. Existing control, for example, is often statically applied on the entire cloud, physical machine, virtual machine or service etc. These can lead to scenarios where the autoscaling approach might result in large overhead or the global benefit is compromised. By leveraging the QoS models, this thesis introduces a dynamic approach that determines the right granularity of control, in such a way that the global benefit is optimised while the overhead is reduced [32][36].

Deciding on the optimal software configurations and hardware resource provisions is an extremely complex task in autoscaling. It becomes even harder when the dynamic, and possibly uncertain trade-offs are required for conflicted objectives, e.g., throughput and cost, and the interfered cloud-based services. However, the most widely applied weighted-sum formulation in autoscaling decision making might lead to some major issues (i) there is only coarse-grained information about the trade-off surfaces and (ii) in some cases, it is very difficult to correctly specify weights, especially for the objectives of interfered cloud-based services. To resolve these issues, this thesis builds on the right QoS models and granularity of control; the thesis then explore a dynamic, weights-free trade-offs decision making approach for cloud autoscaling [34]. The proposed approach optimises for the objectives till the points where trade-offs need to be made, and the resulting decision is guaranteed to achieve ***well-compromised trade-offs***—a large improvements on the majority of the objectives; while causing relatively small degradations to others.

The research activities presented in this thesis have lead to a novel autoscaling framework in the cloud. In particular, to better handle dynamics, uncertainty, QoS interference and trade-offs related to the autoscaling process; while reducing human intervention and the needs of design time knowledge, we leverage the formal principles of self-awareness [18] by investigating the related architectures, methodologies and algorithms. Particularly, we contribute the following research outputs.

1. A self-aware and self-adaptive autoscaling architecture that is mapped to the principles of self-awareness and thus provides fine-grained representation of the required knowledge. This calls for different levels of knowledge and self-awareness capabilities, and hence help to better design and select the underlying algorithms [37][35].

2. A self-aware and self-adaptive QoS modelling approach that correlates the QoS with software configurations, hardware resources, environmental conditions and QoS interference across all the cloud layers. In particular, by acquiring the knowledge of QoS sensitivity, it adaptively determines the most significant model inputs (including QoS interference), the magnitude of models inputs and the best learning algorithms used to tune the model [31][38][33].

3. A self-aware and self-adaptive approach that clusters the objectives into different regions which are optimised independently. It aims to achieve globally-optimal benefit while reducing the overhead in the cloud. This is achieved by knowing the effects of granularity of control to the global benefit [32].

4. A self-aware and self-adaptive decision making approach that dynamically optimises and searches for diversified trade-off decisions for autoscaling, from which well-compromised trade-offs can be reached. This is achieved by knowing the effects of decisions (including QoS interference) on objectives, which permits extensive reasoning and planning in the cloud autoscaling [34].

The remaining of this chapter motivates the need for self-aware and self-adaptive autoscaling in cloud and discusses the assumptions, research questions, objectives, contributions and organisation of this thesis.

## 1.2 Motivation

Consider, for example, a growing company named *Rbay*, which is seeking to deploy numerous services into the cloud for meeting their increasing workloads from the end-users. To reduce management complexity, they have chosen to deploy their services as SaaS on a big cloud provider who is offering both PaaS and IaaS solutions. The cloud provider supports public cloud and it is also currently severing many other consumers. Given the dynamic and uncertain workloads on *Rbay*'s services, their primary concern has been on the QoS of the services running in cloud, and they expect to use as minimal as possible cost to achieve the best QoS possible. This is also the same desire for the cloud provider as they have signed SLA and budget agreement with *Rbay*, which means that improving the QoS with minimal cost is likely to earn better reputation, and save the resources for other consumers. To achieve this, one solution that *Rbay* can facilitate is autoscaling. However, for this to come true, they face several difficulties and challenges:

- *Rbay* are struggling with determining the amount of software configurations and hardware resources that their services need. The difficulty lies in the fact that they lack knowledge about the model and correlation between QoS of their services and the underlying features, including software configurations, hardware resources and environmental conditions. Theses features are referred to as **cloud primitives** throughout the thesis. The QoS models can be heterogeneous depends on the cloud-based services and their QoS attributes, providing that the metrics of QoS attributes and the characteristics of the service tend to be different. More importantly, the challenge is that the dynamic and uncertain interplays among the cloud primitives and their combinatorial influences on the QoS have caused the model accuracy difficult to be preserved. This becomes even more challenging in the presence of QoS interference, because *Rbay* are not able to reason about the likely effects of autoscaling on their services, as well as on those of the other existing consumers.

- When autoscaling in the presence of QoS interference, it is important to consider the interfered cloud-based services, including their QoS and costs during autoscaling. This is actually a challenge for the cloud provider since optimising the objectives for a service in isolation may compromise other services' objectives. However, reaching a right granularity of control in autoscaling becomes a difficulty, providing that the QoS interference is dynamic and uncertain in nature. Existing control, for example, is often statically applied on the entire cloud, physical machine, virtual machine or service etc. The cloud provider must be extremely cautious when designing the autoscaling approach. This is because controlling the entire cloud at a time may achieve global benefits (i.e., QoS and cost) for all cloud-based services, but likely to result in large overheads. On the other hand, a local control (e.g., service level) can often have acceptable overheads while achieving local benefits for some services, but this may come with the consequence that the QoSs of many other services are compromised.

- Like many of the other companies, each of the *Rbay*'s services has different objectives, including QoS attributes and cost. Now, suppose that the autoscaling system has decided to improve the throughput of *Rbay*'s service-instance $S_{ij}$ by provisioning more memory to the underlying VM. Such a decision might not be an issue when the contention is light. However, as the provision increases, eventually it will result in throughput degradation to the other service-instances on the co-hosted VMs, leading to dynamic QoS interference with [26][139][116]. The same issue applies when we increase the number of service threads for a service-instance, where the co-located service-instances on the same VM might be interfered [139][99]. These phenomena imply that there are trade-offs between the throughput of $S_{ij}$ and those of the other service-instances, which might be owned by different cloud consumers. It becomes more complex when we need to consider trade-offs between conflicted objectives, e.g., the throughput and cost of $S_{ij}$. All these facts can lead to a large number of dependent objectives in a decision making process (i.e., more than 4). Given the fluctuated QoS performance of cloud-based service and their interferences, the key challenge is how to dynamically reason about the effects of different trade-off decisions. Among the others, there are decisions that lead to the points where the trade-offs achieve large improvements on the majority of the objectives; while causing relative small degradations to others. These trade-offs, referred to as **well-compromised trade-offs** (a.k.a. knee points), are almost the most preferable ones.

- Typically, current cloud providers provide a limited set of decisions bundles, each of which contains a fixed combination of software configurations and hardware resources provisioning. These predefined bundles can restrict the number of possible trade-off decisions and thus reduce the complexity for human decision making. However, as stated in a recent survey conducted by Galante et al. [61], renting bundles cannot and does not reflect the interests of consumers and the actual demand of their cloud-based services. Therefore considering fine-grained, and arbitrary combinations of software configurations and hardware resources provisioning is an inevitable outcome as the cloud computing paradigm continuous to evolve. This makes the trade-off decision making problem in autoscaling become even more challenging as the possible number of trade-off decisions and their diversity tend to be incredibly large.

Given the dynamic and uncertain nature in cloud, all these challenges and difficulties need to be handled at runtime. This implies that effectively managing autoscaling in the cloud tends to be a task that is far beyond the capability of human analysis and intervention, thus urging for more intelligent foundations and solutions. In this thesis, we argue that the principle of self-awareness can render itself as a neat solution for addressing these challenges. In general, self-awareness is concerned with one's ability to acquire knowledge about one's current state and the environment. Such knowledge permits better reasoning about one's adaptive behaviour. The benefits of self-awareness, as stated by Becker et al. [18], include better adaptivity of a system in handling runtime dynamics, uncertainty, heterogeneity and trade-offs—all belong to the requirements of autoscaling in the cloud.

## 1.3   Research Questions and Objectives

To enable self-aware and self-adaptive autoscaling in the cloud and cope with the aforementioned challenges, one of the key tasks is to understand and accurately model the correlation between QoS attributes and the underlying cloud primitives on-the-fly, taking QoS interference into account. Such a task, namely *self-aware QoS modelling*, is the essential input for the other two subsequent processes for autoscaling, namely *self-aware granularity of control* and *self-aware trade-off decision making*. It is easy to see that, by acquiring knowledge about QoS models, one can easily isolate which QoS or cost objectives tend to influence others (i.e., conflicted or harmonic objectives) while which are the ones that can be considered in independent decision making processes. In this way, the right granularity can be dynamically determined with reduced overhead while not damaging the global benefit. Presumably, the QoS models and the appropriate granularity of control promote better trade-off decision making in autoscaling: the former permits the foundation to dynamically reason about the likely effects of different trade-offs decisions and thus adjust the decision making process accordingly; while the latter, guarantees the global benefit and limits the resulting overhead. The computational self-awareness provides a promising avenue for all these requirements and challenges, and therefore, to systematically design self-aware algorithms and techniques for the three processes, the fundamental task is to study how the general principle of self-awareness can be applied in an autoscaling system. This task, namely *self-aware autoscaling architecture*, is concerned with architecting the self-awareness capabilities and autoscaling in the cloud. As a result, the core research question of this thesis towards self-aware and self-adaptive autoscaling in the cloud is:

> *How can self-awareness and the related algorithms be incorporated into the process of elastically autoscaling cloud-based services, such that the autoscaling system is able to handle runtime dynamics, uncertainties and trade-offs exhibited in the cloud? What are the benefits of self-awareness and to what extent can it be beneficial, when compared to approaches with no or limited self-awareness?*

Specifically, as shown in Table 1.1, more detailed research questions can be discussed in different logical aspects that facilitate autoscaling, i.e., autoscaling architecture, cloud QoS modelling, granularity of control in the cloud and trade-off decision making.

### 1.3.1 Objectives

Driven by the aforementioned research questions, this thesis aims to investigate how to build more intelligent and dependable autoscaling systems in the cloud by leveraging self-awareness and related algorithms. To achieve such, we have identified several objectives, as shown in Table 1.2.

Table 1.1: The Detailed Research Questions of The Thesis.

| Autoscaling architecture | |
|---|---|
| *RQ 1.1.* | How to incorporate and map the self-awareness capabilities to autoscaling in the cloud? |
| *RQ 1.2.* | How to architect self-aware autoscaling system? What are the benefits we can expect from this enriched architecture? |
| **Cloud QoS modelling** | |
| *RQ 2.1.* | How to dynamically select the important, yet uncertain cloud primitives (e.g., software configurations, hardware resources and environmental conditions) when modelling the QoS for cloud-based services. Which cloud primitives tend to be significant while which are the irrelevant ones? When these cloud primitives should be considered in the models? |
| *RQ 2.2.* | How to dynamically model and quantify the uncertain magnitude of cloud primitives in the correlation between them and a QoS attribute. |
| *RQ 2.3.* | How to incorporate the dynamic and uncertain information about QoS interference into the models. |
| *RQ 2.4.* | How to ensure the accuracy of the QoS models. |
| **Granularity of control in the cloud** | |
| *RQ 3.1.* | What are the effects of control granularity on globally-optimal result (i.e., result with respect to QoS and cost of all cloud-based services) and the overhead in cloud? |
| *RQ 3.2.* | Whether local control (e.g., service level) can achieve similar global benefit to global control (e.g., cloud level)? |
| *RQ 3.3.* | How to handle the dynamics and uncertainty associated with the granularity of control in cloud and its effects on the global benefit. |
| **Trade-off decision making in autoscaling** | |
| *RQ 4.1.* | Given the dynamic and uncertain nature of the autoscaling decision making problem in cloud, how to dynamically search and optimise for the uncertain trade-off decisions, considering the naturally conflicted objectives and uncertain QoS interference? |
| *RQ 4.2.* | How to dynamically reason about the effects of decisions on QoS and cost objectives, and the uncertain trade-offs considering their requirements. |
| *RQ 4.2.* | How to quantify the extent of compromises in the trade-off? How to dynamically determine the well-compromised trade-off. |

Table 1.2: The Objectives of The Thesis.

| The Targeted RQ | Objective |
|---|---|
| *RQ 1.1.* | Investigate the self-awareness capabilities and their mapping to the necessary components of autoscaling in the cloud. The mapping should express how various levels of self-awareness and knowledge can be used to address the challenges in autoscaling. |
| *RQ 1.2.* | Drawing on the mapping, blueprint an architecture for self-aware and self-adaptive autoscaling. |
| *RQ 2.1.* and *RQ 2.3.* | Study the sensitivity of the QoS models' accuracy to the selected cloud primitives while considering the QoS interference in cloud. Quantify the relative importance and significance of selected cloud primitives in the correlation. |
| *RQ 2.2.*, *RQ 2.3.* and *RQ 2.4.* | Look into approaches that model the QoS sensitivity for cloud-based services with respect to the cloud primitives. This is concerned with understanding *which* (e.g., are CPU and throughput correlated?), *when* (i.e., at which point in time they are correlated?) and *how* (i.e., the magnitude of primitives in correlation) the primitives correlate with QoS. Particularly, it is necessary to examine and compare the major QoS modelling approaches, which are currently applied in the cloud, in terms of their accuracy and complexity. In this way, we intend to adopt and combine potential approaches for the case of autoscaling with improved accuracy and acceptable complexity. |
| *RQ 3.1.*, *RQ 3.2.* and *RQ 3.3.* | Investigate approaches that dynamically determine the right granularity of control for autoscaling in the cloud through reasoning about the effects of granularity on the global benefit. The ultimate goal is to understand the balance between the effects on global benefit and overhead achieved by the autoscaling system. |
| *RQ 4.1.* and *RQ 4.2.* | Explore approaches that dynamically optimise and search for trade-off decision in autoscaling to reduce the complexity of human intervention; while considering the SLAs and cost requirements. Given the potentially large amount of possible decisions for autoscaling, the approach should be efficient for reasoning about the effects of autoscaling decision on services' objectives and their trade-offs at runtime. |
| *RQ 4.3.* | Identify a method to quantify the extents of compromises in trade-offs and build a mechanism that converges to well-compromised trade-offs without human intervention. |

## 1.4 Scope and Assumptions

In the following, we codify the scope of this thesis through several assumptions.

- In this thesis, cloud-based services can refer to the entire application and system, e.g., Gmail; or any conceptual part within an application, e.g., a payroll reporting service in a large human-resource management application. We assume that applications in the cloud are composed of services, each has different QoS requirements and external environment changes (e.g., changes in workload). The hardware resource can be shared amongst the services (e.g., CPU of the VM); whereas the software configurations are tuneable and can be specific to one service (e.g., threads of a service), as supported by many real-world applications (e.g., the Weblogic [3] and standards (e.g., the JAX-WS standard [4]). This assumption promotes fine-grained differentiation and control for the cloud-based services. More importantly, this assumption about service permit to maximize the flexibility of an autoscaling system such that it can be customized to work on the entire application or any parts of it as necessary.

- Often, multi-tier applications and services in the cloud can have multiple replicas for various purposes, e.g., service differentiation and load balancing etc. Therefore we assume that each tier in a multi-tier application, consisting of concrete services, can have multiple replicas deployed on different VMs or even PMs. In this thesis, we refer to the replicas of concrete services as **service-instances**. The $jth$ instance of the $ith$ cloud-based service is denoted as $S_{ij}$. We aim to optimise the objectives for all those service-instances in the cloud.

- We do not consider the trade-off on scaling actions, i.e., vertical scaling vs. horizontal scaling. The primary concern of this thesis has been on reaching an autoscaling decision that contains the right combination of software configurations and hardware resources for all cloud-based services. In our autoscaling framework, vertical scaling always takes higher priority, providing that modern hypervisors (e.g., Xen [6]) can achieve dynamic vertical scaling with negligible overheads. The resources on a PM are provisioned to the VMs in a first-come-first-serve basis. The horizontal scaling, on the other hand, is only triggered when the resources of the PM tends to be exhausted. We leave the study on the effects of scaling actions for future work.

- Since we tackle the autoscaling problem for the benefits of cloud consumers (i.e., service owners), we do not consider the work that focuses on VM to PM consolidation, VM creation, VM termination, VM migration and VM replication unless they explicitly consider autoscaling as a key contribution of their research.

## 1.5  Research Methodology

The research methodology of the thesis can be seen as similar to the Design Science Research Methodology [109], which is an iterative process. The main processing steps can be discussed as the following:

- *Understanding of the problem:* The first step is to gain conceptual understanding of the problem domain, i.e., autoscaling in the cloud. This is achieved by a large amount of literature review. At the very early stage of the research, the concrete research direction may not be obvious, thus a general understanding of a wide ranges of problem in cloud computing is needed. As the knowledge is built, the research direction is then gradually converge to the related problem for autoscaling in the cloud, which is now the formal problem addressed in this thesis

- *Suggestions and Hypothesis:* In this step, potential issues and gaps in current research are identified. These issues are the keys that should be investigated clearly when reading each related work. Some hypothesis about the possible solution to those issues are made, and they can influence the search key words in the literature review process. For example, when machine learning is identified as one possible solution to the problem of QoS modeling in the cloud, then a new phase of search key words can be added: *machine learning AND cloud computing AND QoS modelling OR performance modelling*

- *Prototyping:* This step requires implementation and development work to realize the concepts developed in the previous stage. It also require close investigation of some related techniques form other fields, e.g., machine learning and optimization algorithms. It is worth noting that from this stage forward, the research can go back to the first step if serious issues (e.g., wrong understanding of the problem) are found or incorrect hypothesis is made in the previous step.

- *Evaluation:* Once the prototype has been completed, qualitative and quantitative evaluation can be carried out. In particular, qualitative evaluation is achieved by examining how well the prototype realize the concepts against design criteria, while quantitative evaluation is based on experiments run in a controlled environments. When the results do not meet with expectations, then it is necessary to iterative back to step 1 or 2 for invalidating the knowledge and hypothesis.

- *Conclusion:* The final step is concerned with formally positioning the contributions and reporting on the results found in the research process.

## 1.6 Contributions

There are numbers of steps that have been taken to tackle the aforementioned research questions and objectives. Collectively, as an ultimate result of this thesis, those steps have leaded to a holistic, self-aware and self-adaptive autoscaling framework that is able to optimise the QoS and costs for all cloud-based services, with limited human intervention and design time knowledge. This framework consists of different components, each deals with a different category of research questions presented in Section 1.3. Particularly, this thesis draws several novel contributions, which are listed in Table 1.3.

Table 1.3: The Detailed Contributions of The Thesis

| Contribution | Addressed RQ |
|---|---|
| 1. A set of common criteria that can be used to assessed and compared existing autoscaling approach in the cloud. Additionally, a taxonomy is produced deriving from these criteria and the corresponding comparisons. Finally, a survey that provides the key background information, strengths, weakness and categorisation of existing approaches for autoscaling in the cloud. | |
| 2. Autoscaling architecture leverages the self-awareness principle and capabilities.<br><br>• A mapping between self-awareness capabilities and the important components in autoscaling by leveraging the general principle of self-awareness. This mapping justifies the need of self-awareness capability at different levels and provides a concise understanding about how self-awareness can be applied to resolve the challenges for autoscaling in the cloud.<br><br>• An autoscaling architecture that is built using self-aware patterns [39]. The proposed architecture not only describes how the self-awareness capabilities are encapsulated into components, but also expresses their potential interactions, which help to better consolidate different levels of self-awareness.<br><br>• By leveraging self-awareness, the autoscaling framework realises *bidirectional* adaptation. That is to say, it is not only able to adapt the underlying cloud-based services and VMs, but also able to further consolidate itself by acquiring the knowledge about itself and the environment through different self-awareness capabilities. | *RQ 1.1.* and *RQ 1.2.* |
| 3. Self-aware and self-adaptive QoS modelling approach for autoscaling in the cloud. | *RQ 2.1.* to *RQ 2.4.* |

- A fine-grained and generic QoS model, which is designed to handle dynamic and uncertain QoS sensitivity; and to incorporate information of the uncertain QoS interference caused by the cloud-based services co-located on a VM and the VMs co-hosted on a PM.

- An in-depth analysis on the correlations of selected cloud primitives to the model accuracy in the cloud; in particular, this analysis shows how the model accuracy can be affected by the selected cloud primitives.

- A self-aware and self-adaptive technique, namely *hybrid dual-learners*, to determine which and the cloud primitives correlates with the QoS on the fly using information theory [130]. This technique has been diverged into four variations, which are experimentally compared and evaluated.

- A suitability analysis of different learning algorithm for modelling QoS against different QoS attributes. Particularly, three widely used machine learning algorithms are examined, including Artificial Neural Network (ANN) [119], Auto-Regressive Moving Average with eXogenous inputs model (ARMAX) [22] and Regression Tree (RT) [117].

- A self-aware and self-adaptive solution, namely *adaptive multi-learners*, to dynamically model how the cloud primitives correlates with the QoS. The proposed solution is not only able to dynamically correlate the selected cloud primitives to the QoS, but also to adaptively select the best learning algorithm and its resulting model during prediction in cloud.

- The QoS modelling approach is experimentally evaluated using RU-BiS [5] benchmark and FIFA 98 workload trend [14]. This is achieved by comparing to various other state-of-the-art approaches; and under four commonly used QoS attributes, these are: Response Time, Throughput, Reliability and Availability. The evaluation criteria have been on accuracy, stability, sensitivity to the online data size and efficiency.

| | |
|---|---|
| 4. Self-aware and self-adaptive mechanism that adapts the granularity of control in autoscaling. <br><br> • A self-aware, self-adaptive and two-phase region clustering mechanism that clusters the QoS and cost objectives into sensitivity independent regions. The basic principle behind the notions of sensitivity independent regions is that it is possible to reach globally-optimal result (with respect to the QoS and cost of all cloud-based services) by asynchronously finding locally-optimal results within each sensitivity independent region. This can eventually shrink the search space and reduce overhead. <br><br> • The mechanism is experimentally evaluated via hypothetical scenarios, which contain different numbers of services. This is achieved by comparing with existing solutions that statically operate on different fixed granularities of control, including cloud-level, PM-level, VM-level and service-level. The achieved globally-optimal result and the produced overhead are assessed. | *RQ 3.1. to RQ 3.3.* |
| 5 Self-aware and self-adaptive trade-off decision making approach for autoscaling in the cloud. <br><br> • In light of many successful applications of metaheuristics algorithms in the cloud, we present self-aware and self-adaptive decision making process where the core is a Multi-Objective Ant Colony Optimisation (MOACO) algorithm that designed to search the optimal (or near-optimal) trade-offs decisions for autoscaling in the cloud. This approach eliminates the need for specifying weights in the objective formulation and is able to handle trade-offs caused by naturally conflicted objectives and QoS interference. In addition, the stochastic nature of MOACO allows it to achieve good coverage in the trade-offs surface, and thus improving diversity in the trade-offs decisions. The search process in our MOACO is similar to conduct many single objective optimisations in one run, which aims to optimise and make trade-offs for a larger number of objectives than the commonly used 2 to 4 objectives (i.e., up to 30 objectives in our experiments). | *RQ 4.1. to RQ 4.3.* |

- A triple mechanism, namely *compromise-dominance*, for finding well-compromised trade-offs based on superiority and fairness of the decisions. The former is measured by pareto-dominance [67], and the latter is achieved via nash-dominance [108] and the distance of decisions measurement. The mechanism is able to dynamically achieve well-balanced improvements and degradations for the objectives, without being guided by weights in the objective formulation.

- The trade-off decision making approach is experimentally evaluated, again, using RUBiS [5] benchmark and FIFA 98 workload trend [14]. This is achieved by comparing our results to those of four widely used approaches for autoscaling: rule-based, single-objective heuristic based, single-objective randomised and multi-objective genetic algorithm based; and under four commonly used QoS attributes, these are: Response Time, Throughput, Reliability and Availability. These approaches are critically examined in terms of their quality of trade-offs, violations on SLA and budget requirements, over-/under-provisioning and overhead.

## 1.7   Publications and Thesis Organisation

The work presented in this thesis has been to a degree or completely derived from the set of papers published during the course of the PhD candidature. These published papers are listed in Table 1.4. Nevertheless, this thesis should be regarded as the definitive account of the work. It is worth mentioning that for all the joint papers, the PhD candidate has contributed to nearly all the content of the work (i.e., more than 95%). The other co-authors are credited for their suggestions, discussions and comments on the proposing ideas and experiments.

Figure 1.2 shows the relationship between the chapters and which logical aspect of the autoscaling they belong to. The remainder of this thesis is organised as follows:

- Chapter 2 presents a taxonomy and survey about autoscaling in the cloud, including detailed background, existing approaches according to the categorisation of research

Table 1.4: The Publications Related to The Thesis.

| Publications | Published | Peer-reviewed | Accepted | Under review |
|---|---|---|---|---|
| T. Chen, R. Bahsoon and G. Theodoropoulos. A Decentralized Architecture for Dynamic QoS Optimization in Cloud-based DDDAS. *In proceeding of International Conference on Computational Science (ICCS)*, Procedia of Computer Science, Elsevier Science, 2013. | ✓ | ✓ | | |
| T. Chen and R. Bahsoon. Self-Adaptive and Sensitivity-Aware QoS Modeling for the Cloud. *IIn proceeding of the 8th International ACM/IEEE Symposium on Software Engineering for Adaptive and Self-Managing Systems (SEAMS), in conjunction with the 35th International Conference on Software Engineering (ICSE)*, San Francisco, CA, 2013. | ✓ | ✓ | | |
| T. Chen, R. Bahsoon and A-R H. Tawil. Scalable Service-Oriented Replication with Flexible Consistency Guarantee in the Cloud. *Information Sciences*, Elsevier, vol. 264, 2014. | ✓ | ✓ | | |
| T. Chen and R. Bahsoon. Symbiotic and Sensitivity-Aware Architecture for Globally-Optimal Benefit in Self-Adaptive Cloud. *In proceeding of the 9th International Symposium on Software Engineering for Adaptive and Self-Managing Systems (SEAMS), in conjunction with the 36th International Conference on Software Engineering (ICSE)*, India, 2014. | ✓ | ✓ | | |
| T. Chen, F. Faniyi, R. Bahsoon, P.R. Lewis, X. Yao, L.L. Minku, and L. Esterle. The Handbook of Engineering Self-Aware and Self-Expressive Systems. Technical Report, Aug. 2014 arXiv:1409.1793 [cs.SE] | ✓ | | | |
| T. Chen, R. Bahsoon and X. Yao. Online QoS Modeling in the Cloud: A Hybrid and Adaptive Multi-Learners Approach. *In proceeding of the 7th IEEE/ACM International Conference on Utility and Cloud Computing (UCC)*, London, UK. 2014. | ✓ | ✓ | | |
| T. Chen and R. Bahsoon. Towards A Smarter Cloud: Self-Aware Autoscaling of Cloud Configurations and Resources, *IEEE Computer*, vol. 48, no. 9, 2015. | ✓ | ✓ | | |
| T. Chen and R. Bahsoon. Self-Adaptive and Online QoS Modeling for Cloud-Based Software Services. *IEEE Transactions on Software Engineering* (This paper is both *accepted* and *under review* because it was accepted subject to revision, which we have submitted.) | | ✓ | ✓ | ✓ |
| T. Chen and R. Bahsoon. Self-Adaptive Trade-off Decision Making for Autoscaling Cloud-Based Services. *IEEE Transactions on Services Computing*, 2015, doi:10.1109/TSC.2015.2499770 | ✓ | ✓ | | |

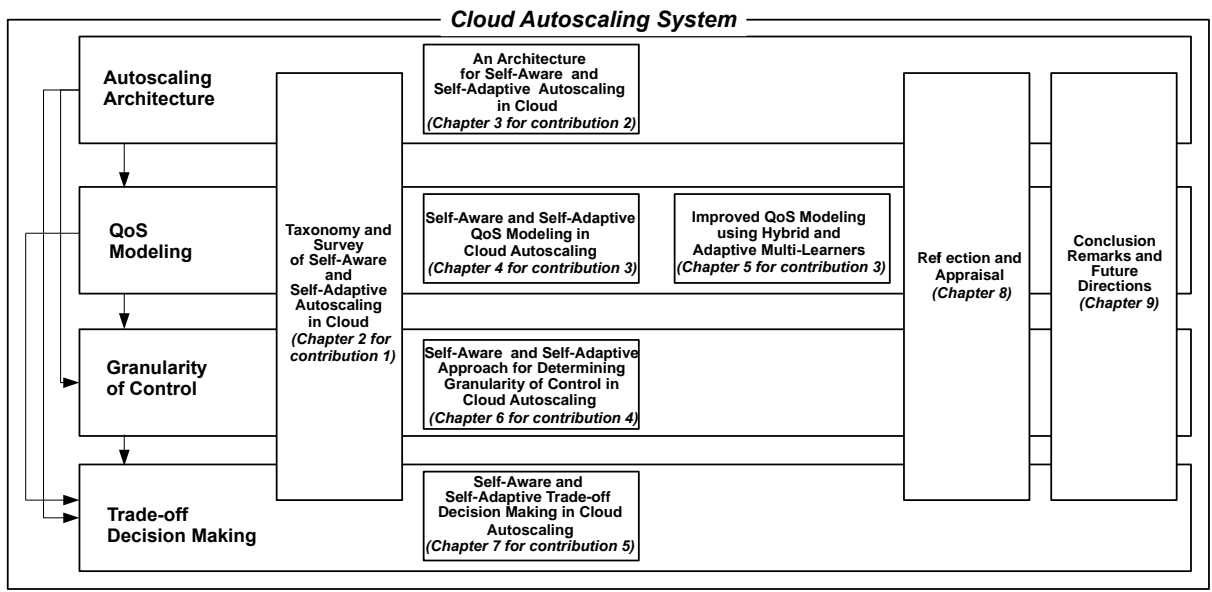

Figure 1.2: Thesis Structure. (the arrows represent dependency between different logical aspects of autoscaling)

questions and the positioning of this thesis in regards to existing work.

- The mapping of self-awareness capabilities to autoscaling in the cloud and the proposed architecture are specified in Chapter 3, which is partially derived from:

  - T. Chen, R. Bahsoon and G. Theodoropoulos. A Decentralized Architecture for Dynamic QoS Optimization in Cloud-based DDDAS. In proceeding of International Conference on Computational Science, Procedia of Computer Science, Elsevier Science, 2013.

  - T. Chen and R. Bahsoon. Towards A Smarter Cloud: Self-Aware Autoscaling of Cloud Configurations and Resources, IEEE Computer, vol. 48, no. 9, 2015.

  - T. Chen, F. Faniyi, R. Bahsoon, P.R. Lewis, X. Yao, L.L. Minku, and L. Esterle. The Handbook of Engineering Self-Aware and Self-Expressive Systems. Aug. 2014 arXiv:1409.1793 [cs.SE]

- Chapter 4 describes the QoS modelling approach for correlating QoS of cloud-based services to the underlying cloud primitives. This chapter is partially derived from:

  - T. Chen and R. Bahsoon. Self-Adaptive and Sensitivity-Aware QoS Modeling for the Cloud. In proceeding of the 8th International ACM/IEEE Symposium on Software Engineering for Adaptive and Self-Managing Systems (SEAMS), in conjunction with the 35th International Conference on Software Engineering (ICSE), San Francisco, CA, 2013.

  - T. Chen and R. Bahsoon. Self-Adaptive and Online QoS Modeling for Cloud-Based Software Services, IEEE Transactions on Software Engineering, accepted with major revision.

- Drawing on experimental observations, the QoS modelling approach is further improved for better accuracy, applicability and less complexity, as described in Chapter 5. This chapter is derived from:

  - T. Chen, R. Bahsoon and X. Yao. Online QoS Modeling in the Cloud: A Hybrid and Adaptive Multi-Learners Approach. In proceeding of the 7th IEEE/ACM International Conference on Utility and Cloud Computing (UCC), London, UK. 2014.

  - T. Chen and R. Bahsoon. Self-Adaptive and Online QoS Modeling for Cloud-Based Software Services, IEEE Transactions on Software Engineering, accepted with major revision.

- Chapter 6 presents the mechanism for dynamically determining the right granularity of control when autoscaling in the cloud. This chapter is derived from:

  - T. Chen and R. Bahsoon. Symbiotic and Sensitivity-Aware Architecture for Globally-Optimal Benefit in Self-Adaptive Cloud. In proceeding of the 9th In-ternational Symposium on Software Engineering for Adaptive and Self-Managing Systems (SEAMS), in conjunction with the 36th International Conference on Software Engineering (ICSE), India, 2014.

– T. Chen, R. Bahsoon and A-R H. Tawil. Scalable Service-Oriented Replication with Flexible Consistency Guarantee in the Cloud. Information Sciences, Elsevier, vol. 264, 2014.

- The trade-off decision making challenges for autoscaling in the cloud are investigated in Chapter 7, which is derived from:

  – T. Chen and R. Bahsoon. Self-Adaptive Trade-off Decision Making for Autoscaling for Cloud-Based Services, IEEE Transactions on Services Computing, accepted as regular paper, to appear.

- A qualitative and reflective evaluation of the thesis with respect to different criteria of cloud autoscaling, including dynamics, uncertainty, scalability, flexibility, complexity of application and practical deployment, is presented in Chapter 8.

- This thesis is concluded in Chapter 9 with detailed remarks and future research directions.

# Chapter 2

# Taxonomy and Survey of Autoscaling in Cloud

## 2.1 Background

### 2.1.1 Self-Adaptivity and Self-Awareness

The broad category of automatic and adaptive systems aim to deal with the dynamics that the system exhibited without human intervention; but this does not necessarily involve uncertainty, i.e., there are changes related to the system but it is easy to know when they would occur and the extent of these changes. Self-adaptivity, being a sub-category, is a particular capability of the system to handle both dynamics and uncertainty. Here, self-adaptive systems refer to the systems that are able to adjust their behaviours according to the perception of the uncertain environment and its own state. According to the adaptive behaviors, self-adaptivity can be regarded as the following four properties, each of which covers a specific set of goals, as explicitly discussed and categorized in many surveys, e.g., [118]:

- **Self-configuring**

The capability of reconfiguring automatically and dynamically in response to changes by installing, updating, integrating, and composing/decomposing software entities.

- **Self-healing**

  This is the capability of discovering, diagnosing, and reacting to disruptions. It can also anticipate potential problems, and accordingly take proper actions to prevent a failure. Self-diagnosing refers to diagnosing errors, faults, and failures, while self-repairing focuses on recovery from them.

- **Self-optimizing**

  This is also called self-tuning or self-adjusting, is the capability of managing performance and resource allocation in order to satisfy the requirements of different users. End-to-end response time, throughput, utilisation, and workload are examples of important concerns related to this property.

- **Self-protecting**

  This is the capability of detecting security breaches and recovering from their effects. It has two aspects, namely defending the system against malicious attacks, and anticipating problems and taking actions to avoid them or to mitigate their effects.

Self-awareness, on the other hand, is concerned with the system's ability to acquire knowledge about its current state and the environment. Such knowledge permits better reasoning about the system's adaptive behaviours. Consequently, self-awareness is often seen as the lowest level of abstraction of self-adaptivity [118], and thus it can improve the perception and self-adaptivity of a system, as surveyed in [92]. Inspired from the psychology domain, Becker et al. [18] have classified self-awareness of a computing system into the following general capabilities (they have used node to represent any conceptual part of a system being managed):

- **Stimulus-aware**

  A node is stimulus-aware if it has knowledge of stimuli. The node is not able to distinguish between the sources of stimuli. It is a prerequisite for all other levels of self-awareness.

- **Interaction-aware**

  A node is interaction-aware if it has knowledge that stimuli and its own actions form part of interactions with other nodes and the environment. It has knowledge via feedback loops that its actions can provoke, generate or cause specific reactions from the social or physical environment.

- **Time-aware**

  A node is time-aware if it has knowledge of historical and/or likely future phenomena. Implementing time-awareness may involve the node possessing an explicit memory, capabilities of time series modelling and/or anticipation.

- **Goal-aware**

  A node is goal-aware if it has knowledge of current goals, objectives, preferences and constraints. It is important to note that there is a difference between a goal existing implicitly in the design of a node, and the node having knowledge of that goal in such a way that it can reason about it. The former does not describe goal-awareness; the latter does.

- **Meta-self-aware**

  A node is meta-self-aware if it has knowledge of its own capability(ies) of awareness and the degree of complexity with which the capability(ies) are exercised. Such awareness permits a node to reason about the benefits and costs of maintaining a certain capability of awareness (and degree of complexity with which it exercises this level).

The benefits that self-awareness introduces for computing systems, including better solution for runtime dynamics and uncertainty, heterogeneity, and trade-offs on objectives, have rendered it as a neat solution for the challenges of cloud autoscaling as we have discussed in Chapter 1.

## 2.1.2   Autoscaling in Cloud

From the literal meaning of the word "autoscaling", it is obvious that the process is dynamic and requires the system to adapt subject to the uncertain, changing state of the services being managed and the environment. In such a way, the cloud-based services can be 'expanded' and 'shrink' according to the environmental conditions at runtime. Given that it is almost impossible to access the low level details of cloud-based services (e.g., their codes and algorithms) at runtime, an autoscaling system often consist of two physical parts: a managing part containing the autoscaling logic and a manageable part including services and VMs running in the cloud. The two physical parts are seamlessly and transparently connected for realising the entire autoscaling process. This characteristic has made autoscaling systems well-suited to the broad category of self-adaptive systems, and the two parts structure of adaptation is known as the external adaptation [40] [118]. Depending on the given QoS attributes and the manageable cloud primitives, an autoscaling system can cover self-configuring, self-healing, self-optimising and self-protecting, or any combination of those, from the notion of self-adaptivity. A recent survey [92] has established the evidences that self-awareness can improve a system that requires self-adaptivity, and therefore achieving self-awareness is a promising way to enable better autoscaling systems in cloud.

The external adaptation of an autoscaling system is shown in Figure 2.1. As we can see, the core of an autoscaling system in the cloud is the autoscaling logic, which can consist of multiple logical aspects. The simplest form of autoscaling system covers monitoring and scaling aspect in its autoscaling logic: the former gathers the service's or application's current state while the latter utilises the information to decide an action. However, such a simplified form of autoscaling system tends to be limited, since it cannot effectively handle the increasing runtime complexity of the cloud environment, including, e.g., dynamic and uncertainty caused by workload, the QoS performance and heterogeneity of cloud-based services. Given the shared infrastructure of cloud, the autoscaling process should be aware of and be able to handle QoS interference. This is because improving the QoS performance of a service may likely to downgrade that of its neighbouring services and VMs, which will negatively affect the overall quality of autoscaling and elasticity.

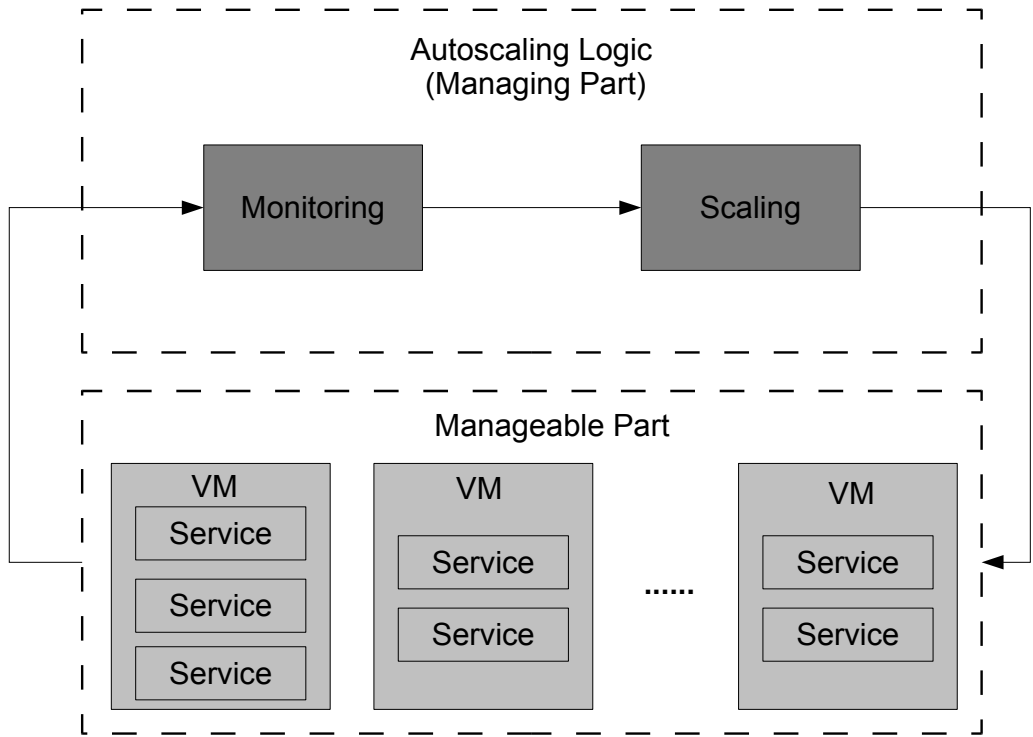

Figure 2.1: The Simplest Form of An Autoscaling System.

To improve the quality of adaptation, modern autoscaling systems often additionally cover other more sophisticated aspects, including *modelling*, *determining granularity of control* and *decision making*. The modelling is concerned with the model of QoS, en-vironment conditions (e.g., workload) and demand of the control knobs (e.g., software configurations and hardware resources). The resulting models are a powerful tool to assist the autoscaling decision making process. Without loss of generality, in this thesis, we term both control knobs and environment conditions in the cloud as *cloud primitives*. In such context, we further decompose the notion of primitives into two major domains: these are **Control Primitive (CP)** and **Environmental Primitive (EP)**. Control Primitives are the internal control knobs and can be either software or hardware, which can be managed by the cloud providers to support QoS. Specifically, software control primitives are software tactics and the key configurations in cloud; such as the number of threads in the thread pool of a service/application, the buffer size and load balancing policies etc. Whereas, hardware control primitives are computational resources, such as CPU and memory. Software and hardware control primitives rely on the PaaS and IaaS layers respectively. In particular, it is non-trivial to consider software control primitives when autoscaling in the cloud as they have been shown to be important features for QoS [26] [139] [99]. On the other hand, Environmental Primitives refer to the external stimuli that cause dynamics and uncertainties in the cloud. These, for example, can be the workload and unpredictable incoming data etc. If the cloud provider is able to control the presence of the stimulus, then these can be considered as control primitives. These models can often assist and improve the autoscaling decision making. It is worth noting that the examples of primitives listed above are not exhaustive, Ghanbari et al. [66] have provided a more completed and detailed list of the possible control primitives in cloud.

Determining granularity of control in the autoscaling logic is essential to ensure the benefit (e.g, QoS and cost objectives) for all cloud-based service. It is concerned with understanding whether certain objectives can be considered in isolation with some of the others. This is because *objective-dependency* (i.e., conflicted or harmonic objectives) often exist in the decisions making process, which implies that the overall quality of autoscaling can be significantly affected by the inclusion of conflicted or harmonic objectives in a decision making process, hence rendering it as a complex task. This is especially true for the shared infrastructure of cloud where objective-dependency exists for both intra- and inter-services. That is to say, objective-dependency is not only caused by the nature of objectives (intra- service), e.g, throughput and cost objective of a service; but also by the QoS interference (inter-services) due to the co-located services on a VM and co-hosted VMs on a PM [26] [139] [99] [116].

The final logical aspect in autoscaling logic is the dynamic decision making process that produces the optimal (or near-optimal) decision, which consists of the newly configured values of the related control primitives, for all the related objectives. In the presence of objective dependency, autoscaling decision making requires to resolve complex trade-offs, subject to the SLA and budget requirements. The trade-off decision can be then executed using either vertical scaling and/or horizontal scaling actions, which adapt the cloud-based services and/or VMs correspondingly.

In the following sections, we provide survey and taxonomy for the most influent and recent work that is related to this thesis. Particularly, we present the review and discussions based on the key logical aspects for autoscaling in the cloud, which are architectural pattern, QoS modelling, granularity of control and decision making. We then position this thesis by discussing how our work differ from those existing approaches.

## 2.2   Architectural Pattern

Autoscaling architecture is the most essential element of an autoscaling system in the cloud. It describes the structure of the autoscaling process, the interaction between components and the modularisation of the important logical aspects in autoscaling. In the following, we survey the key architectural patterns that have been applied for autoscaling in the cloud. In particular, we classify them into three categories based on their basic form;

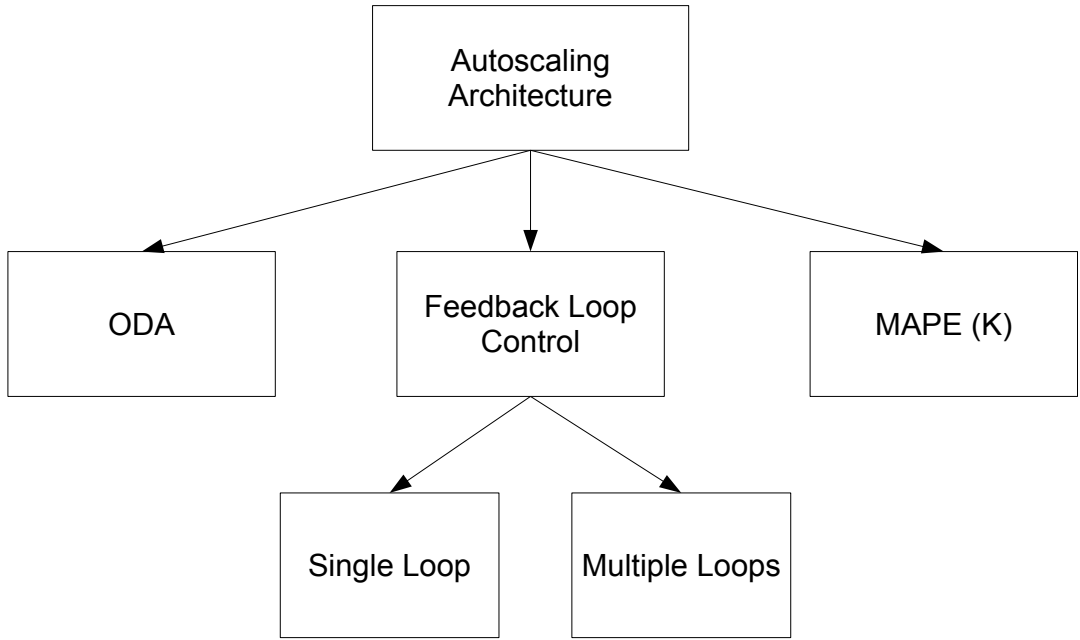

Figure 2.2: The Taxonomy of Architectural Patterns used in Autoscaling.

these are Feedback Loop Control [25], Observe-Decide-Act [74] and Monitor-Analysis-Plan-Execute [76], where the latter two are essentially detailed variations of the former one. The classification of those architectural patterns is the result obtained from literature review. As one of the key outputs, we have identified that those three architectural patterns are predominantly applied for autoscaling system in the cloud. The taxonomy has been illustrated in Figure 2.2.

## 2.2.1 Feedback Loop Control

Feedback Loop Control is the most general architectural pattern for controlling self-adaptive systems, including the autoscaling systems. It is usually a closed-form loop made up of the managing system itself and the path transmitting its origin (e.g., a sensor) to its destination (e.g., an actuator). Here, we further divide the pattern in terms of whether single or multiple loops are used.

### *Single Loop Control*

Single loop control is the simplest, yet the most commonly used pattern for autoscaling in the cloud due to its flexibility. The most common practice with single loop control is to build a feedback loop where the core is the decision making component and an optional QoS modelling component, e.g., Ferretti et al. [56] , CloudOpt [95], SmartSLA [132], Padala et al [107], Kateb et al [51], Jiang et al. [77], CLOUDFARM [106], Grandhi et al. [63]. Some other work has included an additional component for workload or demand prediction based on either offline profiling, e.g., Jiang et al. [78] and Fernandez et al. [55], or online learning, e.g., Kingfisher [121], Gambi et al. [62], Chihi et al. [44] and PRESS [68].

Open feedback loop exists, as presented in Cloudine [60], where the scaling actions are partially triggered by user requests. In particular, they use a centralised *Resource and Execution Manager* to handle all the scaling actions. Apart from the general autoscaling architecture, other efforts are particularly designed upon specific cloud providers [71], [138], [81], [142]. For example, Zhang et al. [138] and Kabir and Chiu [81] propose to use a simple feedback loop for architecting autoscaling system, which is heavily tied to the properties of Amazon EC2 and S3. Other applications of single loop control, which are worth mentioning, include: VScale [137] is a feedback based framework that particularly focus on vertical scaling of VM in the cloud. It is deployed in a decentralised manner where there is a dedicated instance running on each PM. iBoolean [113] is a feedback control approach for autoscaling hardware resource in the cloud. It is designed as a distributed management framework, in which each individual VM initialise its own management.

There are architectures using single loop control where the core is classical control theory [12], [70], [96], [10]. Particularly, Anglano et al. [12] present a fuzzilized feedback control for autoscaling in the cloud. It is a typical controller using fuzzy theory driven by application performance. Guo et al. [70] also present a fuzzy logic based feedback control. However, it only intend to scale the software control primitives.

### *Multiple Loop Control*

Unlike the single loop control approach, it is possible to use multiple loops and controllers for autoscaling in the cloud. Here, multiple feedback loops operate in different levels of the architecture, e.g., one operates at the cloud level while the others operate on each VM. The benefit is that multiple loops provide low coupling in the design of the loops. Notably, multiple loop control can be used to separate global and local controls [82], [15], [16], [104]. Among others, [82] apply a decentralised feedback control for autoscaling CPU in the cloud. Although it aims for individual applications, the controllers actually operate on each tier of an application. Different controllers do not need to interact with each others. ARUVE [15] utilises a global controller in conjunction with the local controller to form multiple feedback loops. The local controllers are decentralised on each PM while the global controller is centralised.

Multiple loop control is also effective for isolating the logical aspects of autoscaling and management in the cloud [131], [134], [85], [43], [52], [126], [135] [26], [141]. For example, Emeakaroha et al. [52] has also used multiple feedback loops. In particular, they use a global feedback loop consisting of three local feedbacks, each of which operate on SaaS, PaaS and IaaS layer. Wang, Xu and Zhao [126] propose a two layer feedback control for autoscaling in the cloud. The first layer, termed guest-to-host optimisation, controls the hardware resources, e.g., CPU and memory. Subsequently, the host-to-guest optimisation adapts the software configuration accordingly.

Overall, existing work adopts feedback loop control for its simplicity and flexibility. However, instead of designing autoscaling with a clear architectural blueprint beforehand, they utilise a bottom-up approach where the design of autoscaling system starts off from the underlying techniques and algorithms. Such design can limit the consideration of re-quired knowledge for the autoscaling system to perform adaptations, or the consideration is rather simple and coarse-grained, as they do not express what level of knowledge is required at which logical aspect of the system, and how they can be beneficial for the adaptation.

### 2.2.2  Observe-Decide-Act

Observe-Decide-Act (ODA) loop [74] is considered as an extended pattern of the feedback loop control, and it is concerned with the system monitoring itself and its environment, making decisions about how to adapt behaviour using a set of available actions. A unique *Decide* component separates it from the feedback loop control, as it explicitly requires to perform reasoning about the effects of adaptation on the system's goals and objectives. While an ODA loop is most commonly applied for self-adaptive systems in general, only few work (e.g., [75]) has included it for the design of autoscaling system in the cloud. This is because one important aspect of ODA is to define the effects of human activities on the adaptive behaviours, which is a difficult practice for autoscaling in the cloud.

SEEC [74], being the very first work to introduce ODA, is a general framework for self-aware and self-adaptive systems. It applies ODA for decoupling the loops to different roles (i.e., application developer, system developer, and the SEEC runtime decision infrastructure) in the development life-cycle, each role focuses on one or more steps in ODA. [122] adopt an ODA loop to manage FPGA-based systems, where the decision and the its translation to actions are conducted by an incorporation of the *Decide* and *Act* steps. Bolchini et al. [21] have used ODA to realise the adaptation for self-adaptive systems because of its simplicity. It observes high-level and raw data from the *Observe* step, such data is then used by *Decide* to know which parts of the system to reason on, and finally, actions are taken depends on the characteristic of the system being managed. Huber et al. [75] also use ODA for self-aware autoscaling resources in the cloud. However, unlike traditional ODA loop, it has an additional *Analysis* step which is used to detect the type of problems that trigger adaptation.

Overall, although the knowledge that a system requirs is sometime discussed in the *Decide* step (e.g., [75]) in ODA, it cannot capture different levels of knowledge in a fine-grained representation as required by the system. This is because ODA is mainly designed for decoupling loops for different human activities, which allows application and systems programmers to separately specify observations and actions, according to their expertise.

### 2.2.3 Monitor-Analyze-Plan-Execute

Another pattern extended from the feedback loop control, namely Monitor-Analyze-Plan-Execute (MAPE), is firstly proposed by IBM for architecting self-adaptive systems. In such pattern, the *Decide* step in OAD is further divided into two substeps, these are *Analyze* and *Plan*, where the former is particularly designed to determine the causes for adaptations, e.g., SLA violation; the latter, on the other hand, is responsible for reasoning about the possible actions for adaptation. MAPE sometime can be extended by a Knowledge component (a.k.a. MAPE-K) which maintains historical data and knowledge used by the system for better adaptation.

MAPE (or MAPE-K) is widely applied for autoscaling in the cloud [94], [30], [23], [101], [28], [91]. For example, the architecture of the FoSII ptoject [23] [101] leverages MAPE-K to realise the self- management interface, which is necessary to devise actions in order to prevent SLA violations in cloud. They also use the additional *Knowledge* (K) component to record cases and the related solutions, which can assist the autoscaling decision making. QoSMOS [28] is designed for service-based systems rather than for cloud specific autoscaling, however, it contains many aspects similar to that of autoscaling in the cloud. To achieve continuous adaptation, it applies MAPE with the focuses on analytical QoS modelling and optimisation of resource allocation. APPLEware [91] is an autoscaling framework which leverages MAPE. In their architecture, the *Analyze* component model the QoS while the *Plan* component conducts optimisation process for autoscaling.

Realising multiple MAPE loops is also possible. Zhang et al. [139] introduce an architecture for autoscaling using two nested MAPE loops. The first loop is responsible for adapting the software primitives while the other loop is used to change the hardware primitives. These two loops run sequentially upon autoscaling, that is, adapting the software primitives before changing the hardware primitives. Similarly, BRGA [7] utilises MAPE to realise a framework for autoscaling in the cloud. Such solution consists of both the local and global view of the cloud-based application. In particular, the *Monitor* and *Execution* phase maintain the global view whereas the *Analyze* and *Plan* phase manage the local view on each PM. The authors claim that such an approach can achieve good global quality with reasonable management overhead.

In conclusion, MAPE can be good for separation of concepts (e.g., *Analyze* and *Plan*) and for expressing the sequential interactions between those concepts. However, although the *Knowledge* component can be considered, there is still no fine-grained representation of the required knowledge for the system. Thus, it is not immediately intuitive that what level of the knowledge is required by each logical aspect of the autoscaling system.

## 2.3 QoS Modelling

QoS modelling, or performance modelling, is a fundamental research theme in cloud computing and it can serve as useful foundations for addressing many research problems in the cloud [98], including autoscaling. The QoS models correlate the QoS attributes to various control primitives and environmental primitives. Clearly, these models are particularly important in cloud autoscaling, as they are a powerful tool that can assist the reasoning about the effects of adaptation on objectives in the autoscaling decision making process. Typically, QoS modelling consists of two phases, namely primitives selection and QoS function training. More precisely, the primitives selection phase determines *which* and *when* the primitives correlate with the QoS; while QoS function training phase identify *how* these primitives correlate with the QoS, i.e., their magnitudes in the correlation. The QoS models can be either static or dynamic, where the former refers to the models' expression and their structure (e.g., the number of inputs and their weights) do not change over time; while the latter permits such changes. Those models can be also applied as online at system runtime, or offline at design phase of the system. In the following, we survey the key work on QoS modelling in the cloud and classify them in terms of the algorithms they apply. The taxonomy has been illustrated in Figure 2.3.

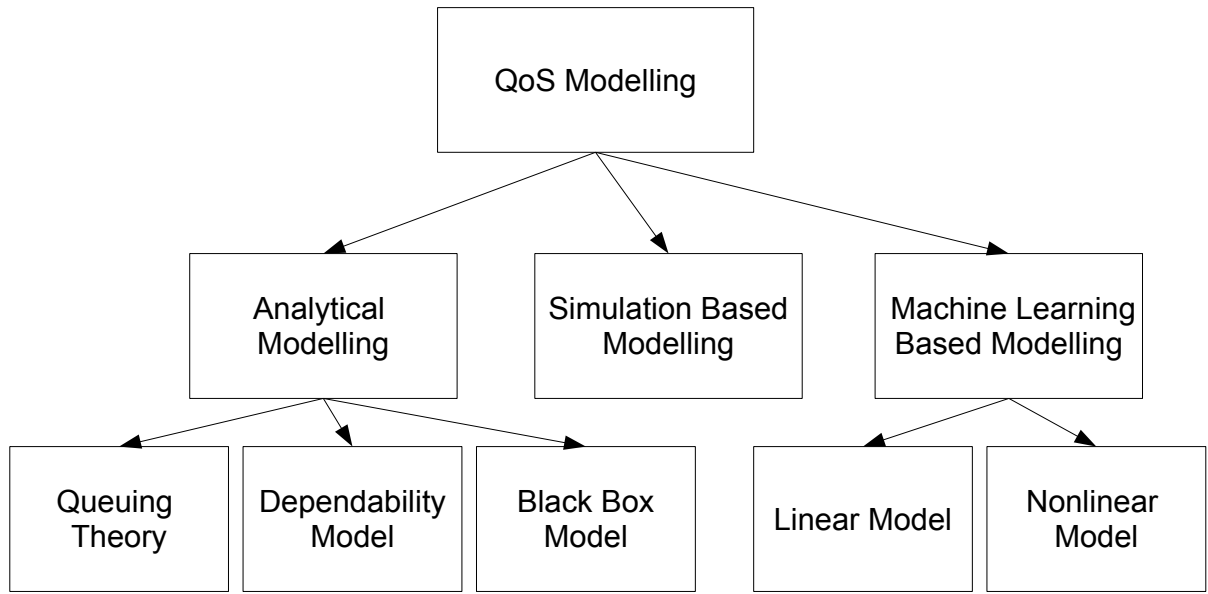

Figure 2.3: The Taxonomy of QoS Modelling in Cloud.

### 2.3.1 Analytical Modelling

Analytical modelling approaches rely on mathematical models that have a closed-form solution to model the cloud-based service. These models are often built offline based on theoretical principles and assumptions. Next, we further divide the analytical modelling approach into queuing theory, dependability models and black box models.

#### *Queuing theory*

Queuing model and queuing network are widely applied for QoS modelling in the cloud. They model the cloud-based services as a single queue or a collection of queues interacting through request arrivals and departures. Specifically, a single queue has been used to model the correlation of response time (or throughput) to CPU, number of VM and workload. For example, depending on the assumption of the distribution on arrival and service rate, the model can be built as M/G/c queue by Zhang et al. [138], M/G/m queue by Jiang et al. [77] , M/M/1 queue by E$^3$-R [124] and JustSAT [125], and M/M/m queue by Jiang et al. [78]. To create more detailed modelling with respect to the internal structure of cloud-based services, multiple queues can be used to create QoS models: Goudarzi and Pedram [69] apply multiple queues to model the response time for cloud-based multi-tiered applications with respect to number of VM and workload. Their work calculates average response time for the queue in the forward direction throughout the tiers. In a similar way, Bi et al. [19] use a queuing network composed of an M/M/c queue and multiple M/M/1 queues to estimate the correlation between response time and number of VMs and workload for cloud-based application. Li et al. [93] apply a single queue to model the correlation between response time and CPU, workload and thread. In particular, the model contains finite capacity regions, which denote the place constraints on the maximum number of jobs circulating in a subnetwork of queues. This is because they are the simplest class of models that offer the features to describe performance scalability as a function of the software threading level and for the number of CPUs.

Unlike classical queuing model and queuing network, the Layer Queuing Network (LQN) additionally model the dependencies arising in a complex workflow of requests to cloud-based services and applications. Chi et al. [41] use LQN (i.e., based on M/M/n queue) to model the QoS of application, which is response time with respect to CPU and workload. CloudOpt [95] relied on LQN as the aggregate QoS model for all the services contained by an application. It models only response time with respect to CPU and workload. Li et al. [94] use LQN for model services in an application. Again, it only captures response time with respect to CPU and workload. Zhu et al. [142] have also used LQN where the authors employ a global M/M/c queue for the entire on-demand dispatcher and then a M/G/1 queue on each tier of an application. The former queue correlates the response time to number of VMs while the latter queue models the relationship between response time and CPU of the VM that contains the corresponding tier.

### *Dependability models*

Dependability models are another widely used technique for QoS modelling in the cloud. This approach focuses on the modelling of stable states for QoS attributes. For example, Copil et al. [46] uses a graph representation to model the dependency between per-service QoS and the necessary primitives. Although the graph can be updated at runtime, the model is essentially analytical. In QoSMOS [28], the authors analytically solve the Markov Models (Discrete-Time Markov Chain and Markov Decision Process) to model the QoS for services in an application. The model correlates QoS attributes with hardware resources and workload. Huber et al. [75] uses Palladio Component Model (PCM) as architecture-level QoS model since it allows to explicitly model different usage profiles and resource allocations. Kateb et al. [51] uses *model@runtime* to correlate QoS attributes with the number of VM. The modelling approach is essentially based on a domain specific language, which does not only able to reason about the system at design time, but is also able to assist decision making during runtime.

### *Black box models*

Black-box models are also popular, in which the QoS is modelled based on empirical knowledge or statistical data of history [52], [23], [106], [54], [53], [8], [81],[7]. Among others, CLOUDFARM [106] uses a empirical QoS model where the correlation between certain QoS values and the required resource is captured (i.e., CPU). In particular, the authors assumed that the magnitudes of resources to the QoS values is known, as specified by the cloud service or application provider. The FoSII project [23] has also applied empirical QoS models such that the correlation between hardware resource (i.e., CPU, memory and bandwidth) and QoS is hard-coded using cases, each of which contains a set of particular values of resource and their resulted utility value. Another work from Emeakaroha et al. [54] propose an empirical model that maps the expected QoS values with CPU, memory, bandwidth and storage. The model relies heavily on the assumption of the system that being managed. Their extended work [53] is also based on a similar approach, where the authors correlate the QoS attributes to different CPU, memory, bandwidth and storage using manual and empirical mapping. The proposed mapping can be as simple as one QoS attribute to one primitives, or a complex form where multiple cloud primitives are associated with a QoS attribute.

Overall, analytical modelling has the advantage of simplicity and interoperability. In particular, such modelling is usually highly intuitive and has negligible overhead when applied for autoscaling in the cloud. However, analytical approaches often require in-depth knowledge about the likely behaviours of the system being modelled. Consequently, their effectiveness is restricted to the assumptions of service's internal operations; such static nature makes these approaches limited in coping with the dynamic and uncertainty at runtime. Finally, both primitives selection and QoS function training phases in analytical approaches are often static and offline. However, for some of the approaches (e.g., [28], [51]), their QoS function training phase can be achieved in a dynamic and online manner.

## 2.3.2 Simulation Based Modelling

Various simulators exist for creating QoS models; here, conducting simulations is usually a complex and expensive process and thus they are used in an offline manner. In practice, simulation is required to be setup by the domain experts, who will often need to analyse, interpret and profile the data collected after simulation runs. Specifically, Fernandez et al [55] have relied on a profiling approach that builds the QoS model for each bundle of VM offline. The process is similar to a simulation modelling approach. CDOSim [57] is a framework that simulates the actual application in the cloud to restrict the search-space for autoscaling and to steer the exploration towards promising decisions. CloudSim [27] is a simulation toolkit that models QoS attributes (of VM) with respect to resource allocation. It supports both single cloud and multiple clouds scenarios. As an extension of CloudSim, CloudAnalysis [128] allows the simulation of QoS attributes for the application deployed on geographically-distributed datacenters. Similarly, DCSim [83] simulates the overall quality of resource autoscaling for the entire cloud.

Overall, simulation can produce good QoS models providing that the scenarios which have been simulated are similar to those that would occur at runtime. However, similar to the analytical approaches, simulation based modelling is also static and restricted by the assumptions made in the simulators, e.g., distribution of workload and the effects of QoS interference. In addition, it can be expensive to use as it often requires heavy human intervention. Commonly, simulation based modelling approach is an offline process, in particular, the QoS function training phases can be dynamic; while the primitives selection phase is static.

### 2.3.3 Machine Learning Based Modelling

The increasing complexity of managing services in the cloud makes the modelling difficulty far beyond the capability of human analysis. To this end, recent works have been leveraging the advances of machine learning algorithms. In the following, we survey the key work that applies machine learning approaches for QoS modelling in the cloud. In particular, we have classified them into two categories, these are: linear and nonlinear modelling.

#### *Linear modelling*

Learning algorithms based on linear models for QoS modelling in the cloud can handle linear correlation between a selected set of inputs (e.g, CPU, memory, number of VM, workload etc) and output (i.e., QoS attributes), and they are sometime very efficient. Diao et al. [48] propose a very early work on QoS modelling using Auto-Regressive and Moving-Average (ARMA) and Multi-Inputs-Multi-Output (MIMO) model on-the-fly. Their work is not cloud specific but it provides insight for many subsequent work on cloud based QoS modelling. Simple linear models most commonly rely on linear regression, where each primitive input is associated with a time-varying weight, e.g., Lim et al. [96], Zhang et al. [139] and Collazo-Mojica et al. [45]. More advanced forms exist, e.g., Padala et al. [107] have used ARMA trained by Recursive Least Squares (RLS). The authors claim that the linear AMRA model is easy to be estimated online and can simplify the corresponding controller design problem. The authors found that the second-order ARMA model can predict the application performance with adequate accuracy. Kalivianaki et al. [82] uses Kalman filter to update the QoS model. The authors claim that the Kalman filter is optimal in the sum squared error sense under the assumptions that the system is described by a linear model, and the process and measurement noise are white and Gaussian.

Linear machine learning algorithms are also commonly used with analytical approaches to form QoS models. Specifically, Grandhi et al. [63] and Zheng et al. [140] have proposed hybrid model: to model the multi-tiered application, they have relied on a modified LQN where there are some time-varying coefficients. The authors then employ the Kalman filter as an online parameter estimator to continually estimate those coefficients. Ghanbari et al. [65] have also followed a similar approach, but through the use of k-mean clustering, they additionally cluster the model into multiple sub-models based on different types of workload. The approach proposed by Xiong et al. [134] has relied on a combined model, where a M/G/1 queue is used to model the correlation between response time and workload; while ARMA is used to model the relationship of response time and CPU.

There is limited work that attempts to capture the information of QoS interference in the linear QoS model and they only focus on the VM-level [91], [116], [127], [99]. As an example, Q-Cloud [116] has explicitly considered QoS interference by using the hardware control primitives of all co-hosted VMs as inputs, rendering it in a MIMO manner. The model itself is a simple linear model and it can be easily trained by using Least Mean Square (LMS).

### Nonlinear modelling

Learning algorithms based on nonlinear models for QoS modelling in the cloud is able to capture complex and nonlinear correlation, in addition to the linear one. However, it can also produce relatively large overhead than the linear modelling. Here, existing work often aim to model the correlation between hardware control primitives (e.g., CPU, memory and bandwidth) and QoS. The nonlinear modelling can be relied on kriging model [62], Regression Tree (RT)[132], Artificial Neural Network (ANN) [90] [105] [87] , Support Vector Machine (SVM) [43] [90] and change-point detection [20]. For example, Gambi et al. [62] utilise kriging model, which is a spatial data interpolator akin to nonlinear and radial basis functions, and it extends traditional regression with stochastic Gaussian processes. SmartSLA [132] employs Regression Tree (RT) and boosting to model the QoS. RT partitions the parameter space in a top-down fashion, and organises the regions into a tree style. The tree is then trained by M5P where the leaves are regression models. The work from Kunda et al. [90] presents sub-modelling based on ANN and SVM for correlating QoS with hardware control primitives in the cloud. Instead of building a single model for a QoS attribute, they train $n$ sub-models, whereby $n$ is determined by performing k-mean clustering based on the similarity between data values of QoS. This is because they observe that large errors were mostly concentrated in a few sub-regions of the output value space, indicating a single model's inability to accurately characterise changes in application behaviour as it moves across critical resource allocation boundaries. The authors claim that the approach can be applied online.

Examples exist for cases where multiple linear and/or nonlinear machine learning algorithms are used together. Zhu and Agrawal [141] use a variant of ARMA and SVM to model the correlation of QoS attributes to software and hardware control primitives. In particular, the ARMA variant, which is trained by SVM, is used to link QoS and software control primitives. Subsequently, another dedicated SVM is used to model the relationship between software control primitives and hardware control primitives (i.e., CPU and memory in the work). Another work from Kousiouris et al. [88] correlates QoS attributes with various primitives using ANN. Additionally, it applies a time-series ANN to predict the workload in conjunction with the QoS models. This aims to provide more accurate information when making prediction online.

### Dynamic primitives selection

All the aforementioned work regards the primitives selection as a manual and offline process, most commonly, they have relied on empirical knowledge and heavy human analysis to select the important primitives as the inputs of QoS models. Although not many, there is some work that explicitly considers dynamic process in primitives selection, which tends to be more accurate and can be easily applied [85], [86], [133]. As an example, vPerfGuard [133] is a framework that correlates QoS attributes with respect to software control primitives, hardware control primitives and environmental primitives. The authors achieve primitive selection based on both filter (relevance based correlation coefficient) and wrapper (i.e., hill-climbing comparison based on algorithms like k-nearest-neighbour and linear regression etc). Linear regression is used as default to train QoS based on the selected primitives.

### *Comparison of different learning algorithms*

Given the various types of machine learning algorithms, it can be difficult to determine which one(s) are the appropriate algorithms for QoS modelling in the cloud, with respect to both accuracy and overhead. There are researches that have conducted detailed comparisons of different possible learning algorithms for QoS modelling in the cloud [103] [97] [42], for example, Lloyd et al. [97] conduct an extensive experiment over various machine learning algorithms (i.e., MLR, MRS and ANN) in cloud. They select the primitives based on manual analysis. The results show that the relevant and useful primitives could be different depending on the characteristics of services and application; and that different machine learning algorithms achieve a variety of accuracy depending on the scenarios.

Overall, machine learning based modelling approaches have the advantage of requiring limited human intervention, and care able to continually evolve themselves at runtime in order to cope with dynamics and uncertainty. Nevertheless, depending on the learning algorithm, the resulting overhead can be high (e.g., the nonlinear ones) and the accuracy is sensitive to the given scenarios (e.g., fluctuation of the data trend). Generally, the machine learning approaches can be applied as offline, online or a mixture of the both. According to the existing work surveyed for QoS modelling in the cloud, we discover that the QoS function training phase is often dynamic; while there is very little work that intends to consider primitives selection phases as a dynamic process (online or offline), and the others have relied on offline and manual analysis. Therefore, we can conclude that majority of the approaches that apply machine learning for QoS modelling are semi-dynamic. In addition, we have also found that only a small amount of existing work intends to consider QoS interference in the modelling; and they only focus on VM-level interference.

### 2.3.4 Comparison of QoS Modelling to Workload and Demand Modelling

Despite the fact that QoS modelling is fundamentally helpful for cloud autoscaling, it can be difficult to achieve given the heterogeneity of possible primitives and the multi-dimensional input space of QoS modelling. Therefore, some existing work on autoscaling (e.g., [121] , [120], [87]) have considered simpler alternatives, i.e., model the workload and demand for assisting autoscaling decision making. In those cases, the modelling is reduced to a single dimension, where the core is to model the trend of the workload or demand using its historical data. The models would be used to predict the likely value at the next interval. However, the single dimension in workload or demand models do not offer the ability to reason about the effects of autoscaling decisions and the possible trade-offs. It is important to note that trade-off decisions making is a critical logical aspect for autoscaling, and it is also one of the cores of this thesis. Therefore, this thesis has explicitly focused on QoS modelling.

## 2.4 Granularity of Control

The ultimate goal of autoscaling is to optimise the QoS and cost objectives, which are referred to as benefit, for all cloud-based services. To this end, the granularity of control in autoscaling plays an integral role, since it determines which and how many objectives should be considered in a decision making process of autoscaling. In the following, we classify existing cloud autoscaling approaches into different categories depending on what level of granularity they tend to operate at. The taxonomy has been illustrated in Figure 2.4.

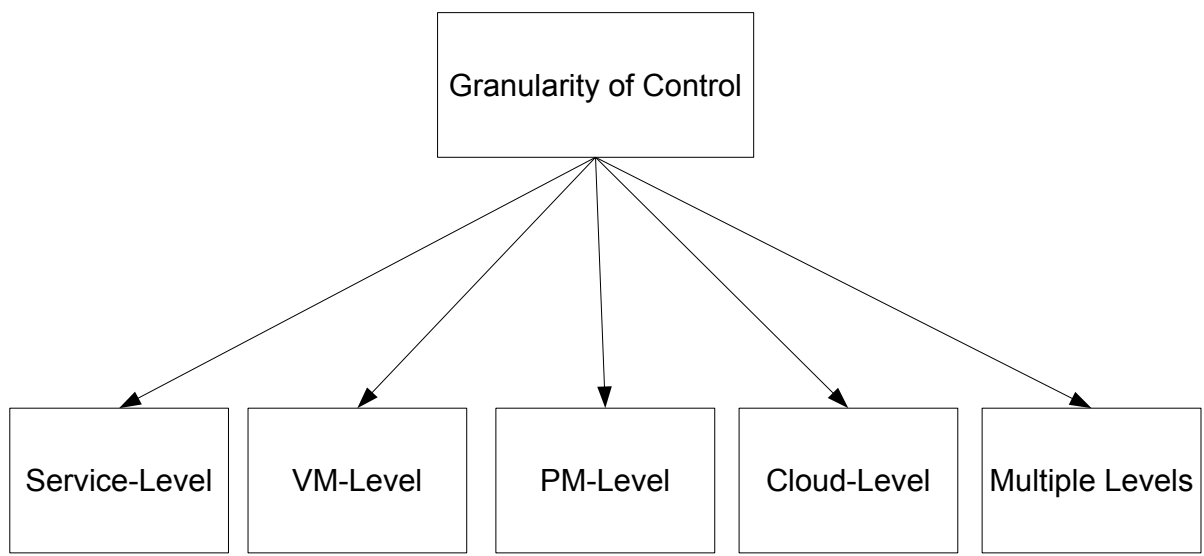

Figure 2.4: The Taxonomy of Granularity of Control for Autoscaling in Cloud.

### 2.4.1 Controlling at Service Level

Service level is the finest level of control in the cloud. It is worth noting that by service, we refer to any conceptual part of the system being managed. As a result, control granularity at the service level may refer to independently controlling/scaling an application, a tier of an application or a cloud-based service.

Specifically, most work has focused on controlling each cloud-based application. These approaches have relied on controlling the QoS and/or cost for each individual application in isolation, and therefore, they regard an application as a service. Examples of such include: [121], [125], [55], [45], [141], [93], [77], [30], [68] and [85]. To describe some of them in detail, Lim et al. [96] control the application and its required VM, in which case an application is regarded as a service. Sedaghat et al. [120] regard application as a service, and considered the required number of VMs and the fixed VM bundles for such service. Jiang et al. [78] control each application, each of which is, again, regarded as a service and therefore it is essentially service-level control. In addition, the authors group VMs into different fixed bundles.

There is also existing work that controls cloud-based service in general, which can be regarded as any conceptual part of a cloud-based system. Copli et al. [46] control the QoS, cost and their elasticity for each service deployed in the cloud. Yang et al. [136] control the cost of individual cloud-based services. The FoSII project [23] controls individual cloud-based service, their QoS and cost. [62] control at the service level, where the controller decides on the optimal autoscaling decision for cloud-based service in isolation. QoSMOS [28] explicitly focuses on each individual service, and adapting it in isolation. Kateb et al. [51] consider many services are encapsulated in the application, and the authors focus on each service in isolation. The $E^3$-R framework [124] and Frey et al. [59] control each service, including their composition and autoscaling.

## 2.4.2 Controlling at Virtual Machine Level

VM level means that the control and decision making operate at each VM. In particular, certain work assumes a one-to-one mapping between application (or a tier) and VM and thus they can be categorised as either service level or VM level granularity. To better separate them from the pure service level granularity of control, these work are regarded as VM level granularity. Specifically, FC2Q [12] regards application tier and VM interchangeably, therefore controlling each tier of an application is equivalent to control each individual VM. Similarly, Kalyvianaki et al. [82] control a tier of an application that resides on a VM, and the authors only focus on CPU allocation of a VM. Wang, Xu and Zhao [126] control the cloud in a per-VM basis, and each VM is adapted in isolation. VScale [137] focuses on vertical scaling only and hence the control granularity is per VM. Zhang et al. [139] assume only one application per VM, and control the deployed VM in isolation correspondingly. Guo et al. [70] control the application deployed on the VM and the authors assume one application per VM. Similarly, Matrix [43] has used VM level control as there is only one application per VM.

### 2.4.3    Controlling at Physical Machine Level

Autoscaling decision making on each PM independently is referred to as PM level control. The primary intention of PM level control is to manage the QoS interference caused by co-hosted VMs. Among the others, Xu et al. [113] control the VMs collectively at the PM level, in this way, it tries to promote better management of QoS interference. The extended work from Xu et al. [135] consider QoS interference at the VM level, therefore the granularity of control is based on each PM. Similarly, Bu et al. [26] considers VM level QoS interference and thus the control is for each PM. Minarolli and Freisleben [105] consider all the co-hosted VM in conjunction with each others and thus its control granularity is at the PM level. Lama et al. [91] control at the PM level in order to handle QoS interference.

### 2.4.4    Controlling at Cloud Level

The most coarse level of control granularity is at the cloud level. The majority of the work achieves autoscaling at the cloud level by using a centralised and global controller, with an aim to manage utility ([41], [15], [60], [8], [106]), profits ([94], [142]) and availability ([52]). Among others, Ferretti et al. [56] control the QoS for all cloud-based services in a global manner. However, the actual deployment can be either centralised or decentralised. Similarly, CRAMP [16] uses a centralized and global controller, it controls the entire cloud for cost and QoS. CloudOpt [95] also controls the entire cloud using centralised control, as the considered optimisation involves all the PM in the cloud. Zhang et al. [138] control the cost of the entire cloud with respect to how the VM instances of Amazon can be utilised. BRGA [7] maintains global view of the entire cloud, and thus it belongs to cloud level control. The FOSII project [102] also controls the entire cloud in a centralised manner, where the goal is to manage the entire cloud at infrastructure level.

Some of the developments have relied on a decentralised manner where a consensus protocol is employed for controlling at the cloud granularity. For example, Wuhib et al. [131] aim to control the entire cloud, and thus the QoS and the overall power consumption of cloud can be collectively managed. In the mean time, they have relied on decentralised deployment, which can reduce the overhead of cloud-level control.

## 2.4.5 Controlling at Multiple Levels

Some work operates at multiple levels, with an aim to better manage the overhead and global benefit. For example, Minarolli and Freisleben [104] combine both PM level and cloud level control, where the PM level is decentralised and the objective is to optimise the utility locally. Similarly, SmartSLA [132] aims to control the resource allocation for all the cloud-based services, therefore it utilises a global, cloud-level control in addition to the decentralised local control on each VM.

In summary, the finer granularity of control implies that it is harder to achieve globally-optimal benefit but likely to generate smaller overhead. On the other hand, globally-optimal benefit can be easier reached with large overhead if the granularity of control is coarser. All of the approaches surveyed operate at static and fixed granularity of control, even for the hybrid ones. As a result, given the time-varying QoS sensitivity and interference in cloud, they can be inflexible for any runtime changes about the effects of control granularity to the global benefit.

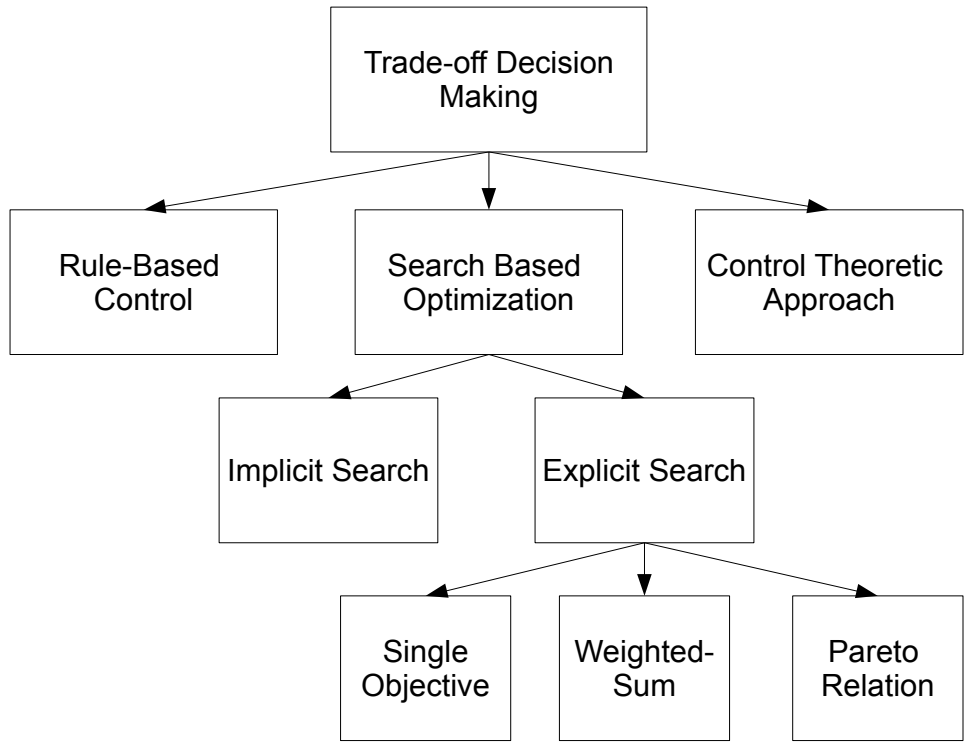

Figure 2.5: The Taxonomy of Trade-off Decision Making for Autoscaling in Cloud.

## 2.5  Trade-off Decision Making

The final important logical aspect in cloud autoscaling is the challenging decision making process, with the goal to optimise QoS and cost objectives. It is even harder to handle the trade-off between possibly conflicting objectives. Such decision making process is essentially a combinatorial optimisation problem where the output is the optimal decision containing the newly configured values for all related control primitives. In the following, we survey the key work on the decision making for cloud autoscaling. In particular, we classify them into three categories, these are Rule Based Control, Control Theoretic Approach and Search Based Optimisation. The taxonomy has been illustrated in Figure 2.5.

## 2.5.1   Rule Based Control

Rule-based control is the most classic approaches for making decision in cloud autoscaling. Commonly, one or more conditions are manually specified and mapped to a decision, e.g., increase CPU and memory by $x$ if the throughput is lower than $y$. Therefore, the possible trade-off is often implicitly handled by the conditions and actions mapping. Specifically, Cloudline [60] allows programmable elasticity rules to drive autoscaling decisions. It is also possible to modify these rules at runtime as required by the users. Copil et al. [46] handle the decision making process by specifying different condition-and-actions mapping for autoscaling in the cloud. In addition, the rules can be defined at different levels, e.g., PaaS and IaaS. Similarly, Ferretti et al. [56] allow to setup mapping between QoS expectation and actions using XML like notations. The autoscaling decision making in the work from Emeakaroha et al. [52] is also based on predefined rules, in addition, a simple heuristic algorithm is used to search for the best available VMs. Rule based control for autoscaling decision making can be also found in other work, e.g., Wuhib et al. [131], Manuer et al. [102], Han et al. [71] and Chazalet et al. [30].

Overall, we discovered that all the rule based autoscaling decision making approaches have considered both vertical and horizontal scaling as the final actuations in the cloud. Generally, rule based control is a highly intuitive approach for autoscaling decision making, and it also has negligible overhead. However, the static nature of the rules requires to assume all the possible conditions and the effects of those decisions that are mapped to the conditions. In addition, the fact that they heavily rely on human intervention and analysis can quickly become an issue. Consequently, they tend to be limited in dealing with the dynamics and uncertainties in cloud, especially when there are complex trade-offs.

## 2.5.2 Control Theoretic Approach

Advanced control theory is another widely investigated approach for autoscaling decision making in cloud because of its low latency and dynamic nature. However, it is difficult to explicitly reason about the effects of possible trade-off decisions in a control theoretic approach.

Among the others, classical controllers (e.g., Proportional-Derivative control [15] [96] [16], Kalman control [82] [63] and Fuzzy control [12] [10] [126] ) are commonly designed as a sole approach to make autoscaling decisions in the cloud. Specifically, ARUVE [15] and CRAMP [16] utilises a Proportional-Derivative (PD) controller, where the proportional and derivative factors do not depend on a QoS model of the application or the infrastructure dynamics, and support proactive resource allocation for the application server tier with dynamic scaling of web applications in a shared hosting environment. Anglano et al. [12] and Albano et al. [10] apply fuzzy control that is updated by fuzzy rules at runtime. The aim is to optimise both QoS, cost and energy by autoscaling hardware resources. Although the authors claim they can cope with any hardware resources, the approach only focus on CPU allocation. They have also assumed that QoS interference rarely occurs. Kalyvianaki et al. [82] and Grandhi et al. [63] use MIMO model and Kalman controller for making autoscaling decisions, and the authors aim at response time by autoscaling CPU on a VM.

Control theoretic approaches can be sometime used with other algorithms to better facilitate the autoscaling decision making [134], [104], [70], [114], [141], [91]. Particularly, the gains in the controllers can be further tuned by optimisation and/or machine learning algorithms, and this is especially useful for Model Predictive Control (MPC). Among others, Zhu and Agrawal [141] utilise a Proportional-Integral-Derivative (PID) and reinforcement learning controller for decision making with respect to adapting software control primitives. Such result is then tuned in conjunction with the hardware control primitives using exhaustive search. The QoS attributes and cost are formulated as weighted-sum relation. The autoscaling decision making process in APPLEware [91] have relied on MPC, which involves optimising a cost function that expresses the local control objectives and resource constraints over a time interval. The current state of the local application and the control decisions made by neighbouring controllers of a VM are taken into account to perform the optimisation using quadratic programming solver.

In summary, we discovery that the majority of the control theoretic work have considered both vertical and horizontal scaling as the final actuations in the cloud. Overall, the control theoretic approaches are efficient for making autoscaling decisions. However, the major drawback of control theoretic approaches is that they require to make many actuations on the physical system, in order to collect the 'errors' for stabilising itself. This means that amateur decisions are very likely to be made. In addition, the trade-offs are only implicitly handled.

### 2.5.3   Search Based Optimisation

A large amount of existing work relies on search-based optimisation, in which the decisions and trade-offs are extensively reasoned in a finite, but possibly large search space. Depending on the algorithms, search-based optimisation for autoscaling decision making in the cloud can be either *explicit* or *implicit*—the former performs optimisation as guided by explicit system models; while this process is not required for the later.

#### *Implicit search*

As mentioned, the implicit and search-based optimisation approaches for autoscaling decision making do not use QoS models. Similar to the control theoretic approaches, the implicit search is also limited in reasoning about the possible trade-offs. For example, the work from Xu et al. [113] , [135] applies a model-free Reinforcement Learning (RL) approach for adapting thread, CPU and memory for QoS and cost. The approach is however implicit, providing that there is neither explicit system models nor explicit optimisation. The authors have considered QoS interference during autoscaling. Similarly, VScale [137] utilises RL for making autoscaling decisions, which are then achieved by vertical scaling. The RL is realised by using parallel learning, that is to say, the authors intend to speed up agent's learning process of approximated model by learning in parallel. Therefore, the agent does not have to visit every state-action pair in a given environment.

The approaches that rely on demand prediction (e.g., the Autoflex [11], PRESS [68], [120], [29], [44] and [79]) are also regarded as implicit search. This is because the autoscaling decision is essentially predicted by the demand models, without the needs of reasoning or optimisation process.

### *Explicit search*

In search-based optimisation category, the explicit approaches for autoscaling decision making rely on the explicit QoS models to evaluate and guide the search process. Depending on the different formulations of the decision making problem for autoscaling in the cloud, explicit search can reason about the effects of decisions and the possible trade-offs in details. In this thesis, we have surveyed the approaches that rely on three the most commonly used formulations, these are single objective optimisation, weighted-sum optimisation, and Pareto optimisation.

It is common to optimise only a single objective (e.g., cost or profit) for autoscaling in the cloud, providing that the requirements of the other objectives are satisfied (i.e., they are often regarded as constraints) [136], [138], [85], [125], [45], [75], [121], [120], [95], [17], [43], [99]. For example, Kingfisher [121] and Sedaghat et al. [120] use Integer Linear Programming (ILP) to optimise the cost for scaling the CPU and memory for VMs of an application while regarding the demand for satisfying QoS as constraint. Sedaghat et al. [120] has additionally assumed fixed VM bundles. Similarly, CloudOpt [95] models the decision making for autoscaling using a weighted-sum of different aspect of cost, which is still regarded as single objective optimisation, and the optimisation is then resolved by mixed integer programming approach. A recent extension of ARUVE [17] formulates the decision making in autoscaling as a single optimisation on cost, which is resolved by using Ant Colony Optimisation (ACO).

To apply search based optimisation for autoscaling in the cloud, the most widely solution for handling the multi-objectivity is to aggregate all related objectives into a weighted (usually weighted-sum) formulation, which converts the decision making process into a single objective optimisation problem. The search based algorithm include: exhaustive search [41] [28] [62] [78], auxiliary network flow model [94], force-directed search [69], binary search [81]. For example, the FoSII project [23] [101] regards the autoscaling decision making as case based reasoning process, where the decision is made by looking for similar cases from the past and reusing the solutions of these cases to solve the current one. The case and solution pairs are linked to aggregative utility values based on the analytical model, thus the reasoning process is essentially an optimisation for the optimal decision using exhaustive search algorithm. Goudarzi and Pedram [69] use a weighted-sum formulation containing QoS and cost for an application tier. The optimisation is resolved using force-directed search, in which an initial solution based on the solution given for the profit upper bound problem is generated. Next, distribution rates are fixed and resource sharing is improved by a local optimisation step.

Some work has relied on more advanced and nonlinear search algorithms, ranging from relatively simple ones: dynamic programming [106] and local-search strategy [8], to more complex forms: grid search [132], decision tree search [55] [99] and quadratic programming [107]. For example, CLOUDFARM [106] addresses the decision making based on a weighted-sum utility function of all cloud-based application and services. The decision making process is formulated as a knapsack problem, which can be resolved by dynamic programming. In SmartSLA [132], the decision making for autoscaling is aimed for optimising SLA penalty, which is essentially based on the aggregation of expected QoS values. The optimisation is resolved using a grid search algorithm. To optimise weighted-sum utilisation, Amiya et al. [99] have explicitly aimed to mitigate QoS interference in the cloud using heuristic based decision tree search, however, they only intend to autoscale software control primitives.

Metaheuristic algorithms are also popular for autoscaling decision making in the cloud, because they can often efficiently address NP-hard problems with approximated results. The most common algorithms include: Tabu Search [142], Genetic Algorithm (GA) [139] [105] [7] , Particle Swarm Optimisation (PSO) [139]. As an example, Zhu et al. [142] formulate the autoscaling decision making as optimise a weighted-sum formulation of response time and cost. To optimise the objectives, the authors apply a hybrid Tabu Search, which, in every iteration, the current matrix is disturbed and a new decision is generated as initial solution of gradient descent. After reaching a particular fixed point, the variation of profit is calculated and the best configuration is returned.

Finally, Pareto relation can explicitly handle multi-objectivity for autoscaling in the cloud without the need to specify weights on the objectives [51], [93], [124], [59]. For example, Kateb et al. [51] formulate the decision making process as Pareto-based multi-objective optimisation, which is solved by Multi-objective Genetic Algorithm (MOGA, e.g., NSGA-II). At each generation, MOGA identifies non-dominating solutions. Crowding distance is used to calculate the distance between an individual and its neighbours. In particular, each generation of the search is evaluated using epsilon dominance which is a relaxed form of the commonly used Pareto-dominance metric. In E$^3$-R [124], the decision making problem is formulated as Pareto front, where it is resolved by using MOGA. In addition, the approach applies objective reduction technique with an aim to remove the objectives, which are not significantly conflicted with the others, from the decision making process. In this way, the author aims to reduce the overhead while not affecting the quality of decisions.

In conclusion, we discover that around two thirds of the surveyed search based approaches have considered both vertical and horizontal scaling as the final actions in the cloud. Overall, the nature of search-based optimisation permits certain level of reasoning during the decision making, and this presumably provides better assurance on the quality of the decisions before conducting the actual scaling actions. As we can see, there are small amount of the approaches surveyed belong to implicit search, which can be efficient as there is no need for QoS modelling. Nevertheless, the absence of QoS models also mean that they cannot explicitly handle the trade-offs. On the other hand, the other approaches make use of the explicit search, however, majority of those work formulate the problem as single objective or rely on weighted-sum of objectives, and hence their search of possible trade-offs decision tend to be limited in terms of both optimality and diversity. A limited amount of effort has considered Pareto relation, however, none of them have considered well-compromised trade-off, i.e., the decisions that have balanced improvements on the related objectives. Finally, QoS interference is often ignored in autoscaling decision making, even if it is considered, there is no explicit solution for handling the related trade-offs.

Table 2.1: Summary of The Key Developments for Cloud Autoscaling

| | Knowledge in Architecture | Considered Entity | Objectives | Objectives Dependency | Primitives | Configured Bundles | QoS Interference | Modelled QoS Attribute | Modelling Online | Primitives Section for QoS | QoS Function Training | Granularity of Control | Decision Making | Well-Compromised Trade-offs |
|---|---|---|---|---|---|---|---|---|---|---|---|---|---|---|
| [121] | implicit | application | cost | no | CPU and memory workload | yes | no | workload | yes | n/a | n/a | service level | ILP | no |
| [41] | no | application | response time and cost | weighted-sum | workload and number of VMs | yes | yes | response time | no | manual | LQN | cloud level | exhaustive search | no |
| [15] | implicit | application | QoS attributes and cost | no | number of VMs | yes | no | n/a | n/a | n/a | n/a | cloud level | PD controller | no |
| [60] | no | application | QoS attributes and cost | no | CPU, memory and number of VMs | yes | no | n/a | n/a | n/a | n/a | cloud level | rules | no |
| [46] | no | service | QoS attributes and cost | no | CPU, memory and disk | yes | no | QoS attributes | no | manual | graph | service level | rules | no |
| [136] | implicit | service | cost | no | hardware resources | yes | no | workload | yes | n/a | n/a | service level | heuristic search | no |
| [12] | no | VM | QoS attributes and cost | no | CPU | no | no | n/a | n/a | manual | n/a | VM level | fuzzy controller | no |
| [82] | no | VM | response time | no | CPU | no | no | response time | yes | manual | kalman filter | VM level | MIMO controller | no |
| [106] | no | application | QoS attributes and cost | weighted-sum | CPU, memory and bandwidth | yes | no | QoS attributes | no | manual | manual | cloud level | dynamic programming | no |
| [56] | no | application | QoS attributes | no | CPU, memory and bandwidth | no | no | n/a | n/a | n/a | n/a | cloud level | rules | no |
| [52] | implicit | application | QoS attributes and cost | no | CPU, storage and bandwidth | no | no | QoS attributes | no | manual | manual | cloud level | rules | no |
| [96] | no | application | CPU utilisation | no | CPU, workload and number of VM | no | no | CPU utilisation | no | manual | linear regression | service level | proportional threshold controller | no |
| [131] | no | application | QoS attributes and power | no | CPU, memory and number of VM | yes | no | n/a | n/a | n/a | n/a | cloud-level | rules | no |

| Ref | | | | | | | | | | | | | | |
|---|---|---|---|---|---|---|---|---|---|---|---|---|---|---|
| [23] | implicit | service | QoS attributes and cost | no | CPU, bandwidth, storage and number of VM | no | no | QoS attributes | no | manual | manual | service-level | CBR | no |
| [16] | no | application | QoS attributes and cost | no | CPU, memory, number of VMs | yes | no | n/a | n/a | n/a | n/a | cloud level | PD controller | no |
| [95] | no | applicator | cost | no | CPU, memory, workload and number of VMs | yes | no | response time | no | manual | LQN | cloud level | mixed integer programming | no |
| [126] | implicit | application | QoS attributes and cost | no | thread, CPU and memory | no | no | n/a | n/a | n/a | n/a | VM level | fuzzy control | no |
| [120] | no | applicator | cost | no | CPU, memory and number of VM | yes | no | demand | no | n/a | n/a | service level | ILP | no |
| [54] | no | application | n/a | n/a | CPU, bandwidth and storage | n/a | no | QoS attributes | no | manual | manual | n/a | n/a | no |
| [62] | implicit | service | QoS attributes and cost | weighted-sum | CPU and memory | no | no | workload and QoS attributes | yes | manual | kriging model | service level | exhaustive | no |
| [94] | implicit | application | QoS attributes and cost | weighted-sum | workload and CPU | no | no | response time | no | manual | LQN | cloud level | FNM | no |
| [135] | no | applicator | QoS attributes and cost | no | CPU, thread and memory | no | yes | n/a | n/a | n/a | n/a | PM level | RL | no |
| [104] | no | application | QoS attributes and cost | weighted-sum | CPU and memory | no | no | QoS attributes | yes | manual | ARMA | PM level and cloud level | quadratic and fuzzy control, hill climbing | no |
| [138] | no | applicator | cost | no | VM type | yes | no | response time | no | manual | queuing analysis | cloud level | dynamic programming | no |
| [69] | no | application | QoS attributes and cost | weighted-sum | number of VM | no | no | response time | no | manual | queuing analysis | service level | force-directed search | no |
| [137] | no | applicator | QoS attributes and cost | no | CPU and memory | no | no | workload | no | n/a | n/a | VM level | RL | no |

| Ref | | | | | | | | | | | | | |
|---|---|---|---|---|---|---|---|---|---|---|---|---|---|
| [85] | implicit | application | QoS attributes and cost | no | software configuration and hardware resources | yes | no | QoS attributes | PCA | n/a | service level | exhaustive | no |
| [8] | implicit | applicator | response time, cost and availability | weighted-sum | CPU and memory | no | no | workload and response time | manual | manual | cloud level | local search heuristic | no |
| [19] | no | application | VM consumption | no | workload and number of VM | yes | no | response time | manual | queue analysis | service level | exhaustive | no |
| [125] | no | applicator | VM consumption | no | workload and number of VM | yes | no | response time | manual | queue analysis | service level | exhaustive | no |
| [71] | no | application | response time and cost | no | CPU, memory and bandwidth | no | no | n/a | n/a | n/a | service level | rules | no |
| [142] | no | applicator | response time and cost | no | workload and number of VM | no | no | response time | manual | queue analysis | cloud level | tabu search | no |
| [78] | implicit | application | response time and cost | no | workload and number of VM | yes | no | workload and response time | manual | queue analysis | service level | tabu search | no |
| [139] | implicit | applicator | QoS attributes | weighted-sum | software CP and hardware CP | no | no | QoS attributes | manual | linear regression | VM level | meta-heuristics | no |
| [28] | implicit | service | QoS attributes and cost | no | hardware CP and EP | no | no | QoS attributes | manual | DTMC and MDP | service level | exhaustive | no |
| [127] | no | applicator | n/a | no | CPU and bandwidth | yes | yes | QoS attributes | manual | fuzzy rules and MIMO | n/a | n/a | no |
| [132] | no | application | SLA panelty | no | CPU, memory, workload and number of VM | no | yes | QoS attributes | manual | RT and boosting | VM and cloud-level | grid search | no |
| [26] | no | applicator | QoS attributes and cost | no | CPU, thread, session, buffer and memory | no | no | n/a | n/a | n/a | PM level | RL | no |
| [70] | no | application | QoS attributes | no | software CP | no | no | n/a | n/a | n/a | VM level | neural fuzzy control | no |

| [ref] | | | | | | | | | | | | | | |
|---|---|---|---|---|---|---|---|---|---|---|---|---|---|---|
| [43] | implicit | applicator | cost | no | software CP | no | no | QoS attributes | yes | manual | SVM | VM level | lagrang algorithm | no |
| [63] | implicit | application | response time and cost | no | CPU, memory, workload and number of VM | no | no | response time | yes | manual | LQN and kalman filter | service level | kalman control | no |
| [105] | no | applicator | QoS attributes and cost | weighted-sum | hardware resources | no | no | QoS attributes | yes | manual | ANN | PM level | GA | no |
| [55] | no | application | QoS attributes and cost | weighted-sum | CPU and memory | yes | no | workload and profiling | no | manual | manual | service level | decision tree search | no |
| [107] | no | applicator | QoS attributes and cost | weighted-sum | CPU, memory and disk | no | no | QoS attributes | yes | manual | ARMA | service level | quadratic programming | no |
| [134] | no | application | cost | no | CPU and workload | no | no | response time | yes | manual | ARMA | service level | lagrange algorithm and PI control | no |
| [114] | no | applicator | QoS attributes and cost | weighted-sum | CPU | no | no | n/a | no | n/a | n/a | service level | fuzzy control | no |
| [45] | no | application | QoS attributes | no | CPU, memory, workload and bandwidth | yes | no | QoS attributes | yes | manual | linear regression | service level | exhaustive | no |
| [113] | no | applicator | QoS attributes and cost | no | CPU and memory | no | yes | n/a | n/a | n/a | n/a | PM level | RL | no |
| [141] | no | application | QoS attributes and cost | weighted-sum | software CP, CPU and memory | no | no | QoS attributes | yes | manual | ARMAX and SVM | service level | PID, RL control and exhaustive search | no |
| [91] | no | applicator | QoS attributes and cost | weighted-sum | CPU and memory | no | no | QoS attributes | yes | manual | fuzzy regression | PM level | quadratic programming | no |
| [20] | no | application | QoS attributes | no | workload and number of PM | no | no | QoS attributes and workload | yes | manual | smoothing splines | service level | n/a | no |
| [86] | no | applicator | n/a | no | hardware CP and BP | no | yes | QoS attributes | no | PCA | linear regression | n/a | n/a | no |
| [90] | no | application | n/a | no | CPU, memory and bandwidth | no | no | QoS attributes | yes | manual | ANN and SVM | n/a | n/a | no |

| Ref | | | | | | | | | | | | | |
|---|---|---|---|---|---|---|---|---|---|---|---|---|---|
| [89] | no | application | n/a | no | CPU, memory and bandwidth | no | QoS attributes | no | manual | ANN | n/a | n/a | no |
| [87] | no | application | n/a | no | hardware CP and EP | no | QoS attributes | yes | manual | ANN | n/a | n/a | no |
| [140] | no | service | n/a | no | hardware CP and EP | no | QoS attributes and demand | yes | manual | LQN, AR and kalman filter | n/a | n/a | no |
| [65] | no | service | n/a | no | CPU and workload | no | QoS attributes and demand | yes | manual | LQN, K-mean and kalman filter | n/a | n/a | no |
| [87] | no | applicator | n/a | no | hardware CP and EP | no | QoS attributes | yes | manual | ANN and TS-ANN | n/a | n/a | no |
| [88] | no | application | n/a | no | software and hardware CP | no | QoS attributes | yes | manual | ARMA and MIMO | n/a | n/a | no |
| [133] | no | applicator | n/a | no | software CP, hardware CP and EP | no | QoS attributes | yes | correlation coefficient and wrapper | linear regression | n/a | n/a | no |
| [116] | no | application | n/a | no | hardware CP | no | QoS attributes | yes | manual | linear regression and mIMO | n/a | n/a | no |
| [53] | no | applicator | n/a | no | CPU, memory, storage and bandwidth | yes | QoS attributes | no | manual | manual | n/a | n/a | no |
| [51] | no | service | QoS attributes and cost | pareto | number of VM | no | QoS attributes | yes | manual | model@runtime | service-level | MOGA | no |
| [81] | no | service | response time and cost | weighted-sum | number of VM | no | response time and workload | no | manual | manual | service-level | linear search | no |
| [10] | no | application | QoS attributes and cost | no | CPU | no | n/a | no | n/a | n/a | service-level | fuzzy control | no |
| [93] | no | applicator | response time and cost | pareto | thread and CPU | no | response time | no | manual | LQN | service-level | SMS-EMOA | no |
| [77] | no | application | response time, throughput, CPU utilisation and cost | weighted-sum | CPU and memory | no | response time and throughput | no | manual | manual | cloud-level | GA | no |

|  |  |  |  |  |  |  |  |  |  |  |  |  |  |  |
|---|---|---|---|---|---|---|---|---|---|---|---|---|---|---|
| [124] | no | service | QoS attributes and cost | pareto | CPU and memory | no | no | response time | no | manual | queue analysis | service-level | MOGA and objective reduction | no |
| [59] | no | service | cost | pareto | number of VMs | no | no | QoS attributes | no | manual | simulation | service-level | MOGA | no |
| [17] | implicit | applicator |  | no |  | yes | no | n/a | n/a | n/a | n/a | service level | ACO | no |
| [102] | implicit | application | QoS attributes, utilisation, number of actions | no | CPU, memory and bandwidth | no | no | QoS attributes | no | manual | manual | cloud level | rules | no |
| [30] | implicit | applicator | QoS attributes | no | thread, CPU and memory | no | no | n/a | no | n/a | n/a | service level | rules | no |
| [99] | implicit | application | utilisation | no | software CP | no | yes | n/a | no | n/a | n/a | service level | heuristics and decision tree | no |
| [75] | implicit | service | QoS attributes, utilisation and cost | no | CPU | no | no | QoS attributes | no | manual | PCM | service level | exhaustive | no |
| Ours | explicit | service | QoS attributes and cost | pareto, nash and distance | CP and EP | yes | yes | QoS attributes and cost | yes | hybrid learners | multiple learners | dynamic | MOACO | yes |

## 2.6   Positioning This Thesis

The number of reviewed papers is a result of an investigation over the key conferences and journals for cloud computing, service computing and self-adaptive systems, such as, SEAMS, IEEE CLOUD, UCC, TSC and TCC etc. They were then carefully selected according to their recentness, relevance, quality and completeness of evaluations, with respect to our research questions for cloud autoscaling. However, the resulting list of papers is not exhaustive, but our investigation and search have tried to improve its conclusiveness as much as possible.

We further extract the surveyed work by removing similar ones from the same research group, this results in 74 papers. Table 2.1 summaries the 74 papers and they are compared using various key criteria of the autoscaling process. The criteria used to compare different approaches are derived from the key aspects that can affect the designing of autoscaling system, the formalization of the problem, and the quality of autoscaling process. Their inclusion is intended to cover the common decisions that need to be made for research in the field of cloud autoscaling, in this way, a general and normalized comparison is made possible. It is worth noting that all the criteria serve as the raw data for producing the taxonomy. That is to say, the classification in this chapter is largely derived from the results presented in Table 2.1. In Table 2.2, we collectively discuss how this thesis is different from this work. Notice that not all the work covers every criteria, e.g., the work on QoS modelling might not cover the aspect of decision making.

Table 2.2: Comparison of the Thesis with Other State-of-the-Art Researches

| Criteria | Comparison |
|---|---|
| Knowledge in Architecture | As we can see from the *Knowledge in Architecture* column, from an architecture perspective, there is a considerable amount of the existing work (i.e., 32 out of 74) that do not intend to discuss the required levels of knowledge and what benefit such knowledge can bring for adaptations. The remaining work, on the other hand, only implicitly discuss the knowledge required in a rather coarse granularity (e.g., system model is needed for optimisation). The absence of explicit consideration for the fine-grained representation of the knowledge in the architecture can results in, e.g., improper inclusion of unnecessary knowledge and/or missing important knowledge that can improve adaptation quality when developing autoscaling systems. Consequently, this fact can mislead the design and application of the underlying algorithms and techniques. In contrast to the other work, this thesis, as we will see in Chapter 3, proposes an autoscaling architecture with fine-grained representation of the required knowledge, which are mapped to the principles of self-awareness [18]. |
| Considered Entity | The *Considered Entity* column describes which level of abstraction that the work intends to model and scale. Only 12 out of 74 work have considered cloud-based service, which is the finest level of abstraction to be scaled in the cloud. The majority of the work (i.e., 60 out of 74) aim at cloud-based application (or application tier) and the rest focus on other levels. This thesis has considered the level of cloud-based service, because it offers better flexibility and the finest modelling granularity in the cloud, as we will discuss in Chapter 4. |
| Objectives | As we can see from the *Objectives* column, certain amount of work (i.e., 18 out of 74) consider only specific objectives (e.g., response time and cost) in autoscaling, which will limit their applicability. In contrast, this thesis aims for any given QoS attributes and cost as the objectives, providing the flexibility of the proposed QoS modelling approach, which will be discussed in Chapter 4 and 5. As we can see, there is also another amount of the related work that can handle arbitrary QoS attributes, i.e., 37 out of 74 work surveyed. However, their underlying algorithms and techniques are different from this thesis. |
| Objectives Dependency | As shown in the *Objectives Dependency* column, most work (i.e., 54 out of 74) do not explicitly consider multiple objectives or they do not intend to handle the objective dependency exhibited by the multi-objectivity. Therefore, they can not produce good trade-off decisions. For the work that does consider multiple, and possibly conflicted objectives, the objective dependency is usually formulated as a weighted-sum relation (i.e., 16 out of 74). Only 4 out of 74 work have used Pareto relation to model multi-objectivity with the aim to provide fine-grained information about the trade-offs. However, these approaches have not considered decisions that achieve well-compromised trade-offs, meaning that the trade-offs can be imbalanced. Unlike that work, this thesis not only models multi-objectivity as Pareto relation, but also intend to search for decisions that achieve well-compromised trade-offs using nash dominance and distance of decisions. |
| Primitives | The *Primitives* column shows what are the cloud primitives that a work has considered. Most work (i.e., 60 out of 74) have only considered specific dimensions (e.g, CPU and memory) of the primitives in the cloud, which will limit their applicability. In contrast, this thesis and the remaning small amount of related work focuses on any given primitives. In addition, this thesis also additionally considers software control primitives and their interplay with the hardware resources, which are the important, but often ignored cause of the fluctuation in QoS. As we can see, there are only 14 out of 74 approaches in the work surveyed have considered software control primitives. However, this thesis applies different algorithms and techniques which offers various benefits, as we will see in the following chapters. |
| Use Bundles | As in the *Use Bundles* column, modern cloud providers (e.g., Amazon) enable autoscaling based on bundles, which is a collection of some fixed configurations (usually at coarse granularity). The benefit is that by limiting the possible autoscaling decision using bundles, the decision making process can be greatly simplified, in which case a simple exhaustive search would be also quite effective. Therefore, 20 out of 74 work has assumed fixed bundles in their autoscaling approaches. However, these fixed bundles cannot meet the increasingly complex demand of cloud-based services, e.g., a cloud-based service may require high CPU but low memory. Consequently, this fact will negatively affect the quality of autoscaling and elasticity. There is a large amount of related work (i.e., 54 out of 74) that does not assume fixed bundles, which are the same as this thesis. However, this thesis applies different algorithms and techniques in the decision making process which offers various benefits, as we will see the the following chapters. |
| QoS Interference | As mentioned previously, QoS interference is an important factor to elasticity in the cloud. Inadequate handling of QoS interference can result in low quality of autoscaling and elasticity. However, as shown in the *QoS Interference* column, only 8 out of 74 work on cloud autoscaling have explicitly modelled and/or handled QoS interference in the cloud autoscaling, but they either ignore or tend to be limited in handling the trade-offs caused by QoS interference. This thesis is different from the related work in the sense that it does not only explicitly model QoS interference, but also effectively resolve the trade-offs between cloud-based services that caused by QoS interference. In addition, we collectively consider QoS interference caused by both the co-located service and co-hosted VM, which has not been well studied in existing work. We will discuss this in the following chapter with greater details. |
| Modelled QoS Attribute | The *Modelled QoS Attribute* column indicates what are the quality attributes that the autoscaling approach intends to model, excluding the cost. Modelling in autoscaling can greatly improve the decision making process. 21 out of 74 work do not explicitly handle and model QoS while 19 work model certain dimensions of QoS only (e.g., response time). Similar to the rest of related work, this thesis models any given QoS attributes. However, as we will see in Chapter 4 and 5, the underlying QoS modelling approach is different. As mentioned previously, apart from modelling QoS, autoscaling approaches can also rely on workload or demand modelling. However, these models cannot be used to reason about the effects of decisions on the objectives. We leave the study about whether and how the workload or demand models can be used to consolidate the QoS model as one of our future work. |
| Modelling Status | The *Modelling Status* column indicates whether the approach can be applied for online modelling. As we can see amongst the approaches that handle QoS modelling, 34 out of of 53 work only focus on offline while the rest can be applied online. The benefit of offline modelling is that there is no need to concern with the modelling overhead, however it is unable to cope with emergent events, e.g., spike workload. On the other hand, online modelling can be used to deal with unexpected scenarios, but its overhead can be a critical issue. This thesis focus on online modelling, however, when offline modelling is beneficial for online scenarios, the proposed approach can be also used offline. |
| Primitives Section for QoS | *Primitives Section for QoS* shows the algorithms and techniques used to select important inputs for QoS models. As we can see, amongst the approaches that model QoS, nearly all (i.e., 50 out of 53) of the related work have relied on manual and offline approach for the selection. However, manual analysis may ignore important features or incorrectly select irrelevant features, which would eventually downgrade the model accuracy. There are only three of the work have considered dynamic primitives selection; however, they focus on the relevance of the inputs while ignoring the redundancy of the inputs which have already selected. This fact, as we will show in Chapter 5, can also negatively affect the model accuracy. This thesis is separated form existing work in the senses that the primitive selection is performed dynamically, and with the consideration of balanced information relevance and redundancy. |
| QoS Function Training | *QoS Function Training* shows the algorithms and techniques used to model the correlation between selected primitives and the QoS. There are work relies on manual and analytical approaches; have used simulation and the rest are machine learning based. However, all the approaches are relied on single algorithms, i.e., only one algorithm is used offline or online. In Chapter 5, we will show that different algorithms perform quite differently under given scenarios, therefore, such a fact means that the selecting the most appropriate single algorithms for QoS modelling in the cloud is mere difficult, if not possible to achieve. In addition, even it can be selected, there is no guarantee that it can be always the best at runtime. Although in the case of modelling workload and demand, we have discovered 2 work that considered multiple learning algorithms, there is still no instance for QoS modelling. As we will see in Chapter 5, this thesis relied on multiple algorithms where the optimal one would be used for modelling and prediction. It is also different from the ones for demand and workload modelling in the sense that it focus on selecting the best algorithm on-the-fly; while [79] aim at ensemble solution at runtime and [72] is an offline approach. |
| Granularity of Control | *Granularity of Control* shows at which granularity level the autoscaling decision would be made. As we can see, 2 out of 74 work have use multiple levels of granularity while the rest work have assumed a single granularity in autoscaling. However, they are static and fixed on the granularity of control. This means that, once the effects of control granularity on the global benefits changes (e..g, due to QoS model change), they would not be able to optimise for the global benefits; or they would have been in a unnecessarily high level of control (e.g., cloud level) that benefit nothing but generating extra overhead. Unlike any of the existing work, this thesis work on dynamic granularity of control, in which the control level can be changed at runtime subject to QoS sensitivity and deployment. We will discuss this in greater detail in Chapter 6. |

| Decision Making | *Decision Making* shows the algorithms and techniques used to perform trade-off decision making for autoscaling in the cloud. We can see that 8 out of 74 work are rule-based approaches and hence they tend to be limited in handling dynamics, uncertainty and trade-offs in cloud. 14 out of 74 work are based on control theory, which can not explicitly handle trade-off decision making. Among the reset search based optimisation for autoscaling in cloud, 9 take implicit approach, e.g., model free RL, which do not guided by explicit QoS model and hence they tend to be limited in reasoning about the trade-offs. 12 of the work assume single objective and apply algorithms that only work on single objective optimisation. A large amount of the work (i.e., 16 out of 74) aggregate multi-objective into a weighted-sum formulation, and resolve the decision making as single objective optimisation. However, it is generally difficult to correctly specify the weights, in addition, weighted-sum formulation can restrict the quality of trade-offs. There are only 4 work have considered pareto relation of the multi-objectivity, and they rely on MOGA, mostly NSGA-II for the optimisation. Different from all the existing work, this thesis formulates the trade-offs decision making problem in autoscaling using pareto relation where the multi-objective optimisation is resolve by Multi Objective Ant Colony Optimisation (MOACO). This is because the commonly applied MOGA, such as NSGA-II, needs pareto-dominance to evaluate the overall quality of decisions for all objectives as the algorithm runs, hence it can be restricted by an inevitably large number of non-pareto-dominated decisions when the number of objectives increases. Such fact has been shown to be limited in optimising and making trade-off for more than 4 objectives [33] while our problem needs to handle larger number as we consider the trade-offs caused by QoS interference. We will discuss this in Chapter 7 with greater details. |
|---|---|
| Well-Compromised Trade-offs | As we can see from *Well-Compromised Trade-offs*, none of the existing work have considered well-compromised trade-offs, which means the achieved decisions may lead to imbalanced improvements on the objectives. In contrast, this thesis explicitly aim to search for decisions that achieve well-compromised trade-offs, as we will show in Chapter 7. |

According to the above table, It may seem that the proposed autoscaling framework introduce various positive points as when compared to existing work. However, this does not come as free. In the following, we discuss some potential negative aspects of our framework:

- *Scalability:* Given the fact that our approach aims to handle dynamics and uncer-

tainty at runtime, making the autoscaling process self-aware is likely to produce extra computational effort, which in turn imposes scalability issue. However, as we will discuss in Chapter 7, our methods and approach has been designed in the way that takes scalability into account. For example, the whole idea of dynamically determining the granularity of control in cloud is aimed to improve scalability. This is because there may be some period of time that the autoscaling system operates at fine-grained control can still yield the optimal benefits, and such control would also produce less overhead, as the number of objectives in a decision making process is reduced. Dynamically switch the system to fine-grained control would potential, in a long term, improve the overall scalability. In Chapter 7, we will discuss scalability of our framework in greater details.

- *Complexity:* The self-awareness in our framework is fundamentally grounded on advance computational intelligence techniques, which may need to be tuned by adjusting their parameters. To achieve such, it does requires some knowledge about the underlying algorithms and some profiling techniques. In Chapter 7, we will discuss the complexity of applying our framework in greater details.

- *Integration with existing autoscaling system:* There may be scenarios where it is difficult to apply the entire autoscaling framework, e.g., when there is a legacy autoscaling system and it is too expensive to replace it. To mitigate this issue, we have designed each component in a way that they can be seamlessly attached to existing modules, in order to achieve certain tasks (e.g., QoS modeling). In Chapter 7, we will discuss practical deployment of our framework in greater details.

## 2.7　Conclusion

In this chapter, we described the background and definitions for cloud autoscaling, self-aware and self-adaptive systems in general. Subsequently, we outlined the major requirements for the key logical aspects of autoscaling in the cloud, and discuss the key state-of-the-art developments proposed for each of the logical aspects. Furthermore, we explicitly discussed the differences of this thesis to the related work and identified the gap in this area of research.

In the next chapter, we propose a self-aware architecture for autoscaling in the cloud, which is enabled and driven by mapping the necessary components to different levels of self-awareness capabilities.

# Chapter 3

# An Architecture for Self-Aware and Self-Adaptive Autoscaling in Cloud

## 3.1 Introduction

Engineering and design autoscaling system in the cloud is a complex process due to the dynamic and uncertain nature of cloud environment. As discussed in Chapter 2, most of the prior work e.g., [141] [23], tend to be limited in one or more logical aspects of the autoscaling process. This can be attribute to the fact that their approaches, especially their architectures, lack to capture the necessary knowledge for optimising the QoS and cost objectives of all cloud-based services. These knowledge can be well-represented and acquired through self-awareness at runtime, and thus render self-awareness as a neat solution to overcome the limitation in existing work.

The autoscaling architecture, being the skeleton of an autoscaling system, is a fundamental element that blueprints the necessary components and their interactions. We argue that incorporating the principles of self-awareness at the architecture level is likely to advance the way we engineer self-aware and self-adaptive autoscaling system.

### 3.1.1 Motivation

Given the definition of autoscaling presented in Chapter 1 and our discussion in Chapter 2, it is clear that autoscaling systems are essentially self-adaptive systems. One general task for architecting self-adaptive systems is to determine what is the necessary knowledge that the system maintains, and how the knowledge can be acquired. Designing autoscaling system in the cloud is not an exception. As mentioned in Chapter 2, the inadequacy of necessary knowledge can resulted in various limitations of the system. This includes, for examples, the absence of accurate QoS models causing the system not be able to reason about the likely effects of adaptation and the possible trade-offs; or ignoring QoS interference can degrade the effectiveness of the autoscaling. The difficulty lies in how to classify the necessary knowledge of an autoscaling system and more importantly, how to acquire such knowledge and how they can improve the adaptation. As we mentioned in Chapter 1, given the heterogeneous, elastic and on-demand nature of cloud, autoscaling in the cloud exhibits high dynamics and uncertainty related to QoS modelling, granularity of control and the trade-off decision making. These unique problems are within the scope of self-awareness principle and its capabilities [18], and hence rending it as a promising solution for autoscaling in the cloud. However, the challenge now becomes how to incorporate and map the principles of self-awareness to autoscaling in the cloud, how the architecture can be enriched and how we can carefully justified benefits of self-awareness capabilities.

As we have extensively surveyed in Chapter 2, existing autoscaling architectures have heavily relied on traditional styles (e.g., MAPE [84] and OAD [74] based) for self-adaptive systems, which are focus mainly on the interactions among different logical aspects of a system and their degree of centralisation. However, these architecture styles lack in cap-turing different levels of knowledge in a fine-grained representation. In particular, the widely adopted MAPE style in autoscaling architecture only model the knowledge in a coarse grained manner. Therefore, the MAPE is only capable to implicitly consider the self-awareness of the knowledge that the system requires. From an architecture perspective, this is not immediately intuitive to deduce which concerns the system addresses in the adaptation. OAD is another style that claimed to be self-aware, however, instead of explicitly model different levels of knowledge, it emphases on a separation or decoupling between the representation of the knowledge and the decision making process. Autoscaling architecture based on feedback loop control that embedded with computational intelligence or control theory techniques are also exist, e.g., [141]. Nevertheless, they either do not consider self-awareness and how it can be used to strength the adaptation or such consideration is implicit. In addition, they are problem specific and do not model the generic concern of knowledge, e.g., with respect to goal, time and interaction. The absence of fine-grained representation of knowledge in the architecture can mislead the design and application of the underlying computational intelligence and/or economic driven techniques. As a result, all those limitations call for novel autoscaling architecture, which should encapsulate fine-grained information about what levels of knowledge are required; how they can be acquired and how they can be beneficial for the adaptation.

### 3.1.2 Contributions

In this chapter, we present an autoscaling architecture leverages on the principles of self-awareness and its mapping to various self-awareness capabilities. By doing so, we obtain an enriched architecture with detailed information about what are the necessary levels of knowledge and their interplay. These levels of knowledge and their interplay promotes better self-adaptivity through *bi-directional* adaptation in the architecture, in which the autoscaling process is not only able to adapt the underlying cloud-based services and VMs, but also capable to further consolidate itself by acquiring the knowledge about itself and the environment. It is worth noting that we only focus on high level view of the architecture in this chapter, the details of each component in the architecture and the related algorithms will be explained in the subsequent chapters.

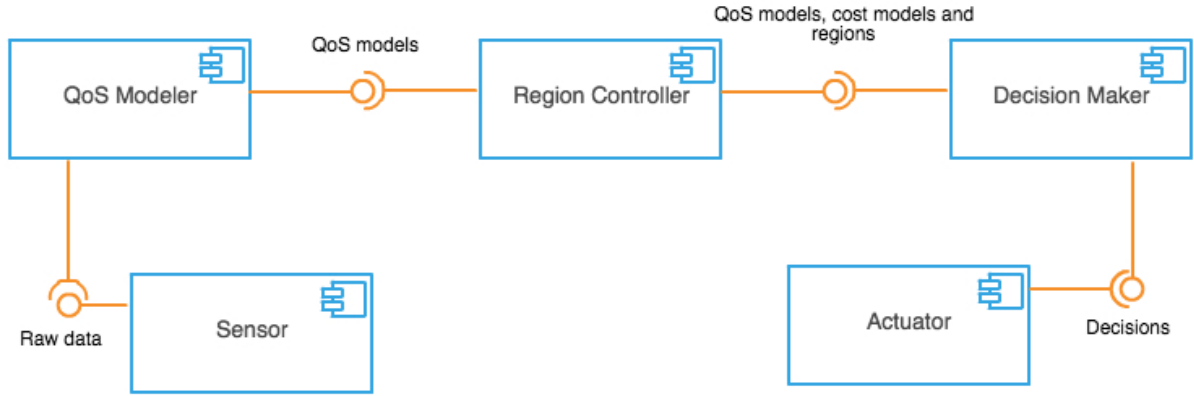

Figure 3.1: The Components of Autoscaling Architecture.

## 3.2 Autoscaling Architecture

As concluded by a recent survey [98], the most common components in the simplest autoscaling system includes a monitor and a scaling unit: the former gathers the service's or application's current state while the later utilise the information to decide an action. However, such approach lacks to capture the necessary levels of knowledge and thus the system can be limited in its adaptation behaviours. To overcome such issue, we propose an autoscaling architecture that offers a unified solution to handle runtime dynamics, uncertainty and trade-offs related to different QoS and cost objectives. As the component diagram depicted in Figure 3.1, the main logical aspects in autoscaling are explicitly modularised into five components, which are described below:

- **Sensor:** The Sensor collects raw data from the underlying service-instances, platform and infrastructure through the corresponding interfaces or logs as offered by the cloud provider. This data includes the currently achieved QoS, environmental primitives, software and hardware control primitives, as well as the agreed constraints in SLA and budgets. In particular, the sensors should sense the data for those service-instances that are likely to have objective-dependency.

- **QoS Modeller:** All historical data from data sensors is analysed in this component using information theory and machine learning algorithms. Its goal is to model the correlation between QoS and the underlying cloud primitives. For each QoS attribute of a service-instance, the *QoS Modeller* filters unnecessary inputs for the QoS and dynamically update their magnitude in the correlation; it is also QoS interference aware by including the primitives of co-located services and co-hosted VMs into the model. Notably, the *QoS Modeller* runs periodically for collecting data and modelling the QoS, which means it is capable for dynamically and continually improving the models for handling runtime uncertainty. The resulted models are particularly useful for reasoning about the effects of autoscaling decisions and their possible trade-offs. Details of the proposed QoS modelling approach are explained in Chapter 4 and 5.

- **Region Controller:** This is the component that rarely exist in current work, as we have surveyed in chapter 2. Such intermediate component between *QoS Modeller* and *Decision Maker* explores full information about QoS sensitivity expressed in the QoS models, and thus enable better global results with reduced overhead. Specifically in *Region Controller*, the most up-to-date QoS models and the cost models are dynamically clustered into regions, in each of which the objectives are dependent (i.e., harmonic or conflicted). This is achieved by examining whether the models have the common inputs that are parts of the autoscaling decision. If these common inputs exist, it means that the objectives are dependent and can be affected differ-ently by the same decision; hence, they need to be considered in conjunction with each other in the decisions making process. On the other hand, the objectives, which are independent, are omitted from the same decision making process as they can benefit nothing but generate overhead. The *Region Controller* can help to dynamically determine the right granularity of control and thus, the autoscaling system can better globally optimise all cloud-based service while resulted in reduced overhead. Details of the proposed region clustering approach are specified in Chapter 6.

- **Decision Maker:** Given the QoS models and the regions, the Decision Maker component is responsible to search for an autoscaling decision that optimises the QoS and cost objectives of the cloud-based services. By leveraging on search-based optimisation, the decision making process can perform exploring and reasoning in a finite, but possibly very large search space; while still produces the optimal (or near-optimal) decision. Such process can also adaptively handle trade-offs on objectives, even in the absence of preference. Details of the search-based decision making process are specified in Chapter 7.

- **Actuator:** Once the final autoscaling decision has been produced, we consider both vertical scaling and horizontal scaling in the *Actuator* component. In our system, vertical scaling always takes higher priority, providing that modern hypervisors (e.g., Xen [6]) can achieve dynamic vertical scaling with negligible overheads. The resources on a PM are provisioned to the VMs in a first-come-first-serve basis. The horizontal scaling, on the other hand, is only triggered when the resources of the PM tends to be exhausted, i.e., when the total upper bounds of all co-hosted VMs for a resource type exceeds the PM's capacity, the last service-instance that requires to increase the upper bound would be migrated/replicated. Likewise, a VM is removed when its provisions and utilisations for all resource types are below thresholds.

This architecture summarises the key components and features for autoscaling system. To better describe the marriage between self-awareness and autoscaling, we have proposed a set of self-aware patterns, which serve as a general guideline for architecting self-aware systems.

## 3.3 Self-Aware Patterns

When faced with the task of designing self-aware computing system, researchers and practitioners need a set of guidelines on how to use the concepts and principle of self-awareness [18]. To this end, we have documented different categories of self-awareness capabilities using 8 architectural patterns, which serve as the guidelines on how to design self-aware computing systems in a principled way. The aim of those patterns is to ensure that, when designing self-aware systems, only relevant types of knowledge are included, and their inclusion justified by identified benefits. To better illustrate the self-aware pattern and its origin, we briefly specify one exampled pattern in the following:

The pattern shown in Figure 3.2, is suitable for situations where goals, time and interactions between machines and processes need to be captured. Here, goal awareness enables the representation of changing runtime goals, so a node can share its knowledge with other goal-aware nodes and the system can adapt to the changed goals. Time awareness allows the representation of temporal knowledge about goal and interaction awareness, enabling capabilities such as forecasting. This pattern also adds meta-self-awareness, enabling the system to manage the trade-offs associated with exercising various self-awareness levels and thereby allowing it to modify goals at runtime. An example of this runtime meta-reasoning is the dynamic selection of the most appropriate learning algorithm for a particular context.

Please refer to Appendix A and our handbook [39] for more detailed specifications of the patterns.

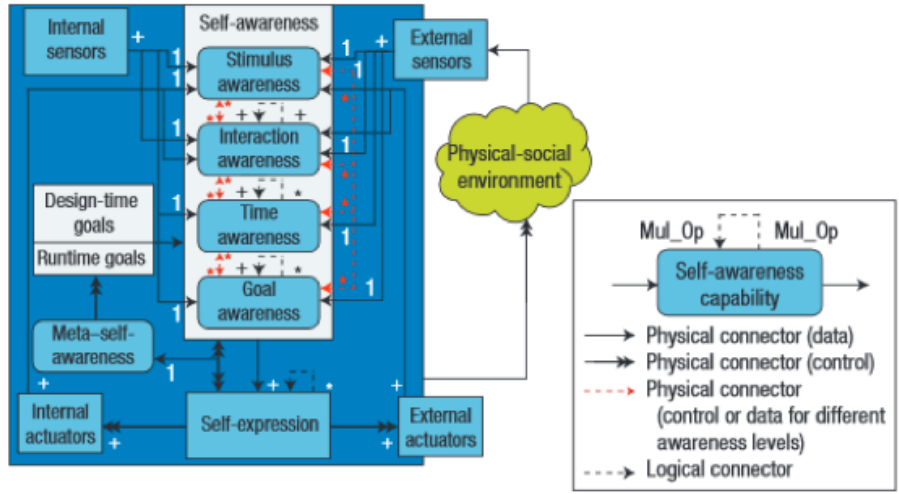

Figure 3.2: The Goal Sharing with Time-Awareness Capability Pattern. There are three types of multiplicity operators (Mul_Op): * expresses that the number of capability of the same type in the interaction is restricted to at least one. 1 indicates that one and only one capability of the same type is permitted. + indicates that zero, one or many of the type specified is permitted in the interaction.

## 3.4 Mapping Between Self-Aware Pattern and Autoscaling Architecture

To apply self-awareness at the architecture level, it is essential to systematically map a selected self-aware pattern to the architecture components of the given problem domain, which in this thesis is autoscaling in the cloud.

Recall that in general term, the 'self' (or node) can be any conceptual part of the system being considered, e.g., process, component or machine. The cloud autoscaling system is typically composed of numeric physical machines, each of which running services benefiting from virtualization. In this context, 'self' is simply the autoscaling process on a physical machine. Self-awareness for an autoscaling process is to aware of the knowledge about its own (or the others') possible impacts on the QoS models of the managed services, the granularity of control and the quality of trade-offs decision for a given runtime scenario. Based on the knowledge acquired via self-awareness, we promote bidirectional adaptations in an autoscaling process. In one direction, adaptation can be concerned with adapting the configurations for its services and virtual machines. On the other direction, adaptation can consolidate its own autoscaling capabilities. This is concerned with building more accurate QoS models, identifying better granularity of control and making better trade-offs decisions. The bidirectional adaptations, through the principles of self-awareness, aim at improving the effectiveness and self-adaptivity of autoscaling.

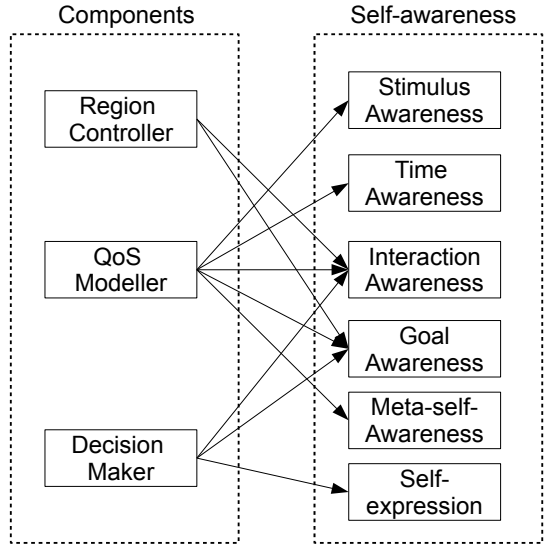

Figure 3.3: The Mapping Between Autoscaling Components and The Self-awareness Capabilities.

With that context in mind, the autoscaling architecture proposed in Section 3.2 is then systemically mapped to the self-awareness capabilities (Figure 3.3) and to a concrete instance of a carefully selected self-aware pattern (Figure 3.4 left). To select the self-aware pattern, we have followed the systematic mapping guideline, which is a result of our Work Package from the EPiCS project [18]. This systematic mapping guideline, which by itself is standalone contribution, has been thoroughly documented in our handbook [39]; and it has been reviewed and evaluated by different partners of the EPiCS project [18]. Additionally, the application of such mapping guideline in the context of cloud

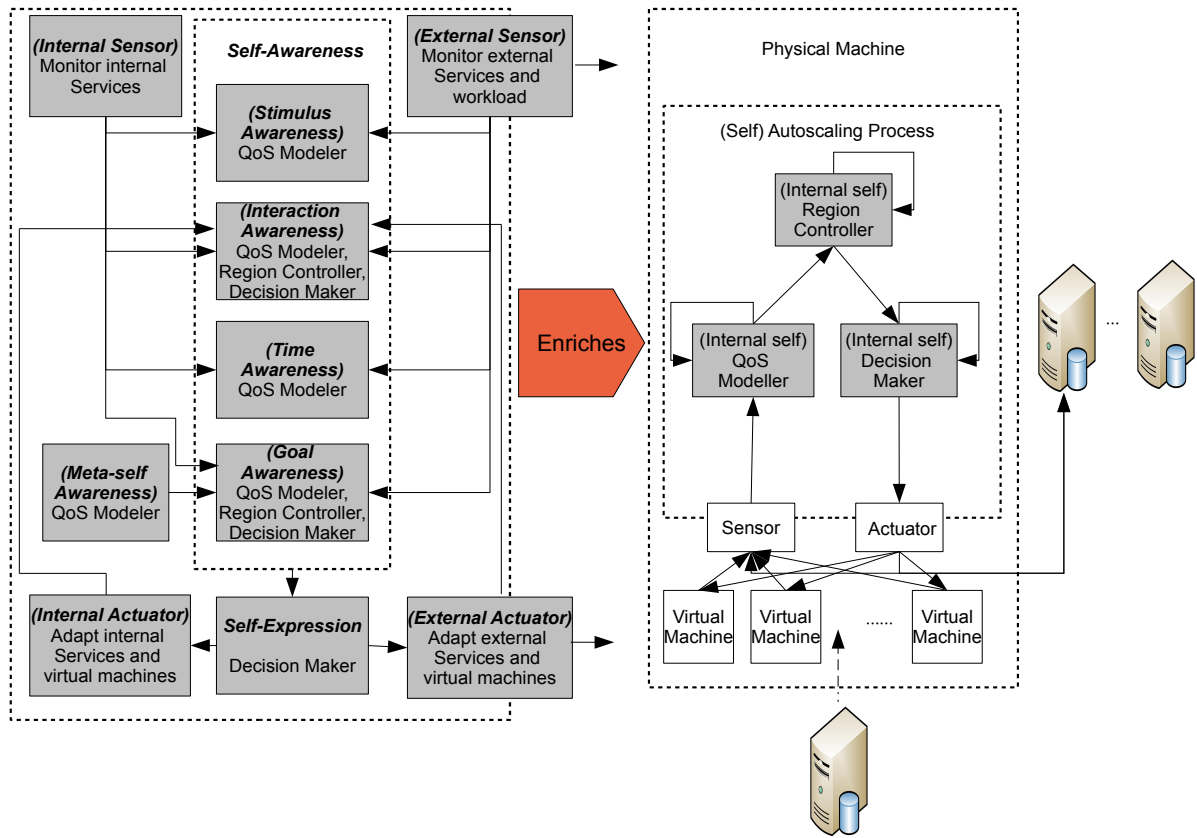

Figure 3.4: Enriching the Self-aware Autoscaling Architecture.

autoscaling has been illustrated as a case study in the handbook.

As we can see from Figure 3.4, the mapping process has identified *Goal Sharing with Time-Awareness Capability Pattern* as a suitable pattern for addressing the challenges of autoscaling. The arrows in Figure 3.4 represent either data or control flow . As we can see from the left figure, different levels of awareness require control or data flow to achieve collaborations. This has provided guidance on how to correctly incorporate self-awareness with autoscaling architecture.The autoscaling architecture is then enriched with self-awareness capabilities, shown as the deployment diagram in Figure 3.4 right. Given that it can be hard to directly achieve self-awareness for a complex autoscaling process, we have modularised the self-awareness capabilities into different internal selves as encapsulated by three components. Consider the autoscaling system of the cloud, which is composed of self-aware autoscaling processes on different physical machines. Here, each autoscaling process would have different internal selves: these are the processes in *QoS Modeller*, the *Region Controller* and the *Decision Maker* component; in this way, we hope to make these processes more intelligent using self-awareness. The autoscaling system is realised as a decentralised instances, where an instance coordinates the selves to realise self-awareness in autoscaling. The components in our autoscaling system, except *QoS Modeller*, are triggered when the system detects violations of the requirements, i.e., violations of SLA and utilisation constraints in case of over-provision. In particular, a requirement is said to be violated only if such violation has been observed for more than $n$ sampling intervals, where $n$ is a variable that controls the trade-offs between stability and adaptivity of our system. The sensors on a PM does not only sense data, but also the QoS models from other PMs. This is because a cloud-based service can, in some cases, be functionally dependent on services running on the other PMs, thus creating the chances for objective-dependency.

In the following sections, we further discuss how the self-awareness capabilities, which have been explained in Chapter 2, are mapped in the context of different internal selves.

### 3.4.1 Mapping Self-Awareness in QoS Modeller

The QoS modelling approach is enriched by self-awareness. At this level, 'self' refers to the QoS modelling process in autoscaling. Self-awareness is concerned with knowing how the QoS modelling can be affected by:

- Features (e.g. workload) of QoS attributes and their changes - **Stimulus-awareness**.

- Possible QoS interference and contention - **Interaction-awareness**.

- Historical modelling errors and data trends - **Time-awareness**.

- Changes in utility functions of QoS - **Goal-awareness**.

- Suitability of learning algorithms - **Meta-self-awareness**.

These levels of knowledge promotes the awareness of QoS sensitivity, i.e., with respect to *which*, *when* and *how* cloud primitives correlate with the QoS. As the knowledge changes, the QoS modelling process can benefit from self-awareness to dynamically self-adapt the selected features and expressions of its QoS models at runtime.

By leveraging on the advances of machine learning algorithms, self-awareness not only able to better handle dynamics and uncertainty, but also eliminates the need for heavy human analysis and prior design time knowledge in the QoS modelling. However, because there is no single learning algorithm which can outperform the others constantly across a range of scenarios, selecting the right algorithm is a challenging task for developers. Self-awareness can even address this issue by using the notion of meta-self-awareness, which performs reasoning at the meta-level, i.e., the modelling process is aware of what is the best algorithm to enable other self-awareness capabilities. In such way, self-awareness help to achieve more accurate and more effective QoS modelling.

### 3.4.2 Mapping Self-Awareness in Region Controller

To improve the global benefit with respect to QoS and cost objectives of all cloud-based services, we dynamically identify the right granularity of control building on self-awareness capabilities [18]. By leveraging on the QoS and cost models, self-awareness helps to adaptively cluster objectives into different regions according to objective dependency ,i.e., we put objectives in one region as long as they have common cloud primitives, which are parts of the final autoscaling decision, in their models. The regions can be different in size and this promotes dynamic granularity of control. We then consider the objectives from each regions in separated decision making processes. A self-aware and self-adaptive process is realised in the *Region Controller*, which maintains different regions. At this level, 'self' refers to the region controlling process in the autoscaling. Self-awareness here is concerned with knowing how such process can be affected by:

- QoS and cost models - **Goal-awareness**.

- objective dependency, i.e., conflicted or harmonic goals - **Interaction-awareness**.

These levels of knowledge promotes the awareness of the effects of control granularity on the global benefit. Henceforth, the region controlling process can better self-adapt its regions and their content to the dynamic changes in knowledge. As such, self-awareness improves global benefits while reducing overhead.

### 3.4.3   Mapping Self-Awareness in Decision Maker

Once the QoS models and regions are defined, we are faced with trade-offs when autoscaling in the cloud. In the *Decision Maker*, we take a multi-objective representation to model the trade-offs of QoS and cost objectives for different cloud-based services, and the problem is resolved by a self-aware and self-adaptive process. At this level, 'self' refers to the decision making process in the autoscaling. Self-awareness here is concerned with knowing how the decision making can be affected by :

- QoS and cost models, as well as their requirements - **Goal-awareness**.

- regions of objectives, which represent the positive, negative or zero interaction between objectives during decision making - **Interaction-awareness**.

Obtaining these levels of knowledge mean that the decision making process is capable to self-adapts its behaviour in the search for better trade-offs decisions. Specifically, self-awareness assists the decision making process in extensively reasoning about the effects of autoscaling decisions on goals and the possible trade-offs. By leveraging on modern search based algorithms, the reasoning serves as a strong assurance about the quality of autoscaling, especially when the search space is incredibly large (i.e., too many combinations of software configurations and hardware resource provisions), which cannot be handled by human decision maker. More importantly, the knowledge acquired from self-awareness helps the process to self-adapt its own search behaviour for better optimality and diversity, e.g., exploring more on the decisions that contain high CPU allocations and more beneficial for certain goals. This permits the *Decision Maker* to handle complex trade-offs even without prior preferences, i.e., achieving well-compromised trade-offs, which largely improving majority of the goals while causing relatively small degradations on others. Given that the regions are dependent to each others, we only make decisions for the regions containing the objectives whose requirement violations have been detected. process is needed.

## 3.5 Conclusion

As cloud computing continuous to evolve, autoscaling requires novel principles and approaches to seamlessly manage the underlying cloud-based services. Self-awareness provides highly promising avenue for improving self-adaptivity and the effectiveness of autoscaling in the cloud. In this chapter, we describe the autoscaling architecture and its mapping to self-awareness capabilities and the related pattern. We focus on discussing how the principle of self-awareness can be beneficial for various logical aspects in autoscaling, including QoS modelling, identifying granularity of control and making trade-offs decisions. In Chapter 4, 5, 6 and 7 we will experimentally evaluate each of these logical aspects in details.

In the next chapter, we will explore the first internal self, namely *QoS Modeller*, in the self-aware autoscaling architecture. Specifically, in such internal self, we propose a self-aware and self-adaptive QoS modelling approach, which is capable to dynamically correlate QoS attributes to various cloud primitives on-the-fly.

# Chapter 4

# Self-Aware and Self-Adaptive QoS Modelling in Cloud Autoscaling

## 4.1 Introduction

The elasticity of cloud has caused a paradigm shift in the way we manage and continually evolve cloud-based software services. However, it would be difficult for software engineers and cloud engineers to predict the wide variation of behaviours that software services can experience when running on a shared and on-demand environment such as the cloud. It is particularly hard to anticipate the dynamic changes in workload and the runtime demands of these cloud-based software services. This fact implies that it becomes more complex to assure the Quality of Service (QoS) when engineering cloud- bases services. The design of offline and manual management strategies for QoS are mere difficult if not impossible exercise to achieve.

With such context in mind, the key problem, which cloud/service providers face is how to manage runtime QoS by autoscaling to the best set of control values on-the-fly. In particular, the fundamental challenge is how to dynamically link QoS with the primitives in cloud, which we address in this Chapter. QoS models allow the use of primitive values as inputs and predict the likely QoS value as outputs. An accurate QoS model in the cloud can serve as a powerful tool that assists software/cloud engineers or other automated agents to profile service characteristics (e.g., CPU intensive services); to diagnose the cause of violation on QoS requirements; and more importantly, to compare and reason about different elastic autoscaling decisions in the cloud.

As we have extensively surveyed in Chapter 2, the majority of the existing approaches for QoS modelling in cloud has been either static (i.e., analytical [77] and simulation based [55]) or semi-dynamic [90]. The former is being static in the sense that the expression of models are fixed, and therefore, they are insensitive to the QoS fluctuations at runtime; this is due to the entire modelling process has relied on manual and offline analysis. On the other hand, the semi-dynamic approaches focus on adaptive and dynamic modelling of the magnitude of primitives in correlation to QoS, which means the model changes with respect to the QoS fluctuations. However, their selection of primitives to determine the feature inputs of models has been manual and offline, resulting fixed inputs for the models. As a result, they suffer limited self-adaptivity.

### 4.1.1   Motivations and Challenges

In this section, we motivate the proposed approach by identifying several important challenges for QoS modelling in the cloud, which have not been or have only been partially considered in previous work.

**_Fine-grained QoS Modelling:_** There can be different cloud-based software services running on a VM, each with its own QoS requirements. Fine-grained QoS modelling is challenging as more heterogeneity (e.g., QoS requirements, their derivatives and service characteristics etc.) need to be considered. However, existing static and semi-dynamic modelling tend to focus on the mean and aggregate QoS of the entire VM. Such coarse-grained analysis suffers from limited sensitivity; it does not apportion sensitivity to changes in QoS of each individual services and the primitives. As a result, the modelling of QoS tend to be inaccurate and limited for individual software service; more accurate and effective modelling needs to be approached from a fine-grained perspective.

***Dynamic and Uncertain QoS Interference:*** QoS modelling in the cloud suffers from the problem of QoS interference. QoS interference refers to scenarios where a software service exhibits wide disparity in its QoS performance that depends on the dynamic behaviours of its neighbours. In particular, we distinguish two major causes of interference, these are: co-located service interference and co-hosted VM interference. In this chapter, we particularly focus on the QoS interference caused by co-located services on a VM; we leave the problem of co-hosted VM interference to the next chapter. Given that the QoS interference tends to be dynamic and uncertain in nature, the challenge lies in the difficulty to dynamically incorporate the information about the related interference in the modelling. Despite the fact that QoS interference is important for QoS modelling in the cloud, there are not many work that target for this challenge. In addition, existing work consider co-hosted VM interference only (e.g., [116]). As a result, such absence of QoS interference in the model can downgrade the accuracy and/or lead to incorrect autoscaling decision.

***Dynamic and Uncertain QoS Sensitivity:*** The core of QoS modelling is how to model its sensitivity with respect to the primitives in cloud. By QoS sensitivity, we are interested in *which* (e.g., are CPU and throughput correlated?), *when* (i.e., at which point in time they are correlated?) and *how* (i.e., the magnitude of primitives in correlation) the primitives correlate with QoS. Given the dynamic and on-demand nature of cloud, QoS sensitivity is dynamic and uncertain, i.e., runtime changes occur in terms of which, when and how primitives correlate with QoS. Specifically, the challenges of QoS sensitivity in the modelling can be attributed to two important phases, namely primitives selection and QoS function training:

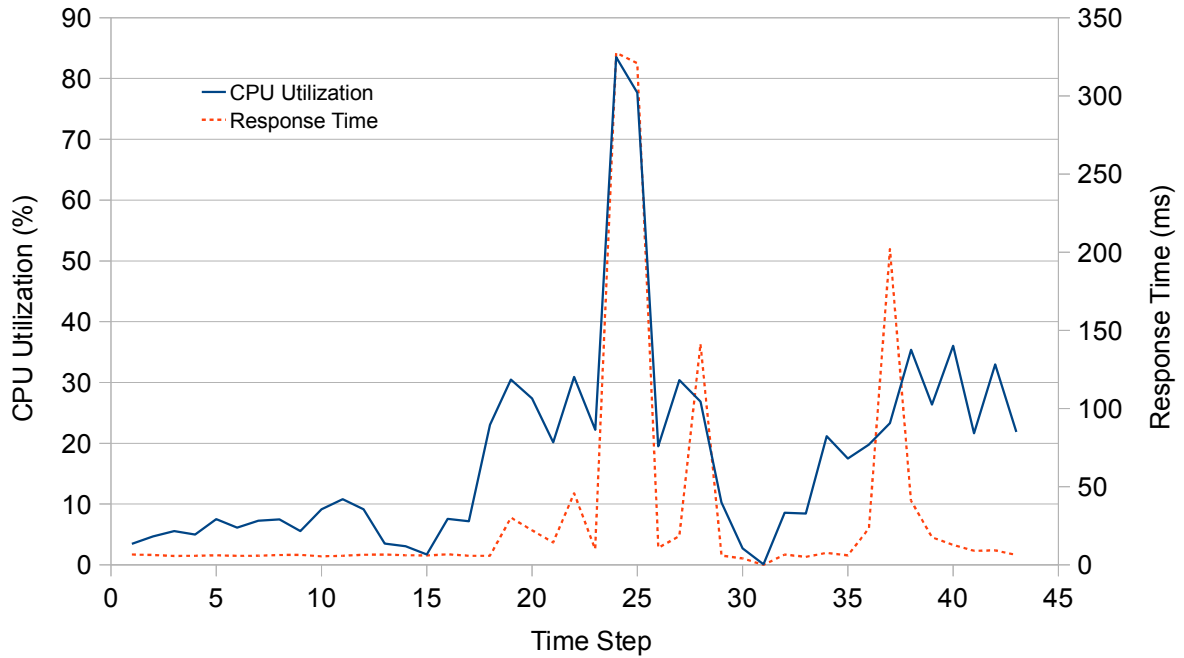

Figure 4.1: The Exampled Correlation Between Response Time and CPU.

- *Primitives Selection:* To model QoS and its sensitivity in the cloud, a fundamental task is to adaptively determine what are the primitives that should be used as feature inputs of the model (i.e., which and when the primitives correlate with QoS). To show a simple example of the dynamics and uncertainties in primitives selection, in Figure 4.1[1], we vary the workload of a service while keeping that of the co-located services and co-hosted VMs unchanged, we can see that the Response Time of the said service tends to be insensitive to CPU at the beginning hence it cannot provide relevant information about the QoS. However, after the *18th* interval, the Response Time gradually become more affected by the CPU as the workload change by time, which is uncertain in nature; this becoming even more true in the cloud when there is uncertain QoS interference, i.e., the workload of neighbour services/VMs changes. Therefore, the primitives selection needs to cope with the dynamics and uncertainties in QoS sensitivity. Given that the selected inputs have a great impact to the model accuracy (as we will show in Section 4.4.1), it is important to select a relevant set of primitives. However, this is a challenge providing the uncertainty of the relevance between QoS and cloud primitives, it is even harder when we take QoS interference into account. Nevertheless, majority of the existing static and semi-dynamic approaches for QoS modelling in the cloud rely on fixed and manual analysis to select the primitives as inputs, which are often offline. In addition, prior work only consider hardware resource while ignoring the software configurations, which can interplay with the hardware provision to influence QoS; these are often the primary causes of QoS violations [26] [139] [99].

---

[1]The data is obtained based on our experiments in Section 4.4

- *QoS Function Training:* Another important task in modelling QoS and its sensitivity is to adaptively determine how the primitives correlate with QoS by means of mathematical function. To show a simple example of the dynamics and uncertainties in QoS function training, we use the aforementioned setup in Figure 4.1. As we can see, from the *18th* interval onwards, the Response Time of the service is becoming more sensitive to CPU till *30th* where the sensitivity is starting to decrease. This shows that the Response Time is always sensitive to CPU for the period, but the magnitude tends to be different depends on the uncertain changes of workload from time to time. Again, this becomes more complex in the cloud when it involves changing workloads of neighbour services/VMs. All These facts imply that the modelling needs to be able to handle the dynamic and uncertain magnitude of primitives in the correlation, which is a challenge. Consequently, the static QoS modelling approaches tends to be insufficient, because the effectiveness of these approaches is restricted by their simplified and fixed assumptions on the environment and service's internal operations [90], which limits them for handling the dynamics and uncertainties of QoS sensitivity in cloud. On the other hand, the semi-dynamic approaches are capable to handle this challenge as they are grounded on sound machine learning algorithms, which tend to be dynamic and self-adaptive in nature. In addition, they rely on no or limited assumptions. Particularly, the online version (e.g., [87]) has been proposed to overcome the inadequacy of the offline version (e.g., [90]) in dealing with the uncertain changes of QoS sensitivity at runtime. However, selecting and efficiently adopting these learning algorithms for QoS modelling in the cloud is a challenge. Each learning algorithm has both advantages and disadvantages, e.g., a complex and nonlinear algorithm may be suitable for handling complex correlation, but the training overhead may be high. On the other hand, a linear algorithm can be efficient, while it may lack in dealing with frequent fluctuation. Given the dynamic and uncertain correlation between QoS and the cloud primitives, it is remain unclear about which are the suitable class of learning algorithms for QoS modelling in the cloud.

All these challenges and limitations of existing work call for novel and fully-dynamic QoS modelling approach in the cloud, with limited or no human intervention.

### 4.1.2 Contributions

In this chapter, we propose a self-aware and self-adaptive QoS modelling approach for each individual cloud-based service. It grounds on sound information theory and machine learning algorithms, which are the key enablers of realising self-awareness for QoS modelling in the cloud. Our approach is fully-dynamic and it is capable to adaptively capture the dynamics of QoS sensitivity by determining *which*, *when* and *how* primitives correlate with QoS at runtime. In particular, we have relied on symmetric uncertainty [130] from information theory to quantify the relevance between two random variables. This is motivated by the fact that it is an highly intuitive and efficient metric supported by strong theoritical foundations. Subsequently, we combine symmetric uncertainty with two learning algorithms: Auto-Regressive Moving Average with eXogenous inputs model (ARMAX) [22] and Artificial Neural Network (ANN) [119] to reach two formulations of the model. The choice of ARMAX and ANN is driven by the fact that they are the most commonly applied algorithm for QoS modeling in the cloud, and that they represent two extreme groups of algorithms in machine learning: ARMAX is essentially linear regression, which is simple and fast; while ANN is capable to handle nonlinear correlation but could lead to high complexity. Experiment results show that our models produce better accuracy when compared with conventional ones. In comparison of the two resulting models, our Sensitivity-aware ANN (S-ANN) can better handle dynamic QoS sensitivity and produce higher accuracy when the fluctuation of measured QoS increases, whereas our Sensitivity-aware ARMAX (S-ARMAX) produces less error when such fluctuation decreases.

## 4.2 Problem Analysis and Models

In this section, we present our assumptions, the system model and the abstract QoS model that drive our design. These assumptions and models will be used in the following chapters of this thesis.

### 4.2.1 Cloud System Model

As mentioned in Chapter 1, we assume that cloud-based applications are composed of services, each has different QoS requirements and external environment changes (e.g., changes in workload). Particularly, we term a replica of a service as **_service-instance_**: the _jth_ instance of the _ith_ cloud-based service is denoted as $S_{ij}$. Unlike most of the existing work, which focus on modelling for the entire application and VM, we aim to create fine-grained QoS models for each service-instance.

It is worth noting that, apart from the co-located services on a VM, QoS interference can also occur due to contention on the functionally dependent services. For instance, $S_{11}$ and $S_{31}$ (both running on different PMs) can be both dependent on $S_{21}$(e.g., a database service). This implies that $S_{11}$ and $S_{31}$ incur QoS interference. However, we discovered that in such case, the primitives of $S_{31}$ tend to be insignificant in the QoS modelling of $S_{11}$ as the same information has already been expressed by the primitives of $S_{21}$, which is also part of the invocation. As a result, we consider the co-hosted services as the primary causes of QoS interference.

## 4.2.2   The Cloud Primitives for Building Models

As mentioned in Chapter 2, the primitives in cloud serve as the fundamental inputs of a QoS model. Without loss of generality, we decompose the notion of primitives into two major domains: these are **Control Primitive (CP)** and **Environmental Primitive (EP)**. Selecting the relevant primitives for QoS is an crucial step in QoS modelling. Specifically, all possible primitives inputs for modelling the QoS attributes of a service-instance form a space, which we call ***possible relevant primitives space***. This space can be defined by:

*Rule 4.1. A primitive belongs to the possible relevant primitives space for modelling the QoS of $S_{ab}$ if it can be classified into one of the following groups:*

1. *It is a software control or environmental primitive of $S_{ab}$.*

2. *It is a hardware control primitive of the VM that runs $S_{ab}$.*

3. *In case of $S_{ab}$ has direct functional dependency[2] on $S_{cd}$, it is a software control or environmental primitive of $S_{cd}$.*

4. *In case of $S_{ab}$ has direct functional dependency on $S_{cd}$, it is a hardware control primitive of the VM that runs $S_{cd}$.*

---

[2]When the completion of a service $S_{ab}$ requires the invocation of another service $S_{cd}$, then it is said $S_{ab}$ has functional dependency on $S_{cd}$. If no intermediate services are required in the invocation between $S_{ab}$ and $S_{cd}$, then it is said $S_{ab}$ has direct functional dependency on $S_{cd}$.

5. *It is a software control or environmental primitive of $S_{cd}$, which is co-located with $S_{ab}$ on the same VM.*

In Section 4.3.1, we will present the solution for selecting the relevant primitives from such possible relevant primitives space.

Another important decision to mention is that, for each control primitive, we need to decide on whether the upper/lower bound of control primitive. However, it is generally impossible to guarantee that the configured value (e.g., CPU cap) can be fully utilised. Such fact obfuscates the sensitivity of QoS to its primitives as using the configuration values to model QoS would take those idle proportions of provisions into account. As a result, using configuration values as inputs is ill-suited in our case. To cope with this issue, we apply the demand values of control primitives (e.g., real-time percentage usage of CPU) as inputs, which better reveal QoS sensitivity. Moreover, modelling QoS with demand values implies that our model is likely to determine the minimal requirement of configurations for achieving certain QoS objectives. This will potentially improve the elasticity of software configuration and hardware provision in cloud, when our modelling approach is used in cloud management. It is worth noting that certain dimensions of control primitives (e.g., thread) can be controlled for each service-instance individually, whereas others (e.g., CPU and memory) are shared on a VM, in which case an identical value would be used for modelling the QoS of all service-instances deployed on such VM.

Instead of using multiple metrics for each primitive and QoS, e.g., CPU percentage and instructions-per-second for measuring CPU of a VM, we follow the state-of-the-art assumption [90] that only one metric is used for each primitive and QoS in the modelling; the proper metric can be chosen by the software/cloud engineers based on certain constraints in the cloud environment (e.g., whether it is supported by the hypervisor). We leave the study of multidimensional metrics as future work.

### 4.2.3 Generic QoS Model

To tackle the aforementioned challenges of QoS modelling in the cloud, we define a generic QoS model. Formally, the model at the *tth* sampling interval is expressed as:

$$QoS_k^{ij}(t) = f_k^{ij}(SP_k^{ij}(t), \delta) \tag{4.1}$$

where $QoS_k^{ij}(t)$ is the *kth* QoS attribute of $S_{ij}$, and its value that used in the modelling is represented by a given metric (e.g., mean Response Time) at $t$. $f_k^{ij}$ is the QoS function for the *kth* QoS attribute of $S_{ij}$, and it changes at runtime using learning algorithms, as we will see in Section 4.3.2. $\delta$ refers to any other inputs (e.g., historical time-series QoS points and tuning variables etc) required by the algorithm to train $f_k^{ij}$ apart from the cloud primitives. We denote the input in (4.1) as the selected primitives matrix of $QoS_k^{ij}(t)$ at $t$, formally depicted in (4.2):

$$SP_k^{ij}(t) = \begin{pmatrix} CP_a^{xy}(t) & \cdots & EP_b^{mn}(t-1) & \cdots \\ \vdots & \ddots & \vdots & \ddots \\ CP_a^{xy}(t-q+1) & \cdots & EP_b^{mn}(t-q) & \cdots \end{pmatrix} \tag{4.2}$$

This matrix contains the primitive inputs of $QoS_k^{ij}(t)$ which are dynamically selected from the possible relevant primitives space for the QoS attributes of $S_{ij}$, as we will see in Section 4.3.1. More concretely, the column entries indicate the selected primitives for the QoS. $CP_a^{xy}(t)$ denotes the *ath* control primitive of $S_{xy}$ and $EP_b^{mn}(t-1)$ means the *bth* environmental primitive of $S_{mn}$ respectively. The actual values of $CP_a^{xy}(t)$ and $EP_b^{mn}(t-1)$ in the modelling are measured by given metrics (e.g., expected CPU % usage and mean request rate) at $t$ and *t-1*, respectively. $q$ determines the number of row entries, which indicates the use of how many historical time-series points of the selected primitives as inputs. During our experiments in Section 4.4, we observed that the best value of $q$ depends on the learning algorithm that trains $f$; in particular, it is better to set $q$ as constant for certain algorithms (e.g., $q=1$ for ANN); however for the others (e.g., ARMAX), we found that $q$ should be determined during training via hill-climbing optimisation, which starts with $q=1$, then automatically increase the number of row entries one by one during training till the accuracy cannot be further improved. To improve numeric stability, we normalised all data values to the range between 0 and 1 before they are used in the modelling.

It is easy to see that (4.1) and (4.2) provide generic and intuitive formulations for modelling QoS in the cloud. Precisely, to model $QoS_k^{ij}(t)$, the objective of our self-aware and self-adaptive modelling approach consists of two-phases: (i) a primitives selection phase that determines the content of $SP_k^{ij}(t)$ at runtime; and (ii) a QoS function training phase that trains function $f_k^{ij}$ on-the-fly.

## 4.3   Designing Self-Aware QoS Modelling

To adaptively build fine-grained, self-aware and self-adaptive QoS models, we implemented our modelling approach as decentralised and independent components. As shown in Figure 4.2, the QoS modelling process is realised as decentralised feedback loops; in particular, a dedicated component instance (CI) is attached to each VM, and it could be deployed on the root domain of a PM (e.g., Dom0 of the hypervisor Xen [6]).

The QoS modelling component encapsulates three sub-components: *Data Collector*, *Primitives Selector* and *QoS Function Trainer*. As we can see in Figure 4.2, the approach is deployed using decentralised component instances, each of which is attached to a VM. The supported QoS attributes and primitive types are provided by the cloud administrators. By leveraging on the cloud providers' own measurement facilities provided at SaaS, PaaS and IaaS layers, an instance of the component monitors the data of each service-instance running on the VM; it also adaptively producing QoS models based on this data. More precisely, the *Data Collector* continually senses QoS values, environmental primitives' values and demand of control primitives from all service- instances on its corresponding VM. It is also responsible for recording all historical data (step 1). The data collectors may need to collect data from the external service-instances, which could be on other VM/PM. This is because these external service-instances may be functionally required by the service-instances running on the VM attached to data collectors. All historical data is then passed to our *Primitives Selector* to determine which and when primitives are correlated with a QoS (step 2). Once the relevant primitives are selected for each QoS, the *QoS Function Trainer* can apply the data set to dynamically train how these primitives correlate with QoS and produce the final QoS models, based on machine learning algorithms (step 3). To capture dynamic sensitivity of QoS, the entire process should be repeated periodically (step 1-4).

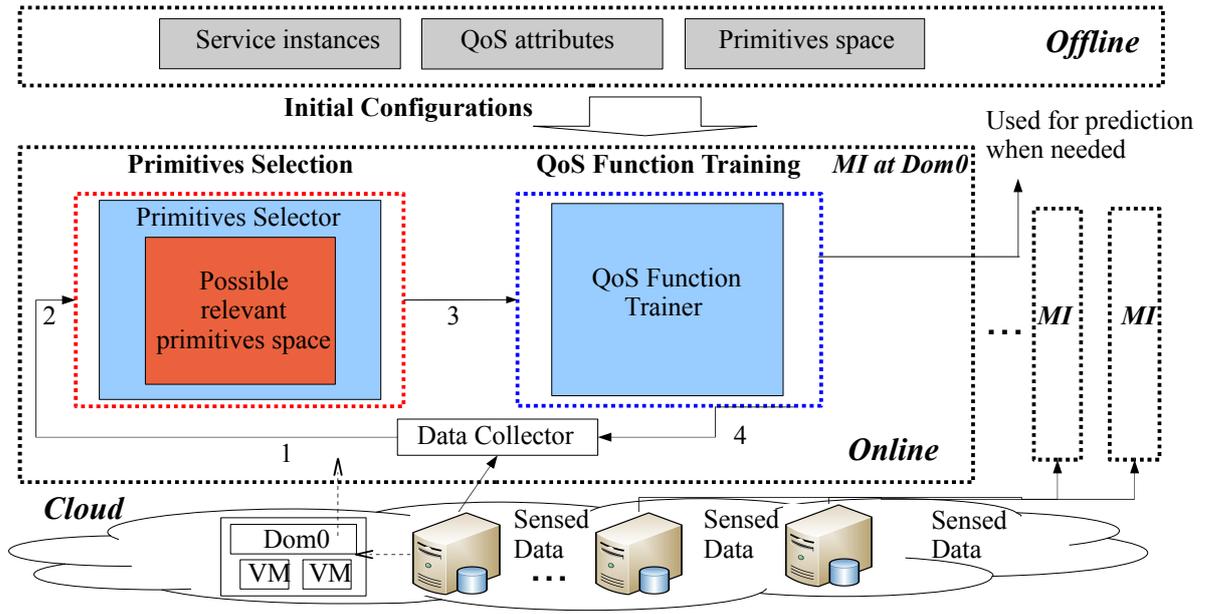

Figure 4.2: The Architecture for QoS Modelling in the Cloud.

For our autoscaling framework, this is the *QoS Modeller* component as described in Chapter 3. Being one of the three internal selves (i.e., *QoS Modeller*, *Region Controller* and *Decision Maker*), self-awareness in *QoS Modeller* is mainly concerned with knowing QoS sensitivity. Table 4.1 shows the mapping between the sub-components of *QoS Modeller* and the self-awareness capabilities. In the following sections, we explain the algorithms and techniques used to achieve self-awareness at the QoS modelling level.

Table 4.1: The Mapping Between Self-Awareness Capabilities and the Sub-Components for QoS Modelling in the Cloud.

| Self-Awareness Capability | Component | Description |
|---|---|---|
| Stimulus-awareness | Primitives Selector and QoS Function Trainer | Knowing how the QoS modeling process can be affected by the features (e.g. workload) of QoS attributes and their changes. |
| Interaction-awareness | Primitives Selector | Knowing how the QoS modelling process can be affected by the possible QoS interference and resources contention. |
| Time-awareness | Primitives Selector and QoS Function Trainer | Knowing how the QoS modelling process can be affected by the historical data points and the overall trends. |
| Goal-awareness | QoS Function Trainer | Knowing how the QoS modelling process can be affected by the utility functions of QoS. |
| Self-expression | Primitives Selector and QoS Function Trainer | Self-adapting the expressions of its QoS models, including the model inputs and function. |

### 4.3.1   Relevance Driven Selection of Cloud Primitives

As shown in (4.1) and (4.2), the first task for modelling $QoS_k^{ij}(t)$ is to explore which primitives should be included as column entries in $SP_k^{ij}(t)$, and determine when is the appropriate interval to consider these primitives. To quantify the relevance of a primitive to the QoS, we have used Symmetric Uncertainty (SU), which is a fundamental concept in information theory [130]. SU measures the degree of relevance between two time series variables by producing a value ranges from 0 to 1, where a greater value implies higher relevance. At one extreme, the value between a QoS attribute and a primitive is 1 indicating that all information of the primitive is correlated with the QoS (and vice versa). At the other extreme, the value of 0 implies that changes in the primitive's behaviour are independent of that of the QoS (i.e., irrelevant primitive). Formally, the symmetric uncertainty for discrete variables is calculated by:

$$U(X,Y) = \frac{2 \times I(X,Y)}{H(X) + H(Y)} \tag{4.3}$$

$$I(X,Y) = \sum_{y \in Y} \sum_{x \in X} p(x,y) \log(\frac{p(x,y)}{p(x) \times p(y)}) \tag{4.4}$$

$$H(X) = -\sum_{x \in X} p(x) \log(p(x)) \tag{4.5}$$

where $X$ and $Y$ are two value vectors of time-series variables (e.g, a QoS attribute and a primitive); $x$ and $y$ are one of these values. $I(X, Y)$ shows the formula for mutual information and $H(X)$ expresses entropy (we have used 2 as the log base); $p(x,y)$ is the joint probability between two values and $p(x)$ is the marginal probability of a value. In the following, we call a primitive as relevant primitive to a QoS attribute if it results in non-zero SU value to such QoS. It is worth noting that at this stage, we do not consider the redundancy in the selected relevance primitives. This issues will be tackled specifically in the next Chapter.

In the *Primitives Selector* component, we adopt the relevance driven technique to determine the relevant primitives of a QoS attribute, the procedure is described as the following: firstly, based on the sensed data, we calculate the symmetric uncertainty between the QoS attributes of a given service-instance and each of the primitives from the corresponding possible relevant primitives space (as described in Rule 4.1). Secondly, we update the corresponding selected primitives matrix by adding the primitives that result in nonzero SU value; while removing primitives that have zero value. To handle runtime dynamics and uncertainty, the selected primitives matrix can be continually updated with newly-measured data.

## 4.3.2 Sensitivity-aware Auto-Regressive Moving Average with eXogenous inputs model

Recall from (4.1), once the $SP_k^{ij}(t)$ is defined by the primitives selector, our next goal of QoS modelling is to determine how those primitives correlate with $QoS_k^{ij}(t)$ by dynamically building the function $f_k^{ij}$. To achieve such goal, we apply two alternative algorithms in QoS function trainer. We have chosen the ARMAX [22] and the ANN [119] as representative of the linear and nonlinear algorithms respectively. Given that they only model the primitives that selected using our relevance driven technique, the final models are termed Sensitivity-aware ARMAX (S-ARMAX) and Sensitivity-aware ANN (S-ANN) respectively. The choice of ARMAX and ANN is driven by the fact that they are the most commonly applied algorithm for QoS modeling in the cloud, and that they represent two extreme groups of algorithms in machine learning: ARMAX is essentially linear regression, which is simple and fast; while ANN is capable to handle nonlinear correlation but could lead to high complexity.

We have considered linear ARMAX [22] as one of the applied learning algorithm for QoS function training. The resulted model of particularly fits our case because it is based on continuous time series. The correlation between primitives and QoSs in our case is unlikely to be linear, however, the behaviour of a service instance can be approximated locally as a linear model [107]. We adopt ARMAX such that the output is the QoS; inputs are historical QoS values and relevant primitives of the said QoS. Formally, based on the generic QoS model and symmetric uncertainty, our Sensitivity-aware ARMAX

(S-ARMAX) is defined as:

$$QoS_k^{ij}(t) = \sum_{z=1}^{q} \alpha_z(t) \times QoS_k^{ij}(t-z) + \sum_{z=1}^{q} \sum_{a=1} \beta_{za}(t) \times CP_a^{xy}(t+1-z)$$
$$+ \sum_{z=1}^{q} \sum_{b=1} \theta_{zb}(t) \times EP_b^{mn}(t-z)$$

(4.6)

$$subject \; to \quad\quad CP_a^{xy}(t+1-z), EP_b^{mn}(t-z) \in SP_k^{ij}(t) \quad\quad (4.7)$$

where $q$ is the number of order, $\alpha_z(t)$, $\beta_{za}(t)$ and $\theta_{zb}(t)$ are the coefficients of corresponding QoS values and relevant primitives at sampling interval $t$. The constraint ensures that any primitives should be selected from $SP_k^{ij}(t)$.

We train the S-ARMAX model using linear Least Mean Square (LMS) approach [129], and the number of order $q$ is determined using hill-climbing algorithm that starts with $q=1$, then automatically increase the number of row entries one by one during training till it reaches good accuracy.

### 4.3.3 Sensitivity-Aware Artificial Neural Network

Artificial Neural Network (ANN) [119] is applied as the second algorithm to build our function $f_k^{ij}$. We chose ANN because it is capable for modelling complex nonlinear correlations. In particular, we adopt ANN with one hidden layer. This is because we observed in our experiments that using two or more hidden layers could exacerbate the problem of local minima, which significantly increases the training time. More precisely, the ANN, which we adopt is a single-output, feedforward and fully connected three layered network, where the inputs are the relevant primitives determined by primitive selector and output is the corresponding QoS. Sigmoid function is chosen as the activation function on each neurone node. We found that the number of order $q$ does not influence ANN's accuracy significantly, therefore we simply set $q$ as 1 (i.e., no time-series information is included). Based on the generic QoS model and symmetric uncertainty, our Sensitivity-aware ANN (S-ANN) model is expressed as:

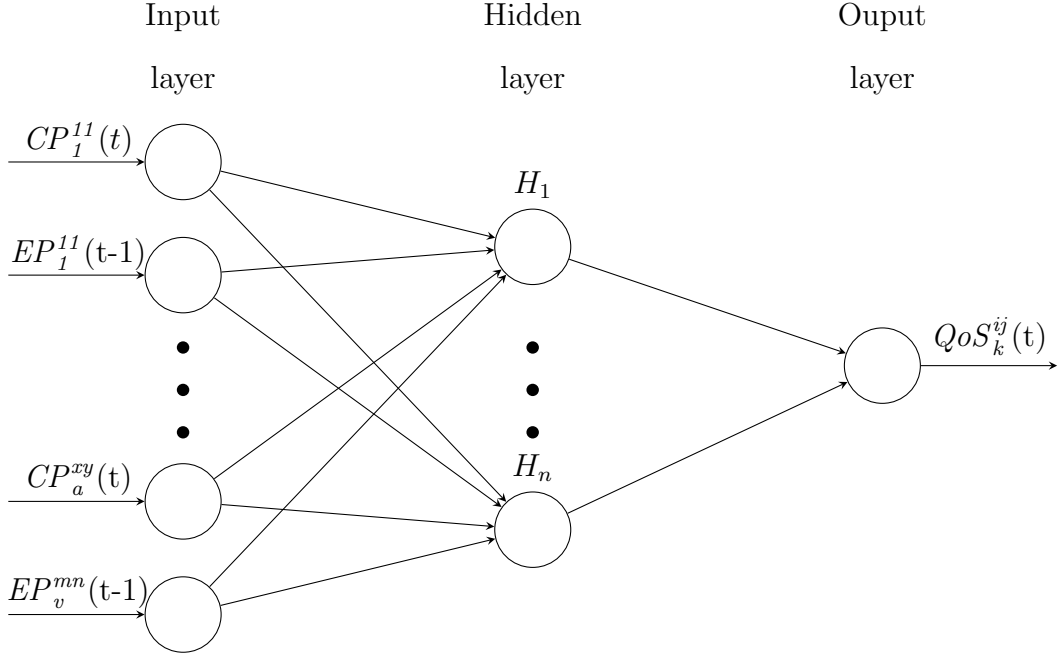

$$subject \ \ to \qquad CP_a^{xy}(t), EP_b^{mn}(t-1) \in SP_k^{ij}(t) \qquad (4.8)$$

The constraint again ensures that any primitives should be selected from $SP_k^{ij}(t)$. ANN model can be trained with arbitrary quality, which reveals the potential accuracy of the model prediction. By model quality, we refer to the degree to which the model is fit with respect to the training data. In this perspective, a good quality model means that the fitness should not be too low or too high; otherwise, the model will suffer from under- and over-fitting. To guarantee model quality, we define minimum and maximum thresholds to represent the *good enough* quality of model. The resulting model should be re-trained immediately if its quality is not good enough.

Similar to S-ARMAX, by the end of each interval, the weights in S-ANN can be retrained with the newly-measured and normalized data. To achieve this goal, we apply the RPROP [115] as the actual training algorithm for the network. This is because RPROP can efficiently reach 'good enough' model quality. To avoid training forever, we have defined a training time threshold such that if this threshold has been reached, the training is concluded with the best ever model. We found that use $q=1$ (i.e., no time series information) can produce the best result; and the right number of hidden neurons is determined using hill-climbing algorithm during training till the accuracy cannot be further improved.

## 4.4 Experiments and Evaluations

The primary intention of our experiments is to evaluate accuracy and effectiveness of the proposed QoS modelling approach with respect to the scale of data. Specifically, we compare the accuracy of the proposed S-ARMAX and S-ANN models with conventional models in continuous time series. We also assess the sensitivity of our models to the size of training data. Finally, we examine training efficiency by looking at the training overhead.

### 4.4.1 Experiments Setup

We conducted experiments on private cloud using a cluster of PMs, each of which has Intel i7 2.8GHz Quad Cores and 4GB RAM. The PMs use Xen v3.0.3 [6] as the hypervisor and the modelling process is running on Dom0. To eliminate the interference caused by modelling, we allocated one CPU core and 1.2GB RAM to Dom0, which tends to be sufficient. Our approach is implemented based on Encog [2] and Apache Mathematics [1] using Java JDK 1.6. To simulate QoS interference caused by the VMs while not exhausting resources, we run three co- hosted VMs on each PM; the remaining resources are evenly allocated to the co-hosted VMs. All VMs run linux kernel v2.6.16.29.

Our experiments leverage on RUBiS [5], which is a cloud-based application consists of 26 co-located software services using the eBay.com model. For simplicity, we have used three RUBiS snapshots, each of which consists of a 2-tiers (i.e., application and database tiers) based RUBiS application; the three RUBiS snapshots differ in terms of the database volume size ranging from 1GB to 5GB data. Each RUBiS snapshot is deployed with a software stack including Tomcat v6.0.28 and MySQL v3.23.58 on each co-hosted VM of a master PM; and we have implemented sensors deployed on each service- instance and VM for sending the online data to *Data Collector*. For each RUBiS snapshot on the master PM, the application tier is replicated to all other servant PMs in the cloud; these replicas are connected to the database on the master PM for handling any database related requests. Finally, each of the three RUBiS snapshots and its replica are linked to its dedicated load balancer. Three client emulators are used and they apply read/write pattern to generate requests for each load balancer.

To simulate a realistic workload within the capacity of our testbed, we vary the number of clients proportionally according to the FIFA98 workload [14], which is compressed in the way that the fluctuation of a day in the trend corresponds to 200s in our case. This setup can generate up to 400 parallel requests, we believe that such compression is realistic and large enough to simulate QoS interference in cloud.

We apply two deployment strategies for the benchmark. The first strategy *D1* assumes that all services of the application are hosted on one VM. The second strategy *D2* involves two VMs for each application replica, where the database server and web/application server are deployed on different VMs. To facilitate the dynamic deployment in the cloud, we switch the deployment from *D1* to *D2* on the fly by live-migrating the database service. To further verify our modelling approach under unusual workload changes, we apply the biding workload pattern for *D1* whereas the browsing pattern is used for *D2*. The entire FIFA 98 workload trend is used for *D1*, and it is repeated again for *D2*. Due to limited space, we only report on evaluation of the QoS models for one service-instance

Table 4.2: The Examined QoS Attributes and Primitives.

| QoS and Primitives | | Description |
|---|---|---|
| Output | Response Time (ms) | The average leaped time between a service-instance receives and replies a request. |
| | Throughput (req/min) | The average rate of completed requests. |
| | Availability (%) | The percentage of time that the average response time above a threshold. (60 ms) |
| CP input | CPU (%) | Observed average CPU utilisation of a VM. |
| | Memory (MB) | Observed average Memory utilisation of a VM. |
| | Thread (no. of req) | Observed maximum concurrent threads of a service-instance. (a modified control knob of Tomcat's *maxThread* property) |
| EP input | Workload (req/min) | Observed average request rate of a service-instance. |

of a concrete service named SEARCHITEMBYCATEGORY.

## 4.4.2 QoS attributes and Cloud Primitives

For the simplicity of exposition, we report on a scenario, which considers the following dimensions: three QoS attributes, two hardware control primitives, one software control primitive and one environmental primitive, as listed in Table 4.2. Considering the fact that there are 26 services in RUBiS benchmark, we need to produce 78 QoS models, each of which can have at most 54 possible relevant primitives and thus this is a non-trivial scale. Additionally, based on the real-life workload and benchmark, this setup sufficiently provides us with valuable insight on the models' behaviour when handling a large stream of live data in complex and dynamic systems.

To sum up, there are three QoS models that need to be adaptively created for each service-instance. The actual inputs to train and predict these three QoSs are the primitives in their selected primitives matrix, which are adaptively determined by the *Primitive Selector* and the used machine learning algorithms as explained in Section 4.3.2 and 4.3.3 at Appendix C. In the case of our experiments, the likely relevant primitives of each QoS would be selected from: CPU and memory demands of the web/application server VM; CPU and memory demands of the database server VM (this is only available when switch to *D2*); threads demands and workload for each of the 26 service-instances on the same VM. To compare S-ANN and S-ARMAX, the *QoS Function Trainer* would simultaneously produce two alternative models for each QoS.

### 4.4.3 Accuracy

To validate the correctness, we measure the accuracy of our QoS modelling approach on the fly. The sampling interval is 30s with the total of 700 intervals. In particular, we examine the accuracy of one interval ahead prediction. That is, by the end of interval *t*, our approach trains QoS models based on historical data up to *t-1* (*t-1* for environmental primitives), the resulting model predicts the QoS at *t* by using historical QoS values (for S-ARMAX), the measured demands of control primitives up to current interval *t* and value of environmental primitives up to interval *t-1*. For all predictions, the accuracy is assessed via Symmetric Mean Average Percentage Error (SMAPE) [58], which is computed as:

$$SMAPE \ = \ \frac{1}{K} \sum_{t=1}^{K} \frac{\left| QoS_k^{ij}(t)' - QoS_k^{ij}(t) \right|}{QoS_k^{ij}(t)' + QoS_k^{ij}(t)} \tag{4.9}$$

where $K$ is the total number of intervals, $SP_k^{ij}(t)$' denotes the measurement of the *kth* QoS of $S_{ij}(t)$ at interval *t* whereas $SP_k^{ij}(t)$ denotes the prediction for the same QoS at the same interval. It has been shown that SMAPE is intuitive, stable and more resilient to outliers than the other metrics [100]. Notably, we regard zero value of QoS as invalid measurement, because it only represent the fact that no one has requested a certain service at a point in time.

To further evaluate the improvement to conventional semi-dynamic approaches, we compare the accuracy of our S-ARMAX and S-ANN models against the conventional ARMAX e.g., [141] and ANN e.g., [90] based models, which only consider limited and fixed hardware control primitives, such as the CPU and memory of the web/application server VM. In addition, they do not cater for QoS interference. We denote these conventional models as C-ARMAX and C-ANN. Given that these models rely on fixed number of relevant primitives, their number of order and hidden neurones are fixed and are obtained by examining given set of measured data (we discovered that in our case set $q$ as 2 for C-ARMAX and 18 hidden neurones for C-ANN could result in the best model). These conventional models predict QoS on per-application basis (denoted as per-app), whereas our models are per-service models. Thereby to eliminate noise caused by granularity, we also compare our models with modified, per-service version of C-ARMAX and C-ANN. To analyse the correlation between model accuracy and the variation of measured QoS trend, we apply Relative Standard Deviation (RSD) to measure how fluctuation of the QoS tends to be in a relative manner, such metric is calculated as: RSD = $\sigma/\mu$, where $\sigma$ is the standard deviation and $\mu$ is the mean of all measured QoS values.

Table 4.3: Comparative Summary of QoS Prediction Accuracy for A Service-Instance of SEARCHITEMBYCATEGORY. (the best is highlighted in bold)

| QoS | SMAPE of Prediction (%) | | | | | | RSD (%) |
|---|---|---|---|---|---|---|---|
| | S-ANN | C-ANN | | S-ARMAX | C-ARMAX | | |
| | | per-service | per-app | | per-service | per-app | |
| Response Time | **6.97** | 12.76 | 31.72 | **11.43** | 15.08 | 34.03 | 120.61 |
| Throughput | **11.14** | 16.88 | 35.28 | **7.99** | 13.22 | 37.82 | 86.41 |
| Availability | 0.96 | **0.38** | 1.36 | **0.01** | **0.01** | 1 | 2.24 |

The accuracies of all the comparative models are summarised in Table 4.3. It clearly indicates that our S-ANN reduces the error from 31.72% to 6.97% for response time; from 35.28% to 11.14% for throughput and from 1.36% to 0.96% for availability, when compared to per-application C-ANN model. In contrast to per-service C-ANN, the S-ANN also reduces 5.79% error (12.76% to 6.97%) for response time and 5.74% error (16.88% to 11.14%) for throughput. The only exception is that the S-ANN tends to produce 0.58% (0.38% to 0.96%) higher error for availability. We believe that this is because the RSD of availability is relatively small and thus the influence caused by dynamic QoS sensitivity tends to be trivial, which could easily cause over-fitting. On the other hand, our S-ARMAX is superior to both per-application and per-service C-ARMAX models. In particular, S-ARMAX reduces the error from 34.03% to 11.43% for response time; from 37.82% to 7.99% for throughput and from 1% to 0.01% for availability, when compared to the per-application C-ARMAX model. In contrast to per-service C-ARMAX, the S-ARMAX also reduces 4.45% error (15.08% to 11.43%) for response time and 5.23% error (13.22% to 7.99%) for throughput. The prediction error for availability remains the same. To conclude, it is clear that both of our S-ANN and S-ARMAX offers better accuracy than the C-ANN and C-ARMAX models.

An interesting discovery is that nonlinear model like S-ANN handles the dynamic QoS sensitivity better when the fluctuation of measured QoS increases (e.g., for Response Time and Throughput with high RSD), whereas the linear S-ARMAX produces less error when such fluctuation decreases (e.g., for Availability with low RSD). This is a useful conclusion as it implies that to better handle the dynamic QoS sensitivity, we shall also adaptively determine the best techniques to train QoS function at runtime.

To provide more detailed view of accuracy when using the proposed modelling approach, Figures 4.3-4.5 illustrate the total of 616 valid measurements of the actual QoS and predicated values produced by S-ANN and S-ARMAX. More precisely, Figure 4.3 demonstrates the trends for response time. Although the figure shows that error tends to increase for some of the peak points, it is obvious that both models still produce good prediction even for *D1* (from interval 1 to 310), where the QoS trend highly fluctuates. Similar observation occurs in Figure 4.4, which illustrates the trends for throughput. As for the availability in Figure 4.5, we can observe that S-ARMAX is better than S-ANN for *D2*. This is because training the ANN model with stable data (e.g., interval 1-174) followed by sudden fluctuation (at interval 175) can easily cause over-fitting, which eventually influences the ANN's prediction accuracy. Nevertheless, we can see that the S-ANN is adaptive enough to correct itself; the prediction becomes better and more stable from interval 230. On the other hand, S-ARMAX obtains perfect prediction fit for availability in all the intervals.

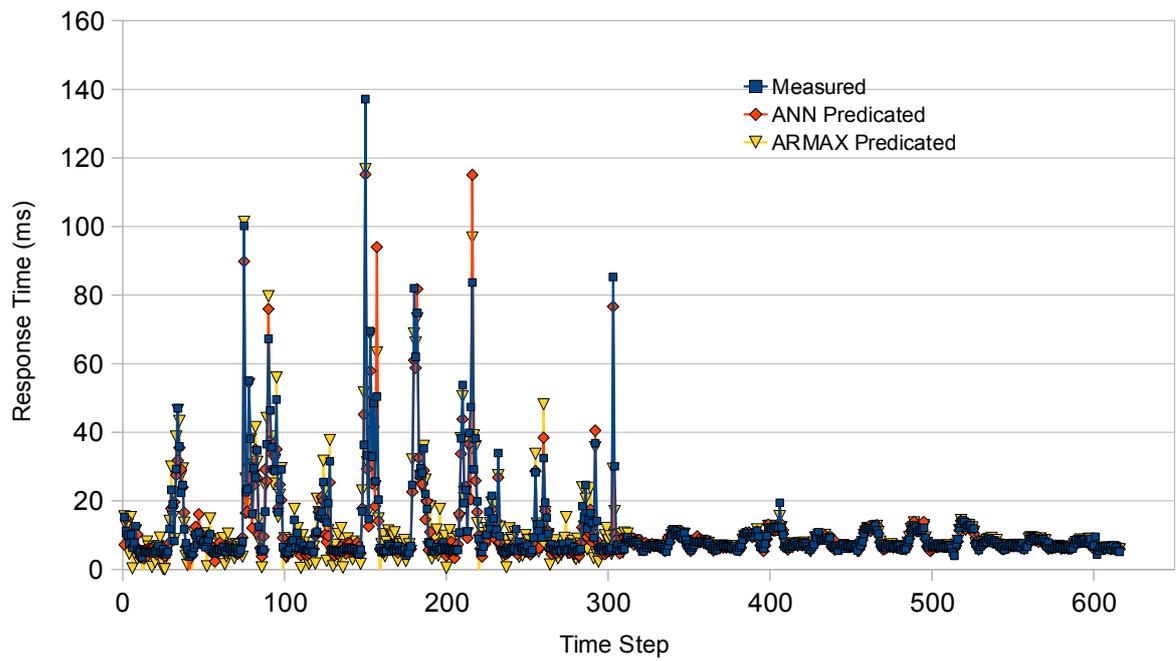

Figure 4.3: Actual and Predicated Response Time.

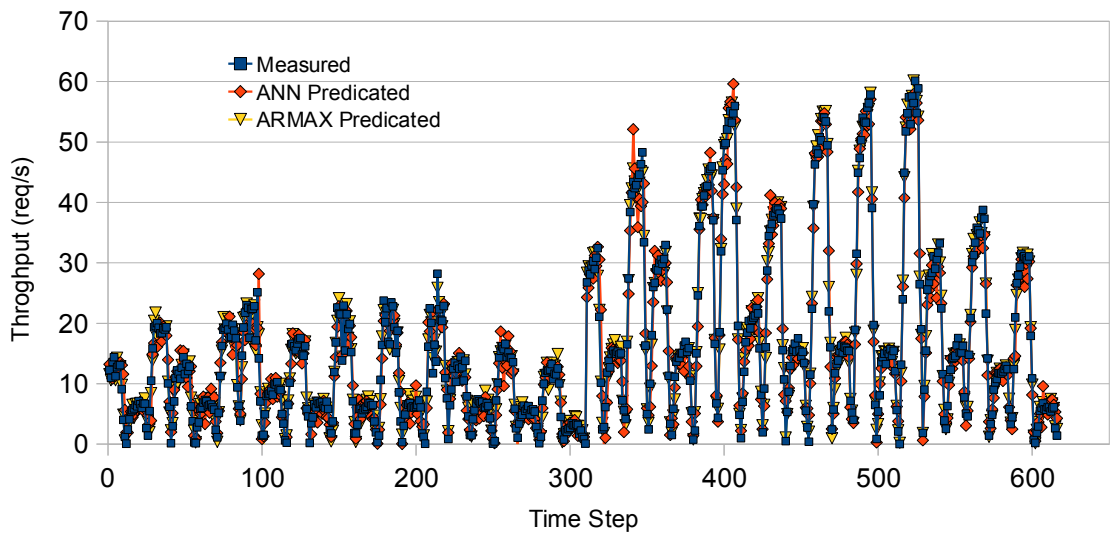

Figure 4.4: Actual and Predicated Throughput.

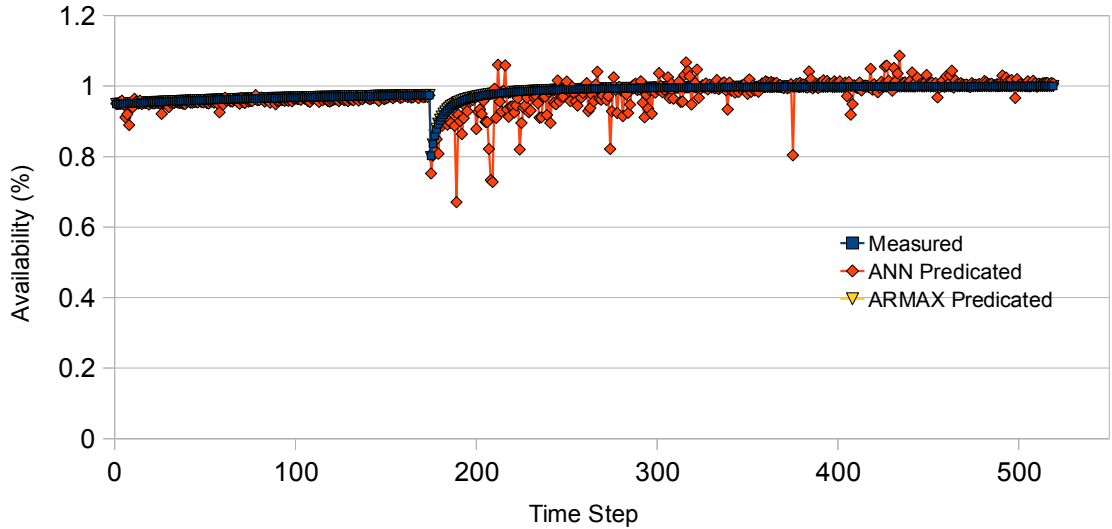

Figure 4.5: Actual and Predicated Availability.

### 4.4.4  Sensitivity of Accuracy to Training Data Size

To understand the sensitivity of model accuracy to the training data size, we divide all the measured intervals into 70% as training data and the rest 30% as testing data. Both S-ANN and S-ARMAX are trained based on a portion of the training data, ranging from 40% to 100%. As shown in Figure 4.6, it is evident that both models improve their accuracy as the size of training data increase. In particular, S-ANN is less sensitive to the training data size and produces acceptable accuracy even under limited data.

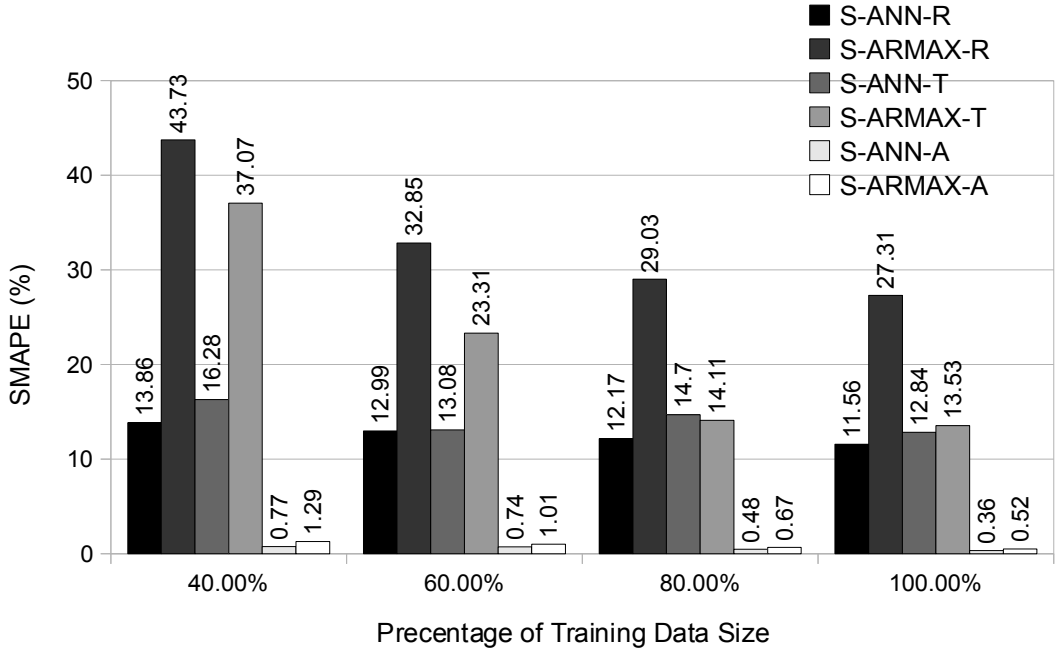

Figure 4.6: Sensitivity of Accuracy to Training Data Size.

### 4.4.5 Overhead

Finally, we examine the overhead to train our sensitivity-aware QoS models with all the training data points. We observed that the time taken to calculate relevance driven primitives selection is rather trivial in our case. From Figure 4.7, we can clearly see that both models have training overhead less than 10s while S-ARMAX is relatively more efficient in all cases as it has simpler structure. Therefore, the training overhead of our models is acceptable within the sampling interval of 30s.

## 4.5 Discussion of Limitations

It is clear that the experiment results have demonstrated the effectiveness of our QoS modelling approach. However, as we observed during the experiment process, there are still limitations for the modelling approach to become fully self-aware and further improve its self-adaptivity. These limitations are discussed as below:

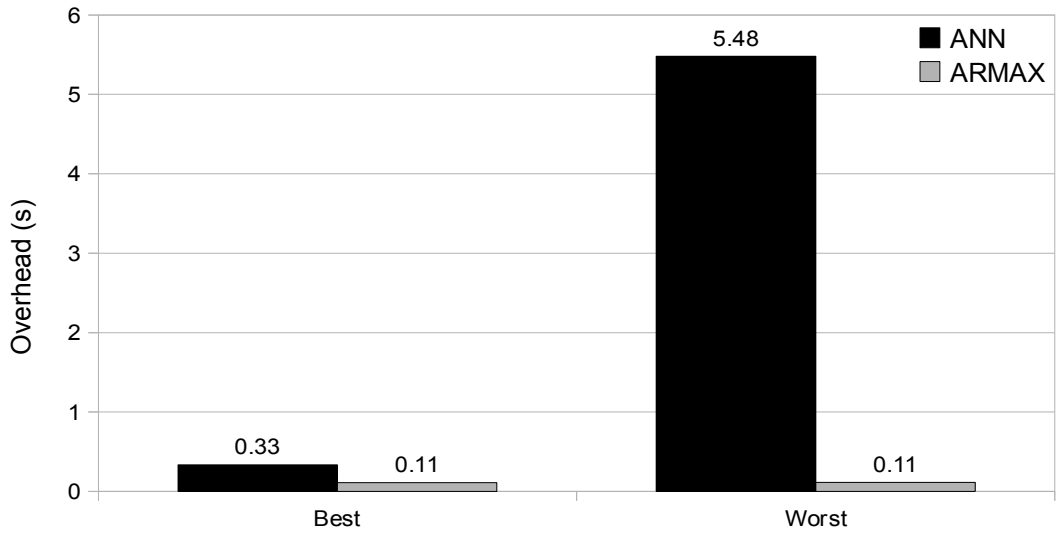

Figure 4.7: Overhead of QoS Modelling.

- Since the proposed QoS modelling approach in this chapter is specifically tailor for handling the QoS interference at the service level, it has not considered the QoS interference caused by co-hosted VM on a PM. This limitation can restrict the application of the approach in scenarios where resource contention becomes a major issue at the virtulization level.

- We have leveraged on relevance driven technique for selecting the primitives as model inputs. However, we subsequently observed that this technique constantly results in a large number of selected primitives in the QoS model. This will result in an overcomplicated model, which can cause the following problems: firstly, it renders the decision for selecting the right elastic autoscaling decision difficult when using the model. Secondly, certain learning algorithms can easily over-fit the model and thus affect its accuracy.

- As we have discussed in Section 4.4.3, the experimental results suggest that dif-

ferent learning algorithms tend to perform quite differently depending on the QoS fluctuation trends and primitives combination. This result indicates that given the generality of the proposed QoS model, the single learning algorithm is limited as we can not determine which algorithm to use without expensive and extensive analysis. In addition, even such process is performed, the offline analysis can still become invalid at runtime.

In the next chapter, we propose an improved QoS modelling approach with the aim to overcome these limitations and the related challenges.

## 4.6  Conclusion

In this chapter, we have proposed a novel self-aware and self-adaptive QoS modelling approach grounded on symmetric uncertainty and two machine learning algorithms, AR-MAX and ANN to reach two formulations of the QoS model. These techniques provide the foundations to enable self-awareness for QoS modelling in the cloud. In this way, the approach can capture the dynamics of QoS sensitivity by determining *which*, *when* and *how* primitives correlate with QoS at runtime. In addition, we cater for QoS granularity, QoS sensitivity and QoS interferences. Our approach considers fine-grained model as well as both software and hardware control primitives. By mapping to the self-awareness capabilities, we have implemented our approach as a independent component that adaptively creates fine-grained QoS models. We have experimentally evaluated our approach with respect to accuracy, sensitivity to data size and efficiency using the RUBiS benchmark and the FIFA 1998 workload trends. The results reveal that our approach is effective and produces better accuracy as when compared with the conventional models in various cases. Experiments also imply that the proposed S-ANN tends to be more accurate than S-ARMAX, when the fluctuation of QoS increases. On the other hand, S-ARMAX tends to be better when the QoS trends is relatively stable.

As mentioned in section 4.5, there are challenges remaining for QoS modelling in the cloud, in the next chapter, we report on an improved approach and demonstrate the necessity and effectiveness of such improvement.

# Chapter 5

# Improved QoS Modelling using Hybrid and Adaptive Multi-Learners

## 5.1 Introduction

Modelling QoS for cloud-based services is an important and necessary, yet challenging task. As mentioned in Chapter 2 and 4, existing *static* (i.e., analytical and simulation based) or *semi-dynamic* modelling approach (i.e., those that rely on manual and static primitives selection) lack in their ability to handle dynamics and uncertainty in the cloud. In particular, they tend to be limited in dealing with QoS interference, primitives selection and QoS function training. As one of the key contributions of this thesis, in Chapter 4, we have proposed and discussed a self-aware, self-adaptive and fine-grained QoS modelling approach in the cloud. This solution and many other existing modelling approaches, e.g., [90] [107], are called single learner-based as they apply single primitives selection technique and learning algorithm to model QoS. We have also demonstrated that the proposed approach in Chapter 4 is able to achieve better accuracy in contrast to existing semi-dynamic modelings. However, our subsequent investigations have revealed several limitations of such approach. Firstly, it ignores the QoS interference caused by VMs co-hosted on the same PM [116]. Secondly, the single learner-based approach can easily result in a QoS model with large numbers of inputs due to its over-sensitive nature when handling QoS interference. This will unnecessarily complicate the model and downgrade the prediction accuracy. Thirdly, we have observed that different learning algorithms (e.g., ANN and ARMAX) can be suitable only for certain QoS trends; however, a single learner-based approach requires the engineers to predetermine the suitable learning algorithm. This can entail manual and extensive investigation rendering it as an expensive process. Moreover, a predetermined approach does not cater for unexpected or envisioned changes in QoS at runtime.

### 5.1.1 Motivations and Challenges

As mentioned in Chapter 4, the improvements of the QoS modelling approach have been motivated by our subsequent investigation from the experiments. Specifically, the motivations and the new challenges can be discussed as the following:

   ***Dynamic and Uncertain QoS Interference at both service- and VM-level:*** We have classified QoS interference into two categories: the co-located service interference and co-hosted VM interference. The former is an inherent issue from the traditional cluster computing, where multiple applications/services running on the same operating system can suffer contention on the shared memory/cache, and therefore cause interference [50]. This is also true for multi-core systems [112]. The latter, on the other hand, is a significant unique problem in cloud computing, where virtualization has been used as the basis. This is because in such scenario, certain aspects of the underlying infrastructure are shared amongst the co-hosted VMs on a machine (e.g., last level cache of CPU and memory bandwidth), henceforth it can result in contention and create the chances for interference, as evident by many recent work [116] [91] [86] [99]. Given that it can be extremely difficult to completely eliminate QoS interference or it can be too expensive to do so [116], it is crucial to consider and handle the interference when modelling QoS in the cloud. In particular, it is important to consider the QoS interference at both service- and VM level as they influence QoS in a *dynamic* and *joint* manner. Here, the challenge lies in the difficulty to dynamically incorporate information about interference in the modelling, especially when the QoS interference is dynamic and uncertain in nature—it is difficult to know when contention would occur and what the degree of such contention is. However, as we have extensively surveyed in Chapter 2, existing work either consider co-hosted VM interference only (e.g., [116]) or completely ignore QoS interference (e.g., [90]) relying on the assumption that interference would never occur, which is unrealistic. As a result, modelling QoS without considering QoS interference at both service and VM level can downgrade the accuracy and/or lead to incorrect autoscaling decision.

***Dynamic and Uncertain Primitives Selection:*** In Chapter 4, we have already discussed the dynamic and uncertain nature of primitives selection. We have proposed a relevance driven technique to cope with this problem. Given that the selected inputs have a great impact to the model accuracy (as we will show in Section 5.4), it is important to select a right set of primitives. In particular, too limited inputs may not provide enough information of relevance to the QoS (i.e., the information that drives the changes in QoS), thus the accuracy may be insufficient. In addition, limited inputs of a model implies that it is not intuitive and can not be used in many scenarios. On the other hand, too many inputs can generate noise in the modelling (as we observed in the previously proposed relevance driven technique), because it introduces irrelevant information and large redundancy in the inputs (i.e., the same information has been provided by more than one selected primitives, thus it becomes noise), this will downgrade the model accuracy [110]. Moreover, it can over-complicate the model, causing difficulties in its application. Though some machine learning algorithms are proven be be resistant to noise e.g., [132], we believe that the benefits gained from primitives selection is vast, e.g., improved accuracy, more intuitive model and faster modelling time. The challenge here is how to dynamically select the ***most significant*** set of primitives as inputs, which provides good model accuracy and adequate complexity. In addition, incorporating the information of QoS interference makes this challenge even more difficult, because the control knobs and environmental conditions of neighbours services and VMs can become possible inputs in the model.

Nevertheless, as we have extensively surveyed in Chapter 2, existing static and semi-dynamic approaches for QoS modelling in the cloud rely on fixed and manual analysis to select the primitives as inputs, which are often offline. It is easy to see that a thoroughly offline and manual analysis would require extensive and careful human intervention by taking every permutation of the primitives into account. Therefore, one trick that has been widely applied is to reduce the possible primitives space based on empirical observations and domain specific assumptions, e.g., most work [116] [90] consider only hardware resources. However, this may mislead the QoS modelling and downgrade accuracy as it can ignore some highly relevant features, e.g., the software configurations, which can interplay with the hardware provision to influence QoS; these are often the primary causes of QoS violations [26] [139] [99]. In addition, ignoring QoS interference can result in inaccurate models. Even though the offline and manual selection is achieved at a good accuracy, the runtime dynamics and uncertainties can become a problem as there is no guarantee that the selected primitives are the best for the entire service life time. Until recently, few techniques [133] [26], including our relevance driven technique presented in Chapter 4, have been proposed for self-adaptive primitives selection in the cloud. However, they implicitly tackle redundancy and regard each primitive equally in the selection. We refer these techniques as single-learner based in the remaining of this thesis. In Section 5.6, we will show why these single-learner based techniques tend to be limited in accuracy.

***Dynamic and Uncertain Suitability of Learning Algorithms:*** As we have surveyed in Chapter 2, the application of machine learning algorithms for QoS function training in many existing approaches, including the approach presented in Chapter 4, are mostly single-learner based, providing that they rely on a single learning algorithm. Nevertheless, as we mentioned in Chapter 4, we observed that a single learning algorithm can be suitable only for certain QoS trends, and such suitability is dynamic and uncertain in nature. Consequently, a significant drawback of these approaches is that, for any given scenarios, they require the engineers to predetermine the suitable learning algorithm. This can entail manual and intensive investigation rendering it as an expensive process. Moreover, a predetermined algorithm does not cater for unexpected or envisioned changes in QoS at runtime. Now, the challenge becomes how to efficiently and dynamically determine the best learning algorithm for different scenarios.

All these challenges exhibit high complexity, runtime dynamics and uncertainties, which urge the need for a fully dynamic, accurate and self-adaptive QoS modelling approach that continually evolves itself in the cloud.

## 5.1.2 Contributions

In this chapter, we propose an improved QoS modelling approach to overcome the above limitations and challenges using hybrid and adaptive multi-learners, which are the key enablers of realising better self-awareness for QoS modelling in the cloud. The approach is capable to dynamically select only the most significant set of primitives as model inputs; while improving the model accuracy. Furthermore, it adaptively selects the most suitable learning algorithm for training the QoS function given a QoS trend and inputs combination. In addition to ANN [119] and ARMAX [22], we further introduce RT [117] as one of the candidate learning algorithm. This is because the tree structure of RT [117] distincts itself from the other two learning algorithms, hence introducing extra variability in the model selection process. All these techniques improve the self-awareness capabilities of the QoS modelling approach. The experiment results reveal that our approach is overall more accurate, more stable and reduce the error quicker than the other approaches; while generating acceptable overhead.

## 5.2   Partitioning of Cloud Primitives

We adopt the same cloud system model, assumptions and the generic QoS model as described in Chapter 4. To improve accuracy and prevent noises, selecting the right primitives as inputs is critical for QoS modelling in the cloud. However, the difficulty is that the primitive inputs, which are relevant and useful for modelling QoS in the cloud tend to be dynamic. In such context, possible inputs of a QoS model can be the primitives that tend to directly influence the QoS (e.g., the threads of the corresponding service-instance); it can also include the primitives that belong to the co-located service-instances and the co-hosted VMs; As discussed in Chapter 4, all possible primitives inputs for modelling the QoS attributes of a service-instance form a space, which we call **possible relevant primitives space**. To incorporate the QoS interference caused by co-hosted VMs, we additionally extend the possible relevance primitives space for each service. The extension is described as follow:

*Rule 5.1. A primitive belongs to the possible relevant primitives space for modelling the QoS of $S_{ab}$ if it can be classified into one of the following groups:*

1. *It is a software control or environmental primitive of $S_{ab}$.*

2. *It is a hardware control primitive of the VM that runs $S_{ab}$.*

3. *In case of $S_{ab}$ has direct functional dependency on $S_{cd}$, it is a software control or environmental primitive of $S_{cd}$.*

4. *In case of $S_{ab}$ has direct functional dependency on $S_{cd}$, it is a hardware control primitive of the VM that runs $S_{cd}$.*

5. *It is a software control or environmental primitive of $S_{cd}$, which is co-located with $S_{ab}$ on the same VM.*

6. *It is a hardware control primitive of the VM, which is co-hosted with the VM that runs $S_{ab}$ on the same PM.*

Rule 5.1 distincts from Rule 4.1 in the fact that it contains an additional condition 6, which aims to capture the information about QoS interference at the VM level. The problem here is how to select on-the-fly a right subset of primitives from the space as the inputs of QoS models. Instead of selecting only the relevant primitives as mentioned in Chapter 4, we intend to select the most significant primitives. The aim is to improve the model's accuracy by taking both relevance and redundancy of the primitives into account. In Section 5.4, we will present detailed analysis and solution for selecting the right primitives.

## 5.3 Improving Architecture for QoS Modelling in the Cloud

Given that we introduce more requirements for self-awareness in the QoS modelling process, the architecture is required to be improved and the related new components needs to be mapped to self-awareness capabilities. As shown in Figure 5.1, the approach is again realised as middleware using autonomic architecture with a feedback loop. The service-instances running on the VMs of a PM are managed by a dedicated Middleware Instance (MI), which is attached to the root domain (e.g., Dom0 [6]) of this PM. Each MI is self-adaptive as the feedback loop runs continually to keep the models updated.

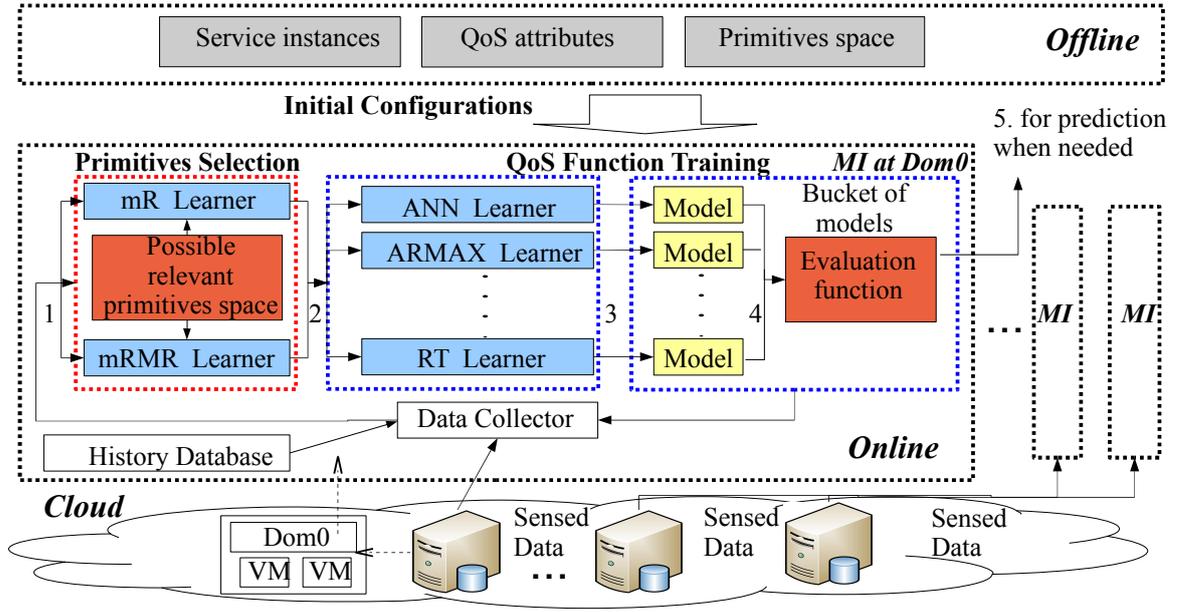

Figure 5.1: The Improved QoS Modelling Architecture.

Our approach is designed for online scenarios; the only offline preparation is to define the current service-instances, their QoS and classification of the primitives in the spaces (i.e., using Rule 5.1). This preparation can be easily done by the cloud engineers and it should be updated accordingly if changes occur. The approach can be also used offline in situations where conducting offline modelling in advance can be beneficial to the online models. Within the feedback loop, *Data Collector* continually monitors and stores sample-values of QoS and primitives from the service-instances/VMs of a PM, and those from the other PMs in the presence of functional dependency. This can be achieved by accessing the cloud sensors or log files. It is worth noting that the modelling interval can be longer than the sampling interval; that is to say, the frequency of data collection do not need to be the same as the frequency of modelling, in which case the sampled data can be stored in a history database and retrieved when needed.

Upon each modelling interval, for each QoS attribute of a service-instance, all historical data is then passed to the primitives selection phase for determining which and when primitives correlate with QoS at runtime (step 1). Here, we have used two learners to select primitives from two sub-spaces as motivated by our analysis in Section 5.4. At step 2, the selected sets of primitives are combined and sent to the QoS function training phase, where multiple learners are used to model how the primitives correlate with QoS online (step 3). At step 4, each QoS attribute is associated with a bucket of models produced by candidate learners and an evaluation function; in addition, the weights in the evaluation function will be updated. This bucket can be then used by, e.g., a *Decision Maker* for performing prediction at any time (step 5). Upon prediction when given a set of inputs, the evaluation function is used to select the best model in the bucket (see Section 5.5).

Table 5.1: The Mapping Between the Additional Meta-Self-Awareness Capability and the Sub-Components.

| Self-Awareness Capability | Component | Description |
|---|---|---|
| Meta-self-awareness | QoS Function Trainer | Knowing the suitability of candidate learning algorithms. |

As one of the extensions to the architecture described in Chapter 4, we incorporating meta-self-awareness in the QoS modelling process. Such self-awareness capability is particularly helpful for reasoning at the meta-level, i.e., to determine which is the best learning algorithm. In addition to Table 5.1, the mapping between the sub-components of the improved *QoS Modeller* and the meta-self-awareness capability is demonstrated in Table 5.1.

## 5.4   Primitives Selection in the Cloud

Recall from (4.1) and (4.2), to dynamically model $QoS_k^{ij}(t)$ at runtime, the first challenge is to determine *which* and *when* the underlying primitives should be included as column entries in $SP_k^{ij}(t)$ for the QoS modelling.

One straightforward solution to the primitives selection problem is to search the best set of primitives using a given learning algorithm that guarantee to produce the best accuracy for the said algorithm; this is regarded as the *wrapper* approach [80]. Nevertheless, given that the learning algorithm needs to be run many times during the selection process, it is clear that such approach can introduce large overheads in terms of both resource and latency. As a result, the wrapper approach is not well-fit for online QoS modelling in the cloud. In this work, we focus on an alternative approach that is more efficient and capable to select primitives independent of the learning algorithms, namely the *filter* [80].

We define two main objectives for the primitives selection, namely, selecting relevant primitives and selecting useful primitives from the possible relevant primitives space. The former try to select all relevant primitives, which can be easily achieved; while the later aims to further select an even better set of primitives that has a right balance between relevance and redundancy, which improves the model accuracy.

In this section, we present a set of experimental analysis on the relevance and redundancy of selected cloud primitives in relation to model accuracy. Subsequently, driven by the observations gained from the conducted analysis, we propose a self-aware, self-adaptive and online solution for primitives selection, namely hybrid dual-learners.

### 5.4.1   Quantifying Relevance and Redundancy

To quantify the relevance of a primitive to the QoS and the redundancy between a pair of primitives, we have used Symmetric Uncertainty (SU), which is a fundamental concept in information theory [130]. As presented in Chapter 4, by using (4.3), it is straightforward to measure the relevance of a primitive to QoS. As for redundancy, we consider it as the relevance between a pair of primitives, which can be also easily quantified via (4.3). It is known that SU can provide correct information about the relative effects of relevance and redundancy for a pair of individual primitives on the model accuracy [110].

Nevertheless, the single SU value and pair-wised comparison are insufficient for selecting useful primitives as they cannot consider both relevance and redundancy simultaneously in the selection. In addition, it cannot properly quantify the effects of combinatorial relevance and redundancy to model accuracy for a whole set of selected relevant primitives. This means given two sets of selected relevant primitives, such comparison cannot determine which set will produce better model accuracy during the selection. Our problem requires a measurement that copes with those issues. As a result, we need to study and select useful primitives by comparing the cumulative representation of relevance and redundancy for any possible sets of selected relevant primitives.

There can be two forms of cumulative representation: firstly we can consider multivariable density function for a given set of selected relevant primitives, in which case (4.3) would be changed into the following formula [110]:

$$U(X_1, X_2...X_n, Y) = \frac{2 \times I(X_1, X_2...X_n, Y)}{H(X_1, X_2...X_n) + H(Y)}, \ X_n \in S \qquad (5.1)$$

where $[X_1, X_2...X_n]$ denotes vectors of $n$ different primitives that has been selected; and $S$ denotes the set of selected primitives. (5.1) expresses both relevance and redundancy as they can be handled by the multivariable density function. However, this method has some serious drawbacks: (i) the number of online data samples can be insufficient for correctly calculating the probability and (ii) the multivariate density estimation often involves computing the inverse of the high-dimensional covariance matrix, which is computationally expensive and thus it is an ill-suited solution in our case. Alternatively, we can compute the cumulative SU values of relevance and redundancy. By cumulative SU values, we refer to the cumulative combination (i.e., total or average) of the single SU values for the primitives in a given set of selected relevant primitives [110]. An example of relevance is shown below:

$$\text{Relevance of a selected set} = \sum_{X \in S}^{n} U(X, Y) \tag{5.2}$$

This cumulative combination involves a bi-variable density function only and thus it is more appropriate for filtering at runtime. In addition, it is highly intuitive and the nature of cumulative combination implies its light computational efforts. The cumulative representation for redundancy can be similarly applied. In this work, we call these cumulative representations as cumulative relevance and cumulative redundancy.

Recall that in selecting useful primitives, we aim to improve the model accuracy by balancing the relevance and redundancy of selected primitives. With this in mind, it is easy to see that even if we incrementally select (add) the relevant primitives one at a time, the validity and usefulness of cumulative relevance and redundancy rely on the following assumption (in the next subsection, we will experimentally verify this assumption):

*Assumption 5.1. For QoS modelling in the cloud, the model accuracy, represented by error, is negative to the difference between cumulative relevance and redundancy if the cumulative relevance is bigger (i.e., the bigger the difference, the smaller error); or being positive to such difference if the cumulative redundancy is bigger (i.e., the bigger the difference, the bigger error).* □

Indeed, if this assumption does not hold, it means that the cumulative SU values cannot correctly differentiate and quantify the effects of some relevant primitives to the model accuracy, and this will significantly mislead the selection process.

In the following, we report on a set of experimental analysis, which are about how the relevance and redundancy of selected primitives affect the model accuracy for QoS modelling in the cloud.

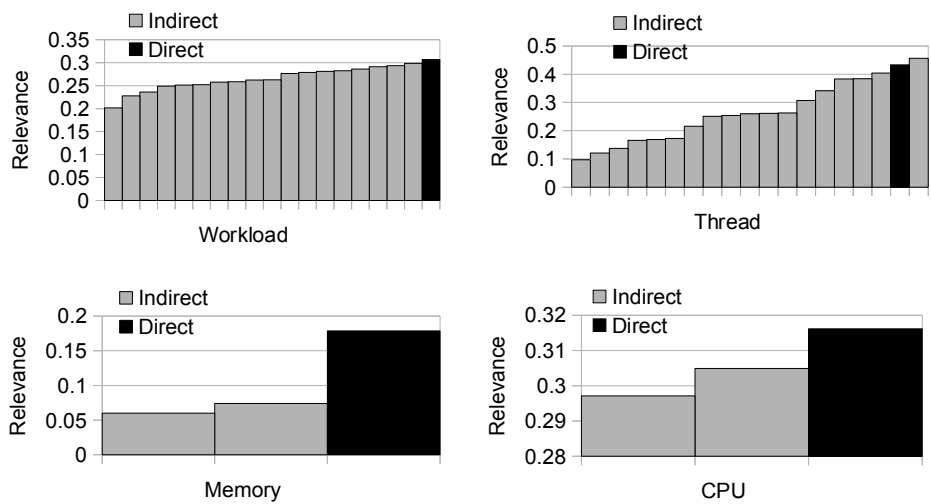

Figure 5.2: The Relevance of Different Cloud Primitives to Response Time for the Exampled Service-Instance.

## 5.4.2 Relevance and Redundancy Analysis on Primitives Selection

To study the correlation of selected primitives to the accuracy for modelling QoS in the cloud, we have conducted several analysis on the relevance and redundancy of selected primitives by means of experiments (see Section 5.6 for the detailed setup). In particular, we have carefully analysed the relevance between possible primitives and QoS from the experiments—we first select the relevant primitives and then we rank them based on their relevance to the QoS. We found that the only constant observation across many QoS attributes and service-instances is that for each feature dimension (i.e., thread, CPU, Memory and Workload), certain primitives are more relevant to the QoS than all or most of the others. As an example, Figure 5.2 shows the relevance (measured by (3)) for Response Time of a service-instance for different feature dimensions, calculated by averaging the values from all 350 intervals in one run. We discovered that the more relevant primitives are the ones that can directly influence the corresponding service (dark bars), e.g., the thread of the service and CPU of the VM; on the other hand, the less relevant primitives are the ones that can only interfere the service and its QoS via contention (light bars), e.g., the thread of co-located service and CPU of co-hosted VM. Such observation indicates that the former is more important to the QoS than the latter as QoS interference can only occur when the contention is quite significant [116]. This fact motivates us to partition the possible relevant primitives spaces into two sub-spaces, namely the direct primitives space and the indirect primitives space. By leveraging the classifications in Rule 5.1, the former is defined by:

*Rule 5.2. A primitive belongs to the direct primitives space for modelling the QoS of $S_{ab}$ if it is in group 1,2,3 or 4 from Rule 5.1.* □

It is clear to see that the direct primitive space contains primitives that can directly influence the QoS, which means they tend to provide different aspects of information. On the other hand, the indirect primitives space contains information about the QoS interference. Consequently, the indirect primitive can be defined as:

*Rule 5.3. A primitive belongs to the indirect primitives space for modelling the QoS of $S_{ab}$ if it is in group 5 or 6 in Rule 5.1.* □

It is worth noting that the indirect primitives space should generally be larger than the direct primitives space as it is sensitive to the number of co-located service and co-hosted VMs, which can be expended largely in the cloud. It is possible that both direct and indirect primitives space have irrelevant primitives, but as mentioned, these can be easily eliminated.

We have experimentally verified the validity of Assumption 5.1 for direct and indirect primitives spaces, as specified in Appendix B. Particularly, we have observed that Assumption 5.1 is true for intra-indirect primitives space. However, for inter- direct and indirect spaces, this assumption does not hold; for intra-indirect primitives space, this assumption is invalid either. We believe the reason being is that Assumption 5.1 can be easily violated when there are certain primitives providing different aspects of information to the QoS (e.g., different dimensions of primitives and QoS interference). However, the assumption is effective for the primitives purely contain information about QoS interference, i.e., those in the indirect primitives space. Therefore, a single-learner based technique (i.e., considering all primitives equally) tends to be insufficient for the primitives selection as it will mislead the selection process. We have also found that the best accuracy is achieved by the combination of direct and indirect primitives, which prove the importance of considering QoS interference in the modelling.

All these facts urge the need for a self-aware, self-adaptive and online primitives selection for modelling QoS in the cloud, which we address in this chapter. Given that the cumulative relevance and redundancy can mislead the selection when Assumption 5.1 does not hold and the fact that it is very difficult to efficiently handle the selection without cumulative representations, we have decided to avoid the misleading selection by partitioning the space. To better tackle the problem of relevance and redundancy in primitives selection, we intend to partition the primitives that provide different aspects of information to the QoS into sub-spaces, and select the useful primitives from each sub-space independently using cumulative relevance and redundancy.

### 5.4.3   The Hybrid Dual-Learners for Primitives Selection

To adaptively and dynamically select primitives as the model inputs online, we design a runtime filtering mechanism based on symmetric uncertainty, which has the advantage to assess the effects of selected primitives on model accuracy without actually training a model. Based on the analysis in Section 5.4.2, we use multi-learners in order to avoid the aforementioned issues caused by single-learner based technique. In particular, we partition the primitives that provide different aspects of information on the QoS into sub-spaces;

this will result in *k+1* partitions, where *k* is equal to the number of primitives in the direct primitives space; while the remaining one partition refers to the indirect primitive space. The objective is to select useful primitives from each sub-space independently using dedicated learners and then produce an ensemble results as the selected inputs for modelling. By doing so, we aim to produce a model with adequate model complexity and improved accuracy.

Inspired by [110], for each sub-space, we formalise a maximal Relevance Minimal Redundancy (mRMR) learner using cumulative relevance and redundancy. The objective of this learner is to continually select the primitives that maximise:

$$max \ \Phi(S, Y), \quad s.t. \ U(X, Y) > 0, X \in S \qquad (5.3)$$

where $X$ corresponds to the value vector of a primitive and $Y$ to the value vector of QoS attribute. $S$ denotes the associated sub-space. Mathematically, the objective function $\Phi$ can have four possible variations, depends on whether we use total or average to represent cumulative SU values; and whether we apply multiplicative or additive formulation to represent the difference between cumulative relevance and redundancy. Specifically, we obtain several variations of the objective function in (5.3):

$$\textit{Total and multiplicative:} \qquad \frac{\sum_{X \in S}^{n} U(X,Y)}{1 + \sum_{X,X' \in S} U(X,X')} \qquad (5.4)$$

$$\textit{Average and multiplicative:} \qquad \frac{\sum_{X \in S}^{n} U(X,Y) \times (n-1)}{n^2 - n + 2 \times \sum_{X,X' \in S} U(X,X')} \qquad (5.5)$$

$$\textit{Total and additive:} \qquad \sum_{X \in S}^{n} U(X,Y) - \sum_{X,X' \in S} U(X,X') \qquad (5.6)$$

$$\textit{Average and additive:} \qquad \frac{\sum_{X \in S}^{n} U(X,Y)}{n} - \frac{2 \times \sum_{X,X' \in S} U(X,X')}{n^2 - n} \qquad (5.7)$$

where $X'$ is the value vector of another primitive. $n$ is the number of primitives, which has been already selected; U is the function of symmetric uncertainty in (4.3). It is clear to see that the constraint filters all the irrelevant primitives and this can be done easily. In this work, we apply incremental random search to optimise these functions for simplicity; however, it can be easily replaced by more sophisticated algorithms (e.g., hill-climbing, integer optimization and evolutionary algorithm). In Section 5.6, we will experimentally compare these variations.

Given that the Assumption 5.1 does not hold in indirect primitive space, we apply dedicated mRMR learner for each sub-space independently. However, because there is only one primitive exist for each $k$ sub-space, the objective here is equivalent to select the relevant primitives from all $k$ sub-spaces, therefore these sub-spaces can be merged into one and the multiple mRMR learners can be simplified to the following single maximal Relevance (mR) learner. The objective of this learner is to continually select the primitives that maximise:

$$max \ \Psi(D, Y), \Psi = \sum_{X \in D}^{n} U(X, Y), \ s.t. U(X, Y) > 0 \qquad (5.8)$$

where $D$ denote the associated direct primitive space all the notations are the same as (6). Again, the constraint filters all the irrelevant primitives.

As for indirect primitives space, we use a mRMR learner for this sub-space, given that the Assumption 5.1 is true and the indirect primitive space tends to provide the same aspect of information to the QoS.

Eventually, we only need to partition the possible relevance primitives space into two sub-spaces, each of which employs learners with different primitives selection techniques (i.e., mR learner and mRMR learner). The final results are combined to form the selected useful primitives. We call this as the hybrid dual-learners technique.

---

**Algorithm 1** Hybrid dual-learners for primitives selection

---

**Inputs:**

given the value vector $Y$ of a QoS attribute $QoS_k^{ij}$ , the associated direct primitives space $D$ and indirect primitives space $ID$

**Declare:**

$C_{direct}$ - the collection of selected direct primitives

$C_{indirect}$ - the collection of selected indirect primitives

**Outputs:**

the column entries of the selected primitives matrix $SP_k^{ij}(t)$

1: **start primitives selection**
2: $C_{direct} := \emptyset, \ C_{indirect} := \emptyset,$
3: $C_{direct} := argmax \ \Psi(D, Y) = \sum_{X \in D}^{n} U(X, Y), \ s.t. U(X, Y) > 0$
4: $C_{indirect} := argmax \ \Phi(S, Y) = \frac{\sum_{X \in S}^{n} U(X, Y)}{1 + \sum_{X, X' \in S} U(X, X')}$ if (5.4) is applied.
5: **end primitives selection**

---

An algorithmic description of the primitives selection phase is illustrated in Algorithm

1. We will show (in Section 5.6) that the proposed hybrid dual-learners technique leads to better accuracy as when compared to other single-learner based and manual solutions.

## 5.5  QoS Function Training in the Cloud

Recall from (4.1), once the primitives in $SP_k^{ij}(t)$ are selected, our next goal for QoS modelling is to determine how those primitives correlate with $QoS_k^{ij}(t)$ in the QoS function $f_k^{ij}(t)$.

Existing work has considered variety of learning algorithms for QoS function training, ranging from simple linear model [107] to complex nonlinear ones [90]. These algorithms are self-adaptive and dynamic in nature thus they are capable to deal with dynamic and uncertain magnitude of primitives in the correlation. In this section and by means of experiments, we study the accuracy of the most widely used single learning algorithms (i.e., ANN, ARMAX and RT.) for QoS modelling in the cloud. We assess the accuracy of the learning algorithms over four different QoS attributes—Response Time, Throughput, Reliability and Availability (see Section 5.6.1 for their detailed definitions). Finally, we present a self-aware, self-adaptive and online solution for QoS function training, namely adaptive multi-learners, to address the issues discovered.

### 5.5.1  Sensitivity-Aware Regression Tree

To increase variability in the QoS function training, Regression Tree (RT) [117] is the final learning algorithm we have considered in this thesis. A possible structure of RT has been shown as below:

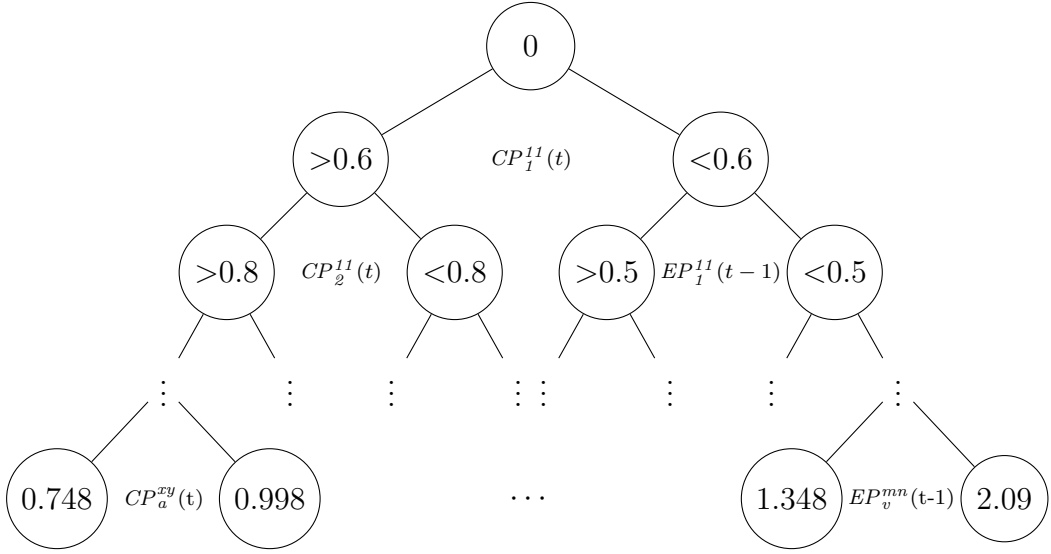

$$subject\ to \qquad CP_a^{xy}(t), EP_b^{mn}(t-1) \in SP_k^{ij}(t) \qquad (5.9)$$

RT is a learning algorithm that maps the relation of primitives and QoS into a tree-like structure, in which leaves represent class labels and branches express conjunctions of features to reach these labels. The tree is trained using the Classification and Regression Trees (CART) technique [24] and we found that use $q=1$ (i.e., no time series information) can produce the optimal results. By using the primitives selected from the primitives selection phase as model inputs and continually updating the RT models online, we achieve a sensitivity-aware RT, which is similar to our design for the sensitivity-aware ANN and ARMAX.

## 5.5.2 Suitability Analysis of Learning Algorithms On QoS Function Training

For simplicity of exposition, we illustrate the results for a service-instance for the three learning algorithms over the four QoS attributes. We have used the variation in 5.4 for primitives selection. To better interpret the result with respect to different trends of QoS attributes, we apply Relative Standard Deviation (RSD) to measure the fluctuation of the QoS in a relative manner, calculated as: RSD = $\sigma/\mu$, where $\sigma$ is the standard deviation and $\mu$ is the mean of all measured QoS values. We can observe from Table 5.2 that the RSD value of the QoS attribute can be sorted by the following ascending order: Availability, Reliability, Throughput to Response Time; this means the trend of Response Time being the most fluctuated one. At the other extreme, the trend of Availability being the most stable one. As shown in Figure 5.3, we can clearly see that the accuracy achieved by a learning algorithm differs significantly from case to case—ANN is the best for Response Time and Throughput while the ARMAX is the best for Reliability and Availability. In particular, the results of ARMAX reduces the error to 0.03% for Reliability and Availability; while ANN tends to be significantly better than ARMAX for Response Time and RT for Throughput. Beside, even though RT perform the worst for most of the cases, it can still largely reduce the error in contrast to ARMAX at the case of Response Time.

An interesting discovery is that, if we interpret the results from Figure 5.3 in conjunction to the RSD of different QoS attributes, we can see that the ANN tends to perform better than ARMAX on Throughput and Response Time where the fluctuations of trend are relatively large; and this improvement tends to increase from Throughput to Response when the trend becomes more fluctuated. On the other hand, ARMAX tends to produce better accuracy than ANN on Reliability and Availability, where the fluctuations of trend are relatively small; and this improvement tends to increase from Reliability to Availability when the trend becomes more stable. These observations reveal that nonlinear model like ANN can better handle the dynamic and uncertain magnitude of primitives in the correlation leading to better accuracy when the fluctuation of the QoS increases, whereas the linear ARMAX produces less error as such fluctuation decreases.

Table 5.2: The Relative Standard Deviation of QoS for A Service-Instance.

| QoS Attribute | RSD |
|---|---|
| Response Time | 4.197056205058521 |
| Throughput | 0.6629050477197764 |
| Reliability | 0.011793827516559515 |
| Availability | 0.010681646537192316 |

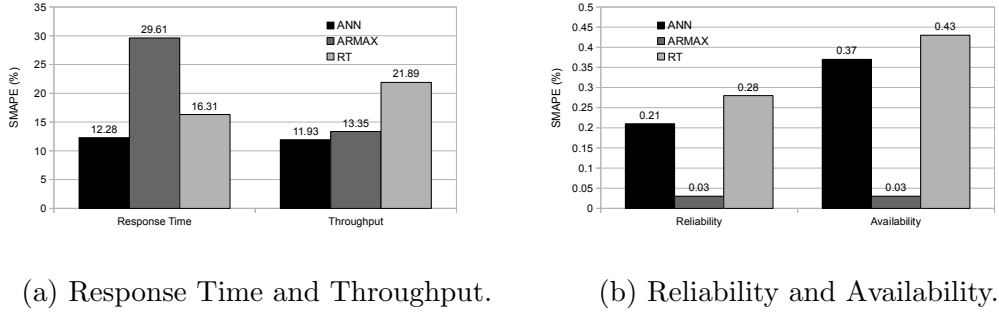

(a) Response Time and Throughput.

(b) Reliability and Availability.

Figure 5.3: The Model Accuracy of Each Learning Algorithm on Difference QoS Attributes.

All these experimental results suggest that the learning algorithms perform quite differently depending on the QoS fluctuation trends and primitives combination; henceforth, we cannot reach a conclusion that a certain algorithm is generally the best learning algorithm for QoS modelling in the cloud. This indicates that given the generality of the proposed QoS model, the single learner is limited as it is difficult to determine which learning algorithm to use without expensive and intensive analysis. In addition, even when such process is performed, the offline analysis can still become invalid at runtime. Therefore, it is desirable to build a self-adaptive mechanism that not only able to adaptively model the magnitude of selected primitives to the QoS, but also dynamically select the suitable algorithm based on the runtime trend of a QoS attribute.

### 5.5.3 The Adaptive Multi-Learners for QoS Function Training

Give the fact that most machine learning algorithms are self-adaptive and dynamic in nature, the crucial challenge here is how to adaptively determine the best learning algorithm for QoS function training. To this end, we propose an adaptive multi-learners technique for updating QoS function on the fly and predicting the QoS values. The technique has two main processes, namely training and prediction. At the training process, we simultaneously apply different learners to train the same QoS function , but each of the learners uses different learning algorithm to build a model. At the prediction process, we evaluate these learning algorithms by comparing the resulted models within the bucket on the fly; the model of the best learning algorithm is used to predict QoS.

One of the most critical design decisions is to determine the evaluation function that compares the models produced by candidate learners. The basic method would be based on global mean error of all historical samples. However as shown by Kundu et al. [90], given a set of primitive values as inputs, the most accurate model using these inputs might not be the one that has the best global error. This is because the accuracy of a model can be sensitive to the local construct of given input values, including the variation of possible combination, scale and granularity, etc. As a result, our evaluation function aims to compare both the local error of a given inputs set produced by a model and the global error of the said model. In this work, we have used SMAPE for measuring the error, but other metrics can be replaced easily.

An algorithmic description of the training process has been shown in Algorithm 2. At the training process, as the collected online data increases, we continually train two QoS models for each learner (line 2-5): (i) A *main-model* that uses 100% of the collected online data; (ii) A *sub-model*, which is trained based on 70% of the total collected data. The *sub-model* is used to test local and global error for its *main-model* of a learner. In particular, it tests the QoS prediction error against the remaining 30% testing data—the

---
**Algorithm 2** Training process in adaptive multi-learners
---
**Inputs:**
given the column entries of $SP_k^{ij}(t)$ from Algorithm 1 and a set of candidate learning algorithms
**Declare:**
$\langle M_{main}, M_{sub}, L \rangle$ - a vector of *main-model*, *sub-model* and the corresponding *local error pattern*
bucket - a collection of model vectors
**Outputs:**
a bucket of model vectors for a QoS attribute $QoS_k^{ij}$

1: **for each** candidate learning algorithm **simultaneously do**
2:     *find* the optimal number of row entries, i.e., the value of $q$ in 4.2 (Chapter 4), for $SP_k^{ij}(t)$ if it has not been predefined for this learning algorithm
3:     *train* main-model $M_{main}$ and sub-model $M_{sub}$ based on the required    inputs defend by $SP_k^{ij}(t)$
4:     *test* the sub-model for building local error pattern $L$
5:     bucket := bucket $\cup \langle M_{main}, M_{sub}, L \rangle$
6: **end for**
---

split of training and testing data follows standard machine learning approach for testing generalisation errors. These generalisation errors and their corresponding samples (i.e., the observed values of all selected primitives and QoS at each interval) within the testing data serve as the *local error patterns* of the *main-model*. Finally, the *main-model*, *sub-model* and local error patterns are put in a bucket.

An algorithmic description of the prediction process has been shown in Algorithm 3. The prediction process is triggered when there is need to perform prediction. In particular, the best main-model in the bucket is used as the final model to predict QoS. To calculate the local error of a *main-model*, we leverage on the prediction error of its *sub-model* for each sample within the testing data, as recorded in the local error patterns (line 3-9). When given a set of inputs (i.e., new values of the selected primitives) for predicting QoS, the local error of a *main-model* is determined by extrapolating the similarity between the given set of inputs and each sample from local error patterns; the error of the most similar sample is used as the local error (line 4-7). To this end, we apply symmetric uncertainty

**Algorithm 3** Prediction process in adaptive multi-learners

**Inputs:**
given a set of inputs $P$ and the bucket from Algorithm 2
**Declare:**
$S$ - the current sample
$S_{selected}$ - the most similar sample to $P$
$d$ - the distance between $P$ and the current sample
$d_{smallest}$ - the smallest distance between $P$ and a sample
$E_{local}$ - the local error of the current *main-model*
$E_{global}$ - the global error of the current *main-model*
$E$ - the final error of the current *main-model*
$E_{smallest}$ - the smallest final error of a *main-model*
$M_{selected}$ - the selected *main-model* for prediction
**Outputs:**
the predicted QoS value of $QoS_k^{ij}$

1: **start prediction**
2: **for each** $\langle M_{main}, M_{sub}, L \rangle$ in the bucket of $QoS_k^{ij}$(t) **do**
3:     **for each** sample $S$ in the *local error pattern* $L$ of $M_{sub}$ **do**
4:         *calculate* distance $d$ between $P$ and $S$ using 5.10
5:         **if** $d_{smallest} > d$ **then**
6:             $d_{smallest} := d$, $S_{selected} := S$
7:         **end if**
8:     **end for**
9:     *get* the error of $S_{selected}$ as the local error $E_{local}$ of $M_{main}$
10:     *get* the global error $E_{global}$ of $M_{main}$
11:     *evaluate* final error $E$ of $M_{main}$ using 5.11
12:     **if** $E_{smallest} > E$ **then**
13:         $E_{smallest} := E$, $M_{selected} := M_{main}$
14:     **end if**
15: **end for**
16: *predict(P)* using the selected *main-model* $M_{selected}$
17: **end prediction**

based Euclidean Distance to measure the similarity. As shown in (5.10), $d$ is the distance of the given set of inputs against a sample in the local error patterns.

$$d = \sqrt{\sum_{x \in X} (SU_x \times (p_x - p'_x)^2)}$$
(5.10)

$p_x$ and $p'_x$ respectively denote the value of $xth$ selected primitive in the given set of inputs and the value of the same primitive in a sample from local error patterns. $SU_x$ is the symmetric uncertainty value between the $xth$ primitive and the QoS attribute. The sample results in the smallest $d$ is the one that we are seeking, then its corresponding error is used as the local error of the *main-model* (line 9).

On the other hand, the global error of a *main-model* is the mean errors of all samples within the 30% testing data produced by its *sub-model* (line 10). Finally, the evaluation function selects the best *main-model* for a given set of inputs by examining on both the local and global error of all *main-models* in the bucket, as formally depicted in (13) (line 11-14).

$$E^i = \alpha \times E^i_{local} + \beta \times E^i_{global}$$
(5.11)

where $E^i$ , $E^i_{local}$ and $E^i_{global}$ denote the final, local and global error of the *ith* main-model respectively. $\alpha$ and $\beta$ are two heuristics expressing the relative importance of local and global errors. We have set the initial value of $\alpha$ and $\beta$ as 0.1, which means the local and global error are equally important from the beginning. The selected main-model and its learning algorithm for a given inputs is the one that has the smallest (line 16).

To capture the right weight of local and global errors, $\alpha$ and $\beta$ are updated via 5.12

when new data is collected.

$$\alpha = \alpha + \Delta\alpha, \quad \beta = \beta + \Delta\beta$$

$$s.t. \begin{cases} \Delta\alpha = e_{\alpha=0,\beta=1} - e_{\alpha=1,\beta=0} & \text{if } e_{\alpha=1,\beta=0} < e_{\alpha=0,\beta=1} \\ \Delta\beta = e_{\alpha=1,\beta=0} - e_{\alpha=0,\beta=1} & \text{if } e_{\alpha=1,\beta=0} > e_{\alpha=0,\beta=1} \end{cases}$$

(5.12)

Specifically, $e_{\alpha=1,\beta=0}$ is the prediction error of new data produced by the selected main-model when $\alpha = 1$ and $\beta = 0$. Similarly, $e_{\alpha=0,\beta=1}$ is the error produced by the selected main-model when $\alpha = 0$ and $\beta = 1$. In this way, the error that is more useful in the selection will gradually gain more importance. This updating process has been illustrated in Algorithm 4.

---
**Algorithm 4** Update $\alpha$ and $\beta$ in the evaluation function
---
**Inputs:**
newly measured values vector $P_{sample}$ and QoS value $y$ for a QoS attribute of selected primitives $QoS_k^{ij}$

1: **start update when newly data is available**
2: $predict(P_{sample})$ using Algorithm 3 when $\alpha = 1, \beta = 0$
3: $predict(P_{sample})$ using Algorithm 3 when $\alpha = 0, \beta = 1$
4: $calculate$ the errors of the predicted values from step 2 and 3 against $y$
5: $calculate$ $\Delta\alpha$ and $\Delta\beta$ using (5.12)
6: $update$ $\alpha$ and $\beta$ using (5.12)
7: **end update**
---

As mentioned in Section 5.5, we employ three different learning algorithms (i.e., AR-MAX, ANN and RT) in the adaptive multi-learners. Our technique is flexible as new algorithms can be added or old algorithms can be removed/substituted if needed.

## 5.6 Experiments and Evaluations

To evaluate our modelling approach, we experimentally benchmark our results against other single-learner and manual techniques. Specifically, the primary intention of the experiments is to validate the approach against the following criteria:

- **Accuracy:** By comparing with various other state-of-the-art modelling approaches, we intend to examine whether the hybrid dual-learners and the adaptive multi-learners can achieve better accuracy.

- **Stability:** We intend to assess the stability of the accuracy achieved by our approach under different scenarios, i.e., different QoS attributes and learning algorithms, in contrast to the other competitors.

- **Sensitivity of accuracy to online data size:** We examine the sensitivity of accuracy of the proposed approach to the available online data size. The purpose is to evaluate how quick the model accuracy changes with respect to the increase in data size.

- **Overhead:** We intend to evaluate the overhead of our approach in terms of the latency in the modelling, for both the primitives selection phase and the QoS function training phase.

In addition to the assessment of accuracy under different QoS attributes and/or learning algorithms using SMAPE, we also intend to examine the overall accuracy and stability for all the considered scenarios. However, given the assumption that the scenarios are equally important, simply calculate the average or sum of all SMAPE can mislead the results. This is because different QoS attributes produce different scale of the prediction error, e.g., the error for predicting Throughput tends to be much larger than that for Reliability; therefore a technique/learning algorithm that performs better for Throughput will more likely to dominate the overall results. To cope with this issue, we use the summation of normalised SMAPE to illustrate the overall accuracy of a competitor, as shown below:

$$Overall\ Accuracy = 100 \times \sum_{i}^{n} \frac{e_i}{e_{i,mean}} \qquad (5.13)$$

whereby $e_i$ is the SMAPE of a competitor for the *ith* QoS attribute and/or learning algorithm and $e_{i,mean}$ is the mean SMAPE of all competitors under such scenario; $n$ is the total number of QoS attribute and/or learning algorithm. In this way, the errors under each scenario are formatted into the same scale where smaller value indicates better overall accuracy. Similarly, we assess the stability of a competitor via the summation of normalised distance to the best competitor under each scenario, formally calculated by:

$$Stability = 100 \times \sum_{i}^{n} \frac{e_i - e_{i,best}}{e_{i,worst} - e_{i,best}} \qquad (5.14)$$

where $e_{i,best}$ and $e_{i,worst}$ are the SMAPE produced by the best and worst competitor respectively, under the *ith* QoS attribute and/or learning algorithm. The remaining notations are the same as 5.13. Again, smaller value indicates better stability across different scenarios.

The experiments setup is similar to what have been described in 4.4.1 of Chapter 4. In addition, we have considered two types of read/write pattern: a read-intensive pattern where read to write ratio is around 9:1; and a write-intensive one, i.e., read to write ratio is around 1:1.

## 5.6.1   The QoS Attributes, Primitives and Evaluation Procedure

The concrete QoS attributes and primitives depend on scenarios. For the simplicity of exposition, we have selected commonly used QoS attributes and primitives in the evaluation, but it is worth noting that our approach is not limited to these dimensions. As listed in Table 5.3, these QoS attributes and primitives are per-service except for CPU and memory as they are shared on a VM. For each service-instance running on a VM of the master PM, a QoS model can at most has 4 direct primitives (i.e., CPU, memory, thread and workload of the said service-instance); and 54 indirect primitives, i.e., 2 (thread and workload)×25 (co-located service-instances)+4 (CPU and memory of another two co-hosted VMs). This combination gives us a maximum of 58 possible relevant primitives for each service-instance.

Table 5.3: The Examined QoS Attributes and Primitives.

| | QoS and Primitives | Description |
|---|---|---|
| Output | Response Time (ms) | The average leap time between a service-instance receives and replies a request. |
| | Throughput (req/min) | The average rate of completed requests. |
| | Reliability (%) | The percentage of requests that being completed less than a threshold. (30 ms) |
| | Availability (%) | The percentage of time that the average response time above a threshold. (60 ms) |
| CP input | CPU (%) | Observed average CPU utilisation of a VM. |
| | Memory (MB) | Observed average Memory utilisation of a VM. |
| | Thread (no. of req) | Observed maximum concurrent threads of a service-instance. (a modified control knob of Tomcat's *maxThread* property) |
| EP input | Workload (req/min) | Observed average request rate of a service-instance. |

We evaluate the prediction accuracy on the fly; and for each experiment run, we examine the accuracy of one interval ahead prediction: by the end of interval $t$, the QoS models are trained based on historical data up to $t$-$1$ (up to $t$-$2$ for environmental primitives), and then we use the observed primitives values at $t$ (at $t$-$1$ for environmental primitives) to predict the QoS value at t, which is finally used to compared with the actual QoS value via SMAPE. The sampling and modelling intervals are both 120 secs with the total of 500 intervals, where the first 150 intervals use a static and stable workload trend aiming at providing some essential data for the modelling; whereas the rear 350 intervals follow the FIFA98 trend. This setup can generate one new sample per interval for updating the model. For all accuracy related experiments, we examine the SMAPE for the rear 350 out of 500 intervals in one experiment run; we calculate the mean accuracy of all service-instances on one VM of the master PM and the reported results are computed by averaging 10 runs.

## 5.6.2 Accuracy of Hybrid Dual-Learners for Primitives Selection

To assess the effectiveness of our hybrid dual-learners technique for primitives selection, we use various criteria, including accuracy, stability and model complexity. To start with, we first compare the four variations of our hybrid dual-learners technique, these are:

- HYBRID-V1 - using (5.8) for mR and (5.4) for mRMR.

- HYBRID-V2 - using (5.8) for mR and (5.5) for mRMR.

- HYBRID-V3 - using (5.8) for mR and (5.6) for mRMR.

- HYBRID-V4 - using (5.8) for mR and (5.7) for mRMR.

We report the results by following the evaluation procedure described in Section 5.6.1. For all cases, we apply three widely used learning algorithms (i.e., ANN, ARMAX and RT) for QoS function training on all the QoS attributes.

Figure 5.4, 5.5 and 5.6 illustrate the results for the overall accuracy, stability and model complexity under two different workload patterns. We can see that for both workload patterns, HYBRID-V1 tends to produce the best accuracy overall, but it has marginal difference to HYBRID-V3 on the write-intensive pattern. As for stability, it is clear that the HYBRID-V1 achieves the best results. We observed that all four variations produce

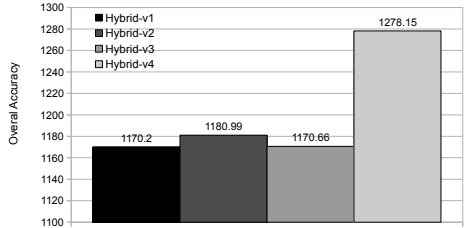 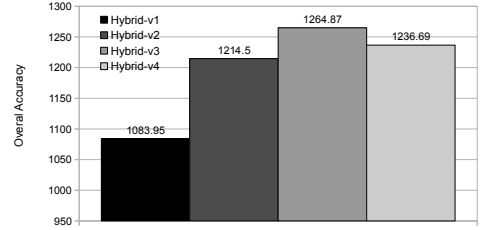

(a) The Write-Intensive Workload.    (b) The Read-Intensive Workload.

Figure 5.4: The Overall Accuracy for Each Variation of the Hybrid Dual-Learners Technique.

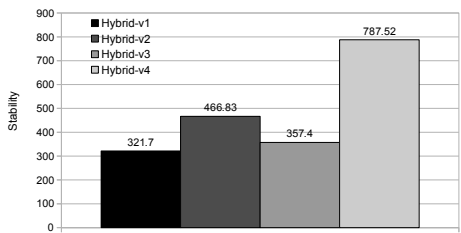 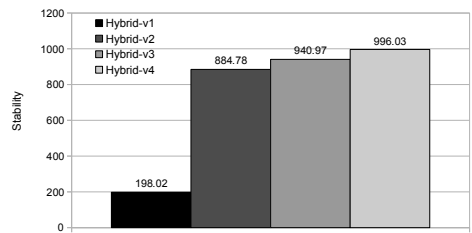

(a) The Write-Intensive Workload.    (b) The Read-Intensive Workload.

Figure 5.5: The Stability for Each Variation of the Hybrid Dual-Learners Technique.

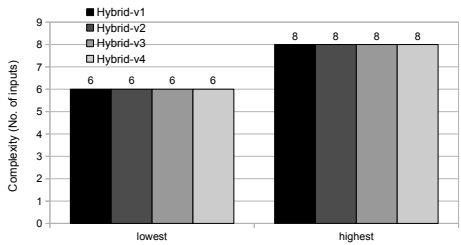 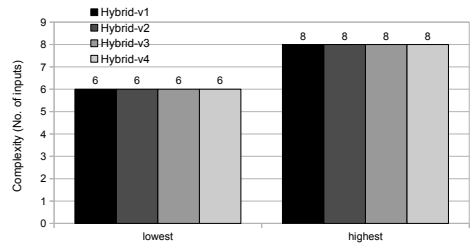

(a) The Write-Intensive Workload.    (b) The Read-Intensive Workload.

Figure 5.6: The Complexity for Each Variation of the Hybrid Dual-Learners Technique.

Table 5.4: The Detailed SMAPE (%) Results of Hybrid Variations for Primitives Selection. (the best is highlighted in bold)

| | | Write-intensive workload | | | |
|---|---|---|---|---|---|
| | | Hybrid-v1 | Hybrid-v2 | Hybrid-v3 | Hybrid-v4 |
| Response Time | ANN | **12.28** | 12.8 | 12.81 | 12.48 |
| | ARMAX | **29.61** | 30.56 | 29.84 | 36.59 |
| | RT | **16.31** | 18.1 | 16.39 | 16.37 |
| Throughput | ANN | **11.93** | 12.29 | 12.67 | 13.59 |
| | ARMAX | **13.35** | 13.55 | 14.02 | 15.12 |
| | RT | 21.89 | 21.8 | **21.14** | 24.2 |
| Reliability | ANN | 0.21 | 0.21 | 0.21 | **0.2** |
| | ARMAX | **0.03** | **0.03** | **0.03** | **0.03** |
| | RT | 0.28 | **0.26** | 0.29 | 0.27 |
| Availability | ANN | 0.37 | **0.36** | 0.36 | 0.43 |
| | ARMAX | 0.03 | **0.02** | 0.03 | 0.03 |
| | RT | 0.43 | 0.45 | **0.41** | 0.55 |
| | | Read-intensive workload | | | |
| Response Time | ANN | **13.51** | 15.44 | 15.14 | 15.35 |
| | ARMAX | 44.85 | 45.62 | **44.36** | 45.68 |
| | RT | **17.42** | 21.19 | 20.26 | 21.58 |
| Throughput | ANN | **13.75** | 15.73 | 15.84 | 15.55 |
| | ARMAX | **15.02** | 17.91 | 17.99 | 17.9 |
| | RT | **22.07** | 24.74 | 25.87 | 26.22 |
| Reliability | ANN | **0.32** | 0.42 | 0.44 | 0.44 |
| | ARMAX | **0.03** | 0.04 | 0.05 | 0.04 |
| | RT | 0.38 | **0.3** | 0.43 | 0.33 |
| Availability | ANN | **0.61** | 0.69 | 0.7 | 0.7 |
| | ARMAX | **0.05** | 0.06 | 0.06 | 0.06 |
| | RT | 0.68 | 0.64 | **0.63** | 0.65 |

the same model complexity. Table 5.4 shows the detailed accuracy results under each of the 12 scenarios. For both workload patterns, HYBRID-V1 produces the best results for most of the cases on Response Time and Throughput; whereas for other two QoS attributes, the best variation tends to be different.

Next, we use HYBRID-V1 (we refer to as HYBRID for simplicity), as the representative of our hybrid dual-learner technique, to compare against various other self-adaptive and online selection techniques that are categorised as single-learner based; and the manual selection technique that has been widely used in existing static and semi-dynamic QoS modelling approaches (e.g., [90] [107]). They are explained as the following:

- HYBRID - using (5.8) for mR and (5.4) for mRMR.

- SINGLE-MR - using (5.8) for mR.

- SINGLE-MRMR - using (5.4) for mRMR.

- MANUAL - fixed and offline selection that statically uses certain primitives (CPU and memory in our case) as inputs e.g., [90] [107] - we modified the model from per-VM to per-service.

- SINGLE-MR-DIRECT - using (5.8) for mR and consider direct primitives only.

Figure 5.7 shows the overall accuracy for the write-intensive and read-intensive workload patterns respectively. From Figure 5.7a, we can see that our HYBRID produces the best accuracy overall. Specifically, in contrast to those single-learner based techniques, HYBRID has better overall accuracy than that of SINGLE-MR-DIRECT because it considers extra information about interference in the modelling, which tends to be important for improving accuracy. In addition, it is also overall more accurate than SINGLE-MR and SINGLE-MRMR, because it is capable to select useful primitives based on both relevance and redundancy while still prevent misleading the selection process. This is achieved by partitioning the possible relevance primitives space. We have also observed that, a carefully designed self-aware, self-adaptive and online selection technique can generally lead to better accuracy than the manual selection; however an inappropriate one (i.e., the SINGLE-MR) might make the accuracy worse off.

As for the read-intensive pattern (Figure 5.7b), SINGLE-MR and SINGLE-MRMR are less accurate than SINGLE-MR-DIRECT; they are even much worse than the manual

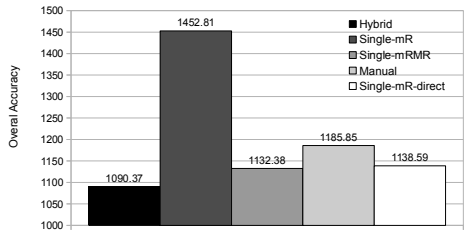
(a) The Write-Intensive Workload.

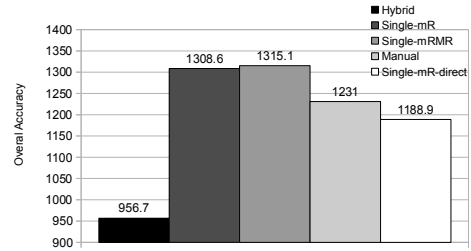
(b) The Read-Intensive Workload.

Figure 5.7: The Overall Accuracy for Each Primitives Selection Technique.

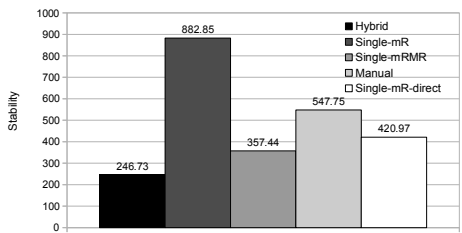
(a) The Write-Intensive Workload.

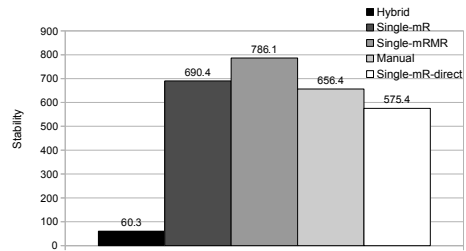
(b) The Read-Intensive Workload.

Figure 5.8: The Stability for Each Primitives Selection Technique.

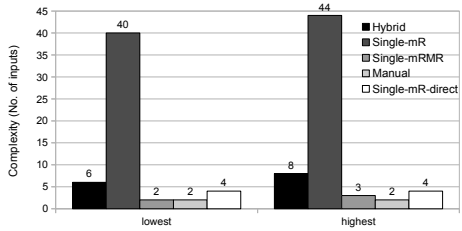
(a) The Write-Intensive Workload.

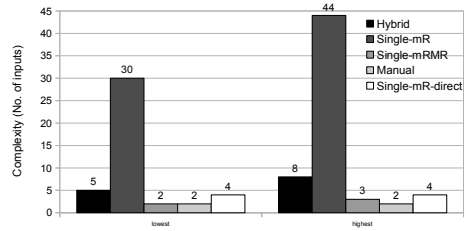
(b) The Read-Intensive Workload.

Figure 5.9: The Complexity for Each Primitives Selection Technique.

technique. This implies that the rich redundancy and the misleading selection cause more serious issues as when compared to write-intensive pattern. For our HYBRID technique, we can note that it again achieves the best accuracy overall, which is a consistent result on both workload patterns.

Figure 5.8 shows the stability of the techniques; it is easy to see that our HYBRID

Table 5.5: The Detailed SMAPE (%) Results of Techniques for Primitives Selection. (the best is highlighted in bold)

| | | Hybrid | Single-mR | Single-mRMR | Manual | Single-mR-direct |
|---|---|---|---|---|---|---|
| **Write-intensive workload** | | | | | | |
| | ANN | **12.28** | 16.12 | 13.03 | 17.8 | 12.92 |
| Response Time | ARMAX | **29.61** | 37.56 | 29.77 | 33.81 | 32.36 |
| | RT | 16.31 | 19.74 | **15.4** | 19.25 | 17.91 |
| | ANN | **11.93** | 13.45 | 16.82 | 14.26 | 12.55 |
| Throughput | ARMAX | **13.35** | 15.29 | 17.53 | 14.17 | 13.6 |
| | RT | 21.89 | 23.52 | 23.17 | **19.88** | 19.94 |
| | ANN | 0.21 | 0.55 | 0.35 | **0.16** | 0.17 |
| Reliability | ARMAX | 0.03 | 0.05 | **0.02** | **0.02** | **0.02** |
| | RT | 0.28 | 0.29 | **0.24** | 0.35 | 0.37 |
| | ANN | 0.37 | 0.39 | **0.34** | 0.36 | 0.36 |
| Availability | ARMAX | 0.03 | 0.04 | **0.02** | 0.03 | 0.02 |
| | RT | **0.43** | **0.43** | **0.43** | 0.55 | 0.56 |
| **Read-intensive workload** | | | | | | |
| | ANN | **13.51** | 15.9 | 21.84 | 30.73 | 15.49 |
| Response Time | ARMAX | **44.85** | 51.44 | **56.47** | 53.46 | 48.52 |
| | RT | **17.42** | 20.21 | 20.01 | 21.09 | 20.73 |
| | ANN | **13.75** | 16.59 | 30.18 | 18.56 | 15.75 |
| Throughput | ARMAX | **15.02** | 17.04 | 17.31 | 17.87 | 17.66 |
| | RT | **22.07** | 24.4 | 30.79 | 25.81 | 24.51 |
| | ANN | **0.32** | 1.2 | 0.45 | 0.36 | 0.55 |
| Reliability | ARMAX | **0.03** | 0.04 | 0.05 | 0.06 | 0.06 |
| | RT | 0.38 | 0.59 | 0.57 | **0.37** | 0.45 |
| | ANN | **0.61** | 0.78 | 0.65 | 0.68 | 0.72 |
| Availability | ARMAX | 0.05 | 0.07 | **0.03** | 0.05 | 0.05 |
| | RT | 0.68 | **0.64** | 1.32 | 0.83 | 0.77 |

technique produces the best result for both workload patterns, meaning that it is the most robust one under different scenarios. As for complexity (Figure 5.9), the HYBRID can be slightly more complex than the others, except for SINGLE-MR. However, the benefit here is that the model's overall accuracy is better and more stable than others with respect to the QoS attributes and the learning algorithms.

Table 5.5 shows the detailed accuracy results for each of the 12 scenarios. Again, we can see that for both workload patterns, the HYBRID produces the best results for most of the cases on Response Time and Throughput, which are highly fluctuate; but the best for Reliability and Availability tend to vary. This is because the Reliability and Availability trends tend to fluctuate less than that of Response Time and Throughput. Therefore, the sensitivity of certain learning algorithms to the number of inputs are amplified; this can easily lead to over-fitting when the model complexity increases, which will significantly influence the model accuracy. However for Reliability and Availability, the differences of accuracy between HYBRID and the best one ranges from 0.01% to 0.05%, which is marginal as when compared to the improvement that HYBRID offers.

According to all these results, we can conclude that although HYBRID does not constantly produce the best accuracy for every learning algorithms and QoS attributes, it tends to produce the best overall accuracy; it is also the most robust and stable technique in the presence of variability introduced by different learning algorithms and QoS trends. In particular, HYBRID provides better accuracy when QoS fluctuates, while leaving the model complexity adequate. It is also worth noting that having a self-aware, self-adaptive and online primitives selection process promotes numerous other benefits, e.g., reduce the needs for complex human analysis and can be easily adapted to many learning algorithms etc.

### 5.6.3 Accuracy of Adaptive Multi-Learners for QoS Function Training

To evaluate our adaptive multi-learners technique (denoted as ADAPTIVE) for QoS function training, we follow the evaluation procedure described in Section 5.6.1. For different QoS attributes, we compare the accuracy and stability of ADAPTIVE with that of the other online learning algorithms that assume single learner (i.e., ANN, ARMAX and RT), which has been widely studied in existing semi-dynamic QoS modelling approaches e.g., [90] [107]. In all the cases, we have used HYBRID for primitives selection.

Figure 5.10 and 5.11 show the overall accuracy and stability results under the two considered workload patterns. We can clearly see that ADAPTIVE produces the best accuracy overall for both workload patterns. It is also the most stable and robust against different QoS attributes. Detailed accuracy results for each of the 4 scenarios has been shown on Table 5.6. Here, we observe similar results on both workload patterns: for Response Time and Throughput, the ANN is the best learning algorithm in contrast to ARMAX and RT; We can see that ADAPTIVE is also much better than ARMAX and RT, but being slightly worse than ANN. These results indicate that although the ADAPTIVE might occasionally produce false positive/negative for selecting the best learning algorithm, it is still able to produce very closed accuracy to the best learning algorithm for a QoS attribute. In cases of Reliability and Availability, we can see that the ADAPTIVE is able to produce the same prediction error as the best learning algorithm, which is ARMAX. This result means that, through the inclusion of meta-self-awareness, the ADAPTIVE successfully determines the best learning algorithm along the QoS trend.

In summary, we can note that although the algorithms behave differently depends on different QoS trends, our adaptive technique can still continually select the suited one to predict QoS and result in good accuracy; it is also the most stable on different QoS trends. Moreover, our self-aware, self-adaptive and online solution eliminates the need of heavy human intervention for identifying the suitable learning algorithm, and hence reduce the errors that can be introduced by human analysis.

### 5.6.4 Detailed Accuracy

Figure 5.12 and 5.13 illustrate examples of the actual and predicted QoS values for all the considered QoS attributes. Due to limited space, we have used an instance of the service named SEARCHITEMBYCATEGORY as the example. We can see that the prediction of the hybrid and adaptive multi-learners approach diverts from the actual QoS scale at some early peak points, e.g., the *30th* interval for throughput. We believe that such inaccuracy is due to the applied FIFA98 trend has limited seasonality, thus the modelling approach can frequently encounter 'new behaviours' of the services at peak points, especially during the early stages of fluctuated trend. However, the figure clearly shows that the multi-learners approach is able to quickly evolve itself and detect most of the change-points in the remaining trend, given that the subsequent predictions are good even for the peak and trough.

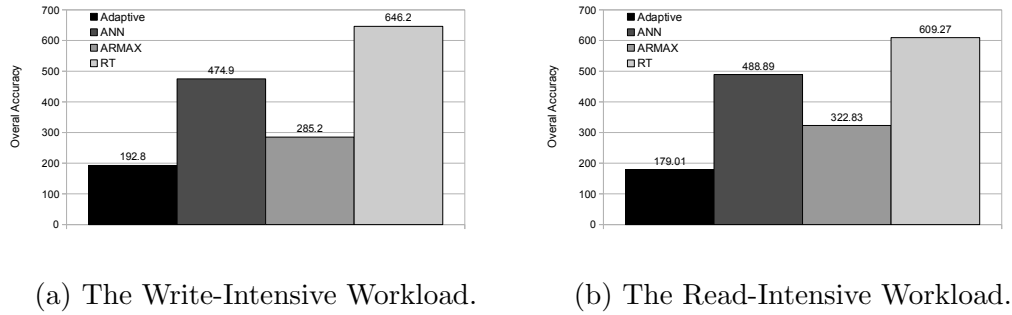

(a) The Write-Intensive Workload.    (b) The Read-Intensive Workload.

Figure 5.10: The Accuracy for Each Learning Algorithm.

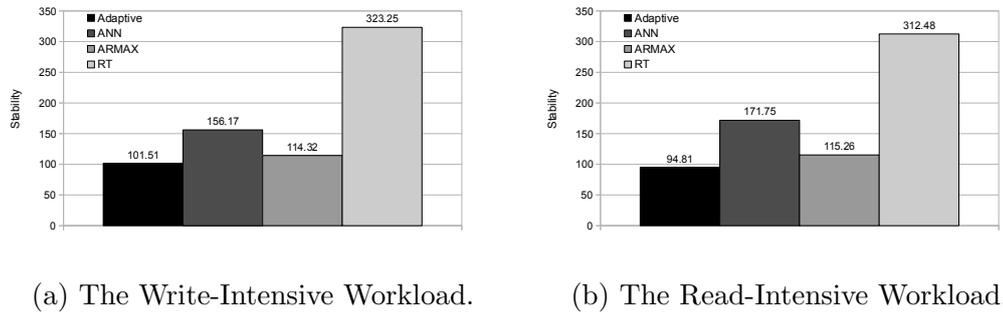

(a) The Write-Intensive Workload.    (b) The Read-Intensive Workload.

Figure 5.11: The Stability for Each Learning Algorithm.

Table 5.6: The Detailed SMAPE (%) Results of Learning Algorithms for QoS Function Training. (the best is highlighted in bold)

| | Write-intensive workload | | | |
|---|---|---|---|---|
| | Adaptive | ANN | ARMAX | RT |
| Response Time | 13.72 | **12.28** | 29.61 | 16.31 |
| Throughput | 12.72 | **11.93** | 13.35 | 21.89 |
| Reliability | **0.03** | 0.21 | **0.03** | **0.28** |
| Availability | **0.03** | 0.37 | **0.03** | 0.43 |
| | Read-intensive workload | | | |
| Response Time | 13.82 | **13.51** | 44.85 | 17.42 |
| Throughput | 14.16 | **13.75** | 15.02 | 22.07 |
| Reliability | **0.03** | 0.32 | **0.03** | **0.38** |
| Availability | **0.05** | 0.61 | **0.05** | 0.68 |

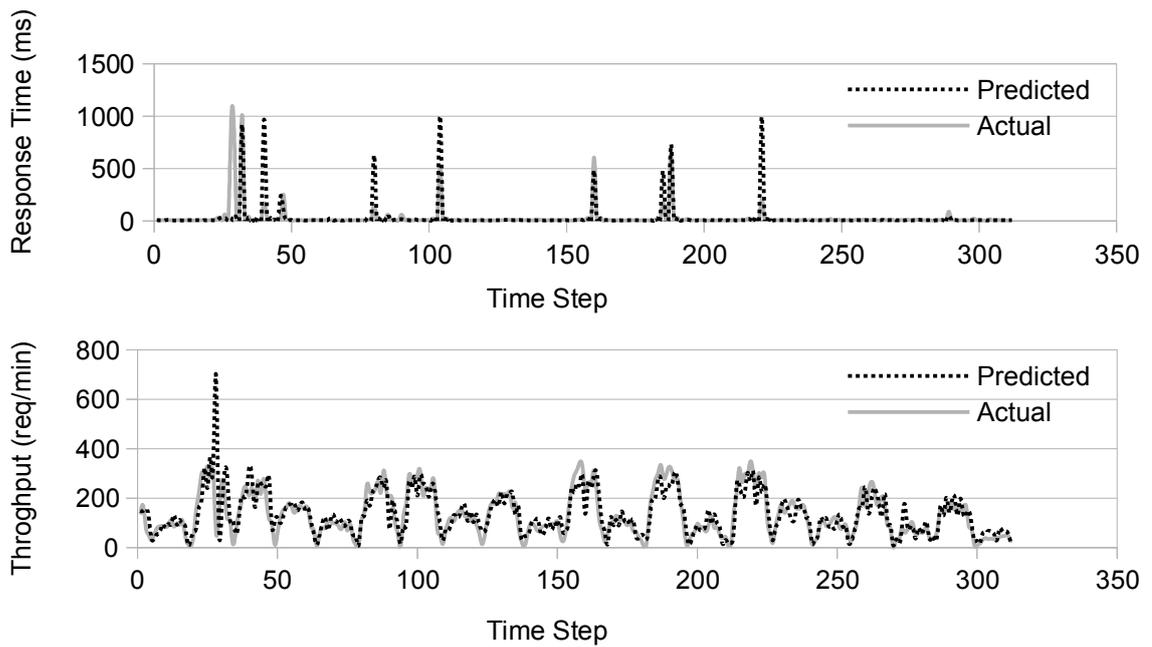

Figure 5.12: The Detailed Accuracy of Response Time and Throughput for SEARCHITEM-BYCATEGORY.

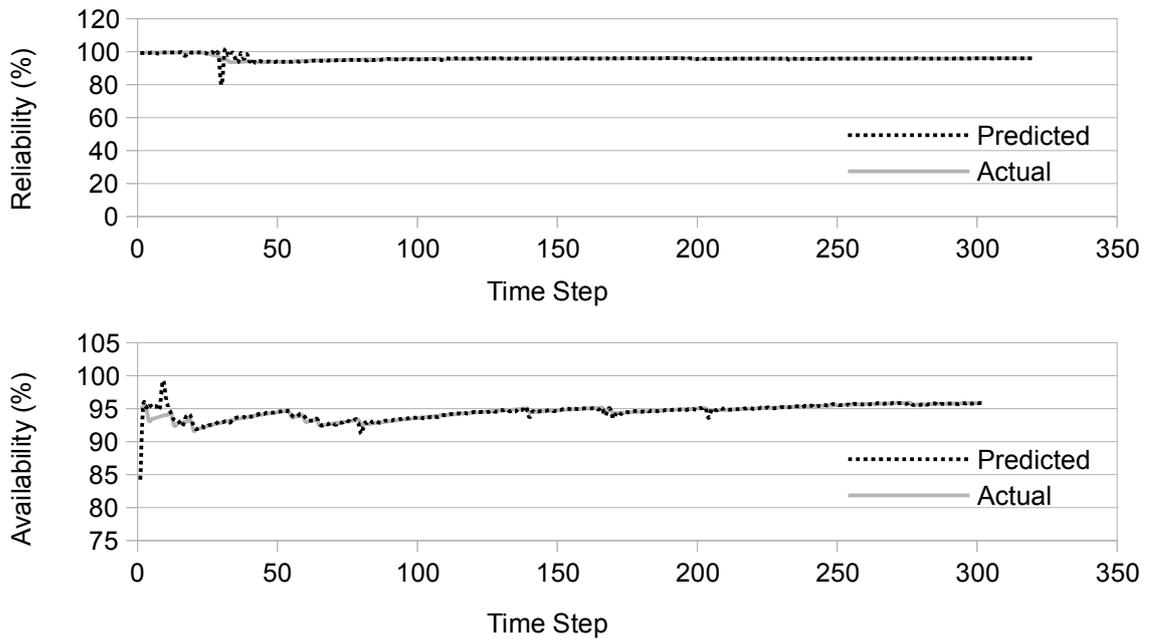

Figure 5.13: The Detailed Accuracy of Reliability and Availability for SEARCHITEMBY-CATEGORY.

### 5.6.5 Sensitivity of Accuracy to Online Data Size

Next, we evaluate the sensitivity of accuracy to the online data size for our approach. This sensitivity expresses how quick accuracy changes as the available data samples increase. Specifically, we split the data size of the entire 350 intervals into the following portions based on the order of time series: 20%, 40%, 60%, 80% and 100%. In the following, we report on the results for the write-intensive pattern. Similar observation has been registered for the read-intensive workload pattern.

Figure 5.14 shows the sensitivity of accuracy to data size for the HYBRID and other single learner-based and manual selection techniques. We note that all primitives selection techniques lead to better accuracy as the data size increases, given the fact that all selected primitives are more or less relevant to the QoS. In most of the cases, the sensitivity of model accuracy to data size has been similar for all the primitives selection techniques. In addition, the comparative accuracy under limited data do not differ much as to what had been reported in Section 5.6.2. However we found that in certain cases (e.g, Figure 5.14a and 5.14e), particularly for fluctuated QoS trends, the accuracy produced by HYBRID clearly has the greatest sensitivity to data size; or being more sensitive than most of the other selection techniques. We also discovered that in these cases, HYBRID tends

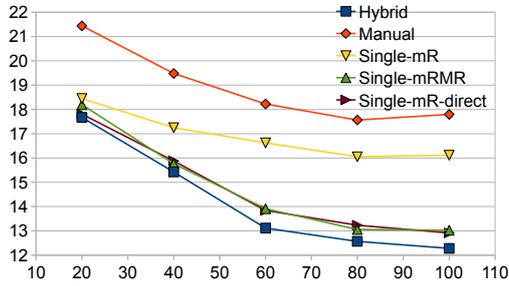

(a) ANN for Response Time.

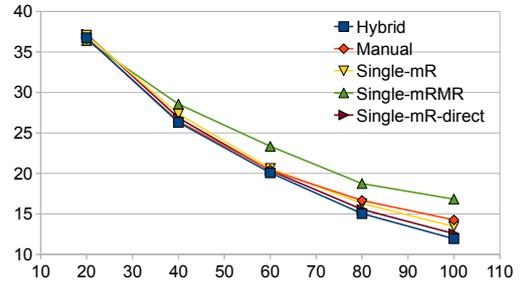

(b) ANN for Throughput.

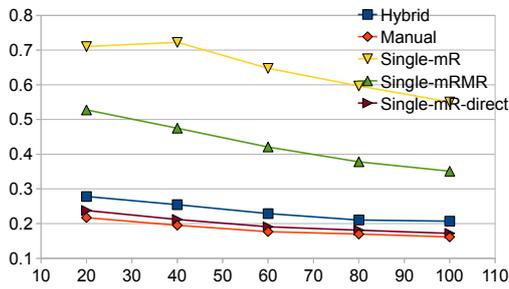

(c) ANN for Reliability.

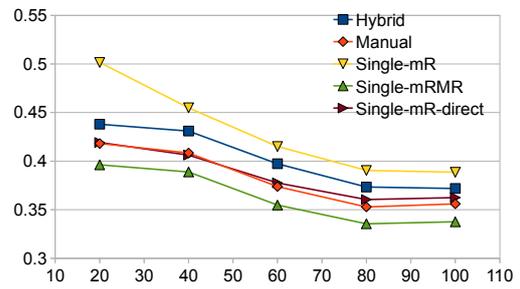

(d) ANN for Availability.

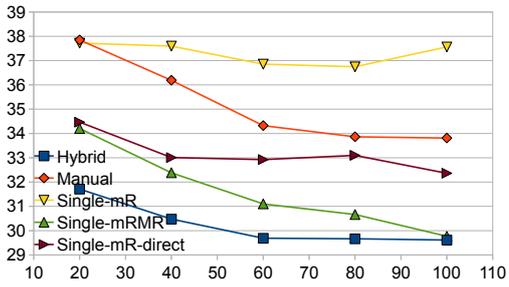

(e) ARMAX for Response Time.

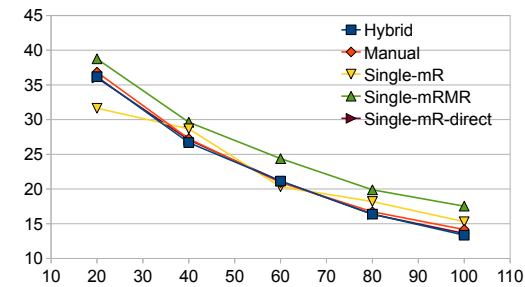

(f) ARMAX for Throughput.

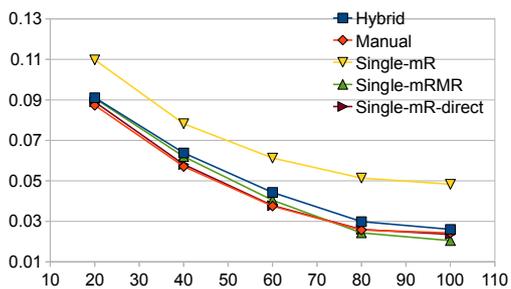

(g) ARMAX for Reliability.

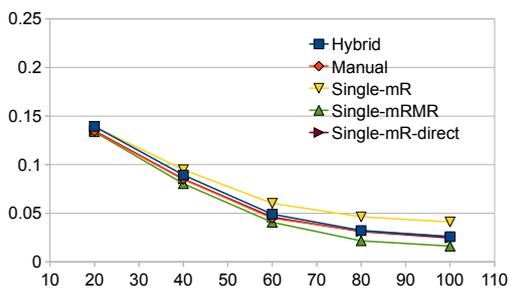

(h) ARMAX for Availability.

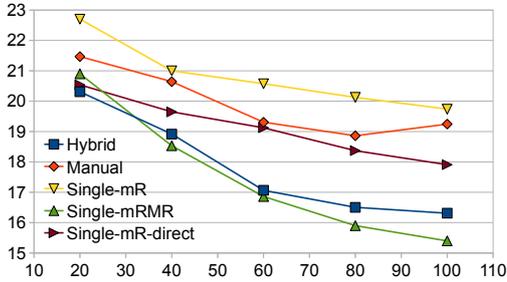

(i) RT for Response Time.

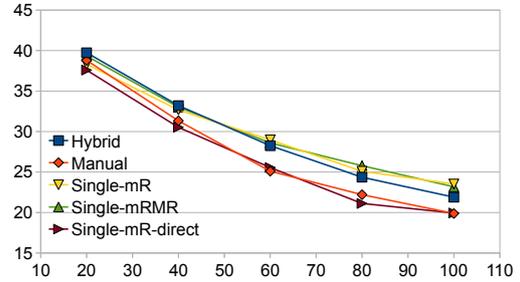

(j) RT for Throughput.

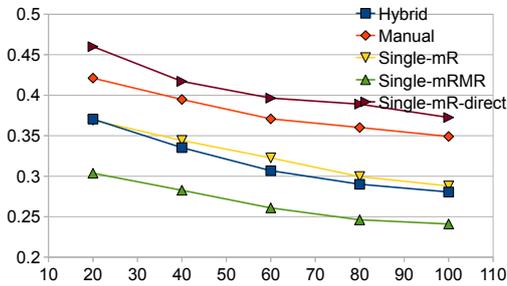

(k) RT for Reliability.

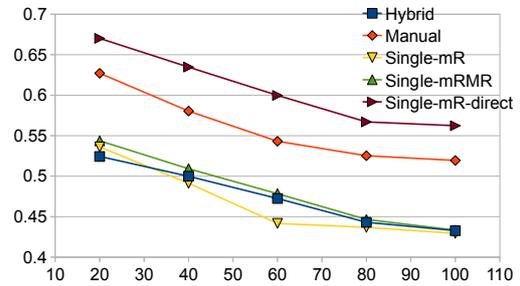

(l) RT for Availability.

Figure 5.14: Sensitivity of Model Accuracy to Online Data Size for Each Primitives Selection Technique. The y-axis is SMAPE (%); x-axis is the online data size (%).

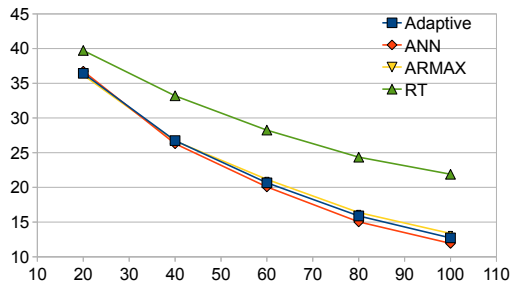

(a) Response Time.

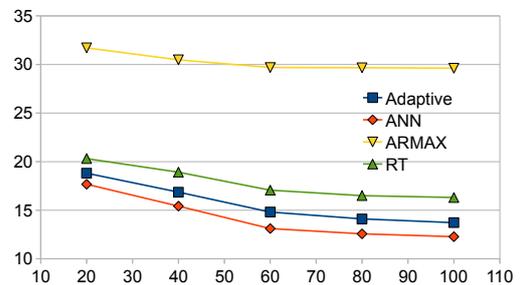

(b) Throughput.

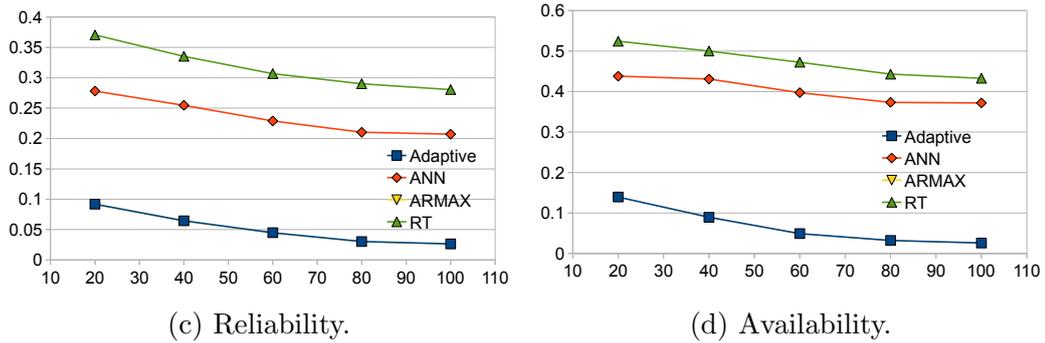

| (c) Reliability. | (d) Availability. |

Figure 5.15: Sensitivity of Model Accuracy to Online Data Size for Each Learning Algorithm. The y-axis is SMAPE (%); x-axis is the online data size (%).

to produce better or similar accuracy in contrast to the other selection techniques, even when the data size is limited. These observations imply that, in contrast to the other approaches, HYBRID can still further improve the accuracy quicker as the data samples increase, while maintaining relatively less or similar error under limited data size.

Figure 5.15 illustrates the sensitivity of accuracy to data size for the ADAPTIVE and other single learner-based learning algorithms. Again, all learning algorithms gradually improve on accuracy as the data size increase. The sensitivity of ADAPTIVE has been similar to most of the others for Response Time and Reliability (i.e., Figure 5.15a and 5.15c). However, for Throughput and Availability (i.e., Figure 5.15b and 5.15d), our ADAPTIVE and the best learning algorithms (i.e., ANN and ARMAX) tends to improve accuracy slightly quicker than the others while maintaining relatively less error under limited data size. We can also observe that, in contrast to the corresponding best single learning algorithm for each QoS attribute, the accuracy of our ADAPTIVE has the same or similar sensitivity to the online data size.

### 5.6.6 Overhead

To assess the overhead of our approach, we compare the latency of HYBRID to other single learner-based techniques, which has been considered in the experiments for primitives selection; we also examine the latency of ADAPTIVE to that of ANN, ARMAX and RT for QoS function training. Because the latency can be varied depends on the characteristics of the service and data size, we have used an instance of the service named SEARCHITEMBYCATEGORY as the example given that it exhibits the most fluctuated workload. The experiments are performed using the rear 10 out of 500 intervals and we report on the average results of all QoS attributes over 10 runs.

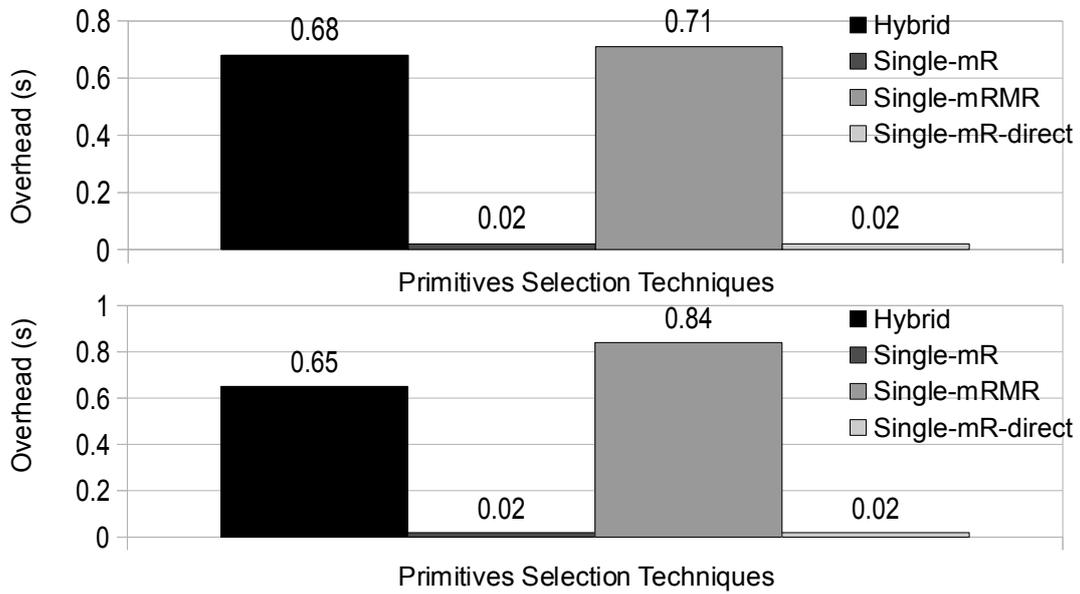

Figure 5.16: Modelling Overhead for Primitives Selection on Write-Intensive Workload (top) and Read-Intensive Workload (bottom).

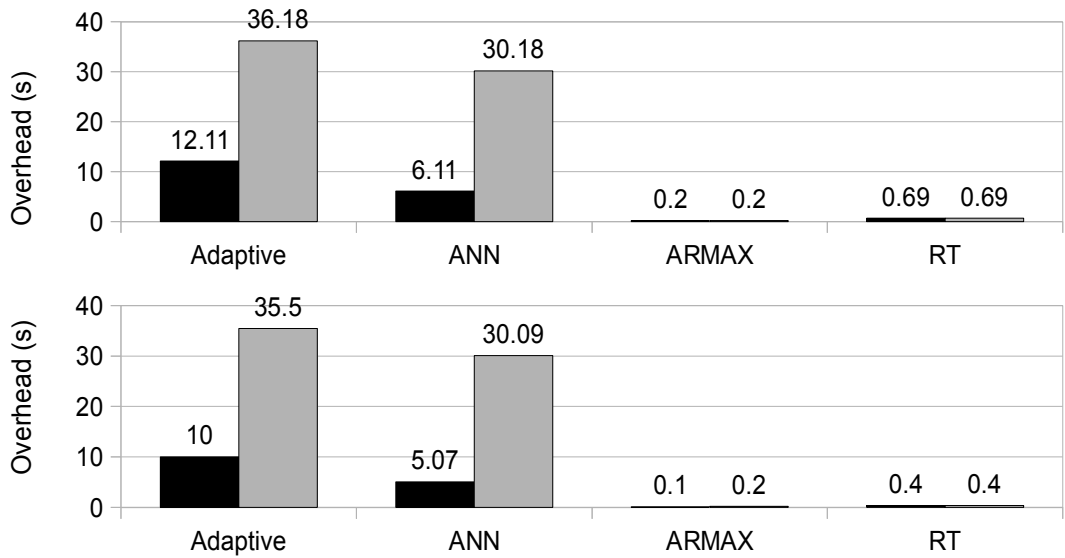

Figure 5.17: Modelling Overhead for QoS Function Training on Write-Intensive Workload (top) and Read-Intensive Workload (bottom).

Figure 5.16 shows the performance overhead for different primitives selection techniques. We can see that under both workload patterns, the HYBRID (0.68s and 0.65s) has relatively bigger overhead as when compared to SINGLE-MR and SINGLE-MR-DIRECT; but it is smaller to that of SINGLE-MRMR. We have observed that this is due to the majority of overhead is caused by the optimisation process in 5.3, which is not part of the process in SINGLE-MR and SINGLE-MR-DIRECT. However, such extra overhead of HYBRID is generally acceptable as it is still less than 1 sec. For the case of QoS function training, Figure 5.17 illustrates the best and worst cases for all learning algorithms. In particular, for both patterns, ANN generally produces bigger overhead as when compared to ARMAX and RT. This is because the ANN is fundamentally more complex than the other two. For both the best and worst cases, the ADAPTIVE has relatively similar overhead to that of ANN; this is expected as the ADAPTIVE needs to wait for the completion of all simultaneously running learning algorithms before determine the best one to use. In conclusion, the overhead of our modelling approach is acceptable under the sampling and modelling interval of 120s, and thus it is efficient enough to be performed online.

## 5.7 Conclusion

In this chapter, we proposed an improved approach for QoS modelling in the cloud. To tackle the dynamics and uncertainties related to QoS sensitivity and interference, we use hybrid dual-learners technique for primitives selection. We have presented a detailed study on how the relevance and redundancy of selected primitives influences the model accuracy, which drives our designs. On the other hand, we have showed that different learning algorithms perform significantly different depends on QoS attributes and their fluctuations. Therefore, we use an adaptive multi-Learners technique for QoS function training. In this way, we aim to dynamically select the best learning algorithms at runtime. All these dual- and multi-learners, as well as the related techniques are the foundations to enable better self-awareness for QoS modelling in the cloud. The experiment results suggest that, in contrast to state-of-the-art QoS modelings, our approach produces better overall accuracy while having acceptable overhead; and it is more stable against the variability introduced by different scenarios. More importantly, the proposed approach eliminates the need for heavy human intervention, which can be complex and error-prone.

The implication of QoS modelling and its dynamic analysis to intelligent adaptation in the cloud are vast: the model can assist autonomic software agents in predicting causes of probable risks leading to QoS violations; reasoning about appropriate mitigation strategies and/or even planning for optimal QoS design and online adaptation strategies. Moreover, it can assist problems related to QoS self-management, self-adaptation, resource utilisation and elastic autoscaling.

We have explored the proposed algorithms and techniques for realising self-awareness in the *QoS Modeller* component. In the following chapters, we will qualitatively and quantitatively demonstrate how the resulted QoS models can be used in the other two important logical aspects (i.e., determining granularity of control and trade-off decision making ) of autoscaling in the cloud; and what are the benefits of these models.

# Chapter 6

# Self-Aware and Self-Adaptive Approach for Determining Granularity of Control in Cloud Autoscaling

## 6.1 Introduction

An effective autoscaling system should be *global-benefit optimised*, with an attempt to optimise both QoS and the required rental cost. The optimal benefit refer to the optimum performance of all QoS attributes with minimal costs for a cloud-based service. If each service in a cloud reaches its optimal benefit, then cloud is said to reach globally-optimal benefit. Achieve globally-optimal benefit in the cloud leads to a win-win situation: the owners of cloud-based services gain better QoS with less rental cost. On the other hand, the cloud provider could better utilise resources and earns better reputation.

The global benefit objective consists of various QoS and cost objectives. In this thesis, we use objectives to refer to various QoS and cost objectives of a cloud-based service.

When making autoscaling decisions in the cloud, the objectives, which we aim to optimise for, can be either conflicting or harmonic due to the presence of overlapping sensitivity (e.g., being sensitive to at least one identical primitive) amongst different QoS attributes and costs; this is referred to as ***objective-dependency***. By sensitivity, we refer to the correlation between the fluctuation of QoS/cost to the stimuli caused by changing primitives. Typically, the cost is based on a fixed model and sensitivity. On the other hand, as we have already discussed in Chapter 4 and 5, QoS models and their sensitivity are generally dynamic (i.e., *which*, *when* and *how* primitives correlate with QoS tends to be dynamic). The objective-dependency could be either intra- or inter-service. Intra-service dependency refers to objectives, which are dependent in nature. This, for example, can be rental cost and throughput of a cloud-based service. The inter-services dependency means the objectives of two services could be dependent on each other because of QoS interference caused by the co-located services on the VM [112] (as resources contention on a VM) and the co-hosted VMs on a PM [116] (as resources contention on a PM). In addition, the loosely coupled nature of cloud-based services implies that dependency might exist between QoS/cost objectives of the services on different PMs, as they are functionally dependent on the same service.

## 6.1.1   Motivation and Challenges

Undoubtedly, objective-dependency has great impact on the achieved global benefits. Therefore, reaching the right granularity of control during the decision making process of autoscaling is extremely important as it needs to cope with the correct objective-dependency. By granularity of control, we refer to which objectives should be considered in the same decision making process for optimisation in the autoscaling. Nevertheless, determining the right granularity of control in autoscaling is a major challenge. One one hand, local optimisation of objectives (e.g., optimise objectives per-VM) might not opti-mise the global benefit due to the presence of objective-dependency caused by overlapping sensitivity. On the other hand, a global optimisation in the cloud is likely to result in large overhead in searching for an autoscaling decision. As a result, there is a trade-offs between global benefit and overhead in the design. In particular, the difficulty is that the objective-dependency tends to dynamic and uncertain in nature. It becomes even harder providing that the QoS models are subject to dynamic changes in their inputs, as we discussed in Chapter 4. The crucial challenge is how can we obtain the correct information about the time-varying objective-dependency, which can be used to determine the effect of a given granularity of control on the global benefits and thus reach the right granularity.

As we have extensively surveyed in Chapter 2, a common problem in existing autoscaling approaches for the cloud is that they are not designed to be self-aware with respect to the dependency of QoS and cost objectives for all cloud-based services. Precisely, they cluster the cloud into fixed regions and granularity of control; optimise for QoS and cost objectives and aggregate the results in each region. For example, existing autoscaling approaches aim at either global optimum in one global region (e.g., cloud-level control) or local optimum in different local regions (e.g., PM-level, VM-level and service-level control) asynchronously and independently. Both solutions ignore QoS and cost sensitivity as their decision making assumes fixed region granularity. Given that the cloud tends to be dynamic and its QoS sensitivity changes at runtime, these approaches can result in inappropriate clustering of regions, which can lead to non-optimal global benefit or large overhead when optimising for the said regions. Both global benefit and the overhead are sensitive to the number of services and their objectives in the decision making process. Therefore, the trade-offs between global benefit and overhead is influenced by the region granularity. Consider now a complicated scenario, where the region granularity is linear to both global benefit and overhead in a given optimisation algorithm: Figure 6.1 shows the likely trend of different fixed region granularities in relation to the global benefit and overhead. Based on the degree of granularity of control in decision making, we classify the approaches for cloud autoscaling into 4 categories, as shown in Figure 6.1. We can see that finer region granularity implies less number of services and the related objectives within each region. This tends to result in worse global benefit but smaller overhead. Consequently, achieving globally-optimal benefit in the cloud call for novel and adaptive approach that is capable to dynamically determine the right granularity of control on-the-fly.

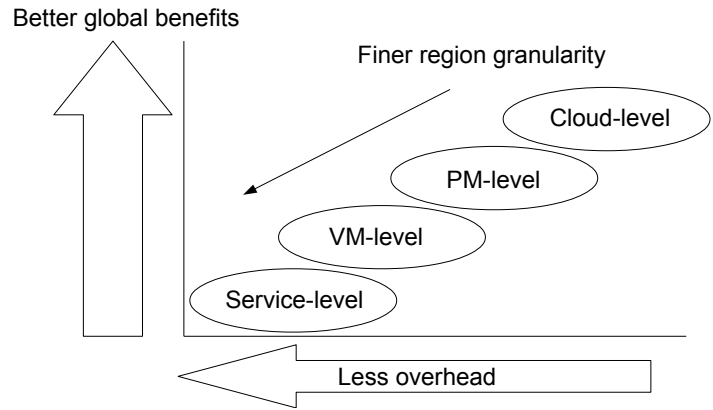

Figure 6.1: Approximated Relationship of Fixed Region Granularities to Global Benefit and Overhead in Cloud.

### 6.1.2 Contributions

The problem, which this chapter addresses is how the autoscaling system can dynamically and efficiently determine an autoscaling decision that produces globally-optimal benefit. To achieve such, we propose a self-aware and self-adaptive approach for determining the right granularity of control, which exploits on the QoS modelling approach from Chapter 4 and 5. Through the awareness of the sensitivity and dependency amongst objectives, this approach can assist the autoscaling process in efficiently produce globally-optimal benefit with reduced overhead. The novelty is that QoS and cost objectives of cloud-based services are dynamically clustered into independent regions where any objectives of a region are independent to those of the other regions, as a result each region can be optimised locally. In particular, we dynamically determine the level of region granularity on-the-fly. This mechanism is the key enabler of realising self-awareness for determining granularity of control. Here, self-awareness provide us with seamless and dynamic clustering of the regions for QoS and cost objectives. To the best of our knowledge, we are the first to consider dynamic granularity of control for autoscaling in the cloud, therefore the proposed approach in this chapter can be seen as a contribution to the fundamentals of elasticity in the cloud enabled by autoscaling. The experiment results reveal that our self-aware and self-adaptive approach is able to produce similar global benefit to the PM-level control, and better than cloud-level, VM-level and service-level controls. On the other hand, it produces smaller overhead than the cloud-level and the PM-level controls; and could be similar to that of the service-level and the VM-level ones. In particular, the achieved global benefit and overhead in our approach tends to be better when it is possible to have more independent regions.

## 6.2    Problem Analysis and Models

We adopt the same cloud system model, assumptions and the generic QoS model as described in Chapter 4. In addition, we do not consider global resources contention caused by shortage in cloud capacity; our approach works for cases where software and hardware resources tend to be available, which is normal in a cloud environment. Henceforth, we assume that the maximum demand of software and hardware resources for all cloud service-instances (e.g., according to their budget) should be satisfied by the capability of the cloud provider. Under such assumption, we eliminate extreme cases where the capacity of cloud provider reaches its limits causing likely global resources contention. This is because the increasing demand of each service-instance would eventually be satisfied by scale up/out as long as the cost does not exceed the budget. We believe this is a reasonable assumption as in realistic scenarios, proper admission control can be applied to restrict the number of cloud-based service-instances. Moreover, in case where the cloud provider actually encounters capacity shortage, the unsatisfied services can be switched to an alternative provider via a cloud selection mechanism, which presumably hold our assumption. However, the design of admission control and selection mechanism is outside the scope of this thesis.

### 6.2.1 Cost and Objective Models

In the context of cloud, utilising control primitives will incur monetary cost, therefore the total costs model for $S_{ij}$ can be represented by the following objective function:

$$Cost^{ij} = \sum_{a=1}^{n} CP_a^{ij}(t) \times P_a \qquad (6.1)$$

where $n$ is the total number of control primitive type that used by service-instance $S_{ij}$ to supports its QoS attributes. $CP_a^{ij}(\text{t})$ is the provision of the $ath$ control primitive for $S_{ij}$ at interval $t$. $P_a$ denotes the corresponding price per unit of the $ath$ control primitive. In this work, we assume that the price of each control primitive type is fixed for all service owners and their service-instances. Given that there can be multiple service-instances running on a VM; while the hardware control primitives (e.g., CPU and memory) can be only provisioned for each VM, we redefine their provision prices to make them suitable for the cost model at the service level.

To achieve globally-optimal benefit in elastic cloud, our approach aims at adaptively and dynamically determine and scale to the configured values of related control primitives, which supports the best of all QoS attributes (4.1 in Chapter 4) with minimal costs (6.1) for all service-instances in the cloud. In this chapter, we apply a linear weighted-sum aggregation to express the global benefit for QoS attributes and costs of different service-instances in the cloud. Formally, at any given interval $t$, we aim to optimise the global objective by maximising the function in 6.2.

$$\sum_{i=1}^{n} \sum_{j=1}^{m} w'_{ij} \cdot (\sum_{a}^{l} w_a \cdot QoS_a^{ij}(t) - \sum_{b}^{r} w_b \cdot QoS_b^{ij}(t) - w_{(l+r+1)} \cdot Cost^{ij}) \qquad (6.2)$$

where $n$ and $m$ are the total number of services and their instances in the cloud respectively; $w'_{ij}$ is the weight for each service-instance. Because the global objective is to maximise (6.2), we need to carefully place the maximised QoS (e.g., throughput) and the minimised ones (e.g., response time); thus $l$ and $r$ are the total number of the maximised and minimised QoS for $S_{ij}$ respectively; $w_a$, $w_b$ and $w_{(l+r+1)}$ are refer to the corresponding weight of the QoS and cost for $S_{ij}$. In addition, the optimisation of (6.2) should be subject to the constraint of budget and SLA.

It is worth noting that the purpose of the approach in this chapter is not to find out the best formalisation of the global benefit and its optimisation algorithms; but to evaluate the effectiveness of our self-aware and self-adaptive approach in handling the granularity of control towards reaching globally-optimal benefit. In the next chapter, we will look at more sophisticated formalisation (e.g., removal of the weights) of the global benefit.

## 6.3 Designing Self-Aware and Self-Adaptive Granularity of Control

Recall that our objective is to optimise the global benefit for QoS attributes and costs of all service-instances. Therefore, from a logical point of view, our basic problem entity in the cloud are different objectives, i.e., to maximise/minimise QoS (4.1) and to minimise cost (6.1), for different service-instances. The objective functions of these objectives are their corresponding QoS/cost models, the $kth$ objective of $S_{ij}$ is denoted by $O_k^{ij}$. In particular, we argue that any two objectives are either dependent (i.e., conflicted or harmonic) or independent (i.e., an objective is neither directly/transitively conflicted nor harmonic with another). With this in mind, we propose a two-phases region clustering, where the first phase clusters the objectives into different independent super-region, which defines the boundary of likely independent objectives for the entire cloud under current deployment. The purpose of super-region is to classify those objectives, which might be independent for now but could become dependent to the others as the QoS sensitivity changes. In other words, the objectives should be clustered into the same super-region as long as they are likely to have objectives-dependency. In the second phase, the objectives within each super-region are further clustered into smaller independent regions where the local optimisation and decision making takes place. By independent regions, we refer to the case where any objective from a region is currently independent to any objective from another region at given time. By doing so, the search space of the global objective function in (6.2) is clustered into $n$ subspaces based on sensitivity, where $n$ is equivalent to the number of regions. The aggregate objective of each subspace (still can be expressed by (6.2), but with smaller search space) is optimised independently and asynchronously.

The basic principle behind our approach is that, we can reach a globally-optimal benefit by asynchronously doing local optimisation and decision making for locally-optimal benefit within different independent regions, which have smaller search space. The clustering of super-region and their regions is a dynamic online process based on the deployment and sensitivity respectively, which are expressed by rules (we will describe in Section 6.3.1 and 6.3.2).

In the following, we use $SR_i$ to denote the *ith* super-region and $R_k^i$ to denote the *kth* region of the *ith* super-region. The clustering should follow the constraints below:

*Constraint 6.1.* $\forall O_c^{xy} \in (R_a^i \cap R_b^j) = \emptyset$ $\qquad\qquad\qquad\qquad\qquad\qquad$ □

*Constraint 6.2.* *if* $(\exists O_a^{ij} \in SR_k)$ *and* $(\exists O_b^{ij} \in SR_l),$ *then* $SR_k = SR_l$ $\qquad\qquad$ □

Constraint 6.1 means that each objective can at most belongs to one region within a super-region. Constraint 6.2 indicates that all objectives of a service-instance should belong to an identical super-region. However, these objectives might belong to different regions within such super-region. The logical view of our two-phases region principle in the cloud is shown in Figure 6.2 where we assume a simple scenario consists of 3 PMs, 4 VMs and 6 service-instances with various QoS/cost objectives. The two solid (red) cycles represent two super-regions. Different colours on the objective entities express different regions within those two super-regions. In addition, there is a functional dependency between $S_{41}$ and $S_{31}$, which means that $S_{41}$ requires the invocation of $S_{31}$ to complete its service.

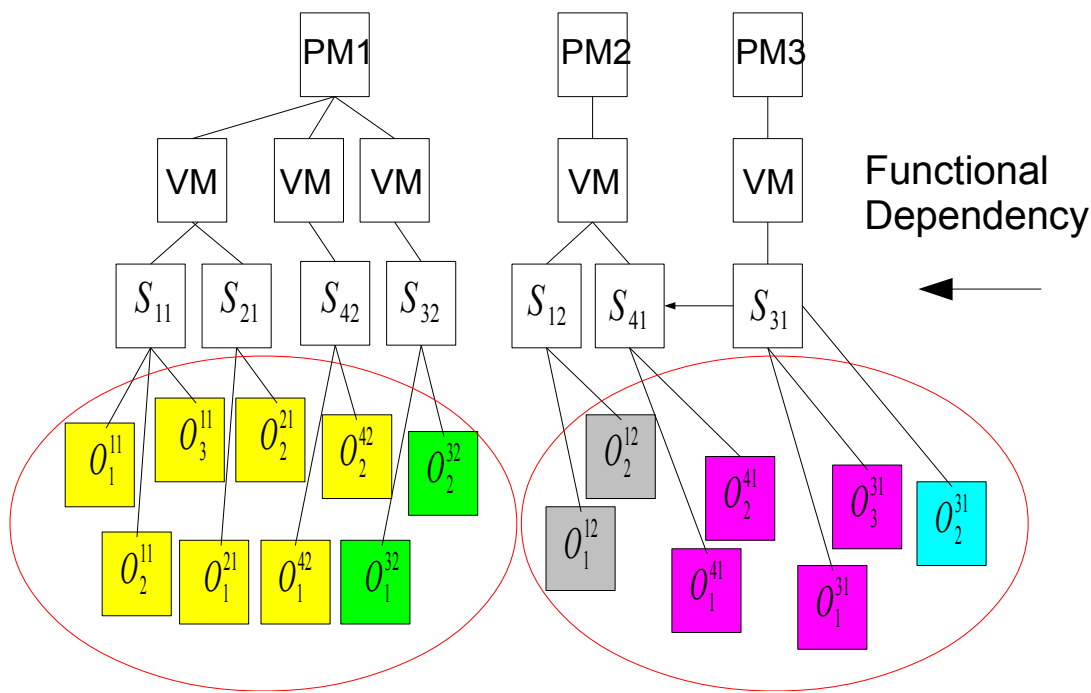

Figure 6.2: Overview of the Notion of Super-Region and Region.

### 6.3.1 Super-Regions

Objectives in the entire cloud can be clustered into different super-regions. Each of the super-region contains the objectives of service-instances that are likely to be directly or transitively dependent. The clustering rule of super-regions is specified as:

*Rule 6.1. Given $S_{ab}$, $S_{cd}$ and $\forall O_i^{cd}$ from $SR_k$, then the jth objective of $S_{ab}$ (i.e.,$O_j^{ab}$)*

*belongs to $SR_k$ if:*

1. *$S_{ab}$ and $S_{cd}$ are deployed on the same VM/PM, or*

2. *$S_{ab}$ has direct functional dependency on $S_{cd}$, or*

3. *$S_{cd}$ has direct functional dependency on $S_{ab}$.*

Rule 6.1 assumes arbitrary service-instances $S_{ab}$ and $S_{cd}$. It also assume that the objectives of $S_{cd}$ are in the super-region $SR_k$. Under these assumptions, objectives of $S_{ab}$ are said to belong to $SR_k$ if and only if it follows any of the above three conditions (either directly or transitively).

Consider the scenario in Figure 3 as an example. The objectives on PM1 are assigned to the same super-region because they satisfy condition 1 in Rule 6.1. On the other hand, the objectives on PM2 and PM3 form another super-region as they satisfy all the conditions. In particular, the objectives of $S_{12}$ and $S_{31}$ are within the same super-region even they do not directly satisfy any of the conditions. This is because $S_{41}$ functionally depends on $S_{31}$, thus they satisfy condition 2. In addition, $S_{41}$ and $S_{12}$ satisfy condition 1. As a result, the $S_{12}$ and $S_{31}$ transitively satisfy the conditions in Rule 6.1 via $S_{41}$.

Given the assumption that shortage in cloud capacity is beyond our concerns, the cluster rule of super-regions are designed based on the fact that the objectives of a service-instance and its functionally dependent service-instances are very likely be dependent under some scenarios (e.g., sequential interaction). In addition, the likely QoS interference can only be caused by the co-located service-instances on a VM and the co-hosted VMs on a PM. Therefore, the objectives from any service-instances that do not directly or transitively satisfy Rule 6.1 can be optimised independently as they would have no way to influence each others.

The clustering of super-region could change at runtime due to the dynamic cloud environment. The super-regions would be re-clustered according to Rule 6.1 upon deployment changes, for examples, VM migration/replication, PM boots-up/shutdown and changes in service compositions etc.

## 6.3.2 Regions

Within each super-region, we further cluster the objectives into different independent regions, where a local optimisation algorithm is running. The clustering of regions could be triggered upon symptoms described in Section 6.3.3. The aim is to further narrow down the number of dependent objectives according to their current sensitivity at a given time. Therefore, the cluster Rule 6.2 of regions are designed based on the sensitivity of QoS and cost models:

*Rule 6.2. Within a super-region $SR_l$, given $O_i^{cd}$ and any $O_j^{ab}$ from $R_k^l$, then $O_i^{cd}$ should also belong to $R_k^l$ if $O_i^{cd}$ has inputs in common to $O_j^{ab}$ and these inputs are parts of the final autoscaling decision, i.e., the control primitives.* $\square$

Concretely, Rule 6.2 expresses that an objective should belongs to a region $R_k^l$ if and only if it has at least one identical primitive input to one or more objectives from $R_k^l$ (meaning that they are dependent and have overlapping sensitivity). If two objectives have neither the common inputs, which are parts of the final autoscaling decision, nor common inputs to the same intermediate objectives, they are said to be independent during optimisation and decision making.

Using the scenario in Figure 6.2 as an example. There are two regions within the left super-region; this is because the objectives of $S_{11}$, $S_{21}$ and $S_{42}$ use certain identical control primitives inputs. On the other hand, the objectives of $S_{32}$ is in an alternative region because it is insensitive to and has no identical inputs to any of those objectives from $S_{11}$ and $S_{21}$ as it suffers limited QoS interference on the co-located services. In particular, suppose that $O_2^{11}$ has identical inputs to $O_1^{21}$ and $O_1^{11}$ ; $O_1^{11}$ and $O_1^{21}$ do not directly satisfy Rule 2. However, all of these 3 objectives are put in the same region because $O_1^{11}$ and $O_1^{21}$ are transitively satisfy Rule 2 via $O_2^{11}$. Similar scenario occurs in the right super-region. In addition, we can see that even $O_1^{31}$, $O_2^{31}$ and $O_3^{31}$ are objectives of the same service-instance, $O_2^{31}$ is put in an alternative region to that of $O_1^{31}$ and $O_3^{31}$. This is a possible scenario: suppose that $O_3^{31}$ is cost objective, $O_1^{31}$ and $O_2^{31}$ are throughput and consistency QoS objective respectively; it is likely that $O_2^{31}$ is only sensitive to an unique control primitive (e.g., ordering error), which is free of charge and henceforth, it is independent on $O_1^{31}$ and $O_2^{31}$.

Similar to the super-regions, the clusters of regions are also subject to dynamic changes. However, region portioning is likely to change more frequently than that of the super-region. This is because it requires updates when changes in QoS sensitivity tend to be significant. Examples of *significant QoS sensitivity changes* could include scenarios, where QoS is becoming sensitive to a new primitive or insensitive to an existing primitive. Insignificant changes on *how* the primitives correlate with QoS cannot trigger re-clustering of the regions.

### 6.3.3  The Workflow

The physical deployment of our approach is shown in Figure 6.3. As we can see the architecture is deployed as decentralised instances, each of which running on a separate VM (e.g., Dom0 on Xen [6]) on every PM in the cloud. For our autoscaling framework, the contributions of this chapter realise self-awareness in the internal self component, namely *Region Controller*. In particular, self-awareness in *Region Controller* is mainly concerned with knowing objective dependency and the effects of control granularity to the global benefit. Table 6.1 shows the mapping between the sub-components of *Region Controller* and the self-awareness capabilities. In such way, self-awareness permits better self-adaptivity, not only at the local level of the region controlling process, but also at the global level of the autoscaling process. In the following, we explain the workflow and those sub-components in greater details.

Table 6.1: The Mapping Between Self-Awareness Capabilities and the Sub-Components for Determining Granularity of Control in Cloud Autoscaling.

| Self-Awareness Capability | Component | Description |
|---|---|---|
| Interaction-awareness | Super Region Cluster and Region Cluster | Knowing how the region controlling process can be affected by the objective dependency. |
| Goal-awareness | Region Cluster | Knowing how the region controlling process can be affected by the QoS and cost models. |
| Self-expression | Super Region Cluster and Region Cluster | Self-adapting its regions and their content. |

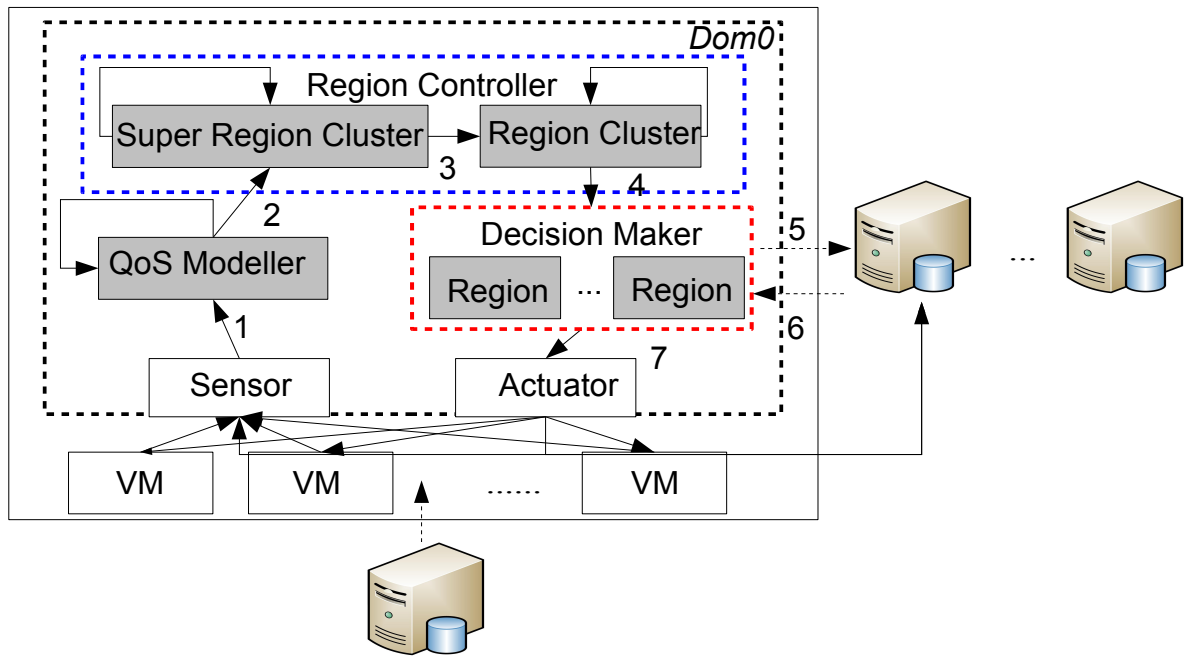

Figure 6.3: Overview of the Architecture for Dynamically Determining Granularity of Control in Autoscaling.

The workflow of our approach has also been demonstrated in Figure 6.3. More precisely, the sensor on each PM collects the data (e.g., QoS values, usages of control primitives and values of environmental primitives) from the underlying VMs and service-instances; and possibly from other PMs due to functional dependency. In addition, the *Sensor* could sense deployment changes and QoS sensitivity changes from other PMs. Next in step 1, the *Sensor* normalises all the raw information. Once the *QoS Modeller* receives both current and historical data after normalisation, this data is then used to build QoS models. At step 2, the QoS models, cost models and the related detected changes are transiting to the *Region Controller*, which realises the region controlling approach proposed in this chapter. The clustering of super-region/region and/or adaptation can be triggered if one or more of the following symptoms is detected:

- *Symptoms 1:* Proactively detect if the QoS of a service-instance is likely to violate SLA constraint by using the QoS models.

- *Symptoms 2:* Reactively detect if the QoS of a service-instance has violated its SLA constraint and/or if the utilisation of a control primitive has violated the constraint.

- *Symptoms 3:* Significant changes in the QoS sensitivity of the objectives in a managed region.

- *Symptoms 4:* Deployment changes occur in a managed super-region.

Symptoms 1 and 2 would trigger the elastic autoscaling of the managed service-instance(s); whereas, symptoms 3 and 4 require the approach to adapt itself by re-clustering the super-regions and/or regions. In particular, to prevent the problem of triggering elastic adaptation too frequently, symptoms 1 and 2 are valid only if the leap time after the previous adaptation for the affected service-instances is more than a threshold $t$. Once we reach the *Region Controller* component, the changes in symptoms 3 and 4 would be addressed separately in a hierarchical stack. Concretely, *Super-Region Cluster* component manages symptom 4 and maintains the super-region on to its PM as only one super-region exists on a PM according to Rule 1. In the lower stack, *Region Cluster* component manages the regions within the aforementioned super-region (step 3) according to Rule 2; it aims to cope with symptom 3. Additionally, it could be triggered by symptom 4 as the cluster of a super-region might change. Once both symptoms 3 and/or 4 are resolved, the propagation goes to the the *Decision Maker* component within each region is designed to address symptoms 1 and 2. This can be done through dynamically searching the best adaptation strategies toward the locally-optimal benefit of region, using the QoS and cost models (step 4). In particular, the *Decision Maker* is triggered independently and asynchronously for each region. Given that functional dependency might exist for service-instances form different PMs, there are cases where a region can be associated with multiple PMs. Therefore in order to ensure that each region is optimised on one PM; the *Decision Maker* can be activated only if the leader of those PMs confirm that the region is not currently being optimised on any other PMs. These processes are expressed as step 5 and 6.

Once the autoscaling decision is determined, the process proceeds to the *Actuator* via step 7. In particular, it is responsible for determining which concrete actions (e.g., scale up/down, in/out and/or VM migration and replication etc) need to be taken in order to fulfil the decision. In this work, we consider both vertical and horizontal scaling and apply a simple solution to determine the actions, i.e., we always try vertical scaling (i.e., scale up/down) first before horizontal scaling (i.e., scale out/in). This is because horizontal scaling is usually more expensive than vertical scaling. As for the VM migration/replication decision, we always choose the one that result in smaller overhead based on a predefined VM profiling pattern.

## 6.4    Experiments and Evaluations

To evaluate global benefit of the autoscaling decisions produced by our approach and the overhead for reaching these decisions, we have conducted an experimental evaluation. We compare our self-aware and self-adaptive approach (we simply refer to as self-aware ap-proach in the following sections) to other 4 non-self-aware approaches and styles that do not cater for sensitivity in the autoscaling process. Each of the 4 approaches assumes different fixed region granularities: service-level, VM-level, PM-level and cloud-level control. Because these 4 styles do not consider symptoms 3 and 4; they trigger elastic adaptation only when symptoms 1 and/or 2 are detected.

### 6.4.1 Experiments Setup

We have implemented the architecture prototype using Java JDK1.6, and we assessed the elastic autoscaling of 8 hypothetical cloud-based service-instances[1] under the control of our prototype. In the experiment setup, each service-instance was deployed on software stack including Apache, Tomcat and MySQL. We simulate a synthetical workload to each service-instance. The workload has been designed in a way that the intensity was sufficient for causing QoS interference on the co-located services and co-hosted VMs. The testbed is a private cloud, where PMs are connected by Gigabit Ethernet and a switch. Xen [6] is used as the underlying hypervisor. The initial deployment and the considered cloud primitives of our experiments are shown in Table 6.2. The scale of each control primitive and their corresponding prices are specified in Table 6.3.

It is worth noting that, unlike the previous chapters, we do not use RUBiS benchmark and the FIFA 98 workload in the experiments. This is because to better evaluate the approach under various objective dependency, it requires a highly customizable testbed to simulate the scenarios of different levels of objective dependency. This is very difficult, if not impossible, to be achieved with the RUBiS benchmark and the FIFA 98 workload, therefore we have used artificial service and workload to better control the scenarios in the experiments.

For simplicity, we assume that each service-instance has only one QoS requirement,

---

[1]We have implemented those service-instances are stateless service. However, stateful services would not affect the experiment results in our problem

Table 6.2: Initial Deployments and the Examined Objectives/Primitives.

| PM | VM | Service-Instance | Objective | Software CP | Hardware CP | EP |
|----|----|----|----|----|----|----|
| PM1 | VM | $S_{11}$ | Throughput and cost | Max threads | CPU and Memory | workload |
| | | $S_{21}$ | Throughput and cost | Max threads | | workload |
| | VM | $S_{31}$ | Throughput and cost | Max threads | CPU and Memory | workload |
| | | $S_{41}$ | Throughput and cost | Max threads | | workload |
| PM2 | VM | $S_{12}$ | Throughput and cost | Max threads | CPU and Memory | workload |
| | | $S_{51}$ | Throughput and cost | Max threads | | workload |
| PM3 | VM | $S_{32}$ | Throughput and cost | Max threads | CPU and Memory | workload |
| | | $S_{61}$ | Throughput and cost | Max threads | | workload |

Table 6.3: Configured Values for Autoscaling and Price of Control Primitives.

| CP | Optional Values | Unit | Price |
|----|----|----|----|
| Max threads | from 5 to 50, 5 unit gap | thread count | $0.8 for each 5 unit per hr |
| CPU | from 1 to 8, 1 unit gap | Compute unit | $2.5 for each 1 unit per hr |
| Max threads | from 0.1 to 2, 0.1 unit gap | GB | $1.5 for each 0.1 unit per hr |

which is throughput and one predefined cost model. To optimise the global objective function in (6.2), we apply random optimisation algorithm with the same number of iterations for each approach. This is because exhaustive algorithms might not be able to produce a decision efficiently due to the large number of possible autoscaling decisions. In addition, we assume that these service-instances and their QoS/cost are equivalently important and thus all weights in the global objective function are set to 1.

Table 6.4: Number of Regions for Each Granularity of Control Under Different Setups of Service-Instances.

| Setup | Number of regions | | | | |
|---|---|---|---|---|---|
| | self-aware and self-adaptive approach | cloud-level | PM-level | VM-level | service-level |
| 2 service-instances | maximum of 1 | 1 | 1 | 1 | 2 |
| 4 service-instances | maximum of 1 | 1 | 1 | 2 | 4 |
| 6 service-instances | maximum of 3 | 1 | 2 | 3 | 6 |
| 8 service-instances | maximum of 4 | 1 | 3 | 4 | 8 |

## 6.4.2 Global Benefits

To examine the global benefit of the autoscaling decision produced by our self-aware and self-adaptive approach, we run 4, 6 and 8 service-instances setups separately for 100 sampling intervals—we omit the case of 2 service-instances here as its result does not differ much as when compared with the case of 4 service-instances. For each of the setup, we collect the quality of global benefit for each autoscaling decision made during the period. The purpose of the different setups is to examine the sensitivity of our approach to the total number of objectives in cloud. Under each setup, we have performed independent runs for each of the five approaches. The global benefit is measured by score, which is the average result calculated by (6.2) for the interval after a previous autoscaling decision point and before the next one. Each of these intervals is referred to as *effect point*. Table 6.4 illustrates the number of regions for each approach, which was observed during the experiments. It is worth noting that unlike the other approaches, the number of regions in our self-aware and self-adaptive approach is subject to dynamic change. Therefore, the result of our approach shown in Table 6.4 is the maximum number of regions that have been observed.

Figure 6.4-6.6 illustrate the results of the global benefit score (y-axis) in relation to each effect point (x-axis). Precisely, Figure 6.4 shows the global benefit of our approach

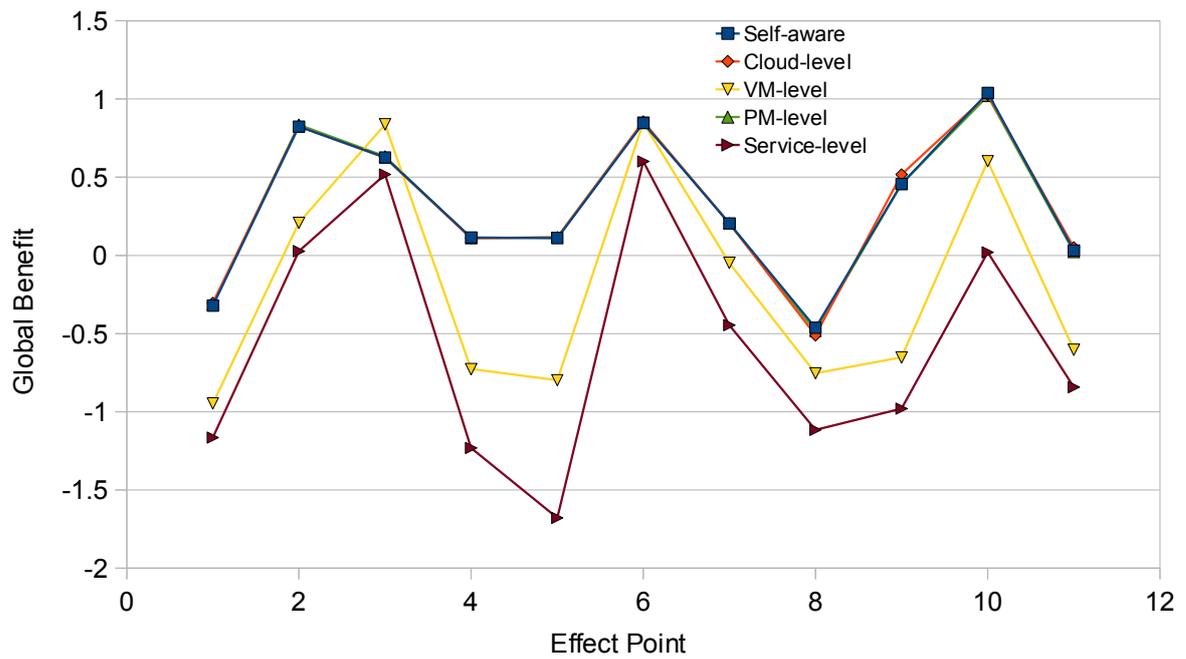

Figure 6.4: Global-Benefit in Case of 4 Service-Instances.

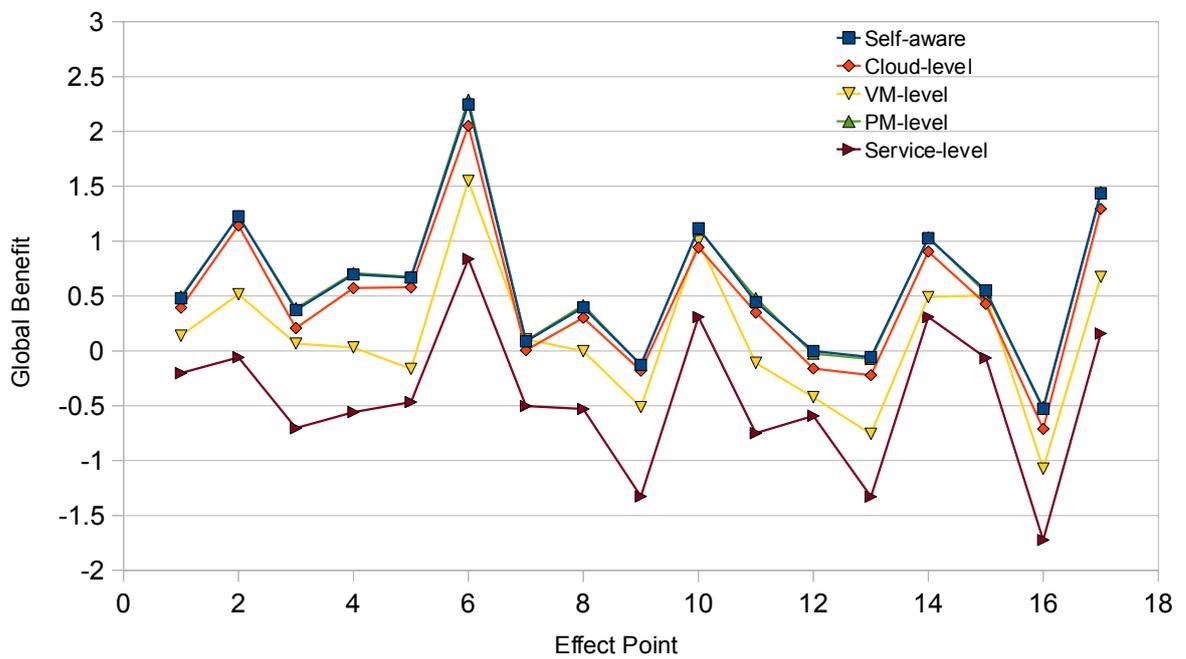

Figure 6.5: Global-Benefit in Case of 6 Service-Instances.

in contrast to the other 4 using setup for service-instances $S_{11}$, $S_{21}$, $S_{31}$ and $S_{41}$. As we can see that the differences in global benefit for the self-aware, the PM-level and the cloud-level approach are marginal. This is because they cluster all the objectives of these 4 service-instances within the same region. Therefore, they perform similarly under such case. In contrast, the service-level and the VM-level control achieve much worse global benefit following the elastic adaptation. This is due to incorrect clustering of the regions as they ignore the sensitivity caused by QoS interferences on co-located services and co-hosted VMs, which are significant in our experiments. Figure 6.5 considers two more service-instances ($S_{12}$ and $S_{51}$) in addition to the ones IN Figure 6.4. We can see that the service-level and the VM-level control performs worse than the other three due to the same reason as the previous case. Surprisingly, although our approach (at most 3 regions) clusters more regions than that of the cloud-level one, its global benefit is better than that of the cloud-level one. We believe that this is because we apply random algorithm in the optimisation and our approach is able to properly cluster the objectives into more regions. This implies that optimising locally and asynchronously on each independent region could result in emergent global benefit using probabilistic algorithms. The PM-level (2 regions) control, on the other hand, also performs better than that of the cloud-level one. We believe that this is because it clusters the objectives per-PM, which similar to the clusters produced by our approach and thus meets the actual sensitivity in the experiments by chance. The self-aware approach performs similarly in contrast to the PM-level control. This is because they produce similar clusters of regions. The only difference is that our approach produces one extra region (we observe only 2 objectives within such region), which is not significant enough to produce emergently better results. However, in the next section we will show that our approach produces much smaller overhead than that of the PM-level one.

Finally, Figure 6.6 illustrates the global benefit for all 8 service-instances. We can see that the service-level and the VM-level control produce the worst results. The gap between their results to the other three is larger than Figure 5 and 6. This is due to the fact that they has incorrectly clustered the regions when introducing more service-instances, and henceforth affecting the global benefit more seriously. Similar to the case of Figure 6.5 , our approach performs slightly better than that of the cloud-level one. The PM-level control performs similar to our approach for the reasons previously explained.

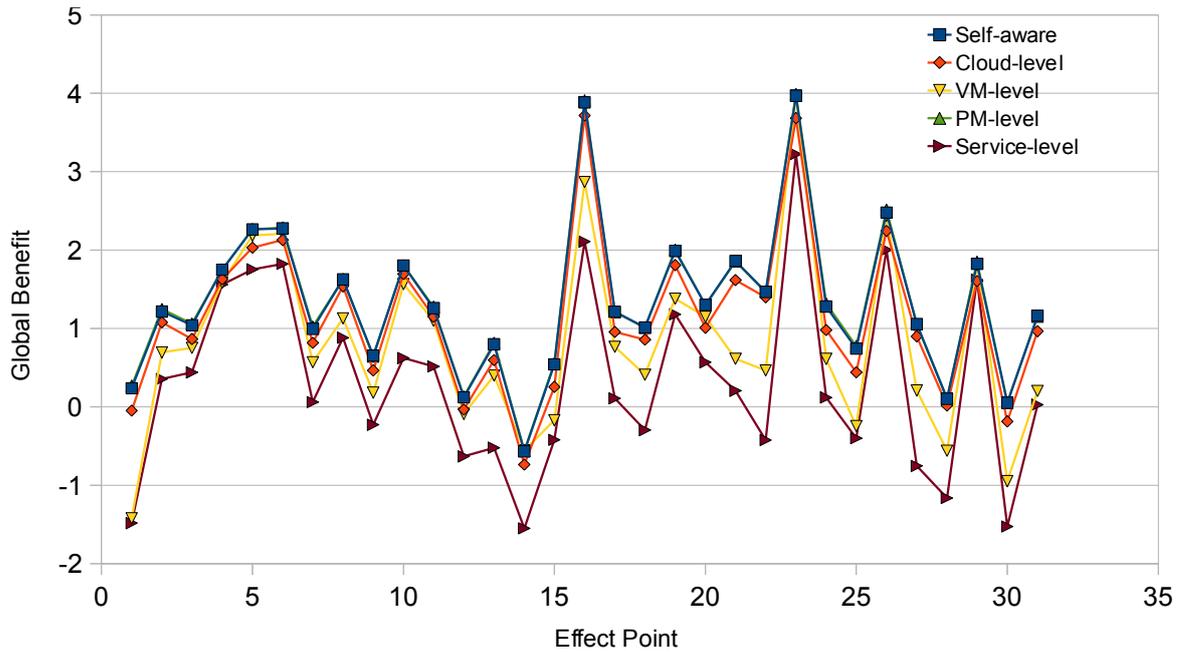

Figure 6.6: Global-Benefit in Case of 8 Service-Instances.

In summary, the elastic adaptations of our self-aware and self-adaptive approach produces much better global benefit than the service-level and the VM-level control under the presence of QoS interferences. In addition, the global benefit produced by our approach are slightly better than that of the cloud-level control and similar to the PM-level control. We observe that the improvement in global benefit tend to be better when having more independent regions. In addition, we believe that our approach can outperform the PM-level one when the number of QoS attributes and/or the number of services on each PM increase.

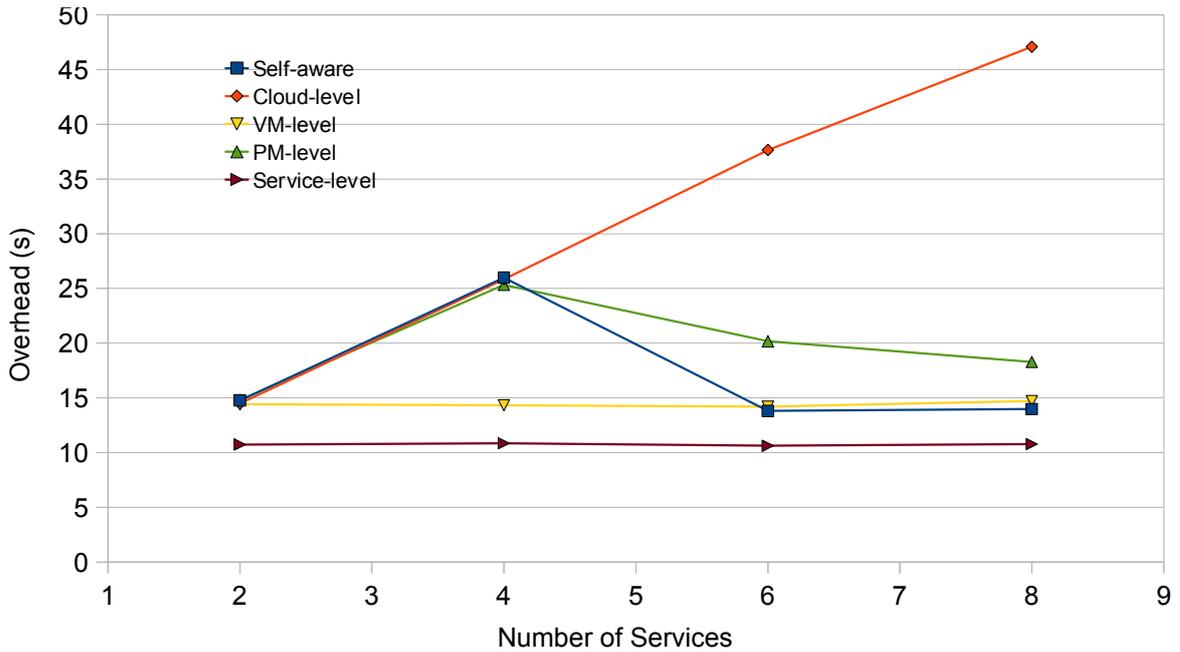

Figure 6.7: Overhead Under Different Numbers of Service-Instances.

### 6.4.3 Overhead

To evaluate the overhead for reaching an autoscaling decision, we compare the average time taken in the decision making processes of our approach to the other 4 competitors, under the setup of 2, 4, 6 and 8 service-instances. In particular, the average time is calculated based on the time taken for reaching all the autoscaling decision within the entire experiment run. As shown in Figure 6.7, which reveals the overhead (y-axis) in relation to the number of service-instances (x-axis), we can see that in case of 2 service-instances ($S_{11}$ and $S_{21}$), the service-level control produces the smallest overhead. This is because it performs optimisation and reaches an autoscaling decision for each service-instance independently. The remaining approaches, on the other hand, produce similar overhead because all the service-instances exist on a single VM.

In the case of 4 service-instances ($S_{11}$, $S_{21}$, $S_{31}$ and $S_{41}$), the differences among the proposed approach, the PM-level and the cloud-level control are marginal. They tend to result in bigger overhead than that of the service-level and VM-level ones. This is because the our approach and the PM-level one only results in one region; they are actually the same as the cloud-level control. In contrast, the service-level and the VM-level style are unaffected by the increasing number of service-instances. In particular, the VM-level control produce bigger overhead than that of the service-level one but better than the other three. This is attributed to the fact that it optimises per-VM, which is coarser-level than the service-level style. As expected, in case of 6 ($S_{12}$ and $S_{51}$ in addition to the case of 4) and 8 service-instances, the overhead of the proposed approach and the PM-level one is becoming better than that of the cloud-level one. This is because our approach and the PM-level control tend to produce more regions (as shown in Table 6.4), which implies that it is able to asynchronously search within a smaller search space for each region with less complexity in contrast to the cloud-level one. On the other hand, the service-level and the VM-level styles remain unaffected. However, we can see that our approach perform similar to the VM-level style and only slightly worse than the service-level style. In contrast to the PM-level style, our approach still performs better. This is attributed to the fact that we further allow clustering within a PM. Consequently, this result in one more regions and thus the search space is further reduced. We can see that even with only one more region, the achieved overhead of our approach gains considerable improvement. We believe that such improvement can be amplified when it is possible to cluster more regions.

Interestingly, we can see that unlike the overhead for cloud-level control, which increases linearly; the overhead of our approach and the PM-level control increase from the cases of 2 to 4 service-instances. They can drop again from the cases of 4 to 6 and remain stable for the case of 8 service-instances. This is because both approaches determine that only one region is allowed for the case of 2 and 4 service-instances. Therefore, it is the same as the cloud-level one and the overhead could also increase in a similar way. When 6 and 8 service-instances exist, both architectural styles result in more than one region. Henceforth, the average result of overhead tends be smaller than previous cases, as there are numbers of autoscaling decisions made for a region with smaller search spaces than the single region in case of 2 and 4 service-instances.

To conclude, our self-aware and self-adaptive approach is able to achieve smaller overhead in contrast to the cloud-level and the PM-level control as the number of region increases. The overhead of our approach is close to that of the service-level and the VM-level style. However, we can observe from Section 6.4.2 that the achieved global benefit are significantly better than these two. In addition, the experiments reveal that the overhead of our approach is sensitive to the number of clustered regions. In particular, the more independent regions are clustered, the smaller overhead is realised.

## 6.5   Conclusion

We have proposed a self-aware and self-adaptive approach that assists the autoscaling process for dynamically guaranteeing globally-optimal benefit in elastic cloud. This is achieved by dynamically clustering the objectives into regions for more efficient and effective decision making, which is the foundation to enable self-awareness for determining granularity of control in cloud autoscaling. In this way, the autoscaling process can adaptively reach to the right granularity of control. Further, the propose approach has been designed leveraging on the principle of self-awareness. Experimentally, we have evaluated our approach with respect to global benefit achieved by the produced autoscaling decisions and the overhead to reach these decisions. We compare the results to other 4 non-self-aware autoscaling approaches, each of which relies on a fixed granularity of control. The results reveal that our approach produces similar global benefit to the PM-level control, and better than the rest approaches. On the other hand, it produces smaller overhead than the cloud-level and the PM-level control; and could be similar to that of the service-level and the VM-level ones. The improvement on global benefit and overhead tends to amplify when it is possible to have more regions.

So far, we have explored the self-aware and self-adaptive process in the *QoS Modeller* and *Region Controller* components of our autoscaling framework. As discussed in this chapter, we have demonstrated the benefit of the QoS models for determining the right granularity of control when autoscaling in the cloud. We have also illustrated that, by incorporating our QoS modelling and the region clustering approach, the global benefits of autoscaling in the cloud can be further improved without heavy human intervention. However, the aforementioned approaches have not explicitly considered the trade-off decision making problem, which is one of the key challenges in autoscaling. In the next chapter, we will refine the objective model and propose self-awareness enabled solution to this problem, building on the approaches described in Chapter 4, 5 and this chapter.

# CHAPTER 7

# SELF-AWARE AND SELF-ADAPTIVE TRADE-OFF DECISION MAKING IN CLOUD AUTOSCALING

## 7.1 Introduction

The core phase in autoscaling is the dynamic decision making process that produces the optimal (or near-optimal) ***decision***, which consists of the newly configured values of control primitives, for all the related objectives. However, as mentioned in Chapter 6, ***objective-dependency*** (i.e., conflicted or harmonic objectives) often exist in the decisions making process, which implies that trade-offs are necessary and the overall quality of autoscaling can be significantly affected by the trade-offs made, hence render it as a complex task. This is especially true for the shared infrastructure of cloud where objective-dependency exists for both intra- and inter-services. That is to say, trade-off is not only caused by the nature of objectives (intra-service), e.g, Throughput and cost objective of a service; but also by the QoS interference (inter-services) due to the co-located services on a VM and co-hosted VMs on a Physical Machine (PM) [112] [116]. This is known as a typical consequence of resources contention in cloud [86] [99] [116]. Therefore, given the presence of complex objective-dependency, it is clear that the decision making for autoscaling in the cloud is very difficult, if not impossible, to be handled by human decision makers; and thus urges the need for self-adaptivity. Among the trade-off decisions that quantified by the commonly used pareto-dominance relation, we are particularly interested in the ones that achieve **well-compromised trade-offs** (a.k.a. knee points). A decision is said to result in well-compromised trade-off, as when compared with its neighbouring decisions, if it largely improves certain objectives; while causing relatively small degradations to others. In other words, the improvements of all dependent objectives tend to be well-balanced. The difficulty lies in how to dynamically and efficiently achieve well-compromised trade-offs for autoscaling in the cloud, which we address in this chapter.

### 7.1.1 Motivation and Challenges

The QoS performance of services and the cloud environment tend to fluctuate; consequently, the QoS interference, the possible trade-off decisions for autoscaling and their effects on the objectives are dynamic and uncertain. Existing work for autoscaling decision making in the cloud can be either static [60] [52] in the sense that the mapping between conditions and decisions are fixed; or dynamic [10] [95] where the runtime conditions and behaviours are used to 'learn' new decisions. Those state-of-the-art approaches often ignore QoS interference and its related trade-offs in autoscaling. Furthermore, they tend to be limited in handling two challenges related to the trade-offs:

- Firstly, most of the work restricts the autoscaling decisions into fixed bundles (e.g., VM instance), which is rather inflexible, and thus it is necessary to consider any combinations of the configured values for control knobs [137]. However, given the potentially large amount of possible combinations of the configured values, finding the optimal decisions and reasoning about their effects on objectives is known to be an NP-hard problem [124] [57]. Henceforth, the key challenge is how to dynamically optimise diversified trade-off decisions and thus produce better coverage of the trade-offs surface.

- Secondly, another challenge is concerned with how to dynamically extract the decisions that achieve well-compromised trade-offs, subject to runtime uncertainty.

Concretely, recall our survey of existing work for making autoscaling decision presented in Chapter 2, the static approaches are insufficient as they are restricted by the simplified assumptions about the conditions and the mapped decisions. Although dynamic approaches have been proposed to address this limitation, most of them, e.g., [69] only focus on optimising a single objective (e.g., cost), where other objectives are treated as constraints. This means that the search process tends to be limited in exploring trade-offs due to the optimisation of single objective. To this end, weighted-sum formulation that aggregates all the objectives into a single one has been widely applied, e.g., [55]. Nevertheless, weighted-sum of objectives requires human intervention to carefully design and tune the weights for the objectives, which is often an extremely complex and error-prone exercise. In addition, finding the right weights in advance is extremely difficult in the presence of QoS interference, as it is difficult to presume the relative importance of the services and their levels of importance. On the other hand, a single aggregation can track the search in a smaller search space and the resulted decisions are driven by coarser and less information about the trade-offs surface. In other words, the optimality and diversity of the resulted trade-offs decisions tend to be limited and therefore causing it difficult to achieve well-compromised trade-offs.

There is a limited amount of work that leverage on the notion of multi-objective optimisation [124] [57] [51] and pareto-dominance [67] based sort. Most commonly, they apply Multi-Objective Genetic Algorithm (MOGA), e.g., NSGA-II [47], to search the trade-offs decisions without explicitly using weights. However, since they do not focus on decisions that produces well-compromised trade-off, the amount of resulted decisions is unavoidably large and can easily lead to imbalanced improvement.

## 7.1.2 Contributions

In this chapter, we propose a self-aware and self-adaptive approach for making autoscaling decision in the cloud without human intervention. This approach grounds on Multi-Objective Ant Colony Optimisation (MOACO) and a new *compromise-dominance* mechanism, which are the key enablers of self-awareness for decision making in cloud autoscaling. In this way, the approach dynamically and adaptively adjust its own behaviours to (i) discover the possible trade-offs decisions at runtime; and (ii) extract the decisions that produce well-compromised trade-offs with respect to all related objectives.

We have chosen MOACO because (i) it has been shown that existing MOGA based autoscaling decision making in the cloud, such as NSGA-II, cannot optimise and make trade-offs for more than 4 objectives [111]; while our problem needs to handle larger numbers as we consider the trade-offs caused by QoS interference, e.g., we have considered 30 objectives in our experiments. (ii) As discussed in [111], the limitation of MOGA for large number of objectives is due to it needs pareto-dominance to evaluate the overall quality of decisions for all objectives as the algorithm runs; henceforth, causing the MOGA to obscure and miss important information about the trade-off surface, which restricts its optimality and diversity when the number of objectives increases. Unlike MOGA, the nature of MOACO allows us to design it in a way that decisions are evaluated against each objective for many single objective optimisations in one run, and thus avoiding the use of pareto-dominance in the optimisation. This is achieved by using aggregative heuristics and different pheromone structures for the objectives. Hence, we only need to evaluate the overall quality of decisions for all objectives (i.e., the compromises) after the optimisation has been competed. By doing so, the optimisation can optimise and make trade-offs for larger number of objectives while ensuring good diversity. (iii) the sequential pareto-dominance sorting of MOGA can incur large overhead; in contrast, MOACO can gain benefits from parallel programming as each ant works in isolation. (iv) In many other domains, e.g., [64], it has been shown that MOACO tends to outperform MOGA in both optimality and diversity.

By separating MOACO and the evaluation of decisions' overall quality for all objectives, the MOACO is encouraged to explore more information about the trade-offs surface while saving computational efforts. This design, as shown in [9], tends to produce better optimised and diversified trade-off decisions. Instead of using pure pareto-dominance [67] to evaluate the overall quality of decisions for all objectives during optimisation, we propose a new mechanism, namely *compromise-dominance*, to search well-compromised trade-offs based on the final result of MOACO. Here, we use pareto-dominance [67] to measure superiority, and a combination of nash-dominance [108] and the distance of decision to measure fairness. In this way, we aim to achieve a well-balanced improvements for the objectives without explicitly weighting them.

To the best of our knowledge, we are the first to address the problem of reaching well-compromised trade-offs for autoscaling in the cloud while considering the trade-offs caused by QoS interference. In particular, we show the effectiveness of the approach for up to 30 dependent objectives, which is significantly larger than what is considered in state-of-the-art work (i.e., 2-4 objectives). The experiment results suggest that our approach produces better trade-offs quality in terms of the numbers of favourable objectives and the extents to which they are optimised; it also produces smaller violation for requirements. Moreover, our approach results in acceptable overheads and has balanced over- and under-provisioning.

## 7.2 Problem Analysis and Models

We adopt the same cloud system model, assumptions and the generic QoS model as described in Chapter 4. It is worth mentioning that dynamic QoS interference is an often ignored, but important factor for decision making in cloud autoscaling. Consider, for example, a scenario where the throughput of a service-instance $S_{ij}$ can be only improved by provisioning more memory to the underlying VM. Such decision might not be an issue when the contention is light. However, as the provision increases, eventually it will result in throughput degradation to the other service-instances on the co-hosted VMs, leading to dynamic QoS interference [26][139][116]. The same issue applies when we increase the number of service threads for a service-instance, where the co-located service-instances on the same VM might be interfered [139][99]. These phenomena imply that there are trade-offs between the throughput of $S_{ij}$ and those of the other service-instances, which might be owned by different cloud consumers. It becomes more complex when we need to consider trade-offs between conflicted objectives, e.g., the throughput and cost of $S_{ij}$. All these facts can lead to large number of dependent objectives in a decision making process (i.e., more than 4). Since it is often too expensive to completely eliminate QoS interference [116], we aim to optimise the services' objectives till the point where QoS interference becomes significant, and then mitigate the effects of QoS interference by making well-compromised trade-offs.

However, well-compromised trade-offs cannot be guaranteed by existing purely pareto-dominance based approaches [67]. Given the large number of dependent objectives caused by QoS interference, quantifying compromise in the trade-offs purely based on pareto-dominance can lead to a large number of trade-off decisions, which also contain the ones that have imbalanced improvements. Consider, for example, two decision A and B, suppose that A leads to 9 significantly better objectives than B; while B can only lead to one slightly better objective than A. These two decisions are regard as indifferentiable in the sense of pareto-dominance. Assuming that both decisions satisfy all the requirements constraints, it is generally the case that A is more preferable than B. However, since they are equivalent in pareto-dominance, B can be selected instead of A, which results in badly compromised trade-offs.

As mentioned in Chapter 6, we cluster the objectives into different regions. As a result, autoscaling in the cloud needs to optimise multiple independent regions, each of which contains different sets of dependent objectives. To cope with multi-objectivity, we revised the weighted-sum objective function (as discussed in Chapter 6) to a multi-objective representation. That is to say, for each region, our goal is to produce an autoscaling decision $d$ that use the minimal costs to achieve the best possible QoSs, shown as the following:

$$\text{Maximise or Minimise } \langle O_1(t), O_2(t)...O_o(t) \rangle \tag{7.1}$$

where $O_o(t)$ denotes the $oth$ objective in the region and it can be either QoS attribute (4.1) or cost (6.1), subject to:

$$QoS_k^{ij}(t) \succeq SLA_k^{ij} \tag{7.2}$$

$$Cost^{ij}(t) \leqslant Budget^{ij} \tag{7.3}$$

$$min_a^{ij} \leqslant CP_a^{ij}(t) \leqslant max_a^{ij} \tag{7.4}$$

whereby (7.2) states that any QoS attribute should meet its Service Level Agreement (SLA). (7.3) denotes that the cost of each service-instance should not exceed its budget requirement on a VM. Finally, (7.4) represents the possible configured values of control primitives must be selected within a given range of the underlying hardware or software. $min_a^{ij}$ and $max_a^{ij}$ are the thresholds to control the range of possible configured values, and they are dynamically updated online, as we will see in Section 7.3.

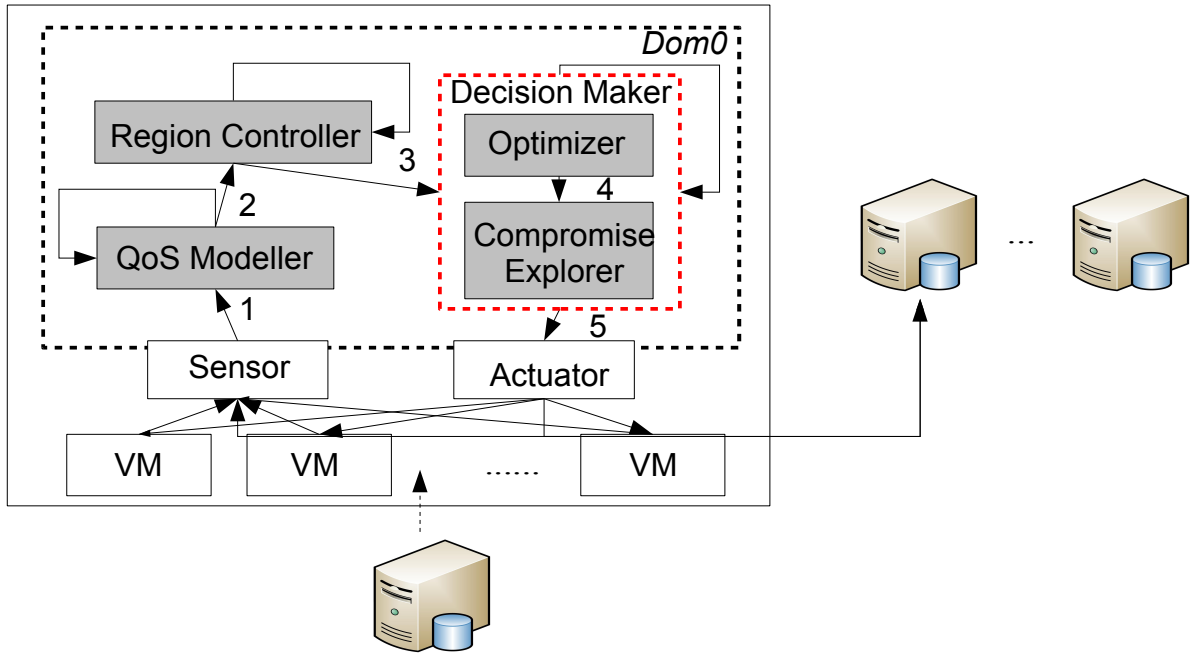

Figure 7.1: Overview of the Autoscaling System with Self-Aware and Self-Adaptive Decision Making Approach.

It is obvious that, by omitting any predefined weights of objectives in (7.1), we render the problem as a discrete multi-objective optimisation problem, which involves multiple trade-offs and is usually NP-hard [124] [57]. In the following sections, we specify our MOACO and compromise-dominance based solution to the problem.

## 7.3 The Architecture for Trade-off Decision Making in Cloud Autoscaling

To enable self-aware and self-adaptive decision making for autoscaling in the cloud, we have designed and implemented an autoscaling system using decentralised feedback loops, which is deployed and running on the root domain of each PM, as shown in Figure 7.1. The components in our system, except *QoS Modeller*, are triggered when the system detects violations of the requirements, i.e., violations of SLA and utilisation constraints in case of over-provision. In particular, a requirement is said to be violated only if such violation has been observed for more than n sampling intervals, where n is a variable that controls the trade-offs between stability and adaptivity of our system. The sensors on a PM does not only sense data, but also the QoS models from other PMs. This is because in some cases, a cloud-based service can be functionally dependent on services running on the other PMs, thus creating the chances for objective-dependency.

Once the QoS models (as explained in Chapter 4 and 5) and regions of objectives (as explained in Chapter 6) are identified, the focus of this chapter is to adaptively produce decision that achieves well-compromised trade-offs with respect to the objectives in each region. To this end, we design the self-adaptive *Decision Maker* component. Specifically, one of the sub-components of *Decision Maker* is the *Optimiser* component, which leverages on MOACO to search and optimise for the possible set of trade-offs decisions. Next, another sub-component, namely *Compromise Explorer*, extracts the decisions that achieve well-compromised trade-offs from the result of Optimiser for autoscaling in the cloud. Theses two sub-components are specified in Section 7.4 and 7.5 respectively. To control the diversity of possible decisions, the possible configured values for each control primitive of a service-instance are bounded within a range, as mentioned in (7.4). The lower bound is set to the maximum value of the predefined value and the latest observed one. On the other hand, after the upper bound is initialised by a predefined value, it is then dynamically adjusted based on the newly decided value and the latest observed one: it is increased by $k\%$ if both values converge to the upper bound; likewise, it is decreased by $k\%$ if both values diverge from the upper bound.

We consider both vertical scaling and horizontal scaling in the actuators. The former refers to change the configurations and provision of control primitives within a PM; the latter refers to boots up/shutdown VMs on the other PMs via migration or replication. In our system, vertical scaling always takes higher priority, providing that modern hypervisors (e.g., Xen [6]) can achieve dynamic vertical scaling with negligible overheads. The resources on a PM are provisioned to the VMs in a first-come-first-serve basis. The horizontal scaling, on the other hand, is only triggered when the resources of the PM tends to be exhausted, i.e., when the total upper bounds of all co-hosted VMs for a resource type exceeds the PM's capacity, the last service-instance that requires to increase the upper bound would be migrated/replicated. Likewise, a VM is removed when its provisions and utilisations for all resource types are below thresholds.

Table 7.1: The Mapping Between Self-Awareness Capabilities and the Sub-Components for Making Trade-Off Decisions in Cloud Autoscaling.

| Self-Awareness Capability | Component | Description |
|---|---|---|
| Interaction-awareness | Optimiser | Knowing how the decision making process can be affected by the objective dependency. |
| Goal-awareness | Optimiser and Compromise Explorer | Knowing how the decision making process can be affected by the QoS and cost models, and their requirements. |
| Self-expression | Optimiser and Compromise Explorer | Self-adapting its behaviours towards well-compromised trade-offs. |

To correctly design the *Decision Maker* component and improve its self-adaptivity at the local level, the proposed approach has been leveraging on the principle of self-awareness. Table 7.1 shows the mapping between each sub-components of *Decision Maker* and the self-awareness capabilities. Here, the self-awareness is concerned with knowing the effects of decisions and the possible trade-offs. By processing the knowledge acquired via the other two internal selves, i.e., *QoS Modeller* and *Region Controller*, the final internal self *Decision Maker* would eventually lead to better self-adaptivity at the global level of the autoscaling process.

From the perspective of our autoscaling framework, the contributions of this chapter realise self-awareness in the internal self component, namely *Decision Maker*. In particular, self-awareness in *Decision Maker* is mainly concerned with knowing the trade-offs in making decisions. In the following, we will specify the techniques and algorithms that achieve self-awareness in *Decision Maker*.

## 7.4 Searching and Optimising Trade-offs Decisions for Autoscaling

Recall that for each region, our aim is to optimise (7.1) subject to the requirements and constraints in (7.2)-(7.4). To this end, we follow the multi-objective ant system described by [9], in which there are one colony and n pheromone structures where n is the number of objectives being optimised. The MOACO relies on probabilistic search-based optimisation that assumes a fixed number of iterations. In each iteration, the ants select different QoS or cost objectives to optimise for. By the end of an iteration, each ant produces an autoscaling decision containing the selected configured values for those cloud control primitives that are inputs of the objectives in (7.1). This is achieved by the use of a probabilistic rule, which expresses the desirability for an ant to choose a particular value for each control primitive. This rule is based on the information about the current pheromone trail, which drives the ants to search better decisions for a particular objective that is selected; and an aggregative heuristic that guides the ants toward choosing better overall decisions with respect to all objectives. Hence, the higher the amount of pheromone and heuristic information is associated with a particular value of the control primitives, the higher the probability is that an ant will choose it. This stochastic nature of the algorithm allows the ants to explore a large space of possible decisions as the search proceeds. Henceforth, it provides more alternatives from the trade-off surface than approaches that based on weighted-sum of objectives.

### 7.4.1 Probabilistic Rule

Suppose that an ant selects the *oth* QoS or cost objective to optimise; the probability for selecting the *xth* configured value of the *ath* control primitive is defined by:

$$p_{x,a,o} = \frac{(\tau_{x,a,o})^\alpha \times (\eta_{x,a})^\beta}{\sum_{y \in S}(\tau_{y,a,o})^\alpha \times (\eta_{y,a})^\beta} \tag{7.5}$$

whereby $S$ denotes the set of possible configured values for the *ath* control primitives; $\tau_{x,a,o}$ is the pheromone for the *xth* configured value of the *ath* control primitives when optimising the *oth* objective. $\eta_{x,a}$ is the heuristic factor for the *xth* configured value of the *ath* control primitive. $\alpha$ and $\beta$ are two parameters that determine their relative importance. It is worth noting that for each ant, the control primitives, which we need to find configured values for, are not restricted to the inputs of the *oth* objective function, but also include those of the other QoS or cost objectives in (7.1).

### 7.4.2 Heuristic Factor

Instead of aggregating the objectives to be optimised, we aggregate the heuristic information to favour the decisions that tend to improve the overall quality of all objectives. In this way, we aim to handle large number of objective while do not require to specify weights on objectives. To this end, we leverage on the normalised, scalar-valued difference between the total improvement and the total degradation for all objectives. This is achieved by comparing the outputs when using a newly configured value to that of the original value, formally expressed as:

$$\eta_{x,a} = \begin{cases} \frac{\sum_{o=1}^m I_{x,a,o}}{1 + \sum_{o=1}^m D_{x,a,o}} & \text{if } \sum_{o=1}^m I_{x,a,o} \neq 0 \\ \frac{\eta_{x,a}^{min}}{1 + \sum_{o=1}^m D_{x,a,o}} & \text{otherwise} \end{cases} \tag{7.6}$$

whereby, for all $m$ objectives that need to be optimised, $\sum_{o=1}^{m} I_{x,a,o}$ is the total improvement over the current setup when using the $x$th configured value of the $a$th control primitive in the decision. Likewise, the total degradation in contrast to the current setup is denoted by $\sum_{o=1}^{m} D_{x,a,o}$. To prevent zero heuristic factor in case where the configured values cannot improve any objectives, we use the minimum non-zero heuristic over all possible configured values, denoted by $\eta_{x,a}^{min}$, as the initial value. It has been shown that these operators promotes better coverage for searching possible trade-offs decisions [9]. In this way, even though the ants select different single objectives to optimise for, the heuristic information still ensure that the configured values, which lead to overall better decisions, are relatively more attractive.

### 7.4.3 Pheromone Update

After all ants complete the search in an iteration, the pheromone trails need to be updated in order to help guiding the search towards better decisions. Unlike the heuristic information, the pheromone is designed to favour the decisions that improve each objective individually. As a result, the pheromone for a particular configured value of a control primitive is specific to an objective. Each pheromone trail is updated by using the rule below:

$$\tau_{a,x,o} = (1 - \rho) \times \tau_{a,x,o} + \Delta\tau_o^{best} \tag{7.7}$$

where $\rho$ $(0 < \rho < 1)$ is a constant that simulates the evaporation of pheromone trails, it determines the speed of evaporation—a larger value implies faster evaporation. Thus, the corresponding configured value becomes unattractive quicker. $\Delta\tau_o^{best}$ is a factor that deposits the pheromone for some favourable decisions. In this work, we follow the MAX-MIN Ant System [9] in which only the configured values that belongs to the iteration's best decision can deposit the pheromone, as defined by the following ($CP_{x,i}$ denotes the *xth* optional value of the *ith* control primitive):

$$\Delta\tau_o^{best} = \begin{cases} \frac{1}{1+h(d_o^{best})^{-1}-h(d_o^{global-best})^{-1}} & \text{if } CP_{x,i} \in d_o^{best}, \text{to argmax } h \\ \frac{1}{1+h(d_o^{best})-h(d_o^{global-best})} & \text{if } CP_{x,i} \in d_o^{best}, \text{to argmin } h \\ 0 & \text{otherwise} \end{cases} \quad (7.8)$$

where $d_o^{best}$ is the best decision for the *oth* objective at the current iteration, and $d_o^{global-best}$ denotes the best ever decision for the same objective; $h$ is the corresponding objective function. Therefore, the configured values in the best decisions would become more attractive; whereas the others, which are not part of the best decisions, will lose pheromone based on the speed of evaporation. In addition, by introducing the best ever decision in the update, we force the search towards the optimal decision for an objective. It is easy to see that, by incorporating the heuristic information and pheromones, the MOACO favours the decisions that do not only benefit all objectives, but also tend to improve each individual objective as much as possible. In this way, the harmonic objectives can be continually optimised in parallel till they reach the point where trade-off needs to be made; while the conflicted objectives would be forced to make trade-offs from the beginning. Consequently, the MOACO is able to produce higher diversity in the trade-off decisions.

Finally, given the fact that only the iterations' best decisions are allowed to deposit the pheromone, the ants may always conclude in the same or similar decisions, which causes the search to be tracked in local spaces. To resolve this issue, we leverage on the solution as used by the MAX-MIN Ant System [9], where the pheromone for the *oth* objective are bounded within a given range, max min denoted as $\tau_o^{max}$ and $\tau_o^{min}$. By the end of an iteration, the bounds are updated using the iteration's best decision:

$$\tau_o^{max} = \begin{cases} \frac{1}{h(d_o^{best})^{-1} \times (1-\rho)} & if \ to \ argmax \ h \\ \frac{1}{h(d_o^{best}) \times (1-\rho)} & if \ to \ argmin \ h \end{cases} \tag{7.9}$$

$$\tau_o^{min} = v \times \tau_o^{max} \tag{7.10}$$

where $h$ is the corresponding objective function. $v$ is a factor that controls the length of the range. By doing so, the diversity in the search is increased and it is more likely to explore a large diversity of different decisions for making trade-offs.

### 7.4.4 Formal Description

Next, we formally explain MOACO in details by means of an algorithmic description. As shown in Algorithm 5, the MOACO takes the models from (7.1) and the possible configured values for each control primitives as inputs. Firstly, MOACO required to take a set of parameters for initialisation; it then compute the heuristic information and set each pheromone using an identical value, meaning that at the beginning, each configured value of a control primitive is equally important for an objective. The search starts off by using numerous working ants (the number of ants should be greater than the number of objectives), each of which selects an objective to optimise for. As we can see from line 6-28, by leveraging on the probabilistic rule, an ant chooses a configured value for each control primitive till it generates a decision. Such decision is then validated by examining whether it meets the SLA and budget in (7.2) and (7.3) for all objectives. if it does then the ant completes its job; otherwise it repeats till the maximum run for finding a satisfied decision (i.e., those that resulted in no violated requirements). It is worth noting that in case no satisfied decision is found, the ant would return the best decision for the selected objective. All discovered decisions are archived regardless of

**Algorithm 5** MOACO for optimising and making trade-offs for autoscaling decisions

**Inputs:**
given a set of objectives to optimise, the associated set of control primitives inputs $CP'$ and $V_a$ ,which denotes the possible configured values for the $ath$ control primitive

**Declare:**
$q$ - the current iteration
$maxIteration$ - the maximum iterations
$ant$ - the current ant
$maxAnt$ - the set of working ants
$r$ - the current run of finding decision for an ant
$maxRun$ - the maximum runs of search for an ant
$d$ - the current decision of an ant for an objective
$d_{ant-best}$ - the best decision of an ant for an objective
$d_o^{best}$ - the best local decision of all ants for an objective
$d_o^{global-best}$ - the best global decision of all ants for an objective so far

**Outputs:**
the set of decisions $D$ for making trade-offs

1: For each configured value of a related control primitives,
2: calculate its heuristic information using (7.6)
3: Initializes pheromone to the same value
4: **for each** $q <maxIteration$ **do**
5:     $q := q+1$
6:     **for each** $ant \in maxAnt$ **do**
7:         select an objective o from (7.1)
8:         $r := r+1$
9:         **for each** $CP_a^{ij} \in CP'$ **do**
10:             *choose* a configured value from $V_a$ via (7.5)
11:         **end for**
12:         **if** $d$ is better than $d_{ant-best}$ for $o$ **then**
13:             $d_{ant-best} := d$
14:         **end if**

```
15:         if not all the objectives in d are satisfied then
16:             if r < maxRun then
17:                 go to line 8
18:             else
19:                 go to line 22
20:             end if
21:         else
22:             d_{ant-best} := d
23:         end if
24:         D := D ∪ d_{ant-best}
25:         if d_{ant-best} is better than d_o^{best} for o then
26:             d_o^{best} := d_{ant-best}
27:         end if
28:     end for
29:     for each objective o from equation (7.1) do
30:         if d_o^{best} is better than d_o^{global-best} for o then
31:             d_o^{global-best} := d_o^{best}
32:         end if
33:         compute τ_o^{min} and τ_o^{max} using (7.9)
34:         for each CP_a^{ij} ∈ CP' do
35:             for each configured value in V_a do
36:                 update τ_{x,a,o} via (7.7) and (7.8)
37:                 if τ_{x,a,o} > τ_o^{max} then
38:                     τ_{x,a,o} := τ_o^{max}
39:                 end if
40:                 if τ_{x,a,o} < τ_o^{min} then
41:                     τ_{x,a,o} := τ_o^{min}
42:                 end if
43:             end for
44:         end for
45:     end for
46: end for
47: return D
```

their quality, and the iteration's best local decision for each objective is determined (line 26). After all ants produce decisions for all the objectives, the best global decision, which results in the best value for an objective so far, is updated (line 31). Next, we compute the pheromone bounds and update the pheromone for each configured value of a control primitive, with respect to each objective (line 33 and 36). To avoid being tracked in local space, a pheromone trail is reinitialised if it exceeds the upper bound; likewise, it is updated accordingly if it fails below the lower bound (line 37-42). Eventually, the search terminates when it reaches its maximum iterations, and returns all the decisions identified.

## 7.5 Identifying Well-Compromised Trade-offs for Autoscaling

It is clear that MOACO is able to search the possible trade-off decisions for autoscaling in the cloud; however, given the large amount of decisions produced, it does not cater for the dynamic and uncertainty of the good compromises in the trade-offs. In this section, we present a simple but efficient mechanism, namely *compromise-dominance*, to adaptively find the decisions that achieve well-compromised trade-offs from the result of MOACO. Specifically, the *compromise-dominance* consists of two phases: superiority phase and fairness phase.

### 7.5.1 Superiority Phase

The first phase in our *compromise-dominance* mechanism is to ensure the superior decisions, which are clearly more favourable than the others. To achieve this, we use the well-known principle of pareto-dominance [67]:

> **Pareto-Dominance:** *A decision $d_1$ pareto-dominates another $d_2$, if and only if, (i) all the objective results achieved by $d_1$ are better than or equivalent to* those achieved by $d_2$; and (ii) the result of at least one objective achieved by $d_1$ is better than the result of the same objective achieved by $d_2$.

It is easy too see that if a decision pareto-dominates another, then it is better than another in terms of the quality of every individual objective and the overall quality for all objectives. In such context, the decisions, which are not pareto-dominated by any others, are called non-pareto-dominated decisions. These decisions are pareto optimal in case no objective can be further improved without making the other objectives worse off. Our aim in this phase is to identify the non-pareto-dominated objectives. If they do not exist, we use the decisions that being pareto-dominated the least.

## 7.5.2 Fairness Phase

After the superior decisions are determined, the second phase aims to ensure the fairness in a decision. That is to say, we are interested in making the trade-offs well-balanced with respect to all objectives. To this end, we leverage on Nash-dominance [108]:

**Nash-Dominance:** *A decision $d_1$ nash-dominates another $d_2$ if and only if there are less objectives that can improve their results by switching from $d_1$ to $d_2$ than vice-versa.*

If a decision nash-dominates another, it means that it is more fair with respect to all objectives, and thus more stable. In particular, the decisions, which are not nash-dominated by others, are called non-nash-dominated decisions. As proven in [108], a non-nash-dominated decision reaches Nash Equilibrium where no objective can be further improved without changing the results of other objectives. It has been shown that, Nash Equilibrium is the most fair state for all objectives in the sense that it exhibits fair competition, or compromise [108]. Here, our aim is to identify the non-nash-dominated objectives; or those that being nash-dominated the least if there is no non-nash-dominated objectives.

However, nash-dominance tends to be limited in reducing the number of decisions when the number of objectives is small, e.g., less than 4 objectives. To this end, we use an additional metric, namely distance of decision, to select well-compromised trade-offs under those cases. Concretely, we select the best value of each objective from all the decisions identified; these values form a theoretically optimal, but unrealistic reference points. We then calculate the normalised Euclidean Distance of the result, which is achieved by each decision, to this reference point. The decision(s), which leads to result that has the minimal distance, is the one(s) that we are seeking.

### 7.5.3 Formal Description

We now formally explain the *compromise-dominance* mechanism in details using an algorithmic description, as shown in Algorithm 6. We can see that the mechanism starts by searching for satisfied decisions from the result generated by MOACO. If it fails to do so, it selects the decisions that result in the least number of violated requirements (line 2-6). This is because in some cases, the violations are inevitable outcomes due to, e.g., heavy conflicts amongst the objectives and/or improper settings of the requirements. In such context, reducing the number of violated requirements takes the highest priority.

Next, we ensure the superiority of decisions by ranking each decision in the set based on the number of other decisions that pareto-dominate it. Smaller number represent higher rank of a decision. Thus, we select a subset of decisions that is ranked the highest in terms of pareto-dominance (line 7-15). Subsequently, by using nash-dominance, the reduced set is further ranked for fairness. The less a decision is nash-dominated, the higher the rank is. Likewise, we select a set of decisions that is ranked as the highest for being nash-dominance (line 16-24). Next, we calculate the reference point and then search a set of decisions that has the smallest distance to that point (line 25-33). Finally, we randomly select one decision from the final set; as at this stage, all decisions tend to

**Algorithm 6** Compromise-dominance based trade-offs for autoscaling decisions

**Inputs:**
given a set of decisions $D$ from Algorithm 5
**Declare:**
$D_{selected}$ -the set of selected decisions
$d_i$ -the $ith$ decision within a set
$n$ -the ranking of nash-dominance
$p$ -the ranking of pareto-dominance
$R$ -the reference point
$dis_i$ -the distance of the $ith$ decision to the reference point
$dis_{smallest}$ -the smallest distance to the reference point
$N$ -the collection of decisions and their ranks of nash-dominance
$P$ -the collection of decisions and their ranks of pareto-dominance
$C$ -the collection of decisions with smallest distance to the reference point
**Outputs:**
the set of decisions $D$ for making trade-offs

1: start
2: $D_{selected} = find$ all satisfied decisions from $D$
3: **if** $D_{selected} = \emptyset$ **then**
4:      $D_{selected} := find$ all decisions that violate the least
5:      number of objectives' requirement from $D$
6: **end if**
7: **for each** $d_i$ in $D_{selected}$ **do**
8:      $p := 0$
9:      **for each** $d_j$ in $D_{selected}$ **do**
10:        **if** $d_i$ is pareto-dominated by $d_j$ **then**
11:          $p := p+1$
12:        **end if**
13:      **end for**
14:      $P := P \cup \langle d_i, p \rangle$
15: **end for**

16: $D_{selected} := find$ the decision(s) with smallest $p$ from $P$
17: $R := find$ the reference point based on the optimal values
18: **for each** $d_i$ in $D_{selected}$ **do**
19:      $n := 0$
20:      **for each** $d_j$ in $D_{selected}$ **do**
21:          **if** $d_i$ is nash-dominated by $d_j$ **then**
22:              $n := n+1$
23:          **end if**
24:      **end for**
25:      $N := N \cup \langle d_i, n \rangle$
26: **end for**
27: $D_{selected} := find$ the decision(s) with smallest $n$ from $N$
28: $R = find$ the reference point based on the optimal values
29: **for each** $d_i$ in $D_{selected}$ **do**
30:      $dis_i :=$ the distance between $d_i$ and $R$
31:      **if** $dis_i \leqslant dis_{smallest}$ **then**
32:          **if** $dis_i < dis_{smallest}$ **then**
33:              $C := \emptyset$
34:          **end if**
35:          $dis_{smallest} := dis_i$
36:          $C := C \cup d_i$
37:      **end if**
38: **end for**
39: **return** randomly selected decision from $C$

be equivalent in terms of superiority and fairness. Whenever there is only one decision left during the process, the algorithm terminates immediately and returns such decision.

## 7.6 Experiments and Evaluations

We integrate our MOACO and the *Compromise-Dominance* (CD) mechanism, denoted as MOACO-CD. To evaluate the proposed approach, we have conducted various quantitative experiments. The primary goal of these experiments is to validate the effectiveness of our approach against other state-of-the-art autoscaling approaches in the cloud, these are:

- **RULE** - A conventional rule-based autoscaling approach that makes decisions using predefined *if-conditions-then-action* mapping e.g., [60] [52]. This approach does not require explicit QoS model as the QoS of a service-instance is assumed to be sensitive to its own control primitives only, e.g., the CPU and thread of the said service-instance. Specifically, violations of QoS would increase all the relevant control primitives to the next higher value; while low utilisation would decrease them to the next lower value.

- **HILL** - A more sophisticated autoscaling approach that relies on our QoS modelling and region controlling techniques, but the decision making process leverages on a weighted-sum formulation of all the dependent objectives e.g., [10] [95] . Here, the approach leverages on greedy and heuristic based solution: the random-restart hill-climbing algorithm for optimisation, in which it starts with an arbitrary decision, then attempts to find a better decision by incrementally and independently changing the values of each control primitives in the models. The algorithm terminates when a maximum iteration has been reached. The best decision, in terms of the weighted-sum formulation, is returned.

- **RANDOM** - Another autoscaling approach that is similar to HILL, but instead

of using hill-climbing, a random optimisation algorithm is applied. This algorithm randomly changes the values of each related control primitive, and terminates when it reaches a maximum number of iterations. The best decision is selected as indicated by the weighted-sum formulation.

- **MOGA** - A most commonly used multi-objective genetic algorithm derived from NSGA-II [124] [57] [51]. We have also designed MOGA to benefit from our QoS mode;ling and region controlling techniques, We configure the optimal population size and number of iterations through careful profiling on our testbed.

Notably, we have configured the approaches to use the identical number of global iterations for the worst case. However, to prevent them from completing with arbitrary latency, we have set a running time threshold (i.e., 75s), which forces the algorithms to terminate and return the best decision found. For HILL and RANDOM, we normalise each objective's result in the weighted-sum of objectives and set all the weights to 1. We use the following 5 criteria to quantify the comparisons:

1. **Coverage of two approaches (C-metric)** [143] - this metric performs pairwise comparison to measure the comparative quality of trade-offs achieved by two approaches. It is calculated using the number of (relatively) better objectives achieved by one approach, divided by the total number of considered objectives. Formally, the C-metric is defined as:

$$C(A, B) = \frac{|r_{o,a} \in A : r_{o,b} \in B, r_{o,a} \succeq r_{o,b}|}{|X|} \quad s.t., \ r_{o,a} = \frac{1}{n} \times \sum_{i=1}^{n} r_{i,o,a} \qquad (7.11)$$

whereby $A$ and $B$ represent two approaches and their corresponding sets of average objective results for all intervals that are being considered. $r_{o,a}$ and $r_{o,b}$ are the average results of the *oth* objective, as achieved by the two approaches; these average results are calculated by averaging the objective values for $n$ intervals, as denoted by $r_{o,a} = \frac{1}{n} \times \sum_{i=1}^{n} r_{i,o,a}$. $|X|$ is the total number of objectives that we consider. $|r_{o,a} \in A : r_{o,b} \in B, r_{o,a} \succeq r_{o,b}|$ counts how many objective results achieved by $A$ are better than those achieved by $B$. Intuitively, the C-metric is an effective method to quantify the quality of trade-offs with respect to the number of the favourable objectives. The greater the value is, the better the approach is. $C(A,B) = 1$ means that the results of all objectives achieved by $A$ are better than those achieved by $B$.

2. ***Generational Distance (G-Distance)*** [123] - this is another intuitive metric that measures the quality of trade-offs. Unlike the C-metric, G-Distance focuses on the generational extents to which the objectives are optimised as achieved by an approach. Formally, it is calculated by:

$$G - Distance = \sqrt{\sum_{o=1}^{|X|} (\frac{1}{r_o^{avg-max}} \times ((\frac{1}{n} \times \sum_{i=1}^{n} r_{i,o,a}) - r_o^{avg-best}))^2} \qquad (7.12)$$

where $r_{i,o,a}$ is the result of the *oth* objective at the *ith* interval, as achieved by the *ath* approach. $r_o^{avg-best}$ and $r_o^{avg-max}$ are the best and the max average result (over all approaches) for the *oth* objective respectively. Smaller value of G-Distance means better results. The remaining notations are the same as (13).

3. ***Violations of Requirements*** - for each approach, we measure the extent to which the requirements (i.e., SLA or budget) of an objective are violated, as defined in:

$$\frac{100}{n} \times \sum_{i=1}^{n} v_i \quad s.t., \ v_i = \begin{cases} \frac{|r_{i,o,a} - t_o|}{t_o} & if \ t_o \succeq r_{i,o,a} \\ 0 & otherwise \end{cases} \qquad (7.13)$$

whereby $v_i$ is the extent of violation at the *ith* interval; $t_o$ is the requirement threshold for the *oth* objective, i.e., SLA or budget; and $n$ is the total number of intervals.

4. **Over- and Under-Provisioning** - for each approach, we quantify over-/under-provision by means of the average difference between the provision and demand for each control primitive type. Formally, the over-provision of a control primitive type is calculated as:

$$\frac{100}{m \times n} \times \sum_{j=1}^{m} \sum_{i=1}^{n} U_{i,j} \ s.t., \ U_{i,j} = \begin{cases} \frac{|u_{i,j} - u'_{i,j}|}{u'_{i,j}} & if \ u_{i,j} > u'_{i,j} \\ 0 & otherwise \end{cases} \quad (7.14)$$

where $u_{i,j}$ is the $jth$ VM (for hardware control primitives) or service-instance (for software control primitives) on a PM. $n$ and $m$ are respectively the number of intervals and VMs/service-instances. $u'_{i,j}$ is the corresponding demand using the highest possible value that we have observed. The calculation of under- provision can be similarly applied.

5. **Overhead** - Finally, we measure the overhead of each approach in terms of the latency in making decisions. In particular, we report on the results for both best and worst case scenarios.

## 7.6.1 Experiments Setup

We conducted experiments on private cloud using a cluster of PMs, each of which has Intel i7 2.8GHz Quad Cores and 4GB RAM. The PMs use Xen v3.0.3 [6] as the hypervisor and the autoscaling process is running on Dom0. To eliminate the interference caused by Dom0, we allocated one CPU core and 600 MB RAM to it, which tends to be sufficient. Our approach and the other competitors are implemented using Java JDK 1.6. To simulate QoS interference caused by the VMs while not exhausting resources, we run three co-hosted VMs on each PM. Initially, we allocate the same amounts of hardware resources for each of the co-hosted VMs, these are 30% cap of a dedicated CPU core and 250 MB RAM. All VMs run linux kernel v2.6.16.29. It is worth noting that the experiments have relied on the same testbed as in Chapter 4 and 5.

Our experiments leverage on RUBiS [5], which is a cloud-based application consists of 26 co-located services using the eBay.com model. For simplicity, we have used three RUBiS snapshots, each of which consists of a 2-tiers (i.e., application and database tiers) based RUBiS application. A RUBiS snapshot is deployed with a software stack including linux kernel v2.6.16.29, Tomcat v6.0.28 and MySQL v3.23.58 on each co-hosted VM of the master PM. The snapshots use heterogeneous database volume size ranging from 1GB to 5GB data. We have implemented sensors and actuators deployed on each service-instance/VM for collecting the online data and scaling the control primitives respectively.

In this work, we have realised vertical scaling actions (a.k.a. scale-up/-down) by using a customised listener on Tomcat and the management module of Xen. As for horizontal scaling actions (a.k.a. scale-in/-out), we leverage on master-salves based replication. Each of the three RUBiS snapshots and its replicas are linked to a dedicated load balancer. Three client emulators are used and they apply read/write pattern to generate requests for each load balancer. To simulate a realistic workload within the capacity of our testbed, we vary the number of clients according to the compressed FIFA98 workload [14]. This setup can generate up to 400 parallel requests, which is large enough to simulate QoS interference.

### 7.6.2 QoS Attributes, Primitives and Configurations

For the simplicity of exposition, we have selected commonly used QoS attributes and primitives in the evaluation. In our experiments, we have used identical setups for all approaches. As listed in Table 7.2, these QoS attributes and primitives are per-service except for CPU and memory as they are shared on a VM. Table 7.3 shows the configurations for each control primitive type. Scale-out occurs if the summed *max* of CPU or memory for all the co-hosted VM exceeds the PM's capacity. The hardware and software control primitives of a new replica VM and service are set as the initial value $i$. Likewise, scale-in occurs if CPU and memory of a VM are provisioned as *min*, and their utilisations are below $u$. By carefully examining the objective-dependency of services based on our QoS modelling and region clustering approaches, we intend to manage and autoscale the services that exhibit the most fluctuated performance, and those that are the most likely to lead to the largest number of dependent objectives in a decision process. We have identified two services on each RUBiS snapshot while leaving the other 24 services as unmanaged, generating interference only. Table 7.4 illustrates the SLA and budget (per interval on a VM) for each managed service-instance. We can see that there are 6 managed service-instance on a PM, each has 5 different objectives. All these setups give us up to 30 dependent objectives in one decision making process. Table 7.5 is the configurations for MOACO.

Table 7.2: The Examined QoS Attributes and Cloud Primitives.

| QoS and Primitives | | Description |
|---|---|---|
| **Output** | Response Time (ms) | The average leaped time between a service-instance receives and replies a request. |
| | Throughput (req/min) | The average rate of completed requests. |
| | Reliability (%) | The percentage of requests that being completed faster than the SLA. (2-4 ms) |
| | Availability (%) | The percentage of time that the average response time above a threshold. (4 ms) |
| **CP input** | CPU (%) | Observed average CPU utilisation of a VM. |
| | Memory (MB) | Observed average Memory utilisation of a VM. |
| | Thread (no. of req) | Observed maximum concurrent threads of a service-instance. (a modified control knob of Tomcat's *maxThread* property) |
| **EP input** | Workload (req/min) | Observed average request rate of a service-instance. |

In each experiment run, the sampling and modelling intervals are both 120s with the total of 70 intervals; and there is one new sample per interval for updating the QoS models. The autoscaling process is triggered when any violations of SLA or low utilisation

Table 7.3: Configurations for Each Control Primitive Type.

|        | i     | u   | step | min   | max   | t   | k   | p      |
|--------|-------|-----|------|-------|-------|-----|-----|--------|
| CPU    | 30%   | 50% | 1%   | 15%   | 40%   | 70% | 10% | $0.01  |
| Memory | 250MB | 50% | 5MB  | 230MB | 280MB | 70% | 10% | $0.002 |
| Thread | 5     | 50% | 1    | 4     | 10    | 70% | 10% | $0.017 |

*i = the initial value; u = the lowest possible utilisation for triggering autoscaling; step = the margin between two neighbour values; min = the minimum value; max = the maximum value; t = the % threshold to trigger change of the max value; k = the % extent to which the max value is changed; p = the price per unit per interval for a service-instance.*

Table 7.4: SLA and Budget for the Managed Service-Instances. (6 service-instances, each has 5 objectives)

|     |          | Response Time (ms) | Throughput (req/min) | Reliability (%) | Availability (%) | Cost ($) |
|-----|----------|--------------------|----------------------|-----------------|------------------|----------|
| VM1 | Service1 | 2                  | 180                  | 85              | 90               | 1.2      |
|     | Service2 | 2                  | 180                  | 85              | 90               | 1.1      |
| VM2 | Service3 | 3                  | 150                  | 85              | 90               | 1.17     |
|     | Service4 | 2                  | 180                  | 85              | 90               | 1.33     |
| VM3 | Service5 | 4                  | 140                  | 90              | 85               | 1.02     |
|     | Service6 | 2                  | 180                  | 90              | 90               | 1.17     |

Table 7.5: Configurations of MOACO.

| $\alpha$ | $\beta$ | $\rho$ | $v$ | maxIteration | maxAnt | maxRun |
|----------|---------|--------|-----|--------------|--------|--------|
| 4        | 1       | 0.1    | 0.5 | 5            | 150    | 100    |

is detected for one interval. Given that the QoS modelling approach requires certain historical data to build the models, we record the achieved QoS and cost of all managed service-instances on the master PM for the rear 50 intervals. We have conducted 10 experiment runs for each approach.

## 7.6.3 Quality of Trade-offs

To evaluate the quality of trade-offs achieved by our approach, we leverage on the afore-mentioned C-metric and G-Distance; the results are plotted in Table 7.6. For C-metric, our MOACO-CD is better than MOGA as the latter is limited in optimising and making trade-off for a large number of objectives; it also does not consider well-compromised trade-offs. MOACO-CD is better than RULE, which does not allow explicit optimisation and trade-off. Finally, MOACO-CD is also better than HILL and RANDOM, because the weighted-sum of objectives in these two has greatly restricted their search into local areas of the search space, henceforth they tend to be limited in improving the diversity of trade-offs decisions. As a result, we can conclude that our MOACO-CD is the best according to C-metric, meaning that it has the best quality of trade-offs in terms of the number of the favourable objectives.

Table 7.6: Quality of Trade-offs. (the best is highlighted in bold)

| Pairwise Comparison | | |
|---|---|---|
| C(**MOACO-CD**, MOGA) : C(MOGA, **MOACO-CD**) | | 0.8 : 0.2 |
| C(**MOACO-CD**, RULE) : C(RULE , **MOACO-CD**) | | 0.73 : 0.27 |
| C(**MOACO-CD**,HILL) : C(HILL, **MOACO-CD**) | | 0.8 : 0.2 |
| C(**MOACO-CD**, RANDOM) : C(RANDOM, **MOACO-CD**) | | 0.73 : 0.27 |

| | MOACO-CD | MOGA | RANDOM | RULE | HILL |
|---|---|---|---|---|---|
| G-Distance | **0.4071** | 1.2707 | 0.9407 | 1.5892 | 1.6958 |

As for G-Distance, we note that our MOACO-CD again achieves the best result, producing the best quality of trade-offs in terms of the extents to which the objectives are optimised. We can see that RANDOM is better than MOGA, RULE and HILL, this is because even though it is restricted by the weighted-sum of objectives, RANDOM tends to largely improve on a few objectives and thus leading to second best G-Distance result. The MOGA is ranked the third, because despite it caters for multi-objective, the inability to handle large number of objective and the limited diversity have caused it to optimise only a small amount of objectives. We can also see that the RULE and HILL are the worst and they exhibit marginal difference. This is because RULE is not capable to perform explicit trade-offs and optimisation; while HILL is greatly affected by high latency due to its greedy nature.

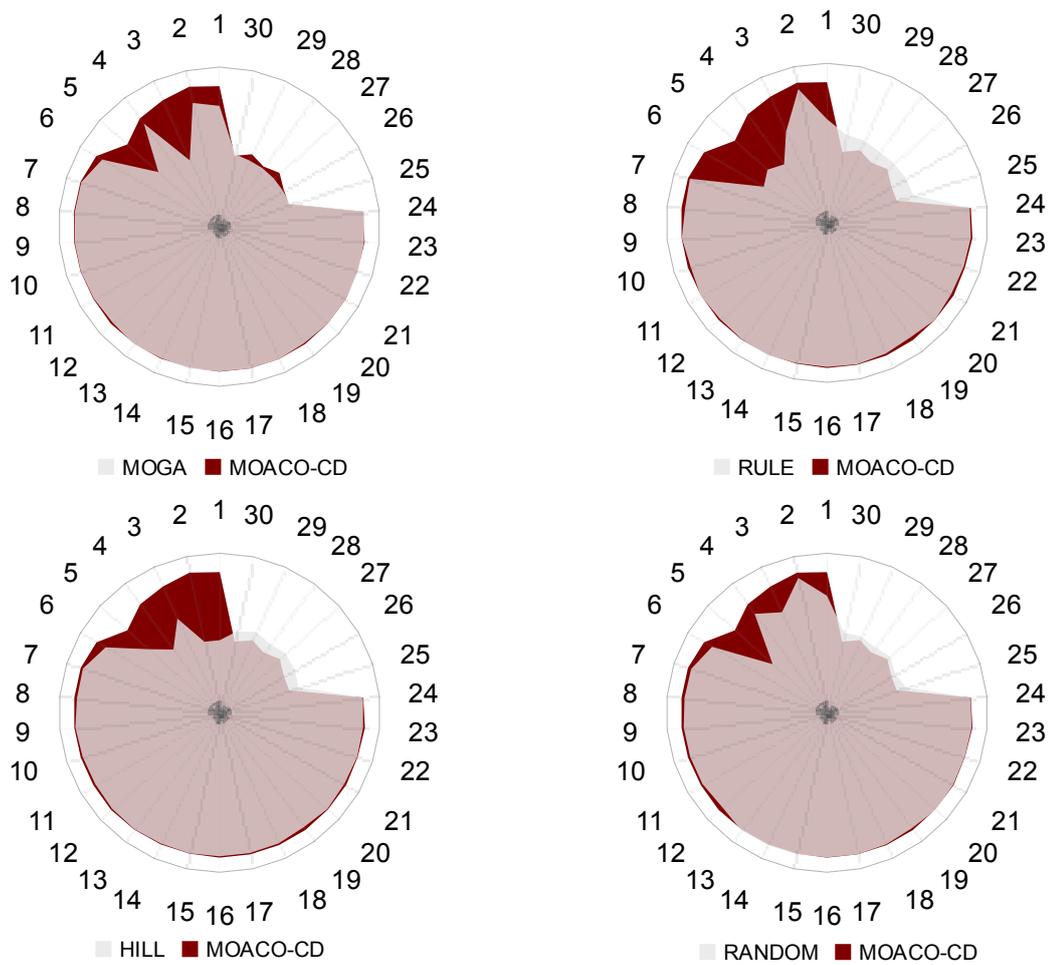

Figure 7.2: Normalised Pair-Wised Comparison Between MOACO-CD and Each of the Other Approaches. (the larger area means better trade-offs; objective number 1-6 denotes Response Time; 7-12 denotes Throughput; 13-18 denotes Reliability; 19-24 denotes Availability and 25-30 denotes cost)

To provide a detailed view of the achieved QoS and cost values, Figure 7.2 shows pairwise comparisons between MOACO-CD and other approaches with respect to each of the 30 objectives that we have considered. For each objective dimension In the figure, a point that closer to the labeled number means better value for that objective. For example, the MOACO-CD produce better result on objective number 1 as when compared with the others. Intuitively, when all 30 objective dimensions are considered, the larger area that cover by an approach means better quality of trade-off is achieved. We can clearly see that in contrast to the other autoscaling approaches, the MOACO-CD covers larger area. In particular, its decisions tend to be significantly better than those of the others on most QoS objectives while slightly worse, mainly on the Availability (against MOGA) or Cost (against RULE, HILL and RANDOM) objectives, which are smaller in number. This means MOACO-CD favours decisions that largely improve on the majority of the objectives; while causing smaller degradation to others.

In conclusion, the MOACO-CD produces better trade-offs than the others in terms of the numbers of favourable objectives and the extents to which they are optimised. This is because it favours the autoscaling decisions that do not only benefit all the objectives, but also tend to improve on each individual objective as much as possible. Therefore, MOACO-CD is capable to perform better optimisation and find trade-offs decisions with higher diversity for large number of objectives. In addition, the compromise-dominance balances the improvements in the objectives, which lead to well-compromised trade-offs. In particular, the possible trade-offs are handled properly, not only for the naturally conflicted objectives (e.g. Throughput and cost objective of a service); but also for the conflicts caused by QoS interference.

### 7.6.4 Violations of Requirements

Next, we examine whether the decisions made by our approach can eliminate runtime violations of the SLA and budget, as listed in Table 7.4. We use (7.13) to assess the extents of these violations when they occur. As shown in Table 7.7, violations do exist, mainly for the Response Time and Throughput objectives. However, we can see that MOACO-CD leads to significantly smaller violations as when compared with the others—it has the best results for 12 out of the 13 cases. In contrast, MOGA is ranked the second as it obtains the second best results for most cases. This proves that MOACO-CD outperforms MOGA in reducing SLA violations, while optimising and making trade-offs for large number of objectives. RANDOM is ranked the third while HILL and RULE do not differ much in terms of the overall violations. In particular, the maximum violation of MOACO-CD is only 102.02%, which is at least 6 times better than the 2307.53% for RULE, the 2364.84% for HILL, the 617.17% for RANDOM and the 614.9% for MOGA. We can also see that MOACO-CD has the smallest standard deviation on the violations for different service-instances, meaning that violations in MOACO-CD are better balanced than the other four. This implies that the trade-offs caused by QoS interference are better compromised; otherwise, it can result in imbalanced scenarios where the QoS attributes of some service-instances are advantaged while those of the others are severely violated, e.g., the cases for RULE and HILL.

Table 7.7: The Average Violations (%). (the best is highlighted in bold)

| | | MOACO-CD | MOGA | RULE | HILL | RANDOM |
|---|---|---|---|---|---|---|
| Service1 | Response Time | **102.02** | 190.42 | 921.62 | 853.32 | 259.94 |
| | Throughput | **10.49** | 13.45 | 14.37 | 14.52 | 14.26 |
| Service2 | Response Time | **86.37** | 316.74 | 2370.53 | 434.88 | 401.09 |
| | Throughput | **19.15** | 20.46 | 21 | 21.86 | 21.06 |
| Service3 | Response Time | **99.52** | 389.12 | 405.33 | 293.55 | 457.53 |
| | Throughput | **39.71** | 39.93 | 39.79 | 41.57 | 40 |
| Service4 | Response Time | **73.79** | 614.90 | 797.53 | 730.25 | 617.17 |
| | Throughput | **19.84** | 21.33 | 20.75 | 21.58 | 19.86 |
| Service5 | Response Time | **0** | 186.71 | 357.58 | 676.56 | 236.49 |
| | Throughput | **13.06** | 13.08 | 13.22 | 16.48 | 14.81 |
| Service6 | Response Time | **16.37** | 560.74 | 214.88 | 2364.84 | 192.83 |
| | Throughput | 61.81 | 61.93 | **60.18** | 62.49 | 62.3 |
| | Availability | **0** | **0** | **0** | 0.02 | **0** |
| Standard Deviation | | **0.52** | 2.69 | 6.62 | 7.55 | 2.49 |

As a detailed example, Figure 7.3 illustrates the fluctuations of QoS and cost for Service 6 in one experiment run. For Response Time (Figure 7.3a and 7.3b), we can clearly see that, in contrast to the others, our MOACO-CD does not only significantly reduce violations, but also produce better and more stable response time when the SLA is complied. This is the same case for Throughput (Figure 7.3c). From Figure 7.3d, we can

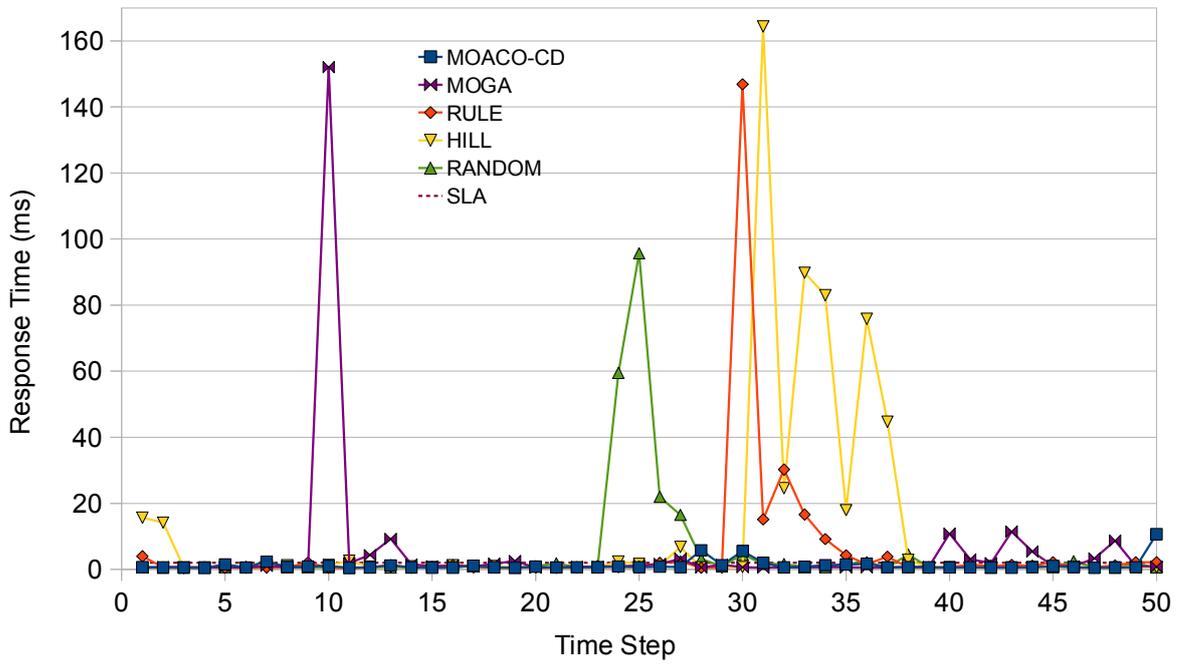

(a) Response Time. (coarse grained)

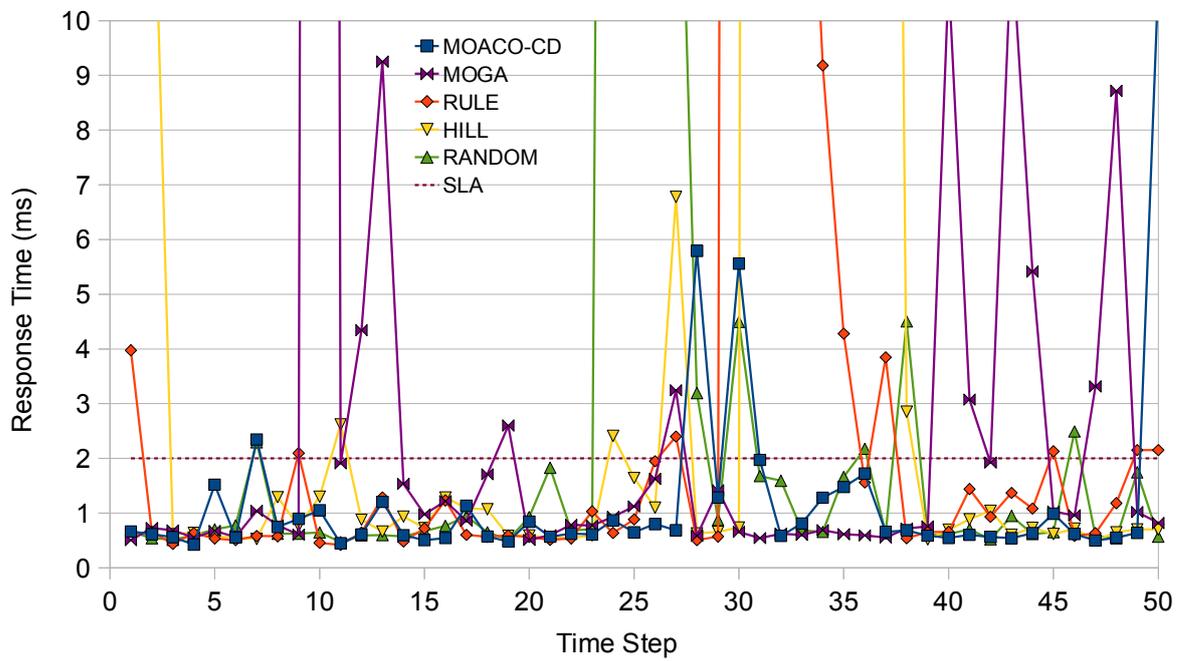

(b) Response Time. (fine grained)

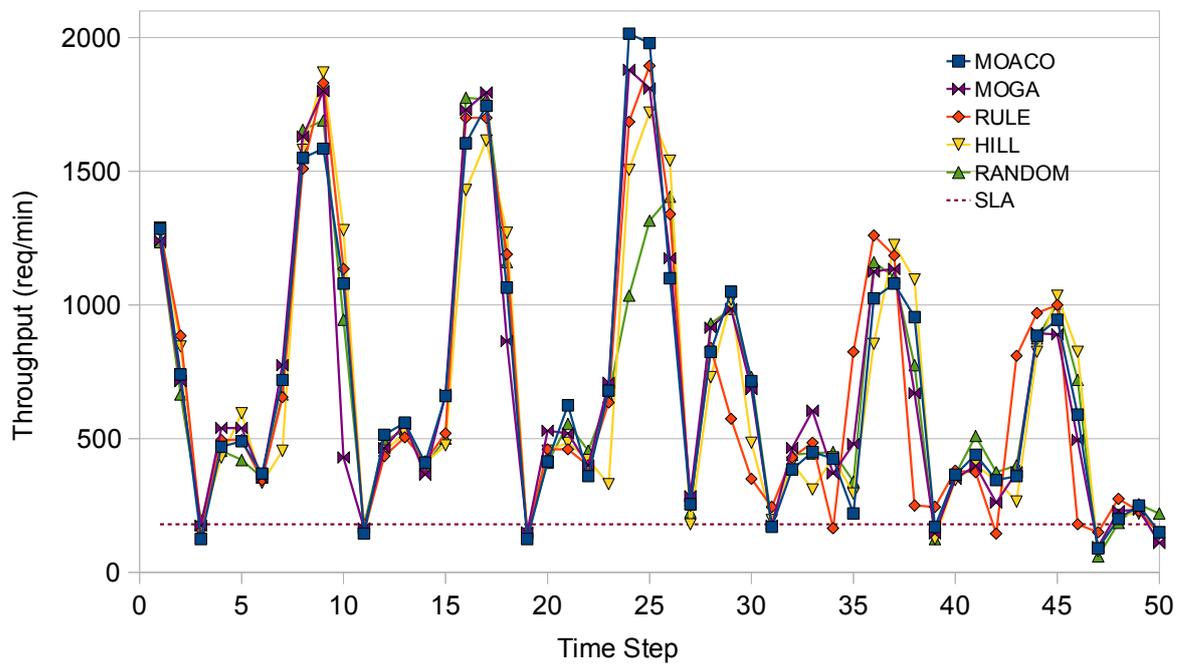

(c) Throughput

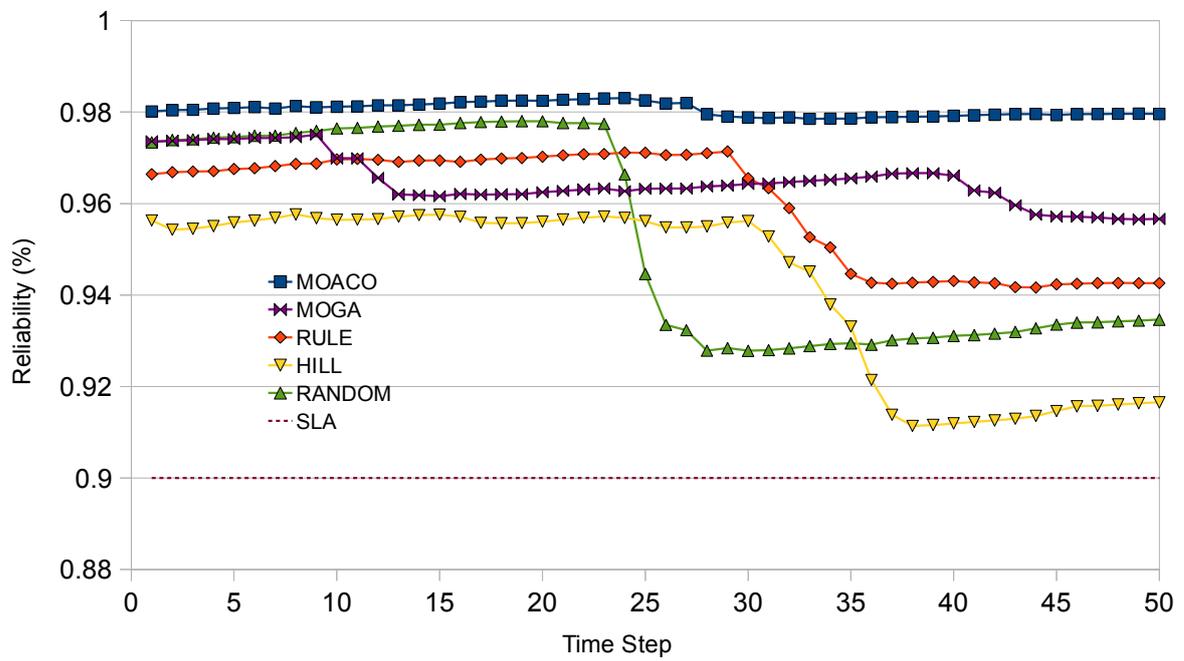

(d) Reliability

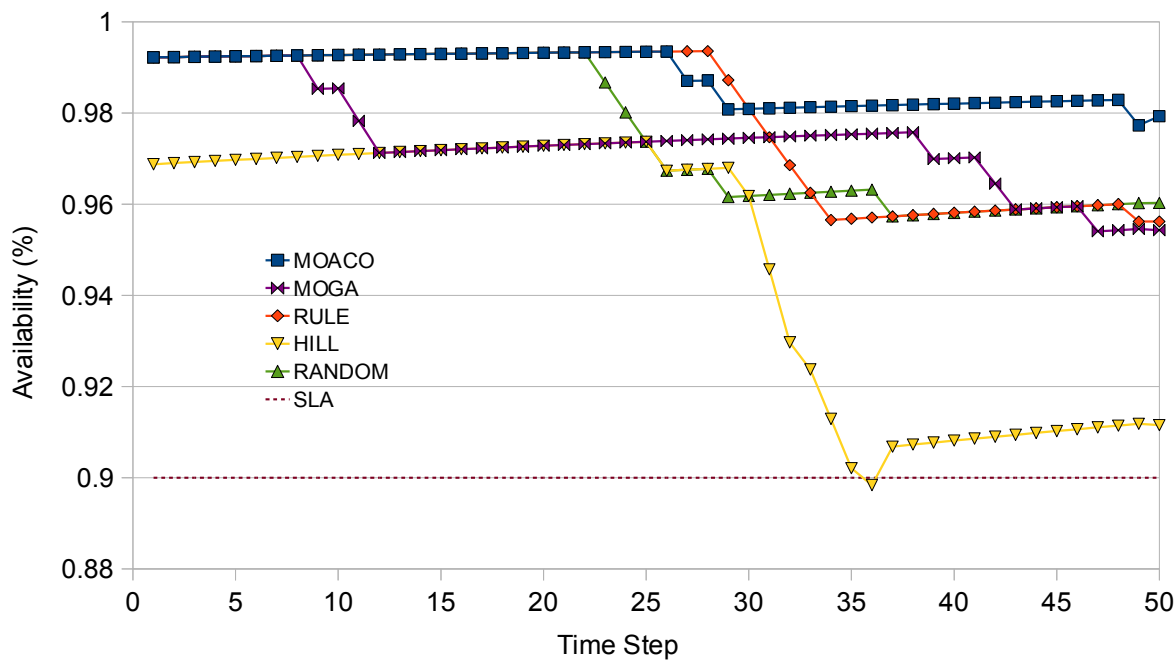

(e) Availability

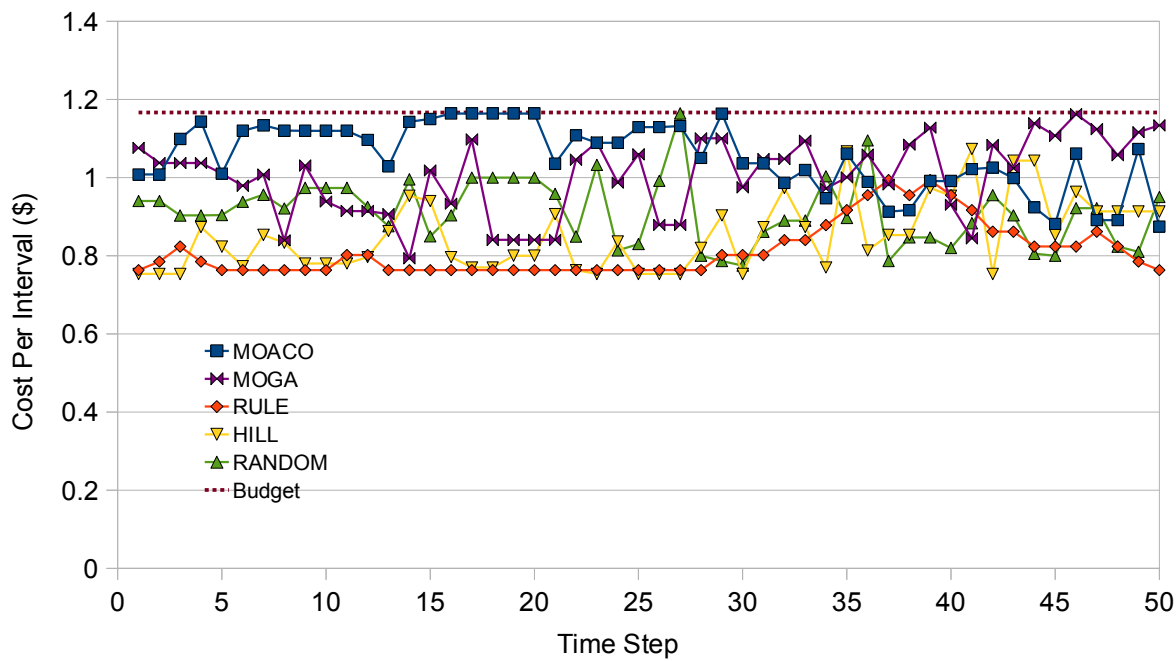

(f) Cost

Figure 7.3: The Achieved QoS Results and Cost for Service 6 in One Experiment Run.

Table 7.8: The Over- and Under-Provisioning (%). (the best is highlighted in bold)

| | | MOACO-CD | MOGA | RULE | HILL | RANDOM |
|---|---|---|---|---|---|---|
| CPU | Over-provision | **6.69** | 15.09 | 9.48 | 20.36 | 15.07 |
| | Under-provision | 14.40 | 9.86 | 11.82 | 13.06 | **9.46** |
| Memory | Over-provision | 10.50 | 10.66 | **0.52** | 0.94 | 3.00 |
| | Under-provision | **2.21** | 2.50 | 19.13 | 20.40 | 11.13 |
| Thread | Over-provision | 22.29 | 37.05 | **21.94** | 32.38 | 40.00 |
| | Under-provision | 22.92 | 15.55 | 39.05 | 27.42 | **9.30** |

observe that there are no violations for Reliability. In such case, MOACO-CD does not only constantly produces the best performance, but also tends to have the most stable results along the trend: we can clearly see that for the other approaches, the achieved reliability drops gradually at around 25-30 time step; whereas the results achieved by MOACO-CD do not fluctuate much. We observe the similar result for Availability (Figure 7.3e). Finally, we can see that the cost incurred by MOACO-CD is similar to MOGA, but slightly higher than the others (Figure 7.3f); however, the extra cost is within the given budget and it is therefore acceptable. That is to say, MOACO-CD comes with some extra costs but it can lead to significantly better performance on many other QoS objectives.

In summary, we can conclude that MOACO-CD performs significantly better than the other approaches for reducing SLA violations on a large number of objectives. In addition, it leads to better and more stable results when the SLAs are complied. This might come with slightly higher cost, yet still comply with the budget requirements. Further, MOACO-CD achieves well-balanced improvement on the QoS attributes for different service-instances, which implies that the trade-offs caused by the QoS interference for both services and VMs are well compromised, even for large number of objectives.

## 7.6.5 Over- and Under-Provisioning

We now evaluate the proposed approach by means of the difference between the provision and demand for each control primitive type. We calculate over- and under-provisioning using (7.14). Table 7.8 shows the average results for all managed service-instances and VMs on the master PM.

For CPU and thread, the results of MOACO-CD do not differ much as when compared to the other four. In addition, the amounts of over-/under-provision are balanced. Interestingly, for memory, we can see that MOACO-CD and MOGA perform significantly better than the others on under-provision, but they are the worst on over-provision with considerable difference. This is because they detect that memory can be the most critical control primitives that significantly influences the QoSs. Moreover, they have assumed that some extra costs can lead to significantly better performance on other objectives. Consequently, both MOACO-CD and MOGA try to avoid under-provisioning by allocating more memory than the actual demand. Indeed, in contrast to MOACO-CD and MOGA, although the other three have better results on over-provision, their bigger under-provision have resulted in significantly worse QoS and SLA violations (especially for RULE and HILL), as evident in Section 6.4. Finally, although MOACO-CD and MOGA obtain similar results for elasticity, we have shown that MOACO-CD outperforms MOGA on the quality of trade-offs and the ability to reduce SLA violation and to optimise QoSs.

In conclusion, our approach results in good elasticity, providing that the amounts of over-/under-provision achieved by MOACO-CD are balanced and acceptable for CPU and thread. Among the others, MOACO-CD tends to have the best under-provision and the second worst, yet acceptable over-provision for memory. However, this is a trade-off between cost and QoS attributes, where the MOACO-CD has assumed that large improvements on the QoS attributes can be achieved by having slightly more costs, which are mainly spent on the memory.

Table 7.9: The Overhead (s). (the best is highlighted in bold)

| | MOACO-CD | MOGA | HILL | RANDOM |
|---|---|---|---|---|
| Best case | **1.2** | 12.3 | 6.8 | 3.5 |
| Worst case | 50.3 | 69.7 | 75.09 | **38.91** |

### 7.6.6 Overhead

Finally, we validate the overhead of our approach by computing latency of the decision making process. Undoubtedly, RULE results in negligible overheads and thus it is omitted from the comparison. As shown in Table 7.9, we can see that for all four approaches, there is a considerable difference between the worst case and best case scenarios. Indeed, their actual overhead can be sensitive to the complexity of the used models (i.e., by the *QoS Modeller*); and the number of objectives that are assigned in the same decision making process (i.e., by the *Region Controller*). This is also the primary reason why we have reported on the best/worst case result instead of using the mean value, which can only provide coarse assessment of the runtime overhead. In contrast, considering results for extreme cases can provide insightful view for decision making in cloud autoscaling.

For the best case scenario, our MOACO-CD has the smallest overhead (1.2s) while the HILL has the biggest. However, the results of all four approaches are acceptable. On the other hand, the RANDOM achieves the smallest overhead (38.91s) in the worst case scenario; while the MOACO-CD, MOGA and HILL report 50.3s, 69.7s and 75.09s respectively. Nevertheless, as we have seen in previous sections, the MOACO-CD is significantly better than RANDOM in terms the quality of trade-offs and its capabilities in reducing SLA violations. For both cases, MOGA has bigger latency than MOACO-CD due to the overhead of pareto-dominance sort during optimisation. Another observation is that HILL is often forced to terminate as it reaches the runtime threshold (i.e., 75s); thus its actual overhead in the worst case scenario can be bigger than 75.09s. This is the main reason that causes its poor performance in the quality of trade-offs and violations. Overall, MOACO-CD has acceptable overhead even for the worst case, providing that the sampling interval is 120s.

## 7.7 Conclusion

The trade-off decision making is undoubtedly a crucial and challenging task for autoscaling in the cloud. In this chapter, we present a self-adaptive approach for autoscaling decision making in the cloud. In particular, it adaptively resolves the trade-offs without human intervention. By leveraging on MOACO, the approach dynamically searches and optimises for possible trade-offs with high diversity. Further, we propose compromise-dominance for adaptively selecting the decision that leads to well-compromised trade-offs. The experiments show that, in contrast to the rule-based, single-objective heuristic based, single-objective randomised and MOGA based autoscaling approaches, our approach produces better trade-offs quality in terms of the numbers of favourable objectives and the the extents to which they are optimised; and much smaller violations of the requirements with large number of objectives. Moreover, it results in acceptable overhead and has balanced elasticity in terms of the over-/under-provision.

By now, we have explored the entire proposed autoscaling framework, not only at a global level (i.e., Chapter 3), but also at local levels of each important components with great details (i.e., Chapter 4, 5, 6 and this chapter). In the next chapter, we will qualitatively evaluate the proposed autoscaling framework against various criteria.

# CHAPTER 8

# REFLECTION AND APPRAISAL

## 8.1 Dealing with Dynamics and Uncertainty

In the context of computing systems, dynamic means continuous changes in the system and the environments. It implies the necessity of self-adaptivity on the system to react on changes in order to assure some of its properties. Uncertainty, on the other hand, is a similar, but still different notion of dynamics. It describes the unpredictability of an object and hence implies the difficulty in realising self-adaptivity on systems.

One can argue that cloud-based services is dynamic but certain in some cases, e.g., there is a fluctuated workload but can exhibit strong seasonality. That is to say, the behaviour of the cloud-based service does change but tend to be predicted to some extent. In such cases, a simple autoscaling system might be a working solution providing that the cloud-based services has undergone a formal profiling process and can be adaptive in response to the changes. However, most commonly, the behaviour of the cloud-based service can exhibit both dynamics and uncertainty, e.g., unexpected spike on workload and QoS interference. In this case, ensuring effective self-adaptivity when designing the autoscaling system becomes mere difficult due to the unpredictability. This thesis provides explicit treatment for scenarios that exhibit both dynamics and uncertainties.

We intend to evaluate how the dynamics and uncertainties of cloud-based services are captured and handled by the proposed autoscaling framework. As we have discussed in previous chapters, dynamics and uncertainty are the fundamental factor in cloud environment and such nature implies that it is hard to presume the QoS performance and cost of cloud-based services. The dynamics and uncertainties can be associated with many factors in cloud autoscaling, including QoS sensitivity, QoS interference, the effects of granularity of control on the global benefit, the effects of decision on objectives and the trade-offs. The proposed autoscaling framework, through using the principle of self-awareness and the related algorithms, have been explicitly designed to cope with those factors.

- *Dynamics and uncertainties in QoS modelling*

  – We have designed a self-aware and self-adaptive QoS modelling approach, which continually learns the knowledge about the significant inputs of QoS models, the magnitude of these inputs, and QoS interference. As the knowledge changes in an uncertain manner, the modelling process can dynamically acquire such knowledge and adapt the expression of the QoS models accordingly.

- *Dynamics and uncertainties in granularity of control*

  – We have used a self-aware and self-adaptive region clustering approach that continually learns the knowledge about the effects of granularity of control on the global benefit and deployment of cloud-based services. In the events of unpredictable and uncertain changes, the approach can aware of when and which regions to adapt because of its ability of acquiring the knowledge.

- *Dynamics and uncertainties in decision making*

  – We proposed a self-aware and self-adaptive autoscaling decision making approach to continually learn the knowledge regarding the effects of decision on objectives and the possible trade-offs, which can be caused by both naturally conflicted objectives and QoS interference. The decision making process is capable to adapt its own behaviours according to the changing knowledge, which is uncertain in nature. In such a way, we aim to adaptively reach the optimal (or near-optimal) decision that achieve well-compromised trade-offs.

Collectively, the previous points lead to a self-aware and self-adaptive autoscaling system, which can handle different forms of dynamics and uncertainties that exhibited by cloud-based services, as we have quantitatively analysed in previous chapters. Therefore, the QoS and cost objective of all cloud-based services can be continually optimised and thus their requirements are better complied, which leads to better elasticity.

## 8.2 Scalability

Scalability refers to the ability of a system to accommodate the growth of data, processes and workload etc. In the context of self-aware and self-adaptive autoscaling system, we are particularly interested in the scalability of the approach with respect to the increasing amount of historical data, the number of cloud primitives, the number of cloud-based services and their objectives.

- **_Scalability with respect to the amount of historical data_**

    - The amount of historical data influence the overhead of the self-aware and self-adaptive QoS modelling. One way to improve scalability is to include only the most significant inputs in the QoS models, as we have achieved in the primitives selection phase. By limiting the inputs in the models, we significantly reduce the training time required for QoS function training because there is less effects caused by the increasing amount of historical data. We have shown that such design also leads to better model accuracy.

– However, although we can limit the inputs for QoS model, the actual amount of historical data does not change and hence they can become extremely large as time goes by. This may not be a major issue for primitives selection process given that the calculation of cumulative relevance and redundancy is very efficient; but it can be a bottleneck for the QoS function training. Therefore a *forgotten strategy* is desired when there is no need to take too much data into account. To achieve such goal, one could set a threshold to the maximum number of historical intervals to be recorded. Once such threshold is exceeded, the QoS function training process can apply cross-validation to examine if dropping data from the oldest intervals would affect the model accuracy. For example, if the reduction in accuracy is less than 1% error then such data can be removed.

- **Scalability with respect to the number of cloud primitives**

  – For self-aware and self-adaptive QoS modelling, the increasing number of cloud primitive can be accommodated by the primitives selection phase, which ensures only the significant primitives are used as inputs. This is achieved by the proposed hybrid-learners approach for primitives selection, where the core is an optimisation for information relevance and redundancy. Increasing the number of primitives can enlarge the search space for primitives selection. To this end, in this thesis, we have applied an efficient randomised optimisation algorithm to this problem, but more sophisticated algorithms can be applied when there is a need. As for the QoS function training process, the primitives selection phase have ensured a less but more useful number of primitives in the model, which have presumably improve the efficiency of the machine learning algorithms in training and calculation of the model.

  – The number of cloud primitives can affect the scalability of the self-aware and

self-adaptive region clustering approach, which is used to determine the granularity of control. However, since the approach performs clustering in a linear manner, the correlation between the overhead and the number of primitives is also linear.

— The search space in autoscaling decision making increases dramatically when the number of cloud primitives increases. However, one of the benefits of the proposed MOACO is that it can efficiently resolve NP-hard problems and achieve good results by exploring in diversified parts of the search space, leading to less effort on computation. In the decision making for cloud autoscaling, the search space consists of $\prod_{i=1}^{n} c_i$ possible decisions, whereby $c_i$ is the number of configured value for the $ith$ control primitives; $n$ is the total number of control primitives for all dependent objectives. MOACO only need to examine $\sum_{i=1}^{n} c_i + i \cdot a \cdot r$ decisions in the worst case, where $i$, $a$ and $r$ are the number of iteration, the number of ants and the number of runs for an ant to find satisfactory decision, respectively. In particular, $\sum_{i=1}^{n} c_i$ decisions are searched when calculating the heuristics information while the remaining $i \cdot a \cdot r$ decisions are examined when updating the pheromones, which is an iterative process throughout the algorithm. Taking the initial setup of the experiment in Chapter 7 as an example, the complete search space has $7.29 \times 10^{11}$ decisions while MOACO at most need to examine $7.5141 \times 10^{4}$, which is only around $1.02 \times 10^{-5}\%$ of the search space.

- **_Scalability with respect to the number of cloud-based services and their objectives_**

    — Increasing the number of cloud-based services and their objectives can cause larger overhead for the self-aware and self-adaptive modelling process on the

root domain (e.g., Dom0). However, we believe that such computational requirements can be fulfilled by modern physical machines. In addition, it is also possible to setup dedicated machine(s) for QoS modelling process. Finally, applying admission control, which restricts the number of services on a PM, can be another possible solution. As for the self-aware and self-adaptive region clustering approach for granularity of control, its scalability tends to be linear to the number of cloud-based services and their objectives.

– The scalability of self-aware and self-adaptive decision making approach is related to the number of dependent objectives, thus it can work in large number of services. This is because increasing the number of services may not influence the approach, as the region clustering approach distributes the independent objectives of these services into different decision making processes, which run independently. In other words, a large number of service may not affect the decision making as long as the number of dependent objectives in a process does not change significantly. If it is known that the number of dependent objective will be largely increased, there are additional mechanisms to improve scalability, e.g., by using admission control to restrict the number of service on a VM and the VMs on a PM, which will limit the number of dependent objectives in one process. However, in cases the additional mechanisms are not applicable, the proposed decision making approach can still be tuned the configurations of MOACO. This can be achieved by profiling with respect to the possible number of dependent objectives. It is worth noting that since we consider the trade-offs caused by QoS interference, we have evaluated up to 30 objectives in one decision making process, which itself is a significantly larger scale as when compared to the small scale (e.g., 2 - 4 objectives) in existing work.

- In particular, our compromise-dominance has similar runtime complexity to the pareto-dominance based sort in NSGA-II, but with some extra overheads on nash-dominance and distance of decisions. However, unlike NSGA-II that sorts in each iteration during the optimisation, we only need to run compromise-dominance once after the optimisation of MOACO completes.

## 8.3   Flexibility

By flexibility, we refer the ability of the autoscaling system to cope with the various heterogeneous scenarios and to incorporate future requirements. In this thesis, the proposed autoscaling framework is designed in a way that aims for the maximal generality. In particular, flexibility can be discussed with respect to the following:

- The framework does not bound to a specific scenarios, environment or cloud vendor.

- We have used control primitives and QoS to refer to the control knobs and quality indicator respectively. Therefore, we do not restrict to any assumptions on these elements.

- The framework promotes flexibility in the way that it can support various architectural styles for the cloud-based services and their requirements.

- The multi-learners approach for QoS modelling can be extended to have more candidate learning algorithms for accuracy, or can be shrinked to less number of learners for efficiency.

## 8.4   Complexity of Application

Conventional autoscaling system requires to define the right conditions and actions mapping. However, by introducing the principle of self-awareness and the related algorithms, these mapping are no longer needed without compromising the effectiveness in our autoscaling system. These observations are evident by various experiments, as we have shown in earlier Chapters.

In our framework, apart from the general configuration of the underlying algorithms, the only offline domain knowledge required are the QoS attributes, cost models, cloud primitives, services and their requirements (i.e., SLA and budget). The domain knowledge is often required to be setup once, it then can be easily maintained and updated online in an automatic manner. This eliminates the need for heavy human intervention, which is can be complex and error-prone.

However, to better ensure the quality of autoscaling, the following offline decisions need to be made for extreme scenarios, e.g., time-critical cloud-based services.

- The self-aware and self-adaptive QoS modelling approach runs periodically in order to capture dynamic QoS sensitivity. Generally, setting the frequency level to minute interval is the common practice for many cloud-based services (e.g., [90] [105] [87]). However, in some extreme scenarios (e.g., real-time critical cloud-based services), if the modelling is too frequent, this may entail large demand on resources for computing the model. In addition, model training may not be completed within its interval. In contrast, too low frequency may fail to capture the actual and evolving diversity of QoS. Consequently, arriving on the right frequency encompasses a trade-off between efficiency and accuracy. The right frequency level can be determined by analysing the characteristics of cloud-based service and or empirically deciding on the frequency level.

- We observed that in some cases when Dom0 suffers contention, the performance of the QoS modelling approach could become worse. However, this can be eliminated by determining the proper amount of provision for Dom0 offline. In the real-world cases, it is still possible to follow the same approach. More precisely, the cloud provider can specify the required computational resources for Dom0 in relation to the total number of service-instances on each PM type. This can be achieved offline by running dummy applications or using historical data. Such decision may influence the VM to PM consolidation strategy, which is out of the scope of this thesis.

- As mentioned, the MOACO can be tuned to improve the quality of optimisation in the decision making process. This can be easily achieved through profiling taking into account the possible number of dependent objectives. In particular, a set of configurations can be applicable for a range types of cloud-based services, providing that they exhibit similar characteristics.

## 8.5   Practical Deployment

Practical deployment is concerned with how difficult the proposed framework can be used in a real world scenarios, and what are the possible ways of deploying the framework in a running cloud environment. Intuitively from Chapter 3, the proposed self-aware and self-adaptive autoscaling framework is designed in a way that it can be seamlessly and transparently deployed on a given cloud scenario. Practically, we strive to enable the maximum flexibility of the framework, that is to say, it can be either fully deployed to form a standalone autoscaling system or be partially deployed to consolidate existing autoscaling system using only one (or more) of the inclusive components. An example of the scenario where the framework is used as a standalone system has been shown in Figure 8.1.

As we have demonstrated in previous chapters, the self-aware autoscaling system can acquire the necessary runtime knowledge and adapt itself to dynamically optimise the QoS and cost for all cloud-based services, which would eventually lead to better elasticity in the cloud.

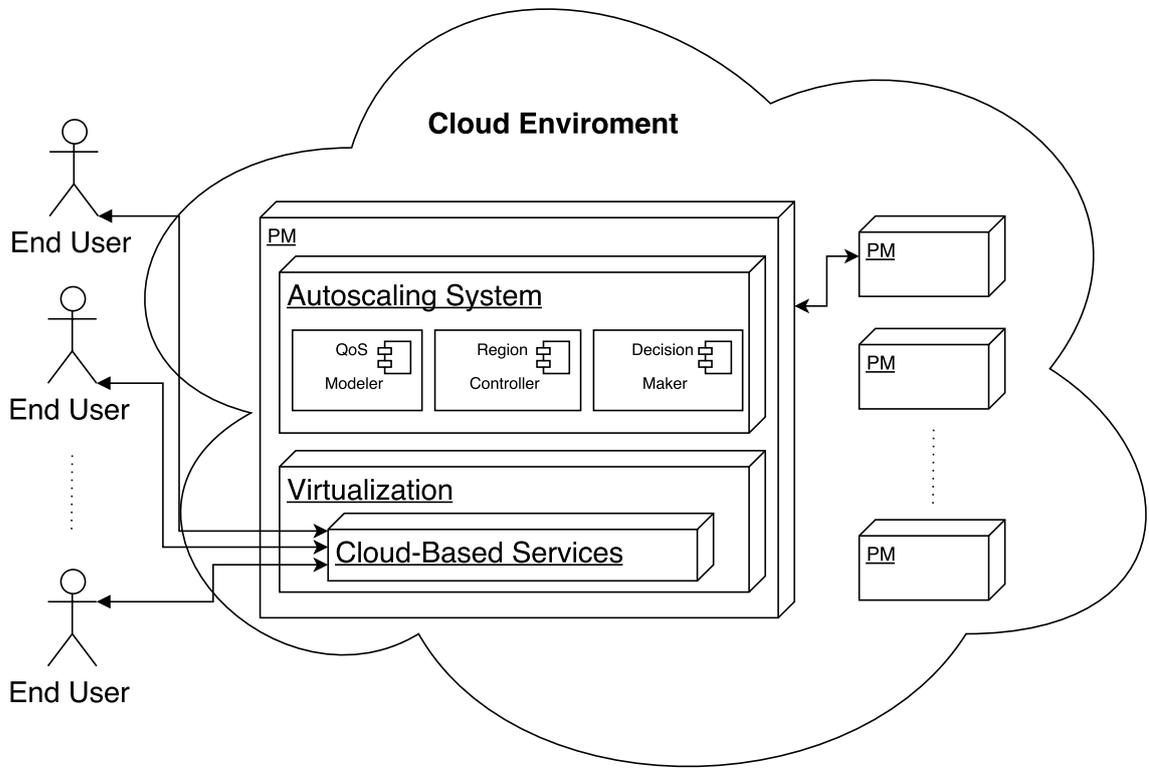

Figure 8.1: The Full Deployment Style of Self-Aware and Self-Adaptive Autoscaling Framework.

There are scenarios where it is difficult to replace the entire autoscaling system due to the transition to self-aware autoscaling is too complex and expensive. For example, suppose that there is a legacy autoscaling system running in the cloud and that the system is currently working fine but require expensive human intervention to maintain. However, it might require a large labour cost in order to replace the system with our self-aware autoscaling framework. In addition, the capacity of cloud infrastructure is rather restricted and thus there is limited additional resources for fully realising self-aware autoscaling. As a result, one solution is to take partial components from our framework and attach them to replace only certain components from the existing system. This is possible as the contributions of this thesis are associated with different independent and internal selves, which are designed as seamlessly connected self-aware components. Two concrete examples are illustrated in Figure 8.2 and 8.3.

As shown in Figure 8.2, the *QoS Modeller* is used to consolidate a simple exhaustive search algorithm for autoscaling decision making. Although such approach may be limited in reasoning about trade-offs, the improved QoS models can still help to improve the decision making by providing more accurate assurance about the effects of decision on objectives. In Figure 8.3, we show another example where the *Decision Maker* is used in conjunction with an analytical modelling approach. In such case, although the QoS models may not be sufficiently accurate, the decision making process can still make the best effort to resolve the trade-offs, and produce the decisions that tend to achieve well-compromised trade-offs.

In summary, we have reviewed the proposed framework and some of the important factors that might affect its effectiveness. We have identified some limitations and discussed the potential solutions to them. Some of those solutions can lead to interesting directions of future work, as we will discuss in the next Chapter.

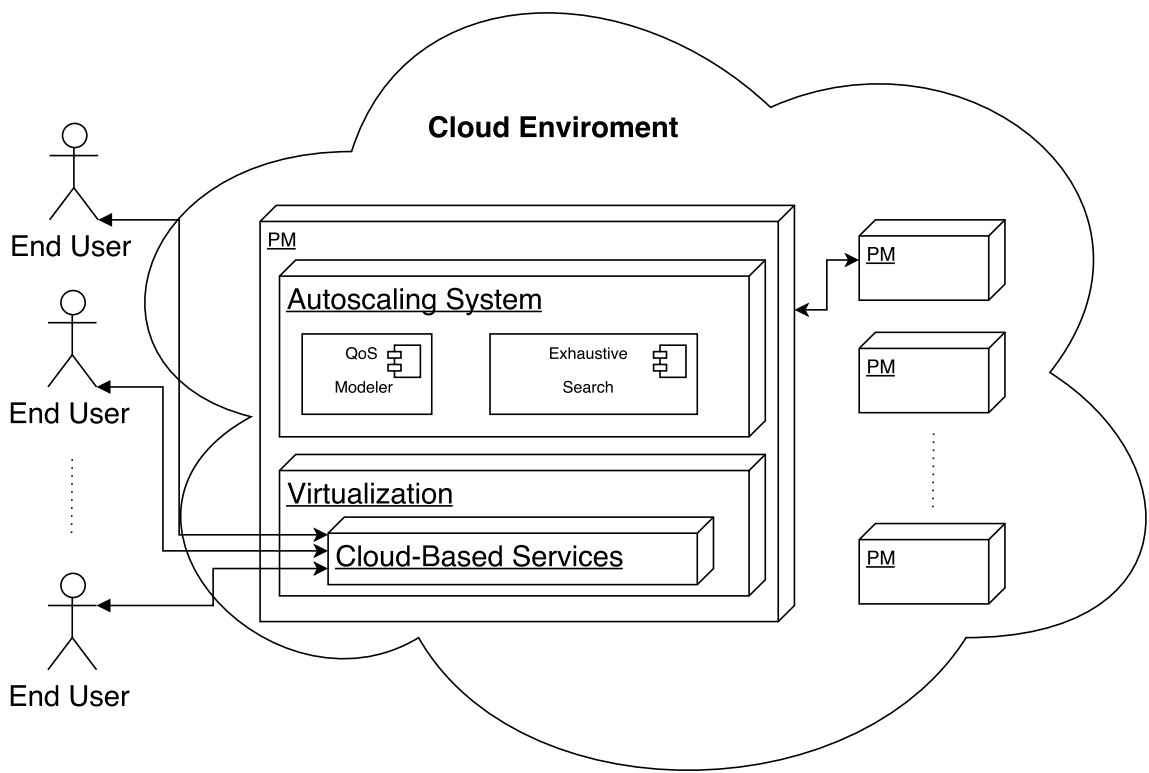

Figure 8.2: The Partial Deployment Style of Self-Aware and Self-Adaptive Autoscaling Framework Containing Only QoS Modeller.

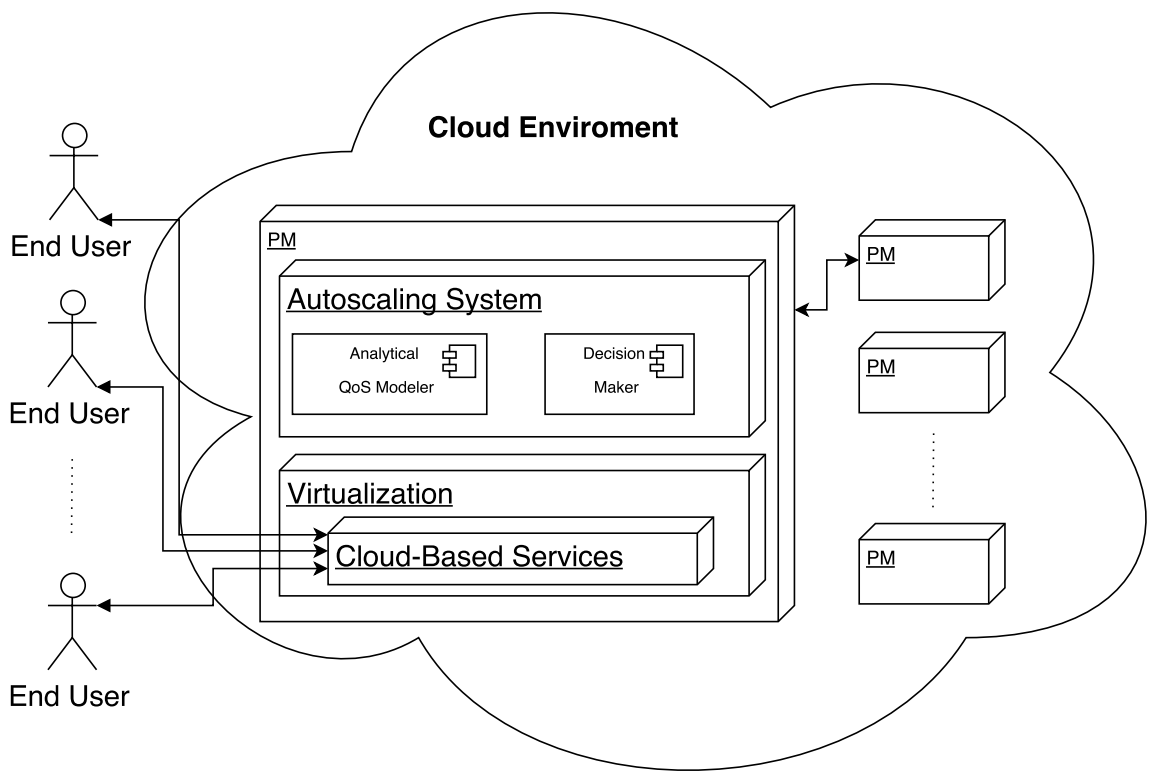

Figure 8.3: The Partial Deployment Style of Self-Aware and Self-Adaptive Autoscaling Framework Containing Only Decision Maker.

# CHAPTER 9
# CONCLUSION REMARKS AND FUTURE DIRECTIONS

## 9.1 How the Research Questions are Addressed

This thesis is driven by numbers of research questions as we have discussed in Chapter 1. In the following, we systematically review how these research questions are addressed throughout the thesis, as summarised in Table 9.1.

Table 9.1: Summary on How the Research Questions are Addressed

*RQ 1.1.* How to incorporate and map the self-awareness capabilities to autoscaling in the cloud?

- In Chapter 3, we have systematically mapped the key components in autoscaling to different self-awareness capabilities, which is considered at different levels, e.g., time-awareness etc. The mapping provides a concise understanding about how self-awareness can be applied to resolve the challenges for autoscaling in the cloud.

*RQ 1.2.* How to architect self-aware autoscaling system? What are the benefits we can expect from this enriched architecture?

- Drawing on the mapping between self-awareness and cloud autoscaling, the autoscaling architecture has been enriched with self-awareness capabilities. The mapping posses high intuition on what are the required level of knowledge at a given logical aspect of the system (e.g., QoS modelling), as we have shown in Chapter 3. This information leads to better design and selection of the underlying algorithms and techniques to enable self-awareness.

The key benefit of the self-aware autoscaling architecture is that it realises bi-directional adaptation. That is to say, it is not only able to adapt the underlying cloud-based services and VMs, but also capable to further consolidate itself by acquiring the knowledge about itself and the environment through different self-awareness capabilities. Such consolidations aim for more accurate QoS models, better granularity of control and better trade-off decisions.

---

*RQ 2.1.* How to dynamically select the important, yet uncertain cloud primitives (e.g., software configurations, hardware resources and environmental conditions) when modelling the QoS for cloud-based services? Which cloud primitive tend to be significant while which are the irrelevant ones? When these cloud primitives should be considered in the models?

- In Chapter 4 and 5, we have quantified the relative importance and significance of primitives to QoS attributes using symmetric uncertainty. We have conducted an in-depth analysis on the correlations between selected cloud primitives and the model accuracy; the results suggest that in general, the direct primitives (i.e., those that directly influence the QoS) tend to be more relevant than the indirect ones (those that only provide information about QoS interference). However, we found no evidence that can tell what dimensions of primitives tends can be constantly significant or at what point in time they can be significant. Instead, the important primitives affecting QoS tend to be dynamic and uncertain.

Therefore, we propose a self-aware and self-adaptive technique, namely hybrid dual-learners, to determine which and when the cloud primitives correlates with the QoS on the fly using information theory [130]. The experiments result show that it can improve the overall accuracy and achieves better stability.

*RQ 2.2.* How to dynamically model and quantify the uncertain magnitude of cloud primitives in the correlation?

- In Chapter 4 and 5, we have demonstrated that machine learning algorithms renders themselves neat solution to this problem. We show, by means of experimental evaluations, that they are capable to produce effective QoS models. However, we have also found that there is no single learning algorithms that can constantly outperform the others across a range of possible scenarios.

*RQ 2.3.* How to incorporate dynamic and uncertain information about QoS interference into the models?

- In Chapter 5, we have codified two classes of QoS interference at both service and VM levels. Given the flexibility of the proposed QoS modelling approach, information about interference can be Incorporated into the models by using the related primitives as inputs. The hybrid dual-learners in primitives selection phases would filter those that are not significant while only keeping the important ones.

*RQ 2.4.* How to ensure the accuracy of the QoS models?

- In Chapter 5, we have considered both information relevance and redundancy when selecting the cloud primitives as model inputs. This design, as we have demonstrated in Chapter 5, leads to better accuracy and stability while keeping the model complexity adequate. In addition, we have used a self-aware and self-adaptive solution, namely adaptive multi-learners, to dynamically model how the cloud primitives correlates with the QoS. The proposed solution can continually guarantee accuracy models by selecting the best learning algorithm and its resulted model during prediction in cloud.

*RQ 3.1.* What are the effects of control granularity on globally-optimal result (i.e., result with respect to QoS and cost of all cloud-based services) and the overhead in cloud?

- In Chapter 6, we have shown that a coarse granularity of control in cloud autoscaling (e.g., cloud level) can lead to global benefit, but the overhead is likely to be high. On the other hand, finer granularity (e.g., service level) reduces the overhead as it usually assume local optimum, however, multiple local optimum may not necessarily imply a globally-optimal benefit.

*RQ 3.2.* Whether local control (e.g., service level) can achieve similar global benefit to the global control (e.g., cloud level)?

- In Chapter 6, we have demonstrated that local control can indeed achieve similar global benefit to the global control, as long as the granularity is divided according to the dependency of objectives. That is to say, multiple local optimums can lead to global optimum if there is no dependency between different local optimums.

*RQ 3.3.* How to handle the dynamics and uncertainty associated with the granularity of control in cloud?

- In Chapter 6, we have used a self-aware and self-adaptive reigning mechanism that dynamically determine the granularity of control through knowing which and how many objectives need to be considered in the same decision making process. Each of the resulted region is regarded as independent decision making process, and they can run simultaneously. The quantitative results show that the approach achieves global optimum (or near-optimum) while reducing the overhead of autoscaling.

*RQ 4.1.* How to dynamically search for the uncertain trade-off decisions, considering the naturally conflicted objectives and QoS interference?

- In Chapter 7, we have formulated the decision making problem as a discrete multi-objective optimisation problem, where the aim is to search for the decisions that optimise different objectives subject to SLA and budget constraints. Given that the possible decisions can form a incredibly large space, this problem is essentially NP-hard. Therefore, we have leveraged on metaherustic algorithm, which is often dynamic and suitable for runtime scenario, to achieve near-optimal solution within polynomial time. Since the QoS models contain information about QoS interference, the decision making process can also handle the trade-offs caused by the QoS interferences.

*RQ 4.2.* How to dynamically reason about the effects of decisions on QoS and cost objectives, and the uncertain trade-offs considering their requirements?

- In Chapter 7, we have proposed a self-aware and self-adaptive decision making approach enabled by Multi-Objective Ant Colony Optimisation (MOACO), which is designed to reason about and optimise the possible trade-offs decisions for autoscaling in the cloud. This approach eliminates the need for specifying weights in the objective formulation and capable to handle trade-off caused by naturally conflicted objectives and QoS interference. MOACO can explore more trade-offs information than the rule, single objective and weighted-sum objectives based decision making approaches. In particular, MOACO distincts from the existing MOGA based approach in the sense that, instead of evaluating the overall quality of decisions for all the objectives during the optimisation, it performs in a way that similar to conduct many single objective optimisations in one run by using aggregative heuristics and different pheromone structures. This design aims for better optimality and diversity for a large number of objectives. Experiments result suggest that, as when compared with state-of-the-art approaches, the MOACO achieves significantly smaller SLA violation and leads to better, more stable QoS performance even when the requirements are complied; while the overhead is still acceptable.

*RQ 4.3.* How to quantify the extent of compromises in the trade-off? How to dynamically determine the well-compromised trade-off?

- In Chapter 7, we design a dynamic triple mechanism, namely compromise-dominance, for finding well-compromised trade-offs based on superiority and fairness of the decisions. The former is measured by pareto-dominance [67], and the latter is achieved via nash-dominance [108] and the distance of decision measurement. The mechanism is a sequential process where a set of decisions is filtered based on their superiority, and then the fairness. Eventually, the resulted set contains the decisions that achieve well-compromised trade-offs. Experimentally, we have shown that it helps to achieve better quality of trade-offs in terms of both the number of objective favoured and the extents to which they are optimised. Notably, by separating MOACO and compromise-dominance, the MOACO is encouraged to explore more information about the trade-offs surface while saving computational efforts.

Recall that the main research question of this thesis is:

*How can self-awareness and the related algorithms be incorporated into the process of elastically autoscaling cloud-based services, such that the autoscaling system is able to handle runtime dynamics, uncertainties and trade-offs*

*exhibited in the cloud? What are the benefits of self-awareness and to what extent can it be beneficial, when compared to approaches with no or limited self-awareness?*

Clearly, the developments of the aforementioned research questions have converged towards the answer of the first half question. The second half question has been addressed through various experimental and quantitative analysis as we discussed in previous chapters. In particular, the key benefit of introducing self-awareness in cloud autoscaling is the ability to handle dynamics, uncertainty and heterogeneity in cloud without heavy human intervention and design time knowledge. Further, we have demonstrated that in contrast to state-of-the-art approaches that have no or limited self-awareness, the improvements in self-aware autoscaling are vast, including more accurate QoS models, better stability of the models against different QoS trends, achieving global optimum with reduced overhead, better quality of trade-offs and better compliance of SLA and budget requirements.

Although we have showed that the overhead associated with self-awareness tends to be acceptable, it is still more computationally expensive than some existing approaches (e.g., rule-based autoscaling). Indeed, there is always a trade-offs between the benefits that self-awareness bring to autoscaling and the extra overhead that it introduces. However, we believe that given the increasingly complex cloud computing environment, making autoscaling self-aware will eventually become an inevitable requirement as it has many potentials to fully unlock the elastic nature of cloud.

## 9.2 Future Directions

This thesis reveals several future directions to further consolidate the effectiveness of cloud autoscaling. They are described as below:

### 9.2.1 Incorporating Workload and Demand Prediction with QoS Modelling in the Cloud

As we have mentioned in Chapter 4 and 5 , QoS modelling offers the fundamentals to reason about the effects of autoscaling decisions and the related trade-offs. Although the QoS models can be used to enable proactive autoscaling, it does not capture the patterns and seasonality related to the workload and demand fluctuations. To this end, demand and workload prediction can be combined with QoS modelling for more accurate, proactive autoscaling.

In addition, incorporating demand and workload prediction can be helpful to reason about whether the autoscaling decision making process should be triggered, and henceforth achieving better sustainability and stability in cloud. This is because in certain cases, the benefits grained by autoscaling is not significant compared with its computational overhead, especially when the spiked changes in the environment are extremely short-term. There has been autoscaling system that relied on workload and demand prediction, e.g., [88], but they rarely combine them with QoS modelling. We advocate that self-awareness can provide useful insights on how to reason about whether the demand and workload prediction are necessary, and what are the added values in addition to the QoS models.

### 9.2.2 Considering Delays of Scaling Actions

Vertical scaling can be achieved with negligible delays, but horizontal scaling often incur larger overhead. Therefore, considering delays in horizontal scaling can further improve the effectiveness of autoscaling. Considering the delays also arise an interesting trade-offs between horizontal and vertical scaling when autoscaling decisions have been made. In this thesis, we have assumed that vertical scaling takes higher priority than the horizontal scaling and the possible configured values are within a range, which is gradually updated. This design aims to handle majority of the cloud scenarios where there is a clear transition when the environment and demand changes. However, in case there is extremely large and sudden changes in the demand, such design may obstruct the self-adaptivity of autoscaling system, since it is often more preferable to go directly to horizontal scaling.

Future research can combine the reasoning about delays of scaling action with the autoscaling decision making process, or alternatively, separate such reasoning based upon the identified decision. This can be better tackled through self-awareness.

### 9.2.3 Combining VM consolidation and Cloud Autoscaling

VM consolidation in the cloud studies the problems of mapping between VMs and PM, as well as the co-hosted VMs. Often, these problems assume that the autoscaling decision has been provided or can be easily obtained by simple profiling process. Clearly, VM consolidation is related to horizontal scaling and it plays an integrals role to the VM-level QoS interference. We believe that, in order to improve the effectiveness of autoscaling, future research for VM consolidation problems, notably their interaction with cloud autoscaling, are necessary.

### 9.2.4 Autoscaling with Cloud Federation

This thesis has explicitly focused on single cloud scenarios, in which we assume that the capacity of cloud is able to handle the full demand of cloud-based services. However, depending on the scale and evolving reputation of the cloud provider, it may be inevitable for switching to alternative cloud providers with larger scale and better reputations. The selection of cloud providers problem has been widely studied, however the linkage between cloud autoscaling and cloud providers selection is remain unclear. Future research should focus on the challenge about how to dynamically reason about when it is needed to scale across multiple cloud providers, instead of autoscaling within the existing one. Investigating how self-awareness can be useful for this problem, especially in the context of cloud market where there is an even larger degree of heterogeneity, is an interesting starting point.

### 9.2.5  Handling Energy Consumption and Economic Profit

Energy consumption is becoming an increasingly important topic for cloud computing. Although not directly resolving energy consumption, this thesis preserves a foundations for achieving energy aware autoscaling. This is because the flexibility of the proposed QoS and cost models, which is often correlated with the required energy. Nevertheless, further studies are required, particularly for modelling the correlation between cost and the consumed energy.

Another interesting future direction is to perform cloud autoscaling with the aim to maximising economic profit of the service providers and cloud providers. This problems can be studied from an economic perspective, where both the buyers (service providers) and seller (cloud providers) aim to optimise their own profit. The challenge for future researches would be how to resolve the trade-off between buyers' profits and that of the seller in such a way that a global economic equilibrium is reached. Achieving self-awareness from an prospective of economic efficiency can be a promising solution.

## 9.3  Closing Remarks

This thesis makes a novel and timely contribution to the field of cloud autoscaling by presenting a self-aware and self-adaptive autoscaling framework. The thesis provides in-depth study and solutions that use the principles of self-awareness and related algorithms to the problems in the absence of closely related work. We believe that the proposed framework can provide many useful insights on how to better engineer self-aware cloud autoscaling system in relation to the architecture, QoS modelling, granularity of control and decision making. The conducted experiments demonstrate the effectiveness of the framework in handling dynamics, uncertainty, QoS interference and trade-offs that are associated with the autoscaling process. The contributions of this thesis have elaborated on different aspects, including software and systems modelling, software architecture and decision making and planning, which eventually advances the understanding of using self-awareness in cloud autoscaling. We hope that our results will motivate further research for more intelligent cloud autoscaling and its interaction with the other problems in the cloud.

# Appendix A
# Specification and Examples of Self-Aware Patterns

The self-awareness capabilities describe the different types of self-knowledge which a system may possess and learn. Subsequently, the presence of these different types of knowledge may lead to different classes of behaviours being possible. This categorisation of self-awareness capabilities as patterns has the possibility to ensure that, when designing self-aware systems, only relevant types of knowledge are included, and their inclusion justified by identified benefits. There is no need for a system to become unnecessarily complex, learning and maintaining knowledge which does nothing to advance the outcomes for that system, generating only overhead. Consequently, design process for self-aware systems will need to take account of the necessity or otherwise of different capabilities of self-awareness. The pattern notation is depicted in Figure A.1.

Two types of connectors are used to express the logical and physical interactions. physical connector means there is a direct interaction between two or more capabilities (from the same or different node), and each capability is required to directly interact with the others. Notably, physical connector (between different levels of awareness), or the red arrow, particularly refers to the interactions for the self-awareness of different types (e.g., goal and time awareness); in contrast, the other black solid arrows represent the interactions for the self-awareness of the same type (e.g., the interaction-awareness from different nodes). On the other hand, the logical connector does not require direct interaction, but rather the data or control in the interaction is sent/received through the other capabilities (e.g., Sensors and Actuators), which have the physical connector. For instance, self-expression might be logically required to reach consensus amongst different nodes, but such interaction is physically realised through Sensors and Actuators. The benefit of additionally introducing the logical connector is that, when design a self-aware capability where the communication protocol (e.g., local/remote function call, multi-cast and broadcast etc) is not needed, the pattern can still show that such capability needs to interact with the others. Thus, this provides the designers with a more precise view about the architecture.

We have used multiplicity operator to represents how many capabilities and their components (a capability can be realised in one or more components), including those from different nodes, are involved in the interaction. There are three types of multiplicity operators (mul_op):

- **\*** expresses that the number of capability of the same type in the interaction is restricted to at least one.

- **1** indicates that one and only one capability of the same type is permitted.

- **0** indicates that zero, one or many of the type specified is permitted in the interaction.

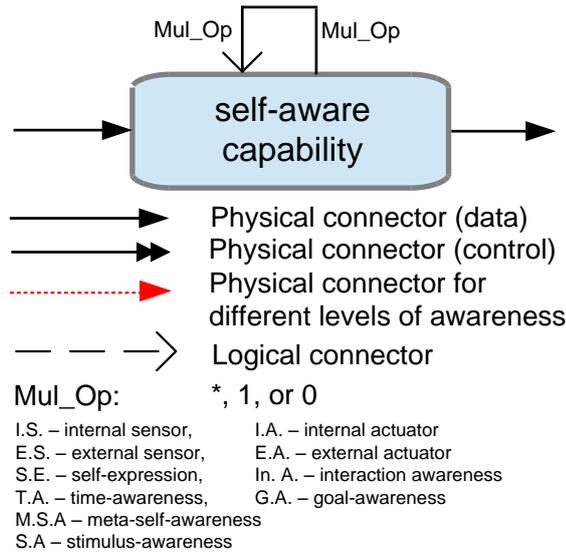

Figure A.1: The Notations for Self-Aware Patterns.

It is worth noting that when the operator is 0, it means that the associated interaction can be removed but does not represent that the corresponding capability can be eliminated. In case a capability is interact with itself, e.g., a \* on both sides of the intra-capability arrow of a capability means that it can interact with the same capability implemented in other nodes. To better clarify the operators, suppose that there is a physical interaction between stimulus awareness and external sensors where the stimulus awareness is associated with 1 whereas the external sensors is associated \*. This means that within the interaction, only one stimulus awareness is permitted whereas the number of external sensors presented in the interaction needs to be one or many. Other multiplicity arrangements can be similarly interpreted.

In the following, we briefly discuss two patterns as examples, full details and the other patterns can be found in our handbook [39].

- **Coordinated Decision-Making Pattern**

  Decisions made by individual self-aware nodes in a group may be suboptimal due to their limited view of the system and its operating environment. In applications requiring near-optimal and consistent global decision making in a cooperative set-ting, a more advanced architectural pattern may be required. In particular, such a pattern should make it possible for nodes to synchronise their adaptation actions.

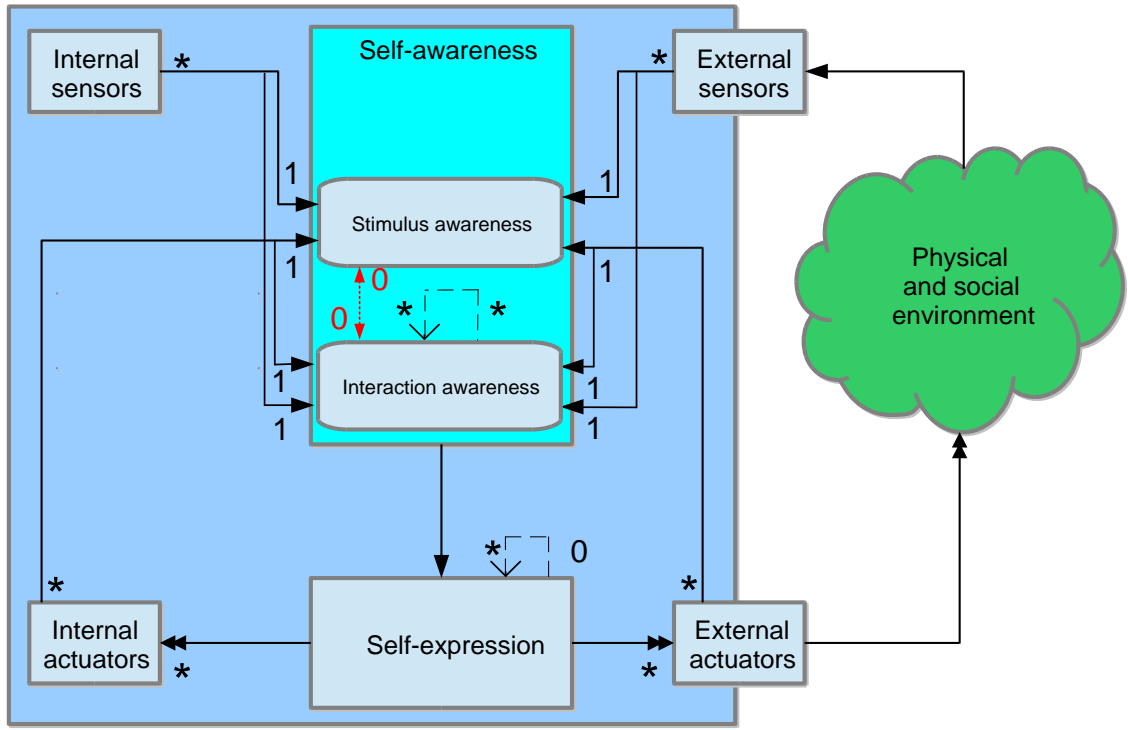

Figure A.2: Coordinated Decision-making Pattern.

The coordinated decision-making pattern provides a means of coordinating actions of multiple, interconnected self-aware nodes. Figure A.2 shows this pattern. It differs from the basic pattern in that self-expressive nodes are linked to one another, such that they are able to agree on *what* action to take. It is clear to see that the coordinated decision-making pattern is a related pattern to the basic information sharing pattern as they only differ on the self-expression capability. However, they are designed to aim for different problems and forces, therefore such separation of concepts paves a better way in pattern selection. The downside of this pattern is that although nodes are able to form clusters and cooperate on *what* action to take, they are unable to decide the timing of such actions, i.e. *when* to act.

- **Temporal Goal Aware Pattern**

The knowledge of goals and time might not necessarily to be shared amongst nodes, especially in cases where the optimisation of local goals could lead to acceptable global optimum. As a result, the presence of interaction awareness capability could cause extra overhead on the system. As shown in figure A.3 the temporal goal aware pattern solves this problem by removing the interaction awareness capability. In this pattern, there is no notion of 'sharing' as the nodes are not aware of any interactions and therefore not aware of the presence of the other nodes. It is worth noting that the absence of interaction awareness does not mean there is no interaction - nodes and the environment could still interact with each other, but the nodes are not aware of it.

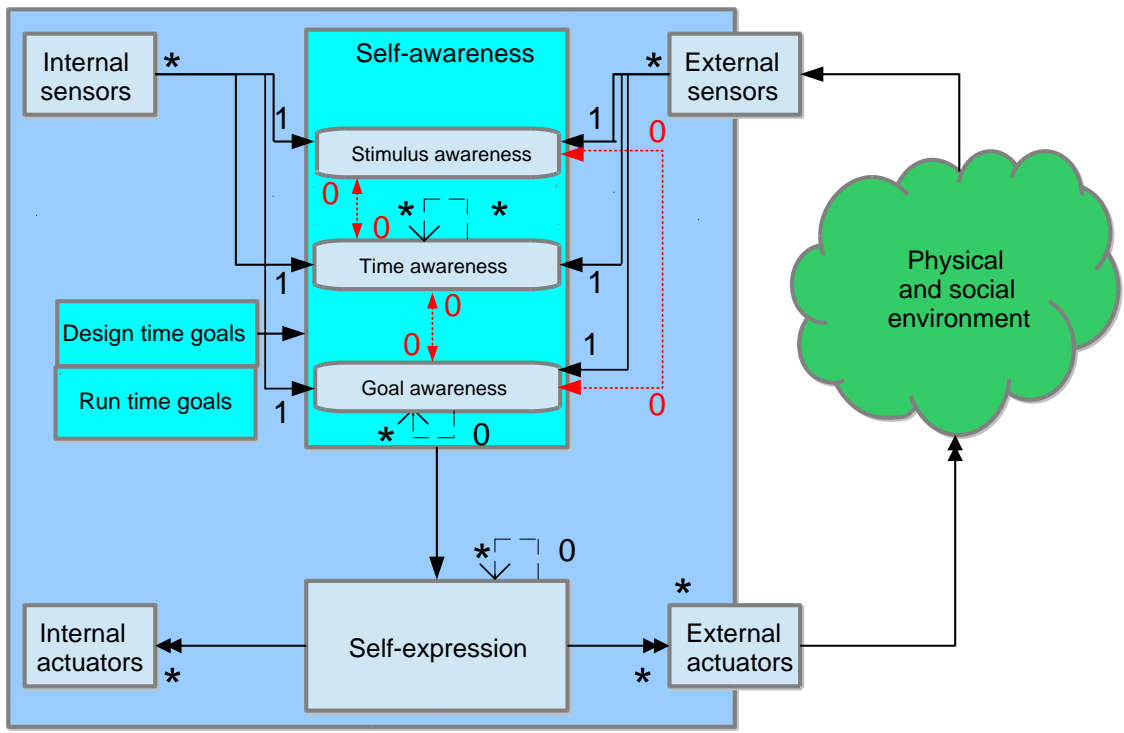

Figure A.3: Temporal Goal Aware Pattern.

The removal of interaction awareness implies that the nodes could be in inconsistent state. The designer should carefully verify that such situation would not result in violations of system requirements. In addition, the self-expression capability could not use any information from other nodes when making decisions.

Meta-self-awareness is useful for managing the trade-off between various levels of self-awareness and for modifying goals at run-time. Since reasoning at the meta level is considered an advanced form of awareness that may be beneficial or necessary in some contexts, we endorse meta-self-awareness as an optional capability for each pattern and this provides the designer with better flexibility.

# Appendix B
# Analysing the Effects of Cumulative Relevance and Redundancy in Selected Primitives to QoS Model Accuracy

To verify whether the Assumption 5.1 is valid for the case of QoS modelling in the cloud, we have conducted a set of analytical experiments to evaluate how the accuracy changes with respect to the changes of cumulative relevance and redundancy. In particular, while keeping the total number of primitives and services unchanged, we gradually add more relevant primitives as the selected inputs (from higher relevance to lower relevance) to the modelling process. For each set of selected primitives, the model accuracy and cumulative values are calculated by averaging the results from all 350 intervals in one run. We have used all the three learning algorithms (i.e., ANN, ARMAX and RT) and assessed the accuracy using SMAPE [58], calculated as shown in (4.9). It has been shown that SMAPE is intuitive, stable and more resilient to outliers than the other metrics [100].

We now explain the process of analysis in details by referring an example to simplify the exposition. In particular, we report on the Response Time of a service- instance, but similar results have been observed on many other instances. To avoid noise caused by the irrelevant primitives, we have considered only relevant primitives in the analysis. Figure B.1 shows how the accuracy tends to change with the cumulative distribution of selected primitives in the modelling. Figure B.2 expresses the changes of the cumulative average of relevance (dash line) and redundancy (solid line) as the number of selected primitives increases. Similarly, Figure B.3 shows the changes of the cumulative total of relevance and redundancy with respect to the number of selected primitives. It is worth noting that, it can be hard to interpret the cumulative relevance and redundancy using cumulative total, as they are on significantly different scales, especially when the number of selected primitives increase. Therefore, we have normalised the data in the way that the scales of both values are in the range between 0 and 1.

We initiate the process by adding the direct primitives before the indirect ones as the former can be relatively smaller in size, which causing minimal noise when the number of primitives increases. In Figure B.1, B.2 and B.3, the trend between 0 and 10% of the x-axis shows the effects of adding direct primitives while the remaining shows the effect of adding indirect ones. From Figure B.2, we can see that the increase of cumulative redundancy tends to be larger than the increase of cumulative relevance, but they become close again as they reach the 10%. We obtained similar results from Figure B.3 for the cumulative total of relevance and redundancy. This means that, if Assumption 5.1 is true from 0% to 10%, then the error is expected to increase gradually and smoothly before it drops slightly toward 10%. Nevertheless, we observed rather contradictory results on the accuracy curve of Figure B.1—for all the three learning algorithms, the error drops almost linearly from 0 to 10%, which means that in the direct primitives space, Assumption 5.1 does not hold.

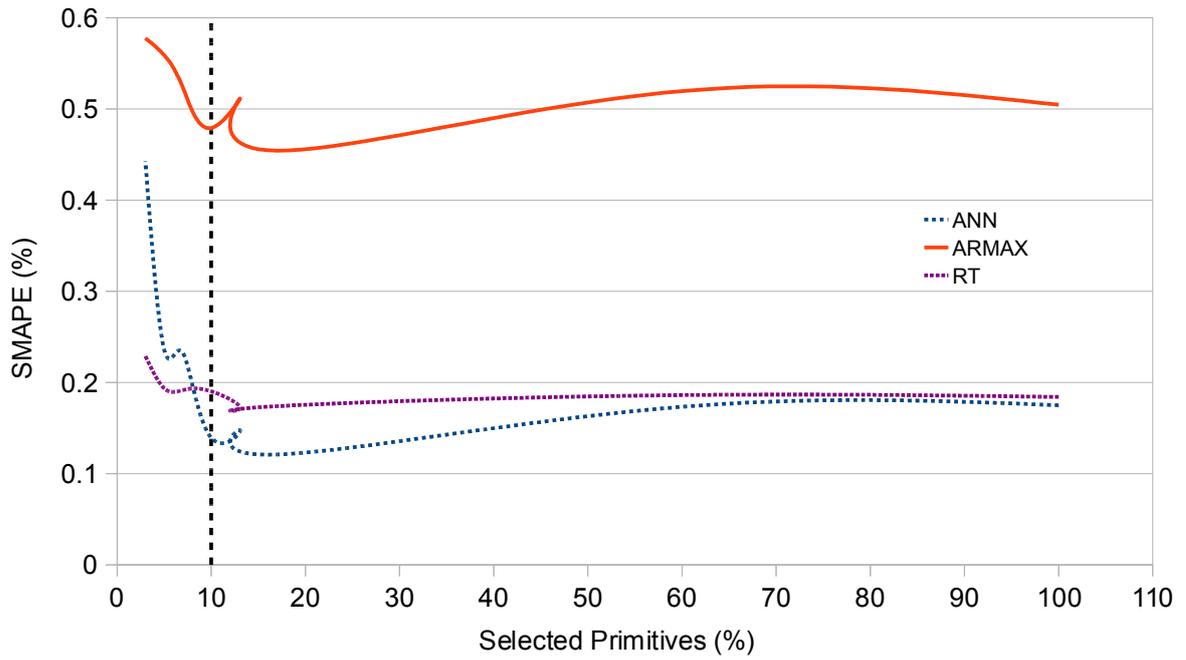

Figure B.1: The Fluctuation on Model Accuracy as the Number of Selected Primitives Increase.

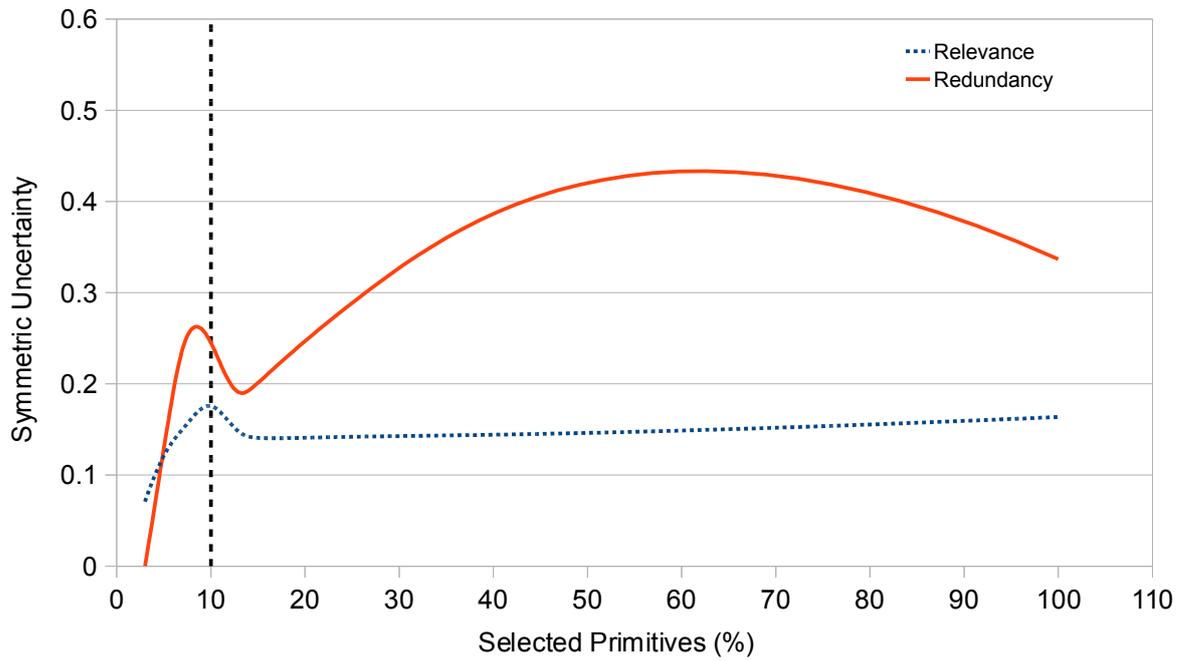

Figure B.2: The Fluctuation on Average Cumulative Relevance and Redundancy as the Number of Selected Primitives Increase.

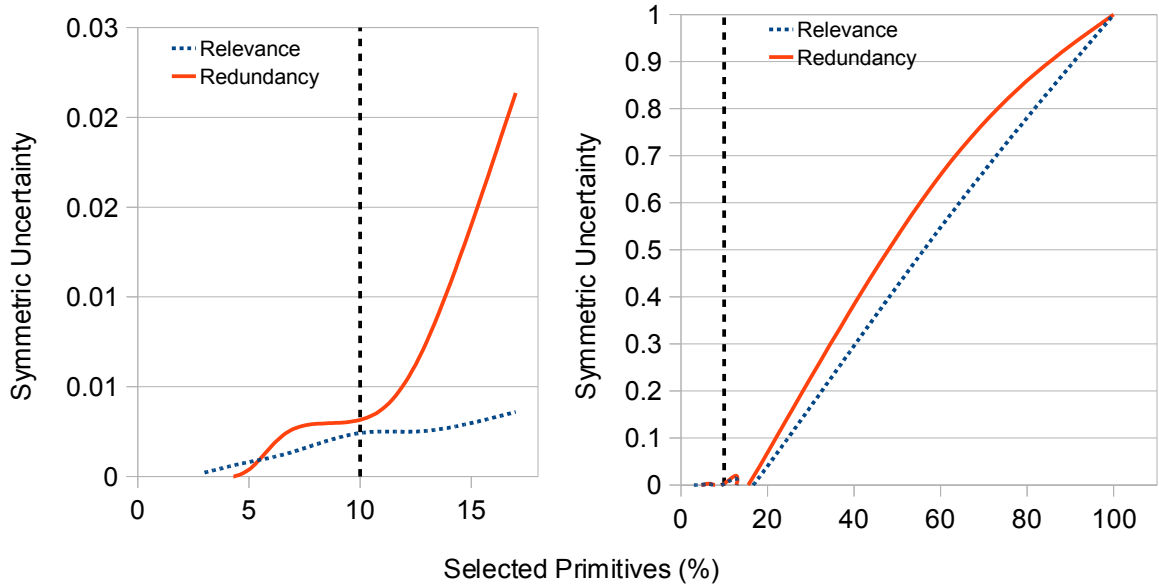

Figure B.3: The Fluctuation on Total Cumulative Relevance and Redundancy as the Number of Selected Primitives Increase.

Next, we can see that similar result also occur at the initial stages when adding the indirect primitives, particularly between 10% and 20% of the x-axis. Precisely, both Figure B.2 and B.3 indicate that from around 13%, the cumulative relevance increase almost linearly and the cumulative redundancy increase following a logarithmic behaviour. This means that if Assumption 5.1 is true, the error is expected to become larger from 13%. This is contradicted with what is shown in Figure B.1—the error continues to drop till it reaches the best point at around 13% to 17%, and the accuracy stabilises up to the 20% x-axis. Given that the number of primitives from both the direct and indirect primitives spaces is close (i.e., 10% of the total number of relevant primitives for each space), theses observations reveal that for the inter direct and indirect primitive spaces, Assumption 5.1 does not hold either. In addition, the accuracy trend implies that a combination of all direct primitives and some indirect ones yields better accuracy as it is important to consider interference in the modelling.

Finally at Figure B.2, we can see that from 20% and onwards, the cumulative relevance increases slightly and linearly whereas the cumulative redundancy tend to exhibit logarithmic and nonlinear behaviour in its increase it increases from 20% and drops by 60%. Similar trend can be observed from Figure B.3—at around 60%, the increasing slope of cumulative redundancy becomes steeper towards the curve of cumulative relevance, which keeps increasing linearly. As a result, if Assumption 5.1 is true, then the error should become larger from 20% to 60%; while from 60% onward, the error should start to drop slightly and smoothly. This is almost what we can observe from Figure B.1 for the three learning algorithms. Since the effects of direct primitives becomes weaker (after 20%) when more indirect primitives are involved in the modelling, the results indicate that in the indirect primitives space, the Assumption 5.1 is indeed valid. Another observation is that the model accuracy when using the direct primitives is generally better than using of indirect primitives.

In summary of the experiments, we have obtained four major observations: (i) within the direct primitives space, Assumption 5.1 does not hold. This is due to the fact that the direct primitives space contains different underlying primitives that directly influence the QoS, hence they can usually provide different aspects of information about a QoS attribute, which cannot be correctly quantified by cumulative SU value. Surprisingly, we also found that (ii) for inter direct and indirect primitives space, Assumption 5.1 does not hold either; (iii) however, within the indirect primitives space, Assumption 5.1 is valid. We believe that the reason for observations (ii) and (iii) is due to the fact that different direct primitives provide different aspects of information about the QoS and they influence the QoS directly. Whereas all the indirect ones can only do so via interference and contention; henceforth, they can only provide information on contention which can be regarded as one aspect of information that influence QoS. Obviously, this aspect of information is different to that in the direct primitive space. These observations also imply that the cumulative SU values can only quantify the effects of primitives to model accuracy, when they provide the same aspect of information. The final observation (iv) is that, although the overall relevance in direct primitives space is smaller than that of the indirect primitives space (as the former is smaller in size), the resulted model accuracy when using direct primitives is generally better than the use of indirect ones. This is a typical consequence of redundancy: the overall redundancy in the indirect primitives space tends to cause more negative effects on model accuracy than that of the direct one. Such observation means that even when redundancy is considered, the direct primitives can be more important than the indirect ones in the modelling. However, we observed that the best accuracy is achieved by the combination of direct and indirect primitives. This means consider proper information of QoS interference in the modelling can be quite beneficial for accuracy.

# Appendix C
# Glossary

Table C.1: The Acronyms in The Thesis.

| Acronyms | Description |
| --- | --- |
| QoS | Quality of Service, this is the non-functional attributes that a cloud-based service contain, e.g., response time, throughput and reliability. |
| CP | Control Primitives, this is the control knobs that realise autoscaling in the cloud, e.g., number of threads, CPU and memory. |
| EP | Environmental Primitives, this is the dynamic and uncertain factors that affect autoscaling in the cloud, e.g., the workload and size of incoming tasks/jobs. |
| VM | Virtual Machine, this is the conceptual unit that contains certain allocation of the resources in the cloud. |
| PM | Physical Machine, this is the machine that running in the cloud infrastructure. |
| SU | Symmetric Uncertainty, this is a metric that used to measure the relevance between two random variables. It is heavily used in Chapter 4 and 5 for the QoS modelling approach. |
| ANN | Artificial Neural Network, this is a machine learning algorithm, derived from biological neural networks, that is capable to model complex nonlinear correlations. It is heavily used in Chapter 4 and 5 for the QoS modelling approach. |
| S-ANN | Sensitivity aware Artificial Neural Network, this is the ANN that improved by primitives selection approach. |
| C-ANN | Conventional Artificial Neural Network, this is the ANN that does not use primitives selection approach. |
| RPROP | Resilient backpropagation, is a learning heuristic for supervised learning in feedforward artificial neural networks. |

| ARMAX | Auto-Regressive Moving Average with eXogenous inputs model, this is a simple, but efficient machine learning algorithm that models linear correlations. It is heavily used in Chapter 4 and 5 for the QoS modelling approach. |
|---|---|
| S-ARMAX | Sensitivity aware Auto-Regressive Moving Average with eXogenous inputs model, this is the ARMAX that improved by primitives selection approach. |
| C-ARMAX | Conventional Auto-Regressive Moving Average with eXogenous inputs model, this is the ARMAX that does not use primitives selection approach. |
| RT | Regression Tree, this is a tree-liked machine learning algorithm. It is heavily used in Chapter 5 for the QoS modelling approach. |
| SMAPE | Symmetric Mean Absolute Percentage Error, this is the metric that measures the percentage error in model prediction. It is also resilient to outliers. |
| RSD | Relative Standard Deviation, this is the metric that measures how fluctuated a QoS trend tends to be. |
| HYBRID | The proposed cloud primitives selection approach that using hybrid dual-learners, as described in Chapter 5. |
| SINGLE-MR | The compared single-learner primitives selection approach that relies on maximal relevance for all the primitives space, as described in Chapter 5. |
| SINGLE-MRMR | The compared single-learner primitives selection approach that relies on maximal relevance and minimal redundancy for all the primitives space, as described in Chapter 5. |
| SINGLE-MRMR | The compared single-learner primitives selection approach that relies on maximal relevance and minimal redundancy for all the primitives space, as described in Chapter 5. |
| MANUAL | The fixed and offline primitive selection approach, as described in Chapter 5. |
| SINGLE-MR-DIRECT | The compared single-learner primitives selection approach that relies on maximal relevance and minimal redundancy for all the direct primitives space only, as described in Chapter 5. |
| ADAPTIVE | The proposed multi-learners approach for QoS function training, as described in Chapter 5. |
| MOACO | Multi-Objective Ant Colony Optimization, this is the proposed search-based algorithm for optimizing autoscaling decisions. |
| CD | Compromise-Dominance, this is the proposed mechanism to search for well-compromised trade-off decisions. |
| MOGA | Multi-Objective Genetic Algorithm, this is the compared algorithm for optimizing autoscaling decisions using MOGA. |

| RULE | This is the compared approach for optimizing autoscaling decisions using predefined *if-conditions-then-action* mapping. |
|------|------------------------------------------------------------------------------------------------------------------------|
| HILL | This is the compared approach for optimizing autoscaling decisions using hill-climbing search algorithm. |
| RANDOM | This is the compared approach for optimizing autoscaling decisions using randomised search algorithm. |